\documentclass[%
a4paper,
twoside,openright,
11pt
]{book}
\usepackage[utf8]{inputenc}
\usepackage{import}
\usepackage{definizioak}
\usepackage{amsmath}
\usepackage{bigints}
\usepackage{soul}
\usepackage{hyperref} 

\def\hmath$#1${\texorpdfstring{{\rmfamily\textit{#1}}}{#1}}

\def\be{\begin{equation}}
\def\ee{\end{equation}}
\def\bseq{\begin{subequations}}
\def\eseq{\end{subequations}}

\def\bea{\begin{eqnarray}}
\def\eea{\end{eqnarray}}

\begin{document}
\counterwithout*{footnote}{chapter}
\pagenumbering{gobble}
         
     \includepdf{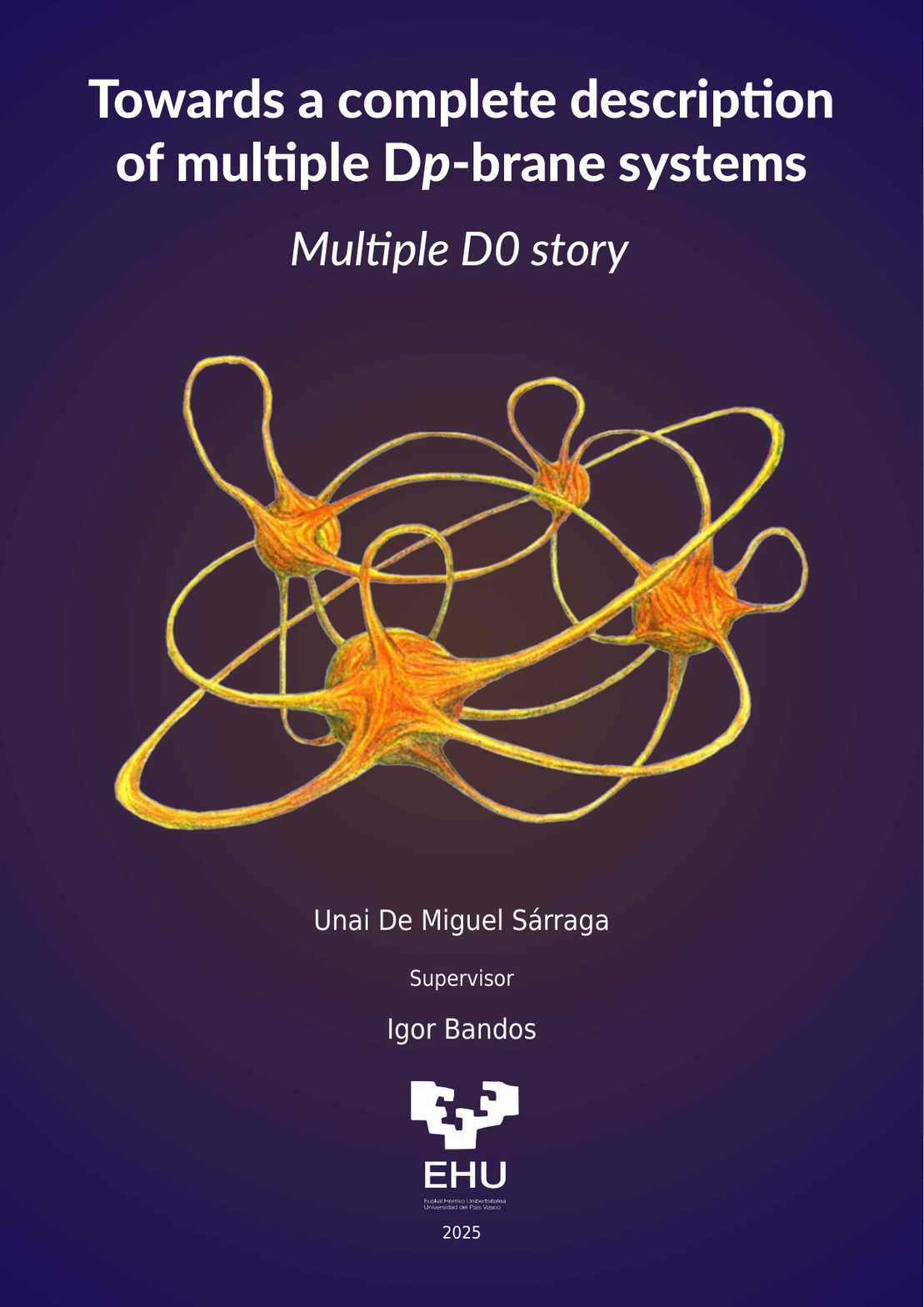} 
    
         \newpage
         \thispagestyle{empty}
         \

 \thispagestyle{empty}

 \begin{center}

 \begin{figure}[h]
 \centering
\includegraphics[width=0.40\textwidth]{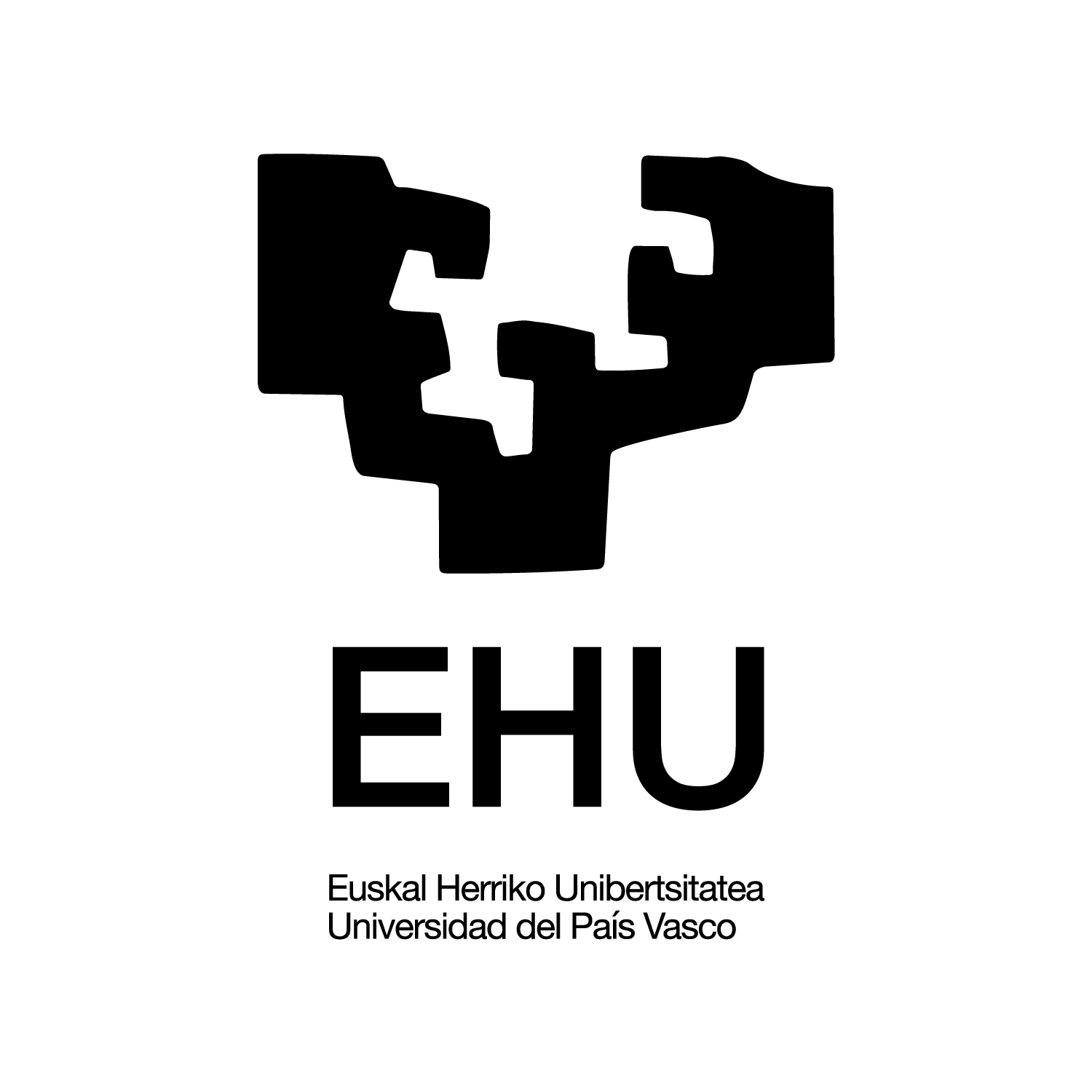}
 \end{figure}

\vspace{0.75cm}
 \vfill
 
 {\large{A thesis submitted in partial fulfilment of the requirements for the\\[2pt]Degree of Doctor of Philosophy}}

\vspace{0.75cm}
\vfill

 \makebox[\linewidth]{\color{chapters}\rule{\textwidth}{1pt}}

 {\sffamily{\textcolor{chapters}{\textbf{\LARGE{{Towards a complete description of multiple\\[12pt] D\textit{p}-brane systems: Multiple D0 story}}}}}}\\
 \vspace{8pt}

 \makebox[\linewidth]{\color{chapters}\rule{\textwidth}{1pt}}

 \vspace{1.5cm}
  \vfill

{\sffamily\LARGE{{{Unai De Miguel Sárraga}}}}

 \vspace{1.5cm}
 \vfill
 
 {{\large{Supervisor}\\[6pt]{\Large{\sffamily Igor Bandos}}}}

 \vspace{1.75cm}
 \vfill

 {\Large{{\textbf{2025}}}}

 \vfill
 \vspace*{1cm}

 \end{center}

\newpage
\thispagestyle{empty}
\

\newpage
\thispagestyle{empty}
\vspace*{1cm}
\vfill

 \begin{changemargin}{-0.04cm}{0cm}
 \begin{center}
     {\texttt{\textsc{\small{The author acknowledges financial support from PIC contract funded by "Gravitation, Cosmology and Fundamental Physics" group (Basque Government) as well as M3 contract funded by "Gravity, Supergravity and Superstrings (GRASS) Project" group (CSIC), which made the development of this thesis possible.}}}}
\end{center}
\end{changemargin}
\vfill
\vspace*{1cm}

\begin{figure}[h]
\centering
\includegraphics[width=0.25\textwidth]{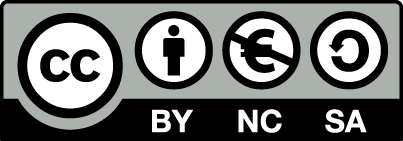}
\end{figure}
\vspace{-16pt}
\begin{center}
{\copyright\hspace{2pt} {\small\texttt{\textsc{Copyright by Unai De Miguel Sárraga, 2025. All rights reserved.}}}}
\end{center}

\newpage
\thispagestyle{empty}
\

     \newpage
\pagenumbering{roman}
\thispagestyle{empty}
\

\vspace*{5cm}
    \begin{flushright}     
         {\sffamily\large\textcolor{chapters}{\textit{Dedication}}}
     \end{flushright}
\begin{flushright}
    A mi familia,
\end{flushright}
por guiarme o dejarme caminar solo cuando el viaje así lo requiere. Por ser el espejo en el que me miro, la viga que me sostiene y el impulso que me anima a seguir adelante cuando el camino se vuelve difícil. En especial...
    \begin{flushright}
    A mi tita por su amor infinito.
    \end{flushright}
    \begin{flushright}
    A mi tito por sus consejos infinitos.
    \end{flushright}
    \begin{flushright}
    A mi hermano por su paciencia infinita.
    \end{flushright}
    \begin{flushright}
    A Sonia por su ser y estar infinitos.
    \end{flushright}
Y por último, y sin embargo en primer lugar...
    \begin{flushright}
    A mi madre por su amor, consejos, paciencia, ser y estar infinitos.
    \end{flushright}

Esta tesis es también para Zor y Lunita, que me regalaron compañía, alegría y amor. Los recuerdo cada día con una sonrisa. Siguen conmigo, como siempre.
 
\newpage
\thispagestyle{empty}
\

     \chapter*{Abstract}
\addcontentsline{toc}{chapter}{Abstract}
\thispagestyle{empty}
\vspace*{-1cm}

\begin{changemargin}{1.0 cm}{0cm}
     \singlespacing\textcolor{cites}{ \small{
                {\begin{flushright}\sffamily {The magic is only in what books say, how they stitched\\ the patches of the Universe together into one garment for us.}\end{flushright}}
     \begin{flushright}     
         {\sffamily  {\textit{Fahrenheit 451}}\\{by {Ray Bradbury}. }}
     \end{flushright}}}
    \end{changemargin}
   \vspace{12pt}

In this thesis, we address the problem of constructing the complete supersymmetric description of systems of $N$ nearly coincident Dirichlet $p$-branes (multiple D$p$-brane or mD$p$). A particularly important result is a completely nonlinear action for the 10-dimensional dynamical system of nearly coincident multiple D$0$-branes (mD$0$) which is doubly supersymmetric, i.e. it is invariant under both spacetime (target superspace) supersymmetry and the worldline supersymmetry; the latter is a counterpart of the local fermionic $\kappa$-symmetry characteristic of a single D$0$-brane (Dirichlet superparticle). This problem is approached in flat superspace using the (spinor) moving frame formalism, which provides us with a geometric framework to the treatment of supersymmetric particles and extended objects ($p$-branes) in higher dimensions. The construction of the complete doubly supersymmetric actions for multiple D$0$-brane system uses this formalism essentially; the counterpart of this action is not known in others approaches.

To approach this goal, we first construct the action for the $\text{D}=4$ $\mathcal{N}=1$ non-Abelian multiwave system (nAmW), which serves as the lower dimensional counterpart of the 11D multiple M$0$-branes (mM$0$ or multiple M-waves), perform its dimensional reduction and thus obtain a nonlinear action having the properties expected for the $\text{D}=3$ $\mathcal{N}=2$ counterpart of the mD$0$ system (the 3D mD$0$ system). The nonlinearity of this (set of) action(s) is due to the presence of a certain function $\mathcal{M}(\mathcal{H})$ of the ($\text{d}=1$ reduction of 3D $\mathcal{N}=2$) supersymmetric Yang-Mills (SYM) Hamiltonian $\mathcal{H}$, constructed from the matrix fields describing relative motion of the constituents of 3D mD$0$ system. Curiously, the action possesses double supersymmetry for any positively definite function $\mathcal{M}(\mathcal{H})$, and not just for the case of particular $\mathcal{M}(\mathcal{H})$ which is obtained by dimensional reduction of 4D $\mathcal{N}= 1$ nAmW system.

Using the insights and procedure refined through the study of these lower dimensional models, we construct a family of complete supersymmetric and $\kappa$-symmetric nonlinear actions with the properties expected for 10D mD$0$-brane system which includes an arbitrary positive definite function $\mathcal{M}(\mathcal{H})$ of the $\text{d}=1$ $\mathcal{N}=16$ SYM Hamiltonian $\mathcal{H}$. We have shown how a particular representative of this family, characterized by a specific nonlinear choice of $\mathcal{M}(\mathcal{H})$, can be derived via dimensional reduction from the action describing the 11D multiple M-wave (mM$0$) system.

Subsequently, we obtain the complete set of equations of motion for the 10D mD$0$ system described by this action, perform a convenient gauge fixing, solve the center of energy equations and establish an interesting correspondence between the relative motion mD$0$ equations and those of maximally supersymmetric SU$(N)$ SYM theory. While this correspondence does not amount to a gauge equivalence, it provides a relation between solutions of the systems. In particular, it implies that all supersymmetric configurations of the mD$0$ equations in the relative motion sector correspond to supersymmetric solutions of the SYM equations.

We have also initiated the quantization program aimed at development of a field theory of multiple D$0$-branes. This quantization in its complete form is expected to lead to a novel supersymmetric field theory formulated on a superspace extended by additional bosonic and fermionic matrix coordinates, whose development may offer significant insights into the deeper structure of String/M-theory. As a first step toward this goal, we construct the Hamiltonian formalism and perform the covariant quantization of the simplest 3-dimensional counterpart of the 10D multiple D$0$-brane system. We conclude with a discussion of some of the properties and solutions of this emerging field theory in an extended (super)spacetime enlarged by bosonic and fermio\-nic matrix coordinates obtained on this way.

\chapter*{Resumen}
\addcontentsline{toc}{chapter}{Resumen}
\thispagestyle{empty}
\vspace*{-1.5cm}
    
    \begin{changemargin}{1cm}{0cm}
    \singlespacing\textcolor{cites}{ \small{
                {\begin{flushright}\sffamily {La magia sólo está en lo que dicen los libros, en cómo unían\\ los diversos aspectos del Universo hasta formar\\ un conjunto para nosotros.}\end{flushright}}
     \begin{flushright}     
         {\sffamily  {\textit{Fahrenheit 451}}\\{de {Ray Bradbury}. }}
     \end{flushright}}}
    \end{changemargin}
    \vspace{12pt}

En esta tesis se aborda el problema de construir una descripción supersimétrica completa para sistemas formados por $N$ Dirichlet $p$-branas casi coincidentes, también conocidos como sistemas de múltiples D$p$-branas (mD$p$). Un resultado particularmente importante es una acción completamen\-te no lineal para el sistema dinámico 10-dimensional de múltiples D$0$-branas casi coincidentes (mD$0$), que resulta ser doblemente supersimétrica; es decir, es invariante tanto bajo la supersimetría del espaciotiempo ($target~superspace$) como bajo la supersimetría de la línea de mundo. Esta última es análoga a la $\kappa$-simetría, una simetría fermiónica local característica de una única D$0$-brana (o superpartícula de Dirichlet). Este pro\-blema se estudia en un superespacio plano mediante el uso del formalismo conocido como ($spinor$)$~moving~frame$, que proporciona un marco geométrico adecuado para tratar con partículas y objetos extendidos ($p$-branas) supersimétricos en dimensiones altas. La cons\-trucción de acciones do\-blemen\-te supersimétricas para sistemas de D$0$-branas múltiples utiliza este formalismo de manera esencial; no se conocen análogos de dichas acciones en ningún otro formalismo.

Con este objetivo, se construye primero la acción del sistema de múltiples ondas no Abe\-lia\-nas (nAmW por sus siglas en Inglés) en $\text{D}=4$ $\mathcal{N}=1$, que actúa como homólogo en dimensiones bajas del sistema de múltiples M$0$-branas en 11 dimensiones (mM$0$ también conocido como múltiples M-ondas). Se analizan sus propiedades y se lleva a cabo su reducción dimensional, obteniendo una acción no lineal con las propiedades esperadas del análogo en $\text{D}=3$ $\mathcal{N}=2$ del sistema de mD$0$ (el sistema 3D mD$0$). La no linealidad de esta acción se debe a la presencia de cierta función $\mathcal{M}(\mathcal{H})$ dependiente del Hamiltoniano $\mathcal{H}$ de la reducción dimensional a $\text{d}=1$ de la teoría de SYM en $\text{D}=3$ $\mathcal{N}=2$ que se construye de los campos matriciales que describen el movimiento relativo de los constituyentes del sistema de mD$0$ 3-dimensional. De forma sorprendente, la doble supersimetría se preserva en el modelo independientemente de la forma de $\mathcal{M}(\mathcal{H})$ si ésta es positivamente definida.

Basándonos en estas ideas y en el procedimiento desarrollado durante el estudio de los mo\-de\-los en dimensiones más bajas, construímos una familia completa de acciones no lineales, supersimétricas y con  $\kappa$-simetría (supersimetría de la línea de mundo), con las propiedades esperadas para el sistema de mD$0$-branas en 10 dimensiones. Esta familia incluye una función arbitraria definida positivamente $\mathcal{M}(\mathcal{H})$. Demostramos que un representante par\-ticular de esta familia, caracterizado por una elección concreta y no lineal de $\mathcal{M}(\mathcal{H})$, puede obte\-nerse mediante reducción dimensional a partir de la acción que describe el sistema de múltiples M-ondas (mM$0$) en 11D.

Asímismo, se obtiene el conjunto completo de ecuaciones de movimiento del sistema de mD$0$ descrita por esta acción; se lleva a cabo una fijación conveniente del gauge, se resuelven las ecuaciones del centro de masas, obteniéndose así una correspondencia interesante entre las ecuaciones del movimiento relativo del sistema mD$0$ y las ecuaciones de la teoría de Yang-Mills supersimétrica (SYM por sus siglas en Inglés) del grupo de gauge SU$(N)$ con supersimetría máxima. Aunque esta corres\-pondencia no implica una equivalencia gauge, sí establece una relación entre las soluciones de ambos sistemas. Entre otros resultados, hemos probado que todas las configuraciones supersimétricas del sistema de mD$0$ en el sector del movimiento relativo se corresponden a soluciones supersimétricas de las ecuaciones de SYM.

Por otro lado, hemos iniciado el programa de cuantización del sistema dinámico con el objetivo de desarrollar una teoría de campos para sistemas de múltiples D$0$-branas. Se espera que, de realizarse en su forma completa, esta cuantización conduzca a la formulación de una nueva teoría de campos supersimétrica definida sobre un superespacio extendido por coordenadas matriciales bosónicas y fermiónicas, cuyo desarrollo podría ofrecer nuevas pers\-pectivas sobre la estructura de la Teoría de Cuerdas/Teoría M. Como primer paso hacia este objetivo, se construye el formalismo Hamiltoniano y se realiza la cuantización covariante del modelo de mD$0$ más sencillo en tres dimensiones, el cual actúa como análogo en dimensiones bajas de un representante del sistema de múltiples D$0$-branas en 10D.  Esta parte concluye con una discusión sobre algunas propiedades y soluciones de la teoría de campos emergente definida en un (super)espaciotiempo extendido por coordenadas matriciales bosónicas y fermiónicas obtenida de esta manera.

     \pagestyle{empty}
\renewcommand{\thepart}{}
{
  \renewcommand{\thispagestyle}[1]{}
{\hypersetup{linkcolor=global}
\singlespacing
\tableofcontents}
}
\newpage
\pagestyle{fancy}

\newpage
\thispagestyle{empty}
\
     
     \chapter*{Resumen General y Marco de Investigación}
\addcontentsline{toc}{chapter}{Resumen General y Marco de Investigación}
\thispagestyle{empty}
\vspace*{-1.5cm}
    
    \begin{changemargin}{1cm}{0cm}
    \singlespacing\textcolor{cites}{ \small{
                {\begin{flushright}\sffamily {Los libros sólo eran un tipo de receptáculo donde\\ almacenábamos un serie de cosas que temíamos olvidar.\\ No hay nada mágico en ellos. La magia sólo está en lo que dicen\\ los libros, en cómo unían los diversos aspectos del Universo\\ hasta formar un conjunto para nosotros.}\end{flushright}}
     \begin{flushright}     
         {\sffamily  {\textit{Fahrenheit 451}}\\{de {Ray Bradbury}. }}
     \end{flushright}}}
    \end{changemargin}
    \vspace{12pt}

El Modelo Estándar (SM por sus siglas en Inglés) de la física de partículas representa uno de los logros más importantes en la historia de la Ciencia, fruto de más de 2000 años de búsqueda de conocimiento. A lo largo del tiempo, uno de los objetivos fundamentales tanto de la Filoso\-fía como de la Ciencia ha sido comprender cuál o cuáles son los constituyentes más pequeños del Universo y cómo interactúan entre sí. Esta inquietud puede resumirse en una pregunta esencial: \textit{``¿De qué está hecha la materia?''}.

Actualmente, en el marco del Modelo Estándar, esta cuestión se responde en términos de partículas elementales\footnote{Se entiende por partículas elementales a aquellas partículas que no están formadas por componentes más pequeños. Por ejemplo, según el conocimiento actual que se tiene sobre la materia (a finales de 2025), el electrón es una partícula elemental, mientras que el protón no lo es, ya que está compuesto por quarks y gluones.}. Es de destacar que toda la materia que nos rodea está formada únicamente por tres de estas partículas fundamentales: los electrones, y los quarks arriba ($up$) y quarks abajo ($down$). La combinación de estos quarks da lugar a protones y neutrones, que a su vez forman los núcleos atómicos.

Pero, ¿es este el final de la historia? La investigación científica actual sugiere que aún es posible ir más allá, es decir, encontrar estructuras aún más fundamentales que las consideradas actualmente como partículas elementales. Sin embargo, antes de adentrarnos en esta posibilidad, conviene describir brevemente cómo el SM explica las interacciones entre los constituyentes fundamentales de la materia.

Desde un punto de vista técnico, el SM se formula dentro del marco matemático de la teoría cuántica de campos (QFT por sus siglas en Inglés), que puede entenderse como la extensión de la teoría clásica de campos relativista al mundo cuántico y representa, hasta el momento, el marco teórico más preciso del que se dispone para describir la Naturaleza. En este marco, el Lagrangiano de la teoría determina su dinámica, donde la materia está formada por partículas fermiónicas (fermiones) mientras que las interacciones entre ellas se describen mediante el intercambio de partículas bosónicas (bosones), las cuales emergen a partir de las llamadas simetrías de tipo gauge. En pocas palabras, una teoría gauge es una teoría de campos cuyo Lagrangiano permanece invariante (en particular) bajo un conjunto de transformaciones locales\footnote{En este contexto, ``local'' significa que los parámetros de simetría cambian de un punto a otro del espa\-ciotiempo; concretamente, las simetrías gauge locales están ``parametrizadas'' por funciones.} que varían punto a punto en el espaciotiempo y que forman ciertos grupos de Lie.

A día de hoy, el SM es la única teoría que describe con éxito tres de las cuatro interacciones fundamentales de la Naturaleza (las interacciones fuerte y débil, y el electromagnetismo) en el marco de una teoría cuántica de campos. Es una teoría gauge invariante bajo la simetría local $\text{SU}(3) \otimes \text{SU}(2) \otimes \text{U}(1)$, donde cada subgrupo se asocia a una interacción distinta.

En particular, la simetría gauge $\text{SU}(2) \otimes \text{U}(1)$ sustenta la teoría electrodébil (EW por sus siglas en Inglés), que proporciona una descripción unificada de las interacciones electromagnética y débil. Esta simetría se rompe espontáneamente a un subgrupo U$(1)$ y, como consecuencia, los bosones gauge $W^{\pm}$ y $Z^0$ adquieren masa a través del mecanismo de Englert-Brout-Higgs-Guralnik-Hagen-Kibble, mientras que el fotón permanece sin masa. El grupo de simetría restante del SM es SU$(3)$, también conocida como simetría de color. Este es el grupo de simetría de la Cromodinámica Cuántica (QCD por sus siglas en Inglés), la teoría gauge que describe la interacción entre quarks y gluones. Los gluones, que actúan como bosones gauge de esta simetría, también están ``cargados'' bajo la interacción fuerte. Curiosamente, QCD presenta dos propiedades clave: el confinamiento de color y la libertad asintótica, las cuales, sin entrar en detalles técnicos, explican la existencia de los hadrones\footnote{Los hadrones son partículas subatómicas compuestas por dos o más quarks.}. 

La unificación de las interacciones electromagnética y débil no constituye el primer caso de unificación en la historia de la física. Una de las ideas más populares (y productivas) de la física teórica ha sido siempre la búsqueda de una descripción unificada. Esta búsqueda puede entenderse como un camino hacia un conocimiento más profundo de la Naturaleza, llegando a verse incluso como un principio filosófico de unidad: la Naturaleza es única y una comprensión fundamental de ella debería revelar una explicación común para fenómenos que, en apariencia, parecen distintos, como lo son las cuatro interacciones fundamentales de la Naturaleza: el Electromagnetismo, las interacciones nucleares Fuerte y Débil, y la Gravedad.

Este camino hacia una comprensión más profunda del Universo comenzó en el siglo XVII, cuando Isaac Newton unificó la mecánica terrestre y la celeste bajo las leyes de la gravitación universal. En el siglo XIX, James C. Maxwell demostró que la electricidad y el magnetismo, hasta entonces considerados fenómenos independientes, podían describirse mediante un único conjunto de ecuaciones que hoy llevan su nombre, logrando así la unificación de la electricidad y el magnetismo en un único marco: el electromagnetismo. Más recientemente, como ya se ha mencionado, el SM unifica las interacciones débil y electromagnética en la teoría electrodébil. Existen además las llamadas Teorías de Gran Unificación (GUT por sus siglas en Inglés), cuyo objetivo es extender esta unificación incluyendo la interacción fuerte. En un sentido amplio, una GUT pretende combinar las interacciones débil, fuerte y electromagnética dentro de un mismo marco teórico. En un sentido más estricto, busca una teoría gauge basada en un grupo de Lie que contenga como subgrupo a $\text{SU}(3) \otimes \text{SU}(2) \otimes \text{U}(1)$. 

El siguiente paso natural sería incorporar la gravedad junto a las otras tres pero, a día de hoy, no está claro que exista una conexión directa entre el SM y la Relatividad General (GR por sus siglas en Inglés). De hecho, una teoría cuántica de campos construida a partir de la acción de GR no es renormalizable, lo que implica que solo puede utilizarse como una teoría de campos efectiva (EFT por sus siglas en Inglés). No obstante, existen sólidos argumentos que sugieren que la gravedad puede (y debe) ser cuantizada. En particular, una teoría cuántica de la gravedad es necesaria para poder estudiar, por ejemplo, las etapas más tempranas del Universo y ciertas propiedades de los agujeros negros.

Entre las alternativas teóricas que existen, la Teoría de Cuerdas se presenta como el marco más desarrollado y prometedor para incorporar todas las interacciones fundamentales. Esta teoría constituye una formulación cuántica de la gravedad (aunque aún no se ha demostrado que sea la gravedad de nuestro Universo), y ofrece un marco unificado para todas las interacciones. Su premisa, aparentemente sencilla, es que las partículas puntuales del SM son, en realidad, excitaciones de objetos unidimensionales llamados cuerdas. Es decir, cada partícula correspondería a un modo vibracional específico de una única cuerda elemental.

Además, la Teoría de Cuerdas incorpora una simetría adicional conocida como Supersime\-trí\-a, que relaciona los bosones y los fermiones. La versión supersimétrica de la teoría (que requiere un espaciotiempo de 10 dimensiones) se conoce como Teoria de Supercuerdas, dentro de la cual existen cinco formulaciones consistentes: Tipo I, Tipo IIA, Tipo IIB, Heterótica SO$(32)$ y Heterótica $\text{E}_8 \otimes \text{E}_8$.\footnote{Además de las cuerdas heteróticas supersimétricas, también ha despertado interés una versión heterótica no supersimétrica pero libre de anomalías, con grupo de simetría $\text{SO}(16) \otimes \text{SO}(16)$.}

Este marco teórico también predice la existencia de otros objetos supersimétricos conocidos como super-$p$-branas. En esta tesis nos centraremos particularmente en las Dirichlet $p$-branas (D$p$-branas), objetos fundamentales en las Teorías de Supercuerdas Tipo II que desempeñan un papel clave en la comprensión de aspectos no perturbativos y de dualidades dentro de este marco teórico. 

Como se anticipa en el título de esta tesis, el interés principal de este trabajo radica en el estudio de los sistemas de múltiples D$p$-branas (mD$p$-branas o simplemente mD$p$), que contienen configuraciones de $N$ D$p$-branas casi coincidentes, con $N^2$ cuerdas abiertas que terminan en estas branas. En estas configuraciones, las cuerdas que se extienden entre di\-fe\-rentes branas pueden describirse por medio de campos vectoriales (casi) sin masa, mientras que aquellas cuyos extremos se sitúan sobre la misma brana se comportan como campos de gauge habituales. En 1995, E. Witten presentó argumentos a favor de la ampliación de la simetría evidente $(\text{U}(1))^{N^2}$ de este sistema a una simetría de gauge no Abeliana U$(N)$. Ade\-más, Witten sugirió que, en el límite de bajas energías (y tras fijar un gauge apropiado, con ruptura de la invariancia de Lorentz), la teoría efectiva que describe el sistema de mD$p$-branas está dada por una acción de Yang-Mills supersimétrica (SYM por sus siglas en Inglés) con simetría gauge U$(N)$. El caso particular de una única brana ($N=1$) corresponde a la acción de SYM Abeliana U$(1)$ con supersimetría má\-xima que coincide con la versión $gauge~fixing$ del límite de campo débil de la acción de una única super-D$p$-brana. Dicha acción viene dada por una combinación de los términos de Dirac-Born-Infeld (DBI) y de Wess-Zumino (WZ); por tanto, se espera que la acción efectiva para mD$p$ incluya una generalización no Abeliana del término de DBI.

Esto condujo naturalmente a un interés más amplio por construir acciones efectivas no li\-nea\-les para los sistemas de múltiples D$p$-branas. La búsqueda de una acción supersimétrica satisfactoria para mD$p$-branas lleva ya más de 25 años en marcha y, aunque existen va\-rios avances significativos y se han desarrollado numerosos resultados e ideas valiosas durante todo este tiempo,  aún no se dispone de una formulación completamente satisfactoria en su caso general. No obstante, las propiedades que se espera que tenga la acción de este sistema de mD$0$ sí están bien establecidas: debe ser invariante bajo transformaciones de Lorentz, poseer supersimetría global en el espaciotiempo, ser invariante bajo una genera\-lización (para los sistemas de múltiples branas) de la simetría local fermiónica llamada $\kappa$-simetría de la acción de una única D$p$-brana (conocida dicha generalización como supersimetría local de la línea de mun\-do, que garantizaría un estado fundamental supersimétrico) y reducirse a la teoría de SYM U$(N)$ en el límite de bajas energías.

Dentro del contexto de estudio de los sistemas mD$p$-brana, resulta especialmente útil abordar el caso de las múltiples M$0$-branas (mM$0$) en 11 dimensiones. Este sistema es particularmente relevante ya que la reducción dimensional de una única M$0$-brana (que se corres\-ponde con la superpartícula sin masa en 11 dimensiones, también conocida como M-onda) reproduce la acción de una D$0$-brana en un espaciotiempo de 10 dimensiones. En consecuencia, es natural suponer que la acción del sistema de múltiples D$0$-branas en 10 dimensiones aparezca como la reducción dimensional de una acción en 11 dimensiones que represente una generalización no Abeliana de la acción de una única M$0$-brana i.e. la acción de múltiples M$0$-branas (mM$0$).

Efectivamente, dicha acción fue construida por I. Bandos en 2012, la cual se obtuvo exten\-diendo la acción de una única M$0$-brana mediante la incorporación de términos de interacción que involucran campos matriciales. Estos campos matriciales que describen la dinámica relativa del sistema coinciden precisamente con los campos de la teoría de SYM SU$(N)$ en $\text{D}=10$ reducida dimensionalmente a $\text{d}=1$ dimensiones. Sin embargo, los intentos iniciales de obtener una acción para el sistema mD$0$ a partir de esta acción en 11D no tuvieron éxito y, al final, en un trabajo de 2018, se acabó construyendo directamente un candidato para la acción de dicho sistema de mD$0$-branas en $\text{D}=10$.

Motivados por estos resultados, en el capítulo~\ref{ch.4D_nAmW} se construye la acción del sistema de multiples ondas no Abelianas (conocido como nAmW por sus siglas en Inglés) en un superespacio plano de $\text{D}=4$ $\mathcal{N}=1$; este modelo sirve como el homólogo en dimensiones bajas del sistema mM$0$ en 11 dimensiones. En el capítulo~\ref{ch.3D_mD0}, se estudia su reducción dimensional, obteniéndose una acción esencialmente no lineal para el análogo 3-dimensional del sistema de mD$0$ en 10 dimensiones.

Siguiendo esta línea de trabajo, en el capítulo~\ref{ch.10D_mD0} se presenta un conjunto completo de acciones no lineales, supersimétricas y con $\kappa$-simetría, para el sistema en $\text{D}=10$ $\mathcal{N}=2$ de $N$ múltiples D$0$-branas casi coincidentes, cuya forma general posee una función arbitraria definida positivamente $\mathcal{M}(\mathcal{H})$, donde $\mathcal{H}$ es el Hamiltoniano de SYM con grupo de gauge SU$(N)$ construido a partir de los campos matriciales que describen el movimiento relativo. Es relevante señalar que todas estas acciones comparten las propiedades esperadas del sistema mD$0$, independientemente de la elección que se haga de la función $\mathcal{M}(\mathcal{H})$. Además, en el capítulo~\ref{ch.11D_origin} se demuestra cómo un representante particular de esta familia de acciones puede derivarse por reducción dimensional a partir del modelo de mM$0$-branas en 11 dimensiones. 

Finalmente, en el capítulo~\ref{ch.quantization} se desarrolla el formalismo Hamiltoniano y la cuantización del representante más simple del sistema en 3 dimensiones que actúa como homólogo del sistema de mD$0$-branas en $\text{D}=10$. Se obtienen soluciones formales para el sistema de ecuaciones de la teoría de campos que aparece como resultado de esta cuantización, especificando su dependencia respecto a las coordenadas fermiónicas del centro de masas. Además, se discuten las configuraciones BPS\footnote{Las siglas corresponden a Bogomol'nyi-Prasad-Sommerfield.} en el sector del movimiento relativo, que describen el estado fundamental del modelo matricial del tipo BFSS\footnote{El acrónimo corresponde a Banks-Fischler-Shenker-Susskind.} para $N$ finito. Todo este análisis constituye un paso necesario hacia una futura formulación completa de la teoría de campos de sistemas de múltiples D$0$-branas en 10D.

Cabe destacar que las acciones de mM$0$ y mD$0$, así como sus homólogos en dimensiones bajas, están escritas utilizando el formalismo de $(spinor)~moving~frame$. Por ello, en el capítulo~\ref{ch.MF_and SMF} se presenta este formalismo y se subraya su importancia a través de un modelo ilustrativo: la M$0$-brana única en un supersespacio plano de 4 dimensiones, como paso previo a la construcción desarrollada en el capítulo~\ref{ch.4D_nAmW}.


\pagenumbering{arabic}  

     \chapter*{Preface}
\addcontentsline{toc}{chapter}{Preface}
\thispagestyle{empty}
\vspace*{-1.5cm}

\begin{changemargin}{4cm}{0cm}
    \singlespacing\textcolor{cites}{ \small{
         \begin{flushright}Our mistake is not that we take our theories too seriously, but that\\ 
         we do not take them seriously enough. It is always hard to realize that\\
         these numbers and equations we play with at our desks\\
         have something to do with the real world.
         \end{flushright}
     \begin{flushright}     
         {\sffamily {\textit{The First Three Minutes}}\\
         {by {Steven Weinberg}. }}
     \end{flushright}}}
    \end{changemargin}
    \vspace{12pt}

The Standard Model (SM) of particle physics represents one of the most significant achievements in the scientific history, spanning over 2000 years of inquiry. Over the course of time, one of the primordial goals of both Philosophy and Science has been to comprehend what is/are the smallest constituent/s of the Universe and their interactions. This quest can be briefly captured by the fundamental question: \textit{``What is the matter made of?''}. 

Nowadays, in the context of the SM, the answer of the above question is given by the \textit{elementary particles}\footnote{Elementary particles are defined as those particles that are not composed from other small constituents. For instance, in the current knowledge that we have of the matter (at the end of 2025), electrons are elementary particles but protons are not because they are composed by quarks.}. Remarkably, all the matter around us is made up of just three of these fundamental particles: electrons, up quarks and down quarks; the combination of these quarks forms the protons and neutrons which, in turn, make up atomic nuclei. 

 In that way, modern physics identifies elementary particles as the most basic constituents, but this idea is not completely new. Ancient thinkers also sought to explain the composition of the world around them. In contrast to alchemical notions\footnote{Alchemists believed that all matter was composed of different combinations of four fundamental elements: water, fire, earth, and air. These ideas, rooted in Aristotelian philosophy, persisted until the time of Antoine Lavoisier, who dropped out the old alchemist ideas as phlogiston and he reformulated the Chemistry in the 18th century~\cite{Lavoisier}.} that emerged with Empedocles and it was strongly rooted by Aristotle~\cite{fourElements}, an alternative perspective was introduced by Leucippus and his disciple Democritus who proposed an indivisible unit called atom (from the ancient Greek word \textit{atomos}\footnote{From $\alpha$ (\textit{a-}, ``not'') and $\tau o \mu o \nu$ of the root of $\tau \grave{\varepsilon} \mu \nu \omega$ (\textit{témn}$\bar{\textit{o}}$, ``cut'')~\cite{atomo_origen}.}, which means ``uncuttable'' or ``indivisible'') as the constituent of matter~\cite{Atomistas}. This concept, diverging from the prevailing beliefs of the time, was dismissed by scholars of the era. It was not until the 19th century, with John Dalton, that the atomic theory was resurrected although his interpretation differed from that of the ancient philosophers. The English chemist noticed that chemical substances seemed to combine in fixed and consistent proportions by weight, leading him to propose that matter was composed of fundamental, indivisible units, which he named atoms~\cite{Dalton}.

The idea of the atoms as elementary particles was predominant until the experiments of Ernest Rutherford in 1911 \cite{Rutherford1}, which demonstrated that atoms consist of a positively charged nucleus surrounded by electrons. Later, in 1919, Ernest Rutherford \cite{Rutherford2} and, in 1932, James Chadwick \cite{Chadwick} discovered that nuclei are composed by protons and neutrons (collectively known as nucleons). In 1969, it was revealed that nucleons themselves are made up of quarks \cite{nucleons1, nucleons2}, which, along with electrons, are currently considered fundamental particles.

Could this be the end of the story? Scientific research suggests that we can still advance a bit more. However, before taking another step forward, let us return to the SM and explore how it describes the interactions between these fundamental constituents.

Technically, the SM is formulated within the mathematical framework of quantum field theory (QFT) \cite{QFT1}, which can be seen as the embedding of the special relativistic classical field theory into the quantum world, being the most accurate framework to describe Nature now. QFT offers a view in which a Lagrangian governs the dynamics of the theory and the ma\-tter is made up of fermionic particles (fermions) and the interactions are described by the exchange of bosonic particles (bosons) which arise from \textit{gauge symmetries}. In simple terms, a gauge theory is a field theory in which the Lagrangian remains invariant (in particular) under a set of local\footnote{The term ``local'' means that the symmetry parameters change as we move between points in the spacetime; specifically, local (gauge) symmetries are ``parametrized'' by functions} transformations that vary from point to point in spacetime and form certain Lie groups \cite{QFT2, LieGroups}.

Today, the SM is the only theory that successfully describes three of four interactions of Nature (the strong, the electromagnetic and the weak interactions) into a QFT description. It is a gauge theory invariant under the local $\text{SU}(3) \otimes \text{SU}(2) \otimes \text{U}(1)$ symmetry \cite{gauge1, gauge2} where different subgroups correspond to different interactions.

In particular, the $\text{SU}(2) \otimes \text{U}(1)$ gauge symmetry underlies the electroweak (EW) theory, which provides a unified description of the electromagnetic and weak interactions \cite{EW1, EW2, EW3}. This symmetry is spontaneously broken down to its certain U$(1)$ subgroup and, as a consequence, the $W^{\pm}$ and $Z^0$ gauge bosons acquire mass through the Englert-Brout-Higgs-Guralnik-Ha\-gen-Kibble mechanism \cite{Higgs1, Higgs2, Higgs3, Higgs4} while photon remains massless. The remaining symmetry group of the SM is SU$(3)$ symmetry, also known as colour symmetry. It is the symmetry group of Quantum Chromodynamics (QCD) \cite{QCD1, QCD2}, the gauge theory that describes the interaction between quarks and gluons. The gluons, acting as the gauge bosons of this symmetry group, are also themselves ``charged'' under the strong interaction. Interestingly, QCD exhibits two key properties: colour confinement and asymptotic freedom, which, without delving into technical details, explain the existence of hadrons\footnote{Hadrons are subatomic composite particles made up of two or more quarks.} \cite{QCD3}.

The unification of the electromagnetic and weak interactions is not the first example of unification in the history of physics. One of the most popular and productive ideas in theoretical physics has always been the pursuit of a unified description\footnote{In some sense, the idea of unification can be seen as a philosophical principle: Nature is unique and a more profound comprehension should reveal a common explanation for seemingly different phenomena.} of the fundamental interactions of Nature. This journey towards a deeper understanding of the Universe began in the 17th century, when Isaac Newton unified terrestrial and celestial mechanics under the laws of universal gravitation \cite{Newton}. In the 19th century, James C. Maxwell demonstrated that electricity and magnetism, previously considered separate phenomena, could be described by a single set of equations that today bears his name. This achievement marked the unification of electric and magnetic disciplines into a single framework: electromagnetism \cite{Maxwell}. More recently, as discussed earlier, the SM successfully unified the weak and electromagnetic interactions into the EW theory.

The ultimate goal of Grand Unification Theory (GUT) is to extend this unification further by incorporating the strong interaction. In a broad sense, a GUT rests of the idea of combine weak, strong and electromagnetism into one theoretical framework. In a narrow sense, it seeks a gauge theory described by a Lie group that contains $\text{SU}(3) \otimes \text{SU}(2)\otimes \text{U}(1)$ as subgroup. Among the most extensively studied GUT are the $\text{SU}(5)$ gauge theory \cite{GUT1} (which is currently excluded) and the $\text{SO}(10)$ gauge theory \cite{GUT2}.

As we have briefly introduced, the development of Physics is marked by successive unifications. However, it is not clear today that there is such a connection between the SM and Ge\-ne\-ral Relativity (GR). Moreover, while we have seen that electromagnetism, the weak and the strong interaction are successfully described by QFT, gravity can not be fitted easily within this framework: a QFT constructed on the basis of GR action is not renormalizable, and thus can be used only as an \textit{effective field theory} (EFT). However, the consistency arguments suggest that gravity can be and should be quantized \cite{quantumGrav1}. In particular, a \textit{quantum theory of gravity} is necessary in order to study, for instance, the very early stages of the Universe \cite{quantumGrav2} and certain properties of black holes \cite{quantumGrav3}.

Of all the models that aim in a unification of SM with a (quantum) GR, \textit{String Theory} stands out as the most developed one: it is a quantum theory of gravity \cite{quantumGrav4} (although it remains unproven whether it describes the gravity of our Universe) and provides a theoretical framework for incorporating all fundamental interactions. Its premise, seemingly simple at first, is that, at most fundamental level, point-like particles of the SM are actually excitations of extended\footnote{The idea of extended objects dates back to earlier works by Paul Dirac and the membrane theory \cite{membranesDirac}.} 1-dimensional objects called \textit{strings}. In other words, each particle would be associated with a specific vibrational mode of a single elementary string, recovering the ancient notion that all matter emerges from a single fundamental constituent. From this subtle but deep idea, the counterparts of all laws of physics can be naturally deduced \cite{quantumGrav5}. So,  String Theory is an extremely ambitious and beautiful project toward a unified understanding of Nature. But, as always, it is Nature itself that will have the last word.

We find it worthwhile to briefly explore some of the key characteristics of String Theory. We begin by examining its origins and the major breakthroughs that have shaped its evolution into the theory we know today. We will see that it is not only a theory with strings; String Theory spectrum contains also extended objects known as \textit{branes} (or $p$-branes, with $p$ the number of spatial dimensions of the worldvolume\footnote{The worldvolume is the path that the $p$-brane follows into the spacetime. Then, $p=0$ corresponds to particle, $p=1$ to string, $p=2$ to membrane and so on.} of the brane) which play an important role in the theory. In particular, we focus on \textit{D}\hmath$p$\textit{-branes} (Dirichlet $p$-branes), a class of these extended objects which are the main characters of this thesis. Notably, one of the most remarkable predictions of String Theory is the existence of \textit{extra dimensions},  extending beyond the three spatial dimensions we observe in Nature in our daily lives. It is precisely because String Theory is formulated in higher-dimensional spaces that D$p$-branes with $p > 3$ can exist.

Thus, after a historial review, we discuss the concept of extra dimensions which must be compactified to reconcile the theory with our 4-dimensional observable Universe. The way these extra dimensions are compactified plays a crucial role in determining the properties of the resulting effective models. Depending on how these models are constructed, two main perspectives arise in String Theory, each offering a different approach to bridging the gap between fundamental theory and observable physics.

\textbf{\underline{A brief story about String Theory}}

During the 1960s, particle accelerators revealed a vast number of so-called elementary particles. In this context, the term ``elementary'' refers to particles were not composed of electrons, protons or neutrons. At that time, most of the discovered particles were hadrons (which experience the strong interaction) rather than leptons, such as the electron, which does not interact via the strong force. In 1964, Murray Gell-Mann~\cite{QCD2} and George Zweig~\cite{QCD4} independently proposed the quark model to explain all this zoo of new elementary particles marking the first step toward QCD, the theory that explains the strong interaction governing hadrons.

However, an alternative approach to describing this interaction was proposed by Gabriele Veneziano~\cite{Veneziano1}, who developed a model that successfully reproduced properties of hadronic scattering amplitudes. Despite this success, the model had many shortcomings as a theory of hadrons. Notably, it required a 26-dimensional spacetime which contributed to its initial lack of recognition~\cite{Veneziano2}. Nevertheless, as it was understood a bit later, it contained a re\-mar\-ka\-ble structure in its core: strings, extended 1-dimensional objects in contrast to point-like particles. These strings can be open, with two free endpoints, or closed, forming continuous loops. The key point here is that if we analyse the oscillation modes of a closed string, we find that one particular mode corresponds to a massless spin-2 particle: the graviton, the particle associated with the gravitational interaction. This discovery, along with the po\-ssi\-bi\-li\-ty of studying the quantization of such objects, positioned the Veneziano's model, what later became known as \textit{bosonic string theory}, as a candidate for a quantum theory of gravity. Ne\-ver\-the\-less, although this model seemed very attractive due to its prediction of the graviton, it still suffered the above mentioned problem, and also the presence of tachyons which implies vacuum instability.

At the beginning of 70th, \textit{Supersymmetry} (SUSY) \cite{SUSY2, SUSY6, SUSY7, SUSY8,  SUSY3, SUSY4, SUSY5} came into play. SUSY, described in~\cite{SUSY1} as \textit{``$\ldots$~one of the most strikingly beautiful recent ideas in Physics''}, is a symmetry between bosons and fermions . In a very schematically way, the generators $\hat{Q}$ of SUSY transformations act as
\begin{equation*}
    \hat{Q}| \text{Fermion} \rangle = | \text{Boson} \rangle ~, \qquad \hat{Q} | \text{Boson} \rangle = | \text{Fermion} \rangle ~.
\end{equation*}
The supersymmetric generalization of Veneziano's model (called \textit{superstring theory}) introduced fermions into the theory and resolved the above mentioned instability of the vacuum and reduced the requirement of 26-dimensional spacetime to 10. This breakthrough led to what is known as the \textit{First superstring revolution}~\cite{1stString}. However, the theory of su\-per\-sy\-mme\-tric string was found to be not unique: there exist five different consistent superstrings mo\-dels~\cite{5SS1} which are generally classified as \textit{Type I}, \textit{Type II} (\textit{IIA} and \textit{IIB})~\cite{5SS2} superstrings and \textit{Heterotic}\footnote{In addition to the supersymmetric heterotic strings, a non-supersymmetric but tachyon-free heterotic string with gauge group $\text{SO}(16) \otimes \text{SO}(16)$ has also garnered interest (see e.g. \cite{nonSUSY1, nonSUSY2}).} (SO$(32)$ and E$_8 \otimes$E$_8$) strings~\cite{5SS3, 5SS4}. 

On the first glance one may think that having five consistent superstring theories could be a problem in the search for a unified theory. However, in the early 1990s, strong evidences emerged that different superstring theories are connected to each other by \textit{dualities} and, in addition, are related to 11D supergravity. These observations suggested the idea that all these models are the different limits of a single, more fundamental underlying 11-dimensional theory: M-theory~\cite{Mtheory}. From this perspective, five superstring theories and 11D supergravity are merely different perspectives of M-theory. This comprehension led to what is known as the \textit{Second superstring revolution} dated by 1995. 

In this period, another crucial development came from Joseph Polchinski~\cite{Polchinski}, who identified D$p$-branes as the missing sources of Ramond-Ramond (R-R) $q$-form gauge fields\footnote{Let us devote a few lines to this assertion: R-R fields appear as some of the field content of the type II superstring theories, which naturally arise in the low energy type II supergravity limit. However, before Polchinski's famous paper~\cite{Polchinski},  it was unclear what objects carried R-R charge.}. Thus, D$p$-branes, which had appeared as the dynamical objects where the fundamental strings can have its ends attached~\cite{Sagnotti, Polchinski_2, Horava}, were found to carry the charges of R-R fields. This knowledge contributed to further breakthroughs, such as the development of the famous AdS/CFT co\-rres\-pon\-den\-ce\footnote{The AdS/CFT correspondence, proposed by Juan Maldacena in 1997, is a duality between a gravitational theory in $(\text{d}$+$1)$-dimensional Anti-de Sitter (AdS) space and a conformal field theory (CFT) without gravity living on its $\text{d}$-dimensional boundary.}~\cite{AdS}, which later evolved into gauge/gravity duality.

\textbf{\underline{Extra dimensions and Landscape vs Swampland perspectives}}

One of the unexpected and interesting properties of String Theory is its prediction of additional dimensions, in contrast to 4-dimensional spacetime used in more conventional models of particle physics and GR. As previously mentioned, the quantum theory of the bosonic string is consistent in a 26-dimensional spacetime, while superstring theory requires a 10-dimensional one. However, although it may seem strange, this idea is not exclusive to String Theory.

Historically, there have been other motivations for considering theories with additional spatial dimensions. For example, Theodor Kaluza attempted to unify electromagnetism with gravity by extending GR to a 5-dimensional theory~\cite{KK}, in which the photon emerged from the extra component of the 5-dimensional metric~\cite{extraDim1}.

Importantly, if extra dimensions are compactified, for instance into circles of small radius~\cite{extraDim2}, they cannot be observed at low energies, providing a plausible explanation for why we perceive only four dimensions in the Universe. Thus, it is necessary to adopt an effective description in order to extract phenomenological particle models from String Theory. In the low ener\-gy limit, the 10-dimensional description can be approximated by 4-dimensional EFTs des\-cri\-bing the dynamics of gravity and matter fields. These EFTs are obtained by compactifying the higher-dimensional models down to four dimensions description, where the structure of the compactification determines the spectrum of the lower-dimensional fields.

Such compactifications imply appearance of a large number of scalar fields, the possible values of which parametrize what is usually called the \textit{moduli space} of the theory. Each point in moduli space  can be associated to the set of vacuum expectation values (vev) of the moduli fields and corresponds to a different vacuum of the theory, leading to different particle masses and coupling constants of the EFT. The collection of all EFTs that arise as low energy limits of String Theory is known as the \textit{Landscape}.

A vast number of possible compactifications from higher dimensions to 4-dimensional theory (EFT) makes the study of the Landscape a challenging task \cite{VAFA2}. However, there are many EFTs with certain properties that cannot be derived from String Theory and, probably, also from any possible quantum theory of gravity. The collection of such theories is called the \textit{Swampland}. The search for some criteria indicating what kind of 4-dimensional EFTs cannot have a String Theory origin, and more generally, origin in any quantum theory of gravity is known as the \textit{Swampland program}~\cite{VAFA1}. 

To conclude this preface, let us note that the journey is far from being over: many challenges remain and solving them is our task as researchers, with Nature as our ultimate and most incorruptible judge.

     \chapter[Introduction]{Introduction}\label{ch.intro}
\thispagestyle{empty}

 \vspace*{-1.5cm}   
\begin{changemargin}{6cm}{0cm}
    \singlespacing\textcolor{cites}{ \small{
         \begin{flushright}We observe without being observed. \\ 
        We learn while remaining a mystery to all.
         \end{flushright}
     \begin{flushright}     
         {\sffamily {\textit{Malazan Book of the Fallen I: Gardens of the Moon}}\\
         {by {Steven Erikson}. }}
     \end{flushright}}}
    \end{changemargin}
    \vspace{12pt}

In the following lines, we briefly introduce \textit{String Theory} (section~\ref{sec:StringTheory}), the theoretical framework underpinning this thesis, as well as \textit{superspace formalism} (section~\ref{sec:diff_SuperForms}), the principal mathematical approach employed throughout. \textit{D}\hmath$p$\textit{-branes}, the central objects of this work, will be introduced in section~\ref{sec:Dp-branes}, where we describe their most relevant features for our purposes.

\section{A brief introduction to String Theory}
\label{sec:StringTheory}
The core of String Theory is that point particles are replaced by strings, 1-dimensional objects whose mass and length scale are given by
\begin{equation}
    m_s = \dfrac{2\pi}{l_s}= \dfrac{1}{\alpha'^{1/2}}~,
\end{equation} 
 where $\alpha'$ is the so-called Regge slope\footnote{This parameter $\alpha'$ characterizes the linear relationship between the squared mass $M^2$ and the spin $J$ of string excitations which is known as a Regge trajectory. Its name provides from the concept of the Regge slope originates from Regge theory of hadronic physics~\cite{Regge_1}, where similar linear relationships were observed between the spin and mass squared of hadrons~\cite{Regge_2, Tong}.} a dimensional parameter of the theory. A crucial point here is that the spectrum of all the particles arises from different vibrational modes of the string; among them the graviton, which is the quantum of the gravitational interaction. Additionally, the string coupling constant\footnote{The string coupling constant controls the perturbative expansion of the theory. In the weak coupling regime, physical quantities can be expressed as power series in $g_s$.} $g_s$, which governs the strength of inte\-ractions between strings, is not a fixed parameter but is determined by the expectation value of the dilaton field $\Phi$ (see e.g. \cite{Tomas, Uranga, Tong} for more details).

As is typical in most introductory texts on  String Theory (see e.g.~\cite{quantumGrav1, quantumGrav5, 5SS1}), we begin by discussing the point particle case as a prelude to bosonic strings since it provides with a useful tool and allows to present some basic concepts that we will use along this thesis.

The material presented here is based on standard textbooks~\cite{quantumGrav1, quantumGrav5, 5SS1, Tomas, Uranga}.

\subsection{Relativistic point particle}
\label{sec:rel_pointParticle}
Let us start by recalling the action for a massive free (not subject to any external force) point (no spatial extension) particle. In non-relativistic classical mechanics, the kinetic energy $T$ of a particle is given by\footnote{Let us consider here the typical 3-dimensional space and the vectorial notation $\vec{x}$.}
\begin{equation}
    T = \frac{1}{2} m v^2~,
    \label{eq:nonRel_T}
\end{equation}
where
\begin{equation}
     \qquad  v^2= \dot{\vec{x}}\cdot \dot{\vec{x}}~, \qquad  \dot{\vec{x}}\equiv \vec{v} = \frac{\text{d}\vec{x}}{\text{d}t}~
\end{equation}
and $\vec{x}= \vec{x}(t)$ is a coordinate function defining the path of the particle as a line in space. The action for a \textit{free massive non-relativistic particle} is then defined by integrating the kinetic energy~\eqref{eq:nonRel_T} over time
\begin{equation}
    S_{\text{nr}} = \int \text{d}t~\frac{1}{2}mv^2~. 
    \label{eq:nonRel_S}
\end{equation}
The equation of motion is derived from variation of the above action,
\begin{equation}
    \frac{\text{d}\vec{v}}{\text{d}t} = 0 \Longrightarrow \vec{v} = \text{constant}~,
\end{equation}
indicates uniform rectilinear motion of the particle.

However, we can check immediately that this formulation allows velocities to exceed the speed of light, thus violating Einstein's second postulate of special relativity\footnote{These postulates were enunciated by Einstein in~\cite{Einstein_1905}, one of his \textit{``annus mirabilis''} papers published in 1905.}~\cite{SR_postulates}, which asserts that the speed of light is the maximum attainable speed for any particle.

Thus, the action $S_{\text{nr}}$ is inadequate for describing relativistic particles. This statement can be followed to the fact that the action~\eqref{eq:nonRel_S} is invariant under Galilean symmetry group including, besides translations in time and in spatial directions, rotations of the space but not pseudorotations mixing space and time (special Lorentz transformations or boosts). This Newtonian mechanics appears as non-relativistic (small velocities) limit of the action invariant under relativistic symmetry, i.e. under Poincaré group transformation (including Lorentz transformations and spacetime translations).

The simplest way to describe relativistic particle is to consider a flat $\text{D}$-dimensional spacetime, the Minkowski space $\mathbb{R}^{1, \text{D}-1}$ with a set of coordinates
\begin{equation*}
    x^\mu = (x^0, x^1,\ldots,x^{\text{D}-1}) = (ct, x^i)~, \qquad i= 1,2,\ldots,\text{D}-1~, 
\end{equation*}
($c$ is the velocity of light), and the distance between points separated by an interval $\text{d}x^\mu$,
\begin{equation}
    \text{d}s^2 = \eta_{\mu \nu}\text{d}x^\mu \text{d}x^\nu~, \qquad \mu,\nu= 0, 1 , \ldots, \text{D}-1~,
    \label{eq:interval}
\end{equation}
defined with metric $\eta_{\mu \nu}$ of indefinite signature
\begin{equation}
    \eta_{\mu \nu} = \text{diag}(+,-,\ldots,-) = \begin{pmatrix}
1 & 0 & \cdots & 0 \\
0 & -1 &  \cdots & 0 \\
\vdots & \vdots & \ddots & \vdots \\
0 & 0 & \cdots & -1
\end{pmatrix}
\label{eq:metric}
\end{equation} 
(mostly-minus in our convections). The interval $\text{d}s^2$ defined in~\eqref{eq:interval} is invariant under pseudorotations of space and time which preserve the metric~\eqref{eq:metric} (Lorentz group SO$(1, \text{D}-1)$).

The action for a~\textit{free massive relativistic particle} is proportional to the total length of the path (worldline $\mathcal{W}^1$) swept by particle in spacetime
\begin{equation}
    S = -\int \text{d}s = -m \int \sqrt{\eta_{\mu \nu}\text{d}x^\mu \text{d}x^\nu}~.
    \label{eq:integral}
\end{equation}
To specify the integral~\eqref{eq:integral} in more detail, following old textbooks, we can choose $t(=x^0/c)$ as the evolution parameter. Then the coordinate functions defining $\mathcal{W}^1$ as a line in spacetime are
\begin{equation}
    x^\mu(t) = (ct, x^i(t))~, \qquad \left[ \text{d}x^\mu(t) = \text{d}t\cdot(c, \dot{x}^i(t))\right]~.
\end{equation} 
Substituting into \eqref{eq:integral}, the action becomes
\begin{equation}
    S = -\int \text{d}s = -m \int \text{d}t \sqrt{c^2-\dot{x}^i\dot{x}^i}= -mc^2 \int \text{d}t \sqrt{1-\frac{\dot{x}^2}{c^2}}~. 
    \label{eq:integral2}
\end{equation}
This action makes no sense when $\dot{x}^i > c$: which reflects the speed of light is the limit for any velo\-city of (massive) particle. If we turn in to the so-called natural units, in which $(\hslash=)~c=1$, the action~\eqref{eq:integral2} reads
\begin{equation}
   S=  -m \int \text{d}t \sqrt{1-\dot{x}^2}~.
\end{equation}
We can compute the canonical momentum
\begin{equation}
    p^i = \frac{\partial L}{\partial \dot{x}^i} = \frac{m \dot{x}^i}{\sqrt{1-\dot{x}^2}}~,
\end{equation}
and, by the Legendre transformation of the Lagrangian
\begin{equation}
    L = -m \sqrt{1-\dot{x}^2}~,
\end{equation}
the corresponding Hamiltonian representing the total energy $E$
\begin{equation}
    H= \dot{x}^i p^i - L = \sqrt{p^i p^i + m^2} =  \sqrt{p^2 + m^2} = E~.
\end{equation}
Although this Lagrangian is physically correct, it treats time $t$ as a parameter rather than a dynamical variable unlike the spatial coordinates $x^i$. This asymmetry obscures the Poincaré invariance of the action. To make this manifest, we can promote $t$ to a dynamical variable (1d field) and parametrize the worldline of the particle by another parameter $\tau$ called proper time. Then, the worldline $\mathcal{W}^1$ (the line ``swept'' by particle in spacetime) is defined with the use of four coordinate functions
\begin{equation}
    x^\mu(\tau) = (x^0(\tau), x^i(\tau)) = (t(\tau), x^i(\tau))~,
\end{equation}
which can be used to write the action~\eqref{eq:integral} in the manifestly relativistic invariant form
\begin{equation}
    S =-m \int \text{d}\tau \sqrt{\eta_{\mu \nu} \dfrac{\text{d}x^\mu}{\text{d}\tau} \dfrac{\text{d}x^\nu}{\text{d}\tau} } =-m \int \text{d}\tau \sqrt{\eta_{\mu \nu}\dot{x}^\mu \dot{x}^\nu } = -m\int \text{d}\tau \sqrt{\dot{x}^\mu \dot{x}_\mu} ~.
    \label{eq:1_massive4Dparticle}
\end{equation}
Besides Poincaré transformations, this action is invariant under reparametrizations of the worldline parameter $\tau$,
\begin{equation}
    \tau \mapsto f(\tau)~,
\end{equation}
where $f(\tau)$ is an arbitrary monotonic function of $\tau$.

We now turn our attention to the case of a \textit{free massless relativistic particle}. The above action does not admit a massless limit, but the following action with additional auxiliary fields $e(\tau)$ does:
\begin{equation}
    S = -\dfrac{1}{2} \int \text{d}\tau \left(e^{-1}\eta_{\mu \nu}\dot{x}^\mu \dot{x}^\nu + e m^2 \right) = -\dfrac{1}{2} \int \text{d}\tau \left(e^{-1}\dot{x}^\mu \dot{x}_\mu + e m^2 \right)~. 
    \label{eq:2_massive4Dparticle}
\end{equation}
The equation of motion of this auxiliary field\footnote{This field $e(\tau)$ can be recognised as an einbein of 1d gravity.} $e= e(\tau)$ is algebraic and substituting its solution into~\eqref{eq:2_massive4Dparticle} we arrive at~\eqref{eq:1_massive4Dparticle}. Thus, the actions~\eqref{eq:1_massive4Dparticle} and~\eqref{eq:2_massive4Dparticle} are physically equivalent. Setting $m=0$ in $\eqref{eq:2_massive4Dparticle}$ we obtain  the action for a massless relativistic particle: 
\begin{equation}
    S = -\dfrac{1}{2} \int \text{d}\tau ~e^{-1}\dot{x}^\mu \dot{x}_\mu~.
    \label{eq:massless4Dparticle}
\end{equation}

\subsection{Relativistic string}
The dynamics of the free relativistic string propagating in a $\text{D}$-dimensional spacetime with metric $g_{\mu \nu}$ is described by the Nambu-Goto action
\begin{equation}
    S_{\text{NG}} = - T \int \text{d}^2\xi \sqrt{|h|}
\end{equation}
where the constant $T$ is the string tension, related to Regge slope via
\begin{equation}
    T = \dfrac{1}{2\pi \alpha'}~,
\end{equation}
$\xi^i$ are the worldsheet coordinates and $|h|$ denotes the determinant of the induced metric on the worldsheet,
\begin{equation}
    |h| = \text{det}(h_{ij})~, \qquad h_{ij} = g_{\mu \nu} \partial_i X^\mu \partial_j X^\nu~, \qquad i,j = 0,1~.
\end{equation}
The fields $X^\mu(\xi)$ are coordinate functions defining the worldsheet of the relativistic string as 2-dimensional surface in spacetime
\begin{equation}
    X^\mu = X^\mu(\xi)~.
\end{equation}
They are scalars with respect to reparametrizations of the worldsheet (general transformations of the worldsheet coordinates $\xi^i \mapsto f(\xi^i)$).

Like in the point particle case, the Nambu-Goto action is invariant under both Poincaré transformation and worldsheet reparametrization. However, the square root in the Nambu-Goto action complicates the quantization of the theory even in flat spacetime. A more convenient, yet classically equivalent, formulation is provided by the so-called Polyakov action~\cite{Polyakov_1, Polyakov_2, Polyakov_3}
\begin{equation}
    S_{\text{P}} = - \dfrac{T}{2} \int \text{d}^2  \xi \sqrt{|\gamma|} \gamma^{ij} g_{\mu \nu} \partial_i X^\mu \partial_j X^\nu~.
    \label{eq:Polyakov_action}
\end{equation}
It contains an auxiliary field $\gamma_{ij} = \gamma_{ij}(\xi) = \gamma_{ji}(\xi)$, the metric of the worldsheet, through its determinant $|\gamma| = \text{det}(\gamma_{ij})$ and its inverse $\gamma^{ij}$. Equations of motion for $\gamma_{ij}$ are algebraic and substituting their solution back into \eqref{eq:Polyakov_action} we reproduce the Nambu-Goto action, thus showing the equivalence between the two formulations.

Importantly, let us note that Polyakov action~\eqref{eq:Polyakov_action} has also Weyl invariance, i.e. invariance under local rescaling of the worldsheet metric
\begin{equation}
    \gamma_{ij} \mapsto \Omega^2(\xi)\gamma_{ij}~
\end{equation}
which plays a crucial role in the quantization of the string~\cite{Polyakov_3}.

Using reparametrization and Weyl gauge symmetries of this system, it is possible to transform the worldsheet metric $\gamma_{ij}$ into the flat 2-dimensional Minkowski metric $\gamma_{ij}= \eta_{ij}$. Then, in the case of flat target spacetime $g_{\mu \nu} = \eta_{\mu \nu}$, Polyakov action~\eqref{eq:Polyakov_action} becomes a free field theory action
\begin{equation}
    S_{\text{P}_{\text{flat}}} = -\dfrac{T}{2} \int \text{d}^2 \xi ~\partial_i X^\mu \partial^i X_\mu~. 
\end{equation}
The equations of motion are obtained by finding the extrema of the action from its variation
\begin{equation}
    \begin{array}{l}
        \begin{split}
            \delta S_{\text{P}_{\text{flat}}} &= - T \int \text{d}^2\xi~ \partial_i X^\mu \partial^i \delta X_\mu = \\
            &~\\
            &= T \int \text{d}^2\xi~ \left(\partial^i \partial_i X^\mu \right) \delta X_\mu - T \int \text{d}^2\xi~ \partial^i\left( \partial_i X^\mu  \delta X_\mu \right)~.
        \end{split}
    \end{array}   
\end{equation}
To deal with the total derivative term $- T \int \text{d}^2\xi~ \partial^i\left( \partial_i X^\mu  \delta X_\mu \right) $, we should consider two types of strings
\begin{equation*}
    \begin{array}{lcl}
         \text{closed:} &~& \xi^0 \in [\tau_0, \tau_f]~, \qquad \xi^1 \in [0, 2\pi)~,\\
         ~\\
         \text{open:} &~& \xi^0 \in [\tau_0, \tau_f]~, \qquad  \xi^1 \in [0, \pi]~, 
    \end{array}
\end{equation*}
where $\xi^1$ is the spatial coordinate of the worldsheet and the time coordinate $\xi^0$ is identified with the proper time $\tau$. Focusing on open strings, the boundary terms from the variation give
\begin{equation}
    T \left[ \int \text{d}\xi^1~ \partial_{\xi^0} X^\mu \delta X_\mu \right]^{\xi^0 = \tau_f}_{\xi^0 = \tau_0} - T \left[ \int \text{d}\xi^0~ \partial_{\xi^1} X^\mu \delta X_\mu \right]^{\xi^1 = \pi}_{\xi^1 = 0} ~.
\end{equation}
We require that $\delta X^\mu = 0$ at the initial and final times, so the first term vanishes. However, the second term does not vanish apparently for open strings. To eliminate it we should require
\begin{equation}
    \partial_{\xi^1} X^\mu \delta X_\mu = 0 \qquad \text{at}~~~\xi^1 = 0,\pi~.
\end{equation}
This can be reached by imposing either Newman boundary conditions,
    \begin{equation}
         \partial_{\xi^1} X^\mu = 0 \qquad \text{at}~~~ \xi^1 = 0,\pi~,
    \end{equation}
which do not restrict $\delta X^\mu$ thus allowing the ends of the string to move freely (at speed of light), or by imposing Dirichlet boundary conditions,
    \begin{equation}
         \delta X_\mu = 0 \qquad \text{at}~~~\xi^1 = 0,\pi~,
    \end{equation}
fixing the end points of the string at some constant positions. In turn, the imposition of Dirichlet boundary conditions along $(\text{D}-1-p)$ constraints the motion of the endpoints of a string within a $(p+1)$-dimensional hypersurfaces, explicitly breaking translations invariance in the transverse directions. As we will see later, these hypersurfaces correspond to the world\-vo\-lu\-me of the dynamical objects of the theory: the D$p$-branes.

As a natural next step, we will explore how supersymmetry fits into this framework. However, instead of discussing superstring model (which is string living in superspace, see textbooks~\cite{quantumGrav1, quantumGrav5, 5SS1, Tomas, Uranga}), here we will introduce the concept of superspace, which extends the notion of spacetime by incorporating fermionic (Grassmann-valued) coordinates alongside the usual bosonic ones. 

\section[Superspace formalism]{Superspace and differential superforms}
\label{sec:diff_SuperForms}
In quantum field theory, supersymmetry (SUSY)~\cite{SUSY2, SUSY3, SUSY4, SUSY5} is the symmetry that exchanges bo\-so\-nic and fermionic fields. As a result, its irreducible representations, known as supermultiplets, necessarily contain both bosonic and fermionic components. A particularly compact and elegant way to encode these supermultiplets is via the superfield approach, also called superspace formalism. This is described in a number of comprehensive references (see e.g.~\cite{superspace_1, superspace_2, superspace_3, superspace_4}), so that here we will briefly discuss some essential elements which are necessary for understanding the content of this thesis.

Superspace is an extension of ordinary spacetime that includes, in addition to the usual bo\-so\-nic coordinates $x^\mu$, a set of extra fermionic coordinates $\theta^\alpha$. The latter are Grassmann variables i.e. anticommutative objects that are nilpotent,
\begin{equation}
    \theta^\alpha \theta^\beta = - \theta^\beta \theta^\alpha \qquad \Longrightarrow \qquad \theta^1 \theta^1 = - \theta^1 \theta^1 =0~, \qquad \theta^2 \theta^2 = - \theta^2 \theta^2 =0~,
    \label{eq:nilpotent}
\end{equation}
with $\alpha, \beta= 1,2$ in simplest $\text{D}=3$ and $\text{D}=4$ cases.

Since the chapter~\ref{ch.4D_nAmW} are devoted to the study of 4D non-Abelian multiwave (nAmW) system which is the counterpart of 11D multiple M$0$-branes in flat $\text{D}=4$~~$\mathcal{N}=1$ superspace, let us describe this superspace in some details.

\subsection[Flat \texorpdfstring{$\text{D}$}{D}=4 \texorpdfstring{$\mathcal{N}$}{N}=1 superspace]{Flat \boldmath\texorpdfstring{$\text{D}$}{D}=4 \boldmath\texorpdfstring{$\mathcal{N}$}{N}=1 superspace}
\label{sec:flatD4N1}
The coordinates of flat $\text{D}=4$~~$\mathcal{N}=1$ superspace $\Sigma^{(4|4)}$ are denoted by
\begin{equation}
    z^M = (x^\mu, \theta^\alpha, \bar{\theta}^{\dot{\alpha}})~,
    \label{eq:superspace4D}
\end{equation}
where $\mu= 0,~1,~2,~3$ is a 4-vector index and $\alpha= 1,~2$ and $\dot{\alpha}= 1,~2$ are Weyl spinorial indices; $M= (\mu, \alpha, \dot{\alpha})$ is referred as \textit{supervector} index. 

The bosonic coordinates commute among themselves
\begin{equation}
    x^\mu x^\nu = x^\nu x^\mu~,
\end{equation}
and with fermionic coordinates $\theta^\alpha,~\bar{\theta}^{\dot{\alpha}}$
\begin{equation}
    x^\mu \theta^\alpha = \theta^\alpha x^\mu~, \qquad x^\mu \bar{\theta}^{\dot{\alpha}} = \bar{\theta}^{\dot{\alpha}} x^\mu ~,
\end{equation}
while the latter anticommute among themselves
\begin{equation}
    \theta^\alpha \theta^\beta = - \theta^\beta \theta^\alpha~, \qquad \bar{\theta}^{\dot{ \alpha}} \bar{\theta}^{\dot{\beta}} = - \bar{\theta}^{\dot{\beta}} \bar{\theta}^{\dot{\alpha}}~, \qquad \theta^{\alpha} \bar{\theta}^{\dot{\alpha}} = - \bar{\theta}^{\dot{\alpha}} \theta^{\alpha}~. 
\end{equation}
Bosonic coordinates are real, $x^\mu = (x^\mu)^*$, while fermionic are complex, $\theta^\alpha = (\bar{\theta}^{\dot{\alpha}})^*$. All these properties can be collected in
\begin{equation}
    z^M z^N = (-)^{\epsilon(M)\epsilon(N)}z^N z^M~,
\end{equation}
where $\epsilon(M) \equiv \epsilon(z^M)$ is the so-called Grassmann parity with $\epsilon(\mu) = 0$, $\epsilon(\alpha) = 1 = \epsilon(\dot{\alpha})$.

Fields living on superspace are called superfields: these are functions of both bosonic and fermionic coordinates. As far as fermionic coordinates are nilpotent, the series expansion of superfields in fermionic coordinates contains only a finite number of terms
\begin{equation}
\begin{array}{l}
     \begin{split}
         F(z)& \equiv F(x,\theta,\bar{\theta}) =f(x) +  \theta \lambda(x)+ \bar{\theta} \bar{\chi}(x) + \theta \theta m(x) + \bar{\theta} \bar{\theta} n(x) + \theta \sigma^\mu \bar{\theta} \phi_\mu(x) +\\
         &+ \theta \theta \bar{\theta} \bar{\lambda}(x) +  \bar{\theta} \bar{\theta} \theta \psi(x) + \theta\theta \bar{\theta} \bar{\theta} g(x)~,
     \end{split} 
\end{array} \label{eq:superfields}
\end{equation}
where the coefficients in this series, called superfield components, are the ordinary spacetime fields and $\theta \theta = \theta^\alpha \theta_\alpha$, $\bar{\theta} \bar{\theta} = \bar{\theta}_{\dot{\alpha}}\bar{\theta}^{\dot{\alpha}}$ and $ \theta \sigma^\mu \bar{\theta}=  \theta^\alpha \sigma^\mu_{\alpha \dot{\alpha}} \bar{\theta}^{\dot{\alpha}}$, where $\sigma^\mu_{\alpha \dot{\alpha}}$ are relativistic Pauli matrices. The Grassmann parity of the superfield $F(z)$ is fixed so that the parity of its component fields is determined by whether they enter multiplied by odd or even powers of $\theta$, $\bar{\theta}$. 

In general, a superfield describes a highly reducible representation of supersymmetry. To reduce its field content to an irreducible representation, one can impose specific constraints on the superfield, formulated in terms of fermionic covariant derivatives which we will be considering later. One distinguishes between on-shell and off-shell constraints:
\begin{itemize}
    \item{On-shell constraints restrict the superfield component content to physical fields and impose their equations of motion.}
    \item{Off-shell constraints, in contrast, do not impose equations of motion on the physical fields. They just reduce the field content to some smaller number of physical and au\-xi\-lia\-ry fields (non-physical components that vanish or are expressed through the phy\-si\-cal fields upon applying the equations of motion) which describe an irreducible re\-presentation of (off-shell) supersymmetry, a supermultiplet. The presence of these auxiliary fields ensures the closure of the supersymmetry algebra off-shell (hence off-shell constraints above\footnote{This means that the commutator of two supersymmetry transformations $\delta_{\epsilon} = i \epsilon^{\underline{\alpha}}Q_{\underline{\alpha}}$, $[\delta_{\epsilon_1}, \delta_{\epsilon_{2}}]$ calculated on all the fields is expressed through other symmetry transformations (from super-Poincaré group) acting on the fields without using the equations of motion.}).}
\end{itemize}

\subsection{Differential superforms and variations}
One of the key advantages of using differential forms is their manifest invariance under coordinate transformations. Let us begin by introducing differentials of the superspace coordinates
\begin{equation}
    \text{d}z^M = (\text{d}x^\mu, \text{d}\theta^{\alpha},\text{d}\bar{\theta}^{\dot{\alpha}})~,
\end{equation}
and their exterior product
\begin{equation}
    \text{d}z^M \wedge \text{d}z^N = - (-)^{(\epsilon(M)+1)(\epsilon(N)+1)}\text{d}z^N\wedge \text{d}z^M~.
\end{equation}
This relation encodes the following identities
\begin{equation}
    \begin{array}{lclcl}
         \text{d}x^\mu \wedge \text{d}x^\nu =  -\text{d}x^\nu \wedge \text{d}x^\mu ~, &~& \text{d}x^\mu \wedge \text{d} \theta^\alpha = - \text{d} \theta^\alpha  \wedge \text{d}x^\mu ~, &~& \text{d}x^\mu \wedge \text{d} \bar{\theta}^{\dot{\alpha}} = - \text{d} \bar{\theta}^{\dot{\alpha}}  \wedge \text{d}x^\mu ~,\\
         \text{d}\theta^{\alpha}\wedge \text{d}\theta^{\beta}= \text{d}\theta^{\beta}\wedge \text{d}\theta^{\alpha}~, &~&  \text{d}\theta^{\alpha}\wedge\text{d}\bar{\theta}^{\dot{\beta}}= \text{d}\bar{\theta}^{\dot{\beta}}\wedge \text{d}\theta^{\alpha}~, &~&
         \text{d}\bar{\theta}^{\dot{\alpha}}\wedge \text{d}\bar{\theta}^{\dot{\beta}} = \text{d}\bar{\theta}^{\dot{\beta}} \wedge \text{d}\bar{\theta}^{\dot{\alpha}}~.
    \end{array}
\end{equation}
Differential form $\Upsilon_q$ of degree $q$ on superspace with coordinates $z^M$ is defined as
\begin{equation}
    \begin{array}{lcr}
         \Upsilon_q = \dfrac{1}{q!} \text{d}z^{M_q} \wedge \ldots \wedge \text{d}z^{M_1} \Upsilon_{M_1 \ldots M_q}(z)~, 
    \end{array}
\end{equation}
where $\Upsilon_{M_1 \ldots M_q}(z)$ is superfield carrying supervector indices.

Since these superfields contracted with exterior product of the differentials $\text{d}z^M$, they have to be graded antisymmetric,$\Upsilon_{M_1 \ldots M_q}(z) = \Upsilon_{\left[ M_1 \ldots M_q \right\rbrace}(z)$. This graded antisymmetry implies invariance under exchanging the pair of fermionic indices and appearance of $(-1)$ when the pair of exchanging indices involves at least one bosonic index, e.g. $\Upsilon_{\left[M_1 M_2 \right\rbrace} = -(-1)^{\epsilon(M_1)\epsilon(M_2)}$ $ \Upsilon_{\left[M_2 M_1 \right\rbrace}$.
Additionally, we observe that, if $\Upsilon_q$ is bosonic, the superfields $\Upsilon_{M_1 \ldots M_q}(z)$ containing an odd number of spinorial indices will have a fer\-mio\-nic behaviour while those with even number of fermionic indices behave as bosonic objects.

We use the exterior derivative, which acts on differential forms \textit{from the right},
\begin{equation}
\begin{array}{c}
\begin{split}
\text{d}\Upsilon_q &= \dfrac{1}{q!}~\text{d}z^{M_q} \wedge \ldots \wedge \text{d}z^{M_1} \wedge \text{d}z^N \partial_N \Upsilon_{M_1 \ldots M_q}(z)\equiv\\
&\equiv  \dfrac{1}{(q+1)!}~\text{d}z^{M_{q+1}} \wedge \ldots \wedge \text{d}z^{M_2} \wedge \text{d}z^{M_1} (q+1) \partial_{[ M_1 } \Upsilon_{ M_2 \ldots M_{q+1} \rbrace}(z)~,
\end{split}
\end{array}
\end{equation}
where $\partial_N = \partial/ \partial z^N$. As we can see, d is an operator that maps $q$-forms into ($q+1$)-forms. As $\partial_{M_1}\partial_{M_2} = (-1)^{\epsilon(M_1) \epsilon(M_2)} \partial_{M_2}\partial_{M_1}$, the exterior derivative is nilpotent
\begin{equation}
    \text{d}\text{d} = 0~.
\end{equation}
It acts on the product of differential forms according to the Leibniz rule
\begin{equation}
    \text{d}(\Upsilon_q \wedge \Upsilon'_p) = \Upsilon_q \wedge \text{d} \Upsilon'_p + (-)^p \text{d}\Upsilon_q \wedge \Upsilon'_p~. 
\end{equation}
The variation of differential forms with respect to superspace coordinates can be calculated using the Lie derivative formula~\cite{LieDerivative}
\begin{equation}
\delta \Upsilon_q = \iota_\delta ({\rm d}\Upsilon_q) +{\rm d} (\iota_\delta\Upsilon_q)\; , \qquad
\label{eq:delta}
\end{equation}
where $\iota_\delta$ is the contraction with variation symbol defined by
\begin{equation}
\iota_\delta \Upsilon_q= \dfrac 1 {(q-1)!} {\rm d}z^{M_q}\wedge \ldots \wedge {\rm d}z^{{M}_2} \delta z^{{M}_1} \Upsilon_{M_1\ldots M_q}(z)\; , \qquad
\label{eq:iota}
\end{equation}
 and hence obeying
\begin{equation}
\iota_\delta(\Upsilon_q \wedge \Upsilon^\prime_p) =\Upsilon_q \wedge \iota_\delta\Upsilon^\prime_p + (-)^p \iota_\delta\Upsilon_q \wedge \Upsilon^\prime_p\; .
\end{equation}
In the context of field theory, the generalization of the Lie derivative expression can be applied to the variation of the Lagrangian d-form ${\cal L}$ of a d-dimensional field theory  with respect to fields,
\begin{equation}
\label{varL}
\delta {\cal L}= \iota_\delta ({\rm d}{\cal L}) +  {\rm d}(\iota_\delta {\cal L}) \; .
\end{equation}
To apply this correctly, the exterior derivative should be considered as acting in a space of more dimensions (ideally in a space where all fields are treated on an equal footing with coor\-dinates, reflecting a kind of coordinate-field democracy). Furthermore, the second term in \eqref{varL} is a total derivative and therefore does not contribute when deriving the equations of motion.

It is often convenient to use a covariant version of the Lie derivative, particularly in the pre\-sence of gauge symmetries. This covariant version reads
\begin{equation}
\label{varL=D}
\delta {\cal L}= \iota_\delta ({\rm D}{\cal L}) +  {\rm D}(\iota_\delta {\cal L}) \; ,
\end{equation}
which is equivalent to \eqref{varL} provided that the connection in the covariant derivative corresponds to a gauge transformation that leaves the Lagrangian form invariant.

\subsection{Supervielbein of flat superspace}
\textit{Supervielbeins} of superspace are 1-forms that define a supersymmetric generalization of local reference frame or vielbein of spacetime (tetrade in $\text{D}=4$ case). We denote the bosonic and fermionic supervielbein 1-forms of $\Sigma^{(4|4)}$ by
\begin{equation}
    E^a(z) = \text{d}z^M E^a_M(z)~, \qquad E^\alpha (z)= \text{d}z^M E^\alpha_M(z)~, \qquad \bar{E}^{\dot{\alpha}}(z) = \text{d}z^M \bar{E}^{\dot{\alpha}}_M(z)~,
\end{equation}
where $a=0,~1,~2,~3$. In our convention, Latin symbols $a,~b, \ldots$ denote flat space $4$-vector indices corresponding to vielbein basis of tangent space, while Greek indices $\mu,~\nu,\ldots$ are reserved for the world 4-vector indices corresponding to the holonomic or coordinate basis.

These supervielbein forms can be collected in a single object with flat supervector index
\begin{equation}
    E^A (z) = (E^a, E^{\underline{\alpha}}) = (E^a, E^\alpha, \bar{E}^{\dot{\alpha}}) = \text{d}z^M E^A_M(z)~.
\end{equation}
Here $\underline{\alpha}= (\alpha, \dot{\alpha})$ can be (roughly) interpreted as Majorana spinor index. 

Superspace torsion 2-forms are defined as the covariant exterior derivatives of the supervielbein 1-forms
\begin{equation}
\begin{array}{l}
   T^a := \text{D}E^a = \text{d}E^a - E^b \wedge w_b^a~, \qquad \qquad \qquad \qquad \qquad \qquad ~
\end{array}
\end{equation}
\begin{equation}
       ~~ T^\alpha := \text{D}E^\alpha = \text{d}E^\alpha - E^\beta \wedge w_\beta^{~~~\alpha}~, \qquad w_\beta^{~~~\alpha}:=\dfrac{1}{4} w^{ab}\sigma_{ab\beta}^{~~~~~~~\alpha}~~,
        \end{equation}
\begin{equation}
        ~~T^{\dot{\alpha}} := \text{D}\bar{E}^{\dot{\alpha}} = \text{d}\bar{E}^{\dot{\alpha}} - \bar{E}^{\dot{\beta}} \wedge w_{~~\dot{\beta}}^{\dot{\alpha}}~~, \qquad ~w_{~~\dot{\beta}}^{\dot{\alpha}}:= \dfrac{1}{4}w^{ab}\tilde{\sigma}^{~~~~\dot{\alpha}}_{ab~~~\dot{\beta}}   ~~~,
\end{equation}
where $w^{ab} = - w^{ba} = \text{d}z^M w^{ab}_M(z)$ is spin connection 1-form. The symbols $\sigma^{ab~\alpha}_\beta = (\sigma^{[a}\tilde{\sigma}^{b]})^{~~\alpha}_\beta$ and $\tilde{\sigma}^{ab~\dot{\alpha}}_{~~~~~~~\dot{\beta}} = (\tilde{\sigma}^{[a}\sigma^{b]})^{\dot{\alpha}}_{~~\dot{\beta}}$ are antisymmetrized products of the relativistic Pauli matrices (see chapter~\ref{ch.MF_and SMF} or section~\ref{sec:4Dto3D_conventions} for more details).

In this thesis we only focus on flat superspace which can be defined by torsion constraints
\begin{equation}
    T^a = \text{d}E^a = -2i E^{\alpha} \wedge \sigma^a_{\alpha \dot{\alpha}} \bar{E}^{\dot{\alpha}}~, \qquad T^\alpha = \text{d}E^{\alpha} = 0~, \qquad T^{\dot{\alpha}} = \text{d} \bar{E}^{\dot{\alpha}} = 0~.
\end{equation}
These constraints admit the following explicit solution in terms of superspace coordinates \eqref{eq:superspace4D}
\begin{equation}
    E^a = \text{d}x^a - i \text{d}\theta^{\alpha} \sigma^a_{\alpha \dot{\alpha}} \bar{\theta}^{\dot{\alpha}} + i \theta^{\alpha} \sigma_{\alpha \dot{\alpha}}^a \text{d}\bar{\theta}^{\dot{\alpha}}~, \qquad E^\alpha = \text{d} \theta^\alpha~, \qquad \bar{E}^{\dot{\alpha}} = \text{d} \bar{\theta}^{\dot{\alpha}}~.
\end{equation}
Having briefly introduced string model and the superspace formalism, we now turn to the central subject of this work: D$p$-branes.

\section[Dirichlet $p$-branes]{Dirichlet \hmath$p$-branes}
\label{sec:Dp-branes}
The study of non-perturbative String Theory experienced tremendous progress during the 90th. Not only the connection between different string theories through dualities~\cite{Mtheory, dualities} was fascinating the scientific world; as mentioned in the preface, the Second superstring revolution ushered in a deeper understanding of the crucial role played by supersymme\-tric extended objects, commonly referred to as \textit{super-}\hmath$p$\textit{-branes} (or simply \hmath$p$\textit{-branes}) within String/M-theory. 

The degrees of freedom of these extended objects are described by $\text{D}-p-1$ coordinate functions which define the embedding of the $\text{d}=p+1$ dimensional $p$-brane worldvolume into a $\text{D}$-dimensional target spacetime~\cite{5SS1}. Examples of $p$-branes with $p=1$ and $p=2$ are the fundamental string(s) (also known as F$1$-brane(s)) in 10 dimensions~\cite{5SS2, WessZumino_term} and supermembrane (M$2$-brane)~\cite{p-branes_6} in 11-dimensional spacetime. However, certain $p$-branes also support additional fields that ``live'' on and propagate along their worldvolume.

In $\text{D}=10$ spacetime, the set of these includes the D$p$-branes of type II String Theory. In type IIA one finds supersymmetric D$p$-branes with $p=0,~2,~4,~6$ and $8$ while type IIB superstring theory admits those with $p=1,~3,~5,~7$ and $9$. Each D$p$-brane carries a $\text{d}=p+1$ dimensional U$(1)$ Abelian gauge field on its worldvolume and (its ground state) can be described by supersymmetric solutions of 10D supergravity equations~\cite{Polchinski}. One of the limiting cases is the D$9$-brane, which ``fills'' all spatial dimensions of the 10D target (super)space. The fundamental superstring ending on \textit{spacetime filling} D$9$-brane is actually an open string whose endpoints move freely across the brane. 

Consistency of bosonic string theory demands it to live in a spacetime of 26 dimensions~\cite{Goddard}, so that the corresponding spacetime filling object is the D$25$-brane. The fundamental bosonic strings have their ends free in empty $\mathbb{R}^{1,25}$ but can be coupled minimally to (and hence carry the charge with respect to) U$(1)$ gauge field on the D$25$-brane worldvolume. Requiring preservation of the gauge invariance (i.e. Weyl anomaly cancellation) for string quantized in the gauge field background fixes the dynamics of this field~\cite{Fradkin} to obey the equations which can be derived from the Born-Infeld (BI) action~\cite{BI}. Thus, the BI action emerges as the effective action of spacetime filling D$25$-brane in bosonic string theory. Similarly, the dynamics of type IIB D$9$-brane is governed by a sum of supersymmetric extension of the 10D BI action and Wess-Zumino term~\cite{p-branes_3, p-branes_4, D-branes_Eric, Akulov_Igor}.

The (theoretical) discovery of lower $p$ objects was delayed largely because the prevailing belief in the scientific community was that only open strings with free endpoints and closed strings preserved Lorentz invariance, SO$(1, \text{D}-1)$ in the $\text{D}$-dimensional target space. Indeed, a string whose endpoints are confined to a $\text{d}=p+1$ dimensional hyperplane in spacetime explicitly breaks  SO$(1, \text{D}-1)$ down to the subgroup  SO$(1,p) \otimes \text{SO}(\text{D}-p-1)$.

A key breakthrough in \cite{Sagnotti, Polchinski_2, Horava} which built upon the insight of Scherk and Schwarz~\cite{Scherk_Schwarz}, who had earlier shown that closed string theory naturally describes a theory of gravity, is as follows. Since open string theories necessarily include closed strings in their spectrum, they implicitly describe gravitational interactions too. Then, in a theory of gravity, fixed hyperplanes cannot be treated as a static configuration. Instead, the $\text{d}=p+1$ dimensional objects on which open string endpoints lie must themselves be dynamical: these are the D$p$-branes. Then, the symmetry breaking
\begin{equation*}
\text{SO}(1, \text{D}-1) \mapsto \text{SO}(1,p) \otimes \text{SO}(\text{D}-p-1)~,
\end{equation*}
by Dirichlet boundary conditions on D$p$-brane is spontaneous. Such spontaneous symmetry breaking is a hallmark of any $p$-brane in String/M-theory~\cite{Polchinski_3}.

Focusing on the endpoints of an open string lie in D$p$-brane and considering now the supersymmetric case, we should conclude that the consistency between the coupling of string ends with the U$(1)$ $\text{d}=p+1$ dimensional supersymmetric gauge theory requires that, in the purely bosonic limit, the gauge field satisfies equations of motion derived from the Dirac-Born-Infeld (DBI) action \cite{Leigh}. This is reflected by the fact that the effective actions describing supersymmetric D$p$-branes \cite{p-branes_1, p-branes_2, p-branes_3, p-branes_4, D-branes_Eric} are the sums of supersymmetric extension of the DBI action and of the so-called Wess-Zumino (WZ) term, which (in the case of nontrivial type II supergravity background) describes the coupling of D$p$-brane to R-R gauge fields (or, more precisely, to their superspace generalization). 

A fundamental feature of the super-D$p$-brane actions is their invariance under \textit{local fermionic} {\boldmath$\kappa$}\textit{-symmetry}. This symmetry plays a crucial role as it eliminates half of the fermionic degrees of freedom, ensuring a balance between bosonic and fermionic modes on the worldvolume. Moreover, $\kappa$-symmetry ensures that the ground state of the system preserves half of the target spacetime supersymmetry (see \cite{Bergshoeff_Tomas, Azcarraga_Igor} for further relevant discussion). Specifically, $\kappa$-symmetry leads to D$p$-brane ground states preserve a half of the original supersymmetries, classifying them as stable $1/2$ BPS\footnote{BPS stands for Bogomol'nyi-Prasad-Sommerfield.} state.

As it is anticipated in the title of this thesis, we are particularly interested in systems of multiple D$p$-branes (mD$p$-brane or simply mD$p$), which correspond to configurations of $N$ nearly coincident\footnote{In the literature, the term ``coincident D$p$-branes'' is often used. Here, we prefer to use ``nearly coincident'' which is more precise particularly when, at low energy, this dynamical system is described by SYM action~\cite{Witten_1996}. This involves nonvanishing scalar matrix fields $\mathbb{X}_A^{I~B}$, where $I = 1,\ldots,\text{D}-p-1$ and $A,B= 1, \ldots,N$. The off-diagonal components of these matrices are generically nonvanishing, e.g. $\mathbb{X}_1^{I~2} \neq 0$ describes the distance in $I$-th direction between D$p$-brane labelled by $1$ and D$p$-brane labelled by $2$. The case of $N$ completely coincident D$p$-branes corresponds to diagonal matrices $\mathbb{X}^I$ and occurs in ground state of mD$p$ system. The supersymmetric solutions of supergravity describing this are correctly referred to as ``coincident D$p$-branes''.} D$p$-branes with $N^2$ open strings ending on these branes. In such configurations, the strings stretching between different branes can be interpreted as (nearly) massless fields, while those with both endpoints on the same brane behave as usual gauge degrees of freedom. Although the manifest symmetry of this system is $(\text{U}(1))^{N^2}$, in~\cite{Witten_1996} E. Witten argued in favour of its enhancement to a non-Abelian $\text{U}(N)$ gauge symmetry.

Furthermore, in this brilliant paper he suggested that in the low energy limit (and after an appro\-pria\-te gauge fixed with breaking Lorentz invariance) the effective theory describing the mD$p$-brane system is given by a $\text{U}(N)$ supersymmetric Yang-Mills (SYM) action. For $N=1$, the $\text{U}(1)$ Abelian SYM action with maximal supersymmetry indeed coincides with the weak field limit of the gauge fixed action of a single super-D$p$-brane; from~\cite{p-branes_1, p-branes_2, p-branes_3, p-branes_4, D-branes_Eric}. It was therefore expected that the mD$p$ effective action should involve a non-Abelian generalisation of the DBI term (or BI term in the case of spacetime filling D$9$-brane).

This expectation naturally led to a broader interest to construct a more complete nonlinear effective action for multiple D$p$-brane systems. The quest for such a supersymmetric mD$p$-brane action has continued for more than 25 years. While significant progress has been made with many valuable results and insights developed along the way~\cite{p-branes_5, Tseytlin, Emparan, Taylor_1, Taylor_2,  Myers, NonAbelian_BI, NonAbelianSUSY_BI, Sorokin_1, Drummond, Janssen_1, Panda, Janssen_2, Janssen_3, Lozano, Howe_1, Howe_2, Howe_3, superembedding_Igor, superembedding, superembedding_Igor_2, McGuirk, Bandos_11D_mM0, Meliveo, Choi_1, Choi_2, 10D_mD0_Igor, Brennan},  \cite{Unai_1, Unai_2, Unai_3, Unai_4}, the problem has not been solved yet in its complete form. However, the expected properties of this mD$p$ action are known. It has to be 10D Lorentz invariance, target space supersymmetry, a counterpart of the local fermionic $\kappa$-symmetry of the single D$p$-brane action (which can be called \textit{local worldline supersymmetry}\footnote{The identity of the $\kappa$-symmetry of superparticle system as worldline supersymmetry was revealed in~\cite{localWlSUSY}.}), which would ensure a supersymmetric ground state, and reduction of the gauge fixed version to $\text{d}= p+1$ $\text{U}(N)$ SYM theory in the low energy limit.

In the bosonic sector, the most widely accepted candidate remains the so-called ``dielectric brane'' action proposed by Myers~\cite{Myers}, which was motivated by consistency with T-duality and exhibits and intriguing dielectric effect\footnote{Whereby lower dimensional mD$p$ systems possess a coupling to higher form gauge fields $C_{q+1}$ with $q>p$ i.e. are polarised with respect to these background fields, which are not coupled to single D$p$-brane.}. This action has been extensively studied and generalised~\cite{Janssen_1, Janssen_2, Janssen_3, Lozano}, and is often taken to represent the bosonic limit of the full mD$p$-brane action. Nonetheless, despite persistent efforts, no Lorentz-invariant or supersymmetric ge\-ne\-ra\-li\-za\-tion of this action is currently known. A similar situation holds for analogous cons\-tructions in M-theory, such as the action for multiple M-gravitons (multiple M-waves) proposed in~\cite{Janssen_1, Janssen_2, Janssen_3, Lozano}, for which supersymmetric completion also remains out of reach.

In the context of mD$0$-brane it is instructive to discuss the 11D multiple M$0$-brane system. This is particularly relevant since, as shown in~\cite{D-branes_Eric}, the dimensional reduction of single M$0$-brane (corresponding to 11D massless superparticle, also known as M-wave~\cite{D-branes_Eric}) reproduces the action of the 10D D$0$-brane. Consequently, it is natural to expect that the action of the 10D mD$0$-brane system should arise as the dimensional reduction of some 11D action, which itself would represent a non-Abelian generalization of the single M$0$-brane action. 

Such a non-Abelian 11D action has been constructed in~\cite{Bandos_11D_mM0} by adding to the M$0$-brane action in its spinor moving frame formulation (which then describes the center of energy sector of mM$0$) an action for matrix fields with interacting terms involving center of energy va\-ria\-bles. These matrix fields, which are precisely the fields of 10D SU($N$) SYM theory dimensionally reduced to $\text{d}=1$ dimension, are understood to govern the relative dynamics of the system. This is in alignment with Witten’s results~\cite{Witten_1996}, where the mD$0$-brane system at very low energies is described by a maximally supersymmetric matrix model, namely the dimensional reduction of 10D SYM to $\text{d}=1$ for D$0$-branes.

The resulting action~\cite{Bandos_11D_mM0} was interpreted as describing the dynamical system of multiple interacting M-waves (multiple M$0$-branes or mM$0$). Its advantages over the bosonic Myers-type 11D action in \cite{Janssen_1, Janssen_2} are that it involves fermions and the action~\cite{Bandos_11D_mM0} is invariant under two types of supersymmetry: the global 11D target space supersymmetry and the local worldline supersymmetry. The latter is especially important in the framework of String/M-theory since it guarantees that the ground state of the system preserves a part (one half) of the target space supersymmetry. This, in its turn, guarantees that the ground state is a $1/2$ BPS state which saturates the BPS bound, ensuring its stability and underscores their importance in the non-perturbative regime of String/M-theory. 

An attempt to derive an action for the 10D mD$0$ system from this 11D mM$0$ action which was performed in~\cite{10D_mD0_Igor} was not successful. Instead a candidate mD$0$ action was constructed in \cite{10D_mD0_Igor} directly, by coupling of 1d reduction of the 10D SU($N$) SYM to the supergravity induced by the embedding of the center of mass worldline into the 10D type IIA superspace. The consistent dimensional reduction of mM$0$ action~\cite{Bandos_11D_mM0} which lead to candidate mD$0$~\cite{Unai_2} and thus is important for interpreting the mM$0$ system as the decompactification limit of the type IIA mD$0$-brane system, is described in this thesis following~\cite{Unai_3}.

Another significant limitation is that both the 11D mM$0$ action constructed in~\cite{Bandos_11D_mM0, Meliveo} and the candidate 10D mD$0$ action proposed in~\cite{10D_mD0_Igor} are currently known only for the case of flat target superspace. To address this issue in a more manageable setting, a 3D counterpart of the mM$0$ action was developed in~\cite{Bandos_3D_mM0}, referred to as the non-Abelian multiwave (nAmW) system which was also generalised to curved AdS superspace~\cite{Bandos_3D_mM0}.

The quantization of the 3D nAmW model was carried out in~\cite{3D_quantization_Bandos}, yielding a system of rela\-tivistic field equations for bosonic and fermionic fields defined on a space involving additional non-commutative bosonic and non-anticommutative fermionic coordinates. In\-te\-res\-tin\-gly, in~\cite{3D_quantization_Bandos} it was suggested that the $\text{D}=4$ counterpart of this quantum 3D model may exhibit a simpler and more transparent structure. This simplification arises from the intrinsic complex structure of $\text{D}=4$~~$\mathcal{N}=1$ superspace, which also appears in $\text{D}=3$~~$\mathcal{N}=2$ supersymmetric theory.

Therefore, constructing and quantizing the $\text{D} = 4$ version of the nAmW system is anticipated to offer a more effective basis for exploring the quantization of both the 11D mM$0$ and 10D mD$0$ models. Such developments could, in turn, shed new light on the structure and properties of String/M-theory\footnote{Moreover, the 4D nAmW system provides a conventional toy model for addressing an important open pro\-blem: the generalization of the 11D mM$0$ action to curved backgrounds. However, this problem is beyond the scope of this PhD thesis.}. 

Motivated by the arguments presented above, in chapter~\ref{ch.4D_nAmW} we construct the action for $\text{D}=4$ nAmW system in flat~~$\mathcal{N} = 1$ superspace; this model serves as lower dimensional counterpart of the 11D mM$0$ system. We demonstrate that the resulting action is invariant under the global 4D target superspace supersymmetry and the local worldline supersymmetry gene\-ra\-li\-zing the $\kappa$-symmetry of the massless $\text{D} = 4~~\mathcal{N} = 1$ superparticle~\cite{kappa_2}. In chapter~\ref{ch.3D_mD0} we study the dimensional reduction of this system and obtain an essentially nonlinear action for $\text{D}=3$~~$\mathcal{N}=2$ counterpart of the 10D mD$0$ system which we call 3D mD$0$-brane abusing a bit the terminology. The general form of this 3D mD$0$ action includes a positive definite function $\mathcal{M}(\mathcal{H})$ of the so-called relative motion Hamiltonian $\mathcal{H}$. This Hamiltonian is itself constructed from the matrix fields that describe the relative motion and interactions among the mD$0$ constituents. Both chapters are based on~\cite{Unai_1}.

In chapter~\ref{ch.10D_mD0} we present a set of nonlinear actions for the $\text{D}=10$~$\mathcal{N}=2$ system of $N$ nearly coincident multiple D$0$-branes which possesses all the expected properties from the mD$0$ system and, like its lower dimensional $\text{D}=3$ counterpart, includes a positive definite function $\mathcal{M}(\mathcal{H})$ with relative motion Hamiltonian $\mathcal{H}$ constructed from the fields of $\text{d}=1$ $\mathcal{N}=16$ SU$(N)$ SYM fields. In chapter~\ref{ch.11D_origin} we will show how one particular representative of this complete set of candidates (characterized by a specific choice of $\mathcal{M}(\mathcal{H})$) can be derived via dimensional reduction from the 11D mM$0$-brane system. There we also derive the equations of motion of this dynamical system and study its properties, particularly its supersymmetric solutions. Both chapters are based on \cite{Unai_2, Unai_3}.

As a first stage towards the field theory of multiple D$0$-branes, in chapter~\ref{ch.quantization} we construct the Hamiltonian formalism and carry out the covariant quantization of the \textit{simplest}\footnote{Here, the word ``simplest'' refers to the choice of the positive definite function $\mathcal{M}(\mathcal{H})=m$, where $m$ the mass of the single D$0$-brane.} three dimensional counterpart of the 10D mD$0$-brane system. This chapter is based on~\cite{Unai_4}.

Since the mM$0$ and mD$0$ actions, and their lower dimensional counterparts, are written using moving frame and spinor moving frame variables, we begin in chapter~\ref{ch.MF_and SMF} by introduction to this formalism and highlighting its significance on the example of a simple model: the single M$0$-brane in $\text{D}=4$~~$\mathcal{N}=1$ flat superspace. This example also plays a key role in the cons\-truction presented in chapter~\ref{ch.4D_nAmW}.

     \chapter[Spinor moving frame formalism]{Spinor moving frame formalism}\label{ch.MF_and SMF}
\thispagestyle{empty}
\vspace*{-1.5cm}   
\begin{changemargin}{4cm}{0cm}
    \singlespacing\textcolor{cites}{ \small{
         \begin{flushright}
        No question; language can free us of feeling, or almost.\\ Maybe that's one of its functions.\\ So we can understand the world without\\ becoming entirely overwhelmed by it.
         \end{flushright}
     \begin{flushright}     
         {\sffamily {\textit{Contact}}\\
         {by {Carl Sagan}. }}
     \end{flushright}}}
    \end{changemargin}
    \vspace{12pt}

The \textit{moving frame} and \textit{spinor moving frame} approach employed throughout this thesis involves the use of some auxiliary variables known as moving frame vectors and their spinor moving frame companions. In this section we will introduce this formalism starting from the first order form of 4D version of the Brink-Schwarz action. Notice that it is also relevant to the so-called \textit{superembedding approach}~\cite{origin_superembedding_1, origin_superembedding_2}, a geometrical description of the dynamics of supersymmetric point-like and extended objects in String Theory (see~\cite{superembedding_1, superembedding_2, superembedding_3} for reviews and references).

Here we present this formalism starting for the case of $\text{D}=4$ $\mathcal{N}=1$ massless superparticle which serves as a prelude of the chapter~\ref{ch.4D_nAmW}, where the 4D counterpart of the 11-dimensional system of multiple M$0$-branes is presented and described. Notice that the action for this latter system is known only in spinor moving frame formulation.


\section{Massless superparticle in different formulations}
In this section we study the action of $\text{D}=4$ $\mathcal{N}=1$ massless superparticle (a counterpart of the 11D single M$0$-brane~\cite{D-branes_Eric}, also known as M-wave) in the framework of spinor moving frame formulation.

\subsection[Massless superparticle in D=4~~\texorpdfstring{$\mathcal{N}$}{N}=1 superspace]{Massless superparticle  in D=4~\boldmath\texorpdfstring{$\mathcal{N}$}{N}=1 superspace}
The first order form of the Brink-Schwarz action for $\text{D}=4$ $\mathcal{N}=1$ massless superparticle~\cite{BS_paper} reads 
\begin{equation}
    S_{\text{BS}} = \int_{\mathcal{W}^1} \left(p_\mu \hat{E}^\mu -  \frac{e}{2}p_\mu p^\mu \right) = \int \text{d}\tau \left(p_\mu \hat{E}^\mu_\tau - \frac{e}{2}p_\mu p^\mu \right)~,
    \label{eq:action_BS}
\end{equation}
where $\mu = 0,1,2,3$ is the vector index of the Lorentz group SO$(1,3)$ and $e=e(\tau)$ is an auxiliary einbein field whose variation imposes the mass shell condition $p_\mu p^\mu = 0$ for the auxiliary momentum variable $p_\mu = p_\mu(\tau)$ and 
\begin{equation}
\hat{E}^\mu = \hat{E}^\mu(\hat{z}) = \text{d}\tau \hat{E}^\mu_\tau = \text{d}\hat{z}^M\hat{E}_M^\mu(\hat{z})~,
\end{equation}
is the pull-back of the bosonic supervielbein of the 4D target superspace to the worldline parametrised by proper time $\tau$. In this thesis, we restrict ourselves by the case of flat target superspace; the bosonic supervielbein in this case (also known as Volkov-Akulov (VA) 1-form \cite{SUSY3,SUSY4}) has the form 
\begin{equation}
    E^\mu = \Pi^\mu = \text{d}x^\mu -i \text{d}\theta \sigma^\mu \bar{\theta} +i \theta \sigma^\mu \text{d}\bar{\theta} = \text{d}x^\mu -i \text{d}\theta^{\alpha} \sigma_{\alpha \dot{\alpha}}^\mu \bar{\theta}^{\dot{\alpha}} +i \theta^{\alpha} \sigma_{\alpha \dot{\alpha}}^\mu \text{d}\bar{\theta}^{\dot{\alpha}} ~,
    \label{eq:VA_4D}
\end{equation}
where $z^M = (x^\mu, \theta^\alpha, \bar{\theta}^{\dot{\alpha}})$ are $\text{D}=4$ $\mathcal{N}=1$ superspace coordinates carrying vector and Weyl spinor indices $\alpha = 1,2$ and $\dot{\alpha}= 1,2$. Here and below we use the relativistic Pauli matrices
\begin{equation*}
    \sigma_\mu = \sigma_{\mu\alpha \dot{\beta}} = (\mathbb{1}, \sigma_1, \sigma_2, \sigma_3)_{\alpha \dot{\beta}} =  \tilde{\sigma}^{\mu \dot{\alpha} \beta}:= \epsilon^{\dot{\alpha}\dot{\beta}} \epsilon^{\beta \alpha} \sigma_{\nu \alpha \dot{\beta}}\eta^{\nu \mu}~,
\end{equation*}
which obey 
\begin{equation}
    (\sigma_\mu \tilde{\sigma}_\nu +\sigma_\nu \tilde{\sigma}_\mu )_\alpha^{~~~\beta} = 2 \eta_{\mu \nu} \delta_\alpha^{~~~\beta}~, \qquad (\tilde{\sigma}_\mu \sigma_\nu +\tilde{\sigma}_\nu \sigma_\mu )^{\dot{\alpha}}_{~~~\dot{\beta}} = 2 \eta_{\mu \nu} \delta^{\dot{\alpha}}_{~~~\dot{\beta}}~, 
    \label{eq:sigmaRelations}
\end{equation}
\begin{equation}
    \sigma_{\mu~\alpha \dot{\beta}}\tilde{\sigma}^{\mu~\beta \dot{\alpha}} = 2 \delta_\alpha^{~~~\beta} \delta^{\dot{\alpha}}_{~~~\dot{\beta}}~.
    \label{eq:matrixRelations}
\end{equation}
The spinorial indices are raised and lowered by the unit antisymmetric tensor
\begin{equation*}
    \epsilon^{\alpha \beta} = -\epsilon_{\alpha \beta} = \begin{pmatrix}
        0 & 1 \\
        -1 & 0
    \end{pmatrix} = -\epsilon_{\dot{\alpha} \dot{\beta}} = \epsilon^{\dot{\alpha} \dot{\beta}}~
\end{equation*}
``acting from the left'', i.e. $\theta^\alpha = \epsilon^{\alpha \beta}\theta_\beta$, $\theta_\alpha = \epsilon_{\alpha \beta}\theta^\beta$ and $\bar{\theta}^{\dot{\alpha}} = \epsilon^{\dot{\alpha} \dot{\beta}}\bar{\theta}_{\dot{\beta}}$, $\bar{\theta}_{\dot{\alpha}} = \epsilon_{\dot{\alpha} \dot{\beta}}\bar{\theta}^{\dot{\beta}}$. 

The pull-back of the Volkov-Akulov (VA) 1-form to the worldline is proportional to differential of proper time $\tau$:
\begin{equation}
     \hat{\Pi}^\mu =\text{d}\tau \hat{\Pi}^\mu_\tau~, \qquad \hat{\Pi}_\tau^\mu = \hat{\Pi}_\tau^\mu(\hat{z})=  \partial_\tau \hat{x}^\mu -i \partial_\tau\hat{\theta} \sigma^\mu \hat{\bar{\theta}} +i \hat{\theta} \sigma^\mu \partial_\tau\hat{\bar{\theta}}~. 
\end{equation}
It is obtained from \eqref{eq:VA_4D} by replacing the coordinates $x^\mu$ and $\theta^\alpha$, $\bar{\theta}^{\dot{\alpha}}$ by bosonic and fermionic coordinate functions $\hat{x}^{\mu}(\tau)$ and $\hat{\theta}^\alpha(\tau)$, $ \hat{\bar{\theta}}^{\dot{\alpha}}(\tau)$, respectively,
\begin{equation}
    z^M = (x^\mu, \theta^\alpha, \bar{\theta}^{\dot{\alpha}}) \quad \mapsto \quad \hat{z}^M(\tau) = (\hat{x}^\mu(\tau), \hat{\theta}^{\alpha}(\tau), \hat{\bar{\theta}}^{\dot{\alpha}}(\tau))~.
\end{equation}
The coordinates functions are used to define parametrically the embedding of the superparticle worldline $\mathcal{W}^1$ into $\text{D} = 4$ $\mathcal{N}=1$ superspace $\Sigma^{(4|4)}$,
\begin{equation}
    \mathcal{W}^1 \subset \Sigma^{(4|4)}: \qquad z^M = \hat{z}^M(\tau)~.
\end{equation}
Below we will omit the hat symbol $\textasciicircum$ from the coordinate functions and the pull-back of the forms in order to simplify equations. Moreover, we prefer to hide \text{d}$\tau$ inside of differential form, in particular to define the Lagrangian 1-form $\mathcal{L}= \text{d}\tau \mathcal{L}_\tau$, and write our actions as integral of this 1-form over the worldline $\int_{\mathcal{W}^1} \mathcal{L}$ rather than as an integral over $\text{d}\tau$ of a Lagrangian, $\int \text{d}\tau \mathcal{L}_\tau$. 

\subsection{Massless superparticle in twistor-like formulation}
The variation of worldline einbein $e(\tau)$ produces the mass shell constraint  $p^\mu p_\mu = 0$. Exploiting this fact it is possible to write the action \eqref{eq:action_BS} in an equivalent form by solving this algebraic equation and substituting the solution back.

Let us introduce the bosonic spinor fields $\lambda$ and its c.c. $\bar{\lambda}$ 
\begin{equation}
    \lambda^\alpha = \lambda^{\alpha}(\tau)~, \qquad \bar{\lambda}^{\dot{\alpha}} = \bar{\lambda}^{\dot{\alpha}}(\tau)~.
\end{equation}
The mass shell constraint $p^\mu p_\mu = 0$ can be easily solved using these bosonic spinors\footnote{Indeed, as bosonic spinors $\lambda^\alpha \lambda^\beta = \lambda^\beta \lambda^\alpha~,$ we find
\begin{equation*}
     \lambda^\alpha \lambda_\alpha = \epsilon_{\alpha \beta} \lambda^{\alpha} \lambda^{\beta} = \epsilon_{[\alpha \beta]} \lambda^{\left(\alpha \right.} \lambda^{\left.\beta\right)} \equiv 0~
\end{equation*}
and then\begin{equation*}
p_\mu p^\mu     = \lambda^\alpha \bar{\lambda}^{\dot{\alpha}} \lambda^{\beta} \bar{\lambda}^{\dot{\beta}} \sigma_{\mu \alpha \dot{\alpha}} \sigma^{\mu}_{\beta \dot{\beta}} = \lambda^\alpha \bar{\lambda}^{\dot{\alpha}} \lambda^{\beta} \bar{\lambda}^{\dot{\beta}} (2\epsilon_{\alpha \beta} \epsilon_{\dot{\alpha}\dot{\beta}}) = 2\epsilon_{[\alpha \beta]} \lambda^{\left(\alpha \right.} \lambda^{\left.\beta\right)} \epsilon_{[\dot{\alpha} \dot{\beta}]} \bar{\lambda}^{\left(\dot{\alpha} \right.} \bar{\lambda}^{\dot{\beta})} \equiv0~.
\end{equation*}} 
\begin{equation}
    p_\mu = \lambda^\alpha \sigma_{\mu \alpha \dot{\alpha}} \bar{\lambda}^{\dot{\alpha}}~.
    \label{eq:p_mu_lambda}
\end{equation}
Substituting this solution back into \eqref{eq:action_BS}, we arrive at a new equivalent form of the superparticle action
\begin{equation}
    S_{\text{FS}} = \int_{\mathcal{W}^1} \lambda^\alpha \sigma_{\mu \alpha \dot{\alpha}} \bar{\lambda}^{\dot{\alpha}} \Pi^\mu~,
    \label{eq:action_FS}
\end{equation}
which is known as twistor-like or Ferber-Shirafuji formulation~\cite{Ferber, Shirafuji} of the massless superparticle action.

\subsection{Moving frame formulation of the massless superparticle}
\label{sec:SMF4D_action}
In this section we present the action of the same system in a one more equivalent form. This new form is quiet close to the Ferber-Shirafuji formulation but it provides a much convenient basis for the study of the multiple branes systems, the main subject of this thesis.

Let us begin by considering a solution of the mass shell condition $p_\mu p^\mu = 0$ in a special reference frame, 
\begin{equation}
    P_{(a)} = (1,0,0,-1)\rho ~.
    \label{eq:P_a}
\end{equation}
Clearly the vector field $P_{(a)} = P_{(a)}(\tau)$ is light-like and $\rho = \rho(\tau)$ describes the energy of the massless particle. The solution of the mass shell condition in an arbitrary frame can be obtained from~\eqref{eq:P_a} by making a Lorentz transformation
\begin{equation}
    p_\mu := U_\mu^{(a)}P_{(a)} = U_\mu^{(a)}(1,0,0,-1)\rho =: u_\mu^{=}\rho^{\#}~.
    \label{eq:p_mu}
\end{equation}
Here
\begin{equation}
    U_\mu^{(a)}:= \left(\frac{u^{\#}_\mu+u^=_\mu}{2},  U^{i}_\mu, \frac{u^{\#}_\mu-u^=_\mu}{2}\right) \in \text{SO}(1,3)
    \label{eq:4D_MF_matrix}
\end{equation}
is a Lorentz group valued matrix field whose elements are known as \textit{moving frame variables} (also referred to as \textit{Lorentz harmonics}~\cite{Igor_Russian}; see section~\ref{sec:MF_and_SMF} for more details on the connection between these terms) and $i=1,2$. Eq.~\eqref{eq:4D_MF_matrix} implies that the elements of $U_\mu^{(a)}= U_\mu^{(a)}(\tau)$ obey the constraints~\cite{harmonic, lightcone}
\begin{equation}
    U_\mu^{(a)} U^{\mu (b)} = \eta^{ab} = \text{diag}(1,-1,-1,-1)~,
    \label{eq:U_term}
\end{equation}
which can be split into
\begin{equation}
\begin{array}{rcrcr}
u^{=}_\mu u^{\mu =} = 0~, &~&  u^{\#}_\mu u^{\mu \#} = 0~, &~& u^{=}_\mu u^{\mu \#} = 2~,\\
~~u^{=}_\mu U^{\mu i} = 0~, &~& ~u^{\#}_\mu U^{\mu i} = 0~, &~& U^{i}_\mu U^{\mu j} = - \delta^{ij}~.
\end{array}
\label{eq:4D_MF_properties}
\end{equation}
The relations~\eqref{eq:4D_MF_properties} are invariant under $\text{SO}(\text{D}-2)= \text{SO}(2)$ symmetry action on $i,j$ indices and SO$(1,1)$ symmetry action on light-like vectors as indicated by their sign indices $u_\mu^= = u_\mu^{--}$, $u_\mu^{\#} = u_\mu^{++}$. In \eqref{eq:p_mu} we have supplied the energy variable $\rho$ of \eqref{eq:P_a} with $\#$ index, indicating its transformations with respect to SO$(1,1)$ symmetry defined in such a way that~\eqref{eq:p_mu} remains invariant. Let us stress that, as $p_\mu = p_\mu(\tau)$ was a dynamical variable in the superparticle action \eqref{eq:action_BS}, such is the Lorentz group matrix valued $U_\mu^{(a)} = U_\mu^{(a)}(\tau)$ in~$\eqref{eq:p_mu}$. 

Substituting $\eqref{eq:p_mu}$, we can rewrite the Brink-Schwarz action \eqref{eq:action_BS} in the following equivalent form
\begin{equation}
    S_{\text{M}0} =  \int_{\mathcal{W}^1} \rho^\# u_\mu^= \Pi^\mu  = \int_{\mathcal{W}^1} \rho^\# \text{E}^=~,
    \label{eq:action_M0_MF}
\end{equation}
which corresponds to the action of massless superparticle in \textit{moving frame formalism}. In the second form of~\eqref{eq:action_M0_MF} we have introduced the notation
\begin{equation}
    \text{E}^==u_\mu^= \Pi^\mu
\end{equation}
which will be useful below.

For our purposes it is interesting to go one step further: the above analysis shows that the superparticle action in this moving frame formalism ``hides'' a more useful \textit{spinor moving frame formulation}.

\subsection{Moving frame and spinor moving frame for massless superparticle in D=4}
\label{sec:MF_and_SMF}
In $\text{D}=4$ the spinor moving frame is constructed from a pair of nonvanishing complex spinors $v^{\pm} = (\bar{v}_{\dot{\alpha}}^\mp)^*$ that satisfy the normalization conditions
\begin{equation}
    v^{\alpha -}v_\alpha^+=1\; ,  \qquad \bar{v}^{\dot{\alpha} -}\bar{v}_{\dot{\alpha}}^+=1\; .
    \label{eq:v-v+=1}
\end{equation}
These constraints ensure that the matrix formed by the columns $v_\alpha^+$, $v_\alpha^-$ (and its c.c. formed by $\bar{v}_{\dot{\alpha}}^+$, $\bar{v}_{\bar{\alpha}}^-$) is unimodular and thus an element of the SL$(2,{\mathbb C})$ group\footnote{The spinor frame variables are complexification of the
harmonic variables of ${\cal N}=2$ harmonic superspace approach to the off-shell description of the
${\cal N}=2$ SYM, matter and supergravity in terms of unconstrained superfields \cite{Galperin2, Galperin4, Galperin5}. Hence the name of Lorentz harmonics used in \cite{Igor_Russian}.}
\begin{equation}
    \epsilon^{\alpha\beta}v_\alpha^+v_\beta^-=1\qquad \Longleftrightarrow \qquad V_\alpha^{(\beta)}:=(v_\alpha^+, v_\alpha^-)\; \in\; \text{SL}(2,{\mathbb C})\; .
    \label{VinSL}
\end{equation}
This matrix is referred to as the \textbf{spinor (moving) frame matrix}, and it can be regarded as a kind of~~``square root'' of  a Lorentz vector frame, described by the Lorentz group valued \textit{moving frame matrix}~\eqref{eq:4D_MF_matrix}
\begin{equation*}
        U_\mu^{(a)}:= \left(\frac{u^{\#}_\mu+u^=_\mu}{2}, U^{i}_\mu, \frac{u^{\#}_\mu-u^=_\mu}{2}\right) \in \text{SO}(1,3)
\end{equation*}
in the following sense. The real light-like vectors $u^=_\mu$ and $u^\#_\mu$ of \eqref{eq:4D_MF_matrix} can be constructed from $v_\alpha^-,\bar{v}_{\dot{\alpha}}^-$ and $v_\alpha^+, \bar{v}_{\dot{\alpha}}^+$, respectively, by the Cartan-Penrose representation similar to~\eqref{eq:p_mu_lambda},
\begin{equation}
\label{u--=v-sv-}
u^=_\mu =    v^-\sigma_\mu \bar{v}^-\; , \qquad u^\#_\mu =     v^+\sigma_\mu \bar{v}^+\; . 
\end{equation}
Then, to proceed further, it is convenient to express also orthogonal and normalized space-like vectors $U^{(i)}_\mu = (U^1_\mu, U^2_\mu)$ in (\ref{eq:4D_MF_matrix}) in terms of complex light-like vectors
\begin{equation}
\label{u-+=v-sv+}
    u_\mu^{+-} = \frac{1}{2} (U^1_{\mu} + iU^2_{\mu})~, \qquad u_\mu^{-+} = \frac{1}{2} (U^1_{\mu} - iU^2_{\mu}) = (u_\mu^{+-})^*~.
\end{equation}
For these, the relation with spinor moving frame variables are given by
\begin{equation}
\begin{array}{ccc}
u^{+-}_\mu= v^{\alpha +} \sigma_{\mu \alpha \dot{\alpha}} \bar{v}^{\dot{\alpha} -}~, &~& u^{-+}_\mu= v^{\alpha -} \sigma_{\mu \alpha \dot{\alpha}} \bar{v}^{\dot{\alpha} +} = (u_\mu^{+-})^*~.
\end{array}
\end{equation}
It is easy to check that the vectors~\eqref{u--=v-sv-} and~\eqref{u-+=v-sv+} are light-like
\begin{equation}
    u_\mu^= u^{\mu =} = 0~, \qquad  u_\mu^\# u^{\mu \#} = 0~, \qquad u_\mu^{-+} u^{\mu -+} = 0~,\qquad u_\mu^{+-} u^{\mu +-} = 0~  
\end{equation}
and, as a result of~\eqref{eq:v-v+=1}, are normalized by
\begin{equation}
u^=_\mu u^{\mu\#}=2\; ,  \qquad u^{+-}_\mu u^{\mu\,-+}=- 2\;  .
\end{equation}
All other contraction of these vectors vanish, so that~\eqref{eq:U_term} and hence~\eqref{eq:4D_MF_matrix} hold
\begin{equation}
    \begin{array}{c}
       U_\mu^{(a)} \in \text{SO}(1,3)\qquad \Longleftrightarrow \qquad U_\mu^{(a)}U^{\mu(b)}=\eta^{ab}= \rm{diag} (+1,-1,-1,-1)\; ,  \\ 
       ~\\
       u^\#_\mu = U_\mu^{(0)} - U_\mu^{(3)}\; , \qquad u^=_\mu = U_\mu^{(0)} + U_\mu^{(3)}\; , \qquad u^{\mp\pm}_\mu = -U_\mu^{1} \pm  i U_\mu^{2}\; . \qquad
    \end{array}
\end{equation}
Actually, the light-like moving frame vectors form a complete light-like tetrad of the Newman-Penrose formalism \cite{Newman} (see also \cite{Newman_Igor}).

It is important that \eqref{u--=v-sv-} and \eqref{u-+=v-sv+} can be written in an equivalent matrix form relations (see~\eqref{eq:matrixRelations})
\begin{equation}
    \begin{array}{lcl}
       ~u_{\alpha \dot{\beta}}^= = u^=_\mu {\sigma}^\mu_{\alpha\dot{\beta}}= 2v_\alpha^- \bar{v}_{\dot{\beta}}^-~,  & ~ & ~~u_{\alpha \dot{\beta}}^\# = u^\#_\mu {\sigma}^\mu_{\alpha\dot{\beta}}= 2v_\alpha^+ \bar{v}_{\dot{\beta}}^+ ~,  \\
       ~\\
        u_{\alpha \dot{\beta}}^{-+} = u^{-+}_\mu {\sigma}^\mu_{\alpha\dot{\beta}}= 2v_\alpha^- \bar{v}_{\dot{\beta}}^+ ~, &~& ~u_{\alpha \dot{\beta}}^{+-} = u^{+-}_\mu {\sigma}^\mu_{\alpha\dot{\beta}}= 2v_\alpha^+ \bar{v}_{\dot{\beta}}^-= (u_{\beta\dot\alpha}^{-+})^*~.
        \label{eq:u_v_complete}
    \end{array}
\end{equation}
Substituting the expression from~\eqref{u--=v-sv-} for $u_\mu^=$ in~\eqref{eq:action_M0_MF}, we arrive at the massless superparticle action in its spinor moving frame formulation~\cite{Igor_Russian}
\begin{equation}
   S_{\text{M}0} = \int_{\mathcal{W}^1} \rho^\# u_\mu^= \Pi^\mu = \int_{\mathcal{W}^1} \rho^\# v^{\alpha -}\sigma_{\mu \alpha \dot{\alpha}} \bar{v}^{\dot{\alpha}-} \Pi^\mu~.
    \label{eq:action_M0_MF_SMF}
\end{equation}
It is not difficult to see that a simple change of bosonic spinor variables
\begin{equation}
    \lambda_\alpha = \sqrt{\rho^\#} v_\alpha^-~, \qquad \bar{\lambda}_{\dot{\alpha}} =  \sqrt{\rho^\#} \bar{v}_{\dot{\alpha}}^-~
    \label{eq:lambdas}
\end{equation}
converts~\eqref{eq:action_M0_MF_SMF} into the Ferber-Shirafuji action~\eqref{eq:action_FS},
\begin{equation*}
    S_{\text{FS}} = \int_{\mathcal{W}^1} \lambda^\alpha \sigma_{\mu \alpha \dot{\alpha}} \bar{\lambda}^{\dot{\alpha}} \Pi^\mu~.
\end{equation*}
At first glance, this reformulation of the Ferber-Schirafuji action~\eqref{eq:action_FS} might appear to be trivial: the only difference seems to be the inclusion of the Stückelberg field $\rho^\#$, which introduces a scale gauge symmetry. However, the spinor moving frame approach goes far beyond this. It is rather a method to deal with Ferber-Schirafuji-like formulation of superparticle, allowing for systematic generalization to higher dimensions, to extended objects such as $p$-branes, and for the construction of complete doubly supersymmetric actions for systems like the mM$0$ and mD$0$.

The group-theoretical structure beyond the spinor moving frame variables, in the present case including $v_\alpha^-$ and its partner $v_\alpha^+$ obeying $v^{\alpha-} v_\alpha^+ = 1$, is essential to this generalizations, and the gauge invariance under SO$(1,1)$ subgroup of SO$(1, \text{D}-1)= \text{SO}(1,3)$ implemented in~\eqref{eq:action_M0_MF_SMF} is essential to treat these as generalized homogeneous coordinates of celestial sphere $\mathbb{S}^{\text{D}-2} = \mathbb{S}^2$.

Notice also the possibility to supplement the action~\eqref{eq:action_M0_MF_SMF} by a kind of bosonic Wess-Zumino term constructed from the forms $v^{\alpha-}\text{d}v_\alpha^+$ and $\bar{v}^{\dot{\alpha}-}\text{d}\bar{v}_{\dot{\alpha}}^+$ which have the properties of connection under SO$(1,1) \otimes \text{SO}(2)$ gauge transformation~\cite{Igor_Russian}. These described the (super)helicity of massless superparticle at the (quasi-)classical level~\cite{Igor_Russian, Igor_2}.

\subsection[\texorpdfstring{Local fermionic $\kappa$}{K}-symmetry]{\boldmath\texorpdfstring{Local fermionic $\kappa$}{K}-symmetry}
\label{sec:kappa-symmetry}
To derive the equation of motions as well as to analyse the symmetries of a dynamical system, one studies the variation of the action with respect to all independent variables~\cite{variations}. While we will not delve into the general details here, we are particularly interested in a specific local fermionic symmetry characteristic for supersymmetric particles and extended objects. Let us consider the variations $\delta_\kappa$ obeying $\iota_\delta \Pi^\mu = \delta x^\mu -i \delta \theta \sigma^\mu \bar{\theta} +i \theta \sigma^\mu \delta \bar{\theta} = 0$,\footnote{$\iota_\delta \Pi^\mu$ corresponds to \eqref{eq:VA_4D} but with replacement $\text{d} \mapsto \delta$.} which leads to the following transformation of the bosonic coordinate function in terms of the fermionic ones,
\begin{equation}
    \iota_\kappa \Pi^\mu \equiv \iota_{\delta_\kappa} \Pi^\mu = 0 \qquad \Longrightarrow \qquad \delta_\kappa x^\mu = i\delta_\kappa \theta \sigma^\mu \bar{\theta} - i \theta \sigma^{\mu} \delta_\kappa \bar{\theta}~.
\end{equation}
Furthermore, imposing  $\iota_\kappa \Pi^\mu = 0$ in the variation of the action and requiring vanishing of this latter determines the complete set of $\delta_\kappa$ transformations for the remaining variables. The local fermionic gauge symmetry thus obtained is known as $\kappa$-symmetry\footnote{The $\kappa$-symmetry was discovered in~\cite{kappa_2} and~\cite{kappa_1}, and was shown to be equivalent with worldline supersymmetry in~\cite{localWlSUSY}.}.

Such a variation (with  $\iota_\kappa \Pi^\mu = 0$) of 4D Brink-Schwarz action \eqref{eq:action_BS} gives
\begin{equation}
\delta_\kappa S_{\text{BS}} = \int \text{d}\tau \left[ \delta_\kappa p_\mu \left(\Pi_\tau^\mu - ep^{\mu} \right) - \frac{1}{2}\delta_\kappa e p_\mu p^\mu - 2i p_\mu \partial_\tau \theta \sigma^\mu \delta_\kappa \bar{\theta} + 2i p_\mu \delta_\kappa \theta \sigma^\mu \partial_\tau \bar{\theta} \right]~.
\label{eq:4D_Svar}
\end{equation}
Studying~\eqref{eq:4D_Svar} one observes that, for $\delta_\kappa \theta^{\alpha} = \bar{\kappa}_{\dot{\alpha}} \bar{\sigma}^{\nu \dot{\alpha} \alpha} p_\nu =\bar{\kappa} \bar{\sigma}^\nu p_\nu$, the contribution from variation of coordinates functions can be compensated by some $\delta_\kappa e (\tau)$. This implies that~\eqref{eq:action_BS} is invariant under followed local fermionic $\kappa$-symmetry transformations
\begin{equation}
    \begin{array}{lclcr}
        \delta_\kappa x^\mu = i\delta_\kappa \theta \sigma^\mu \bar{\theta} - i \theta \sigma^{\mu} \delta_\kappa \bar{\theta}~, & ~ & \delta_\kappa \theta^\alpha = \bar{\kappa} \bar{\sigma}^\nu p_\nu~, & ~ &  \delta_\kappa \bar{\theta}^{\dot{\alpha}} = \kappa \sigma^\nu p_\nu~,  \\
        \delta_\kappa e = 4i( \bar{\kappa}\text{d}\bar{\theta} - \text{d}\theta \kappa)~, & ~ & \delta_\kappa p_\mu = 0~.
    \end{array}
    \label{eq:kappa1}
\end{equation}
This symmetry is important because it guarantees that the ground state of the system is invariant under $1/2$ of spacetime supersymmetry (see \cite{1/2_1, 1/2_2} and refs. therein) and a stable so-called BPS state. Moreover, the $\kappa$-symmetry can be used to remove half of the components of $\theta$ (and its c.c. $\bar{\theta}$) \cite{kappa_2} which equates the number of bosonic and fermionic degrees of freedom\footnote{The equality of degrees of freedom is a necessary condition for linearly realized supersymmetry in $d>1$, so that massless superparticle action can be written in (super)space of any (bosonic) dimension $\text{D}$. This balance however is necessary for superstring and higher $p$-branes.}.

To see this, one should observe that $\kappa$-symmetry of Brink-Schwarz action is \textit{infinitely reducible}. Indeed, with $p_\mu p^\mu = 0$ one can check that $\kappa_\alpha$ and $\kappa_\alpha + p_\mu\sigma^\mu_{\alpha \dot{\alpha}} \bar{\kappa}^{(1)\dot{\alpha}}$ (with an arbitrary $\bar{\kappa}^{(1){\dot{\alpha}}}= \bar{\kappa}^{(1)\dot{\alpha}}(\tau)$) generate the same $\kappa$-symmetry transformations~\eqref{eq:kappa1}. This is usually expressed by saying that $\delta_\kappa$ has a null vector $\bar{\kappa}^{(1)\dot{\alpha}}$ and hence that $\kappa$-symmetry is \textit{reducible}. But moreover, it  can be easily observed that $\bar{\kappa}^{(1)\dot{\alpha}}$ and $\bar{\kappa}^{(1)\dot{\alpha}} + p_\mu \bar{\sigma}^{a \dot{\alpha} \beta}\kappa^{(2)}_\beta$ (with an arbitrary $\kappa_\beta^{(2)}= \kappa_\beta^{(2)}(\tau)$) lead to the same variation of the parameter $\kappa_\alpha$, indicating a second order of reducibility for this $\kappa$-symmetry (existence of null-vector for null-vector). Furthermore, this process can be continued infinitely many times, the fact which is referred to as \textit{infinitely reducible} nature of the $\kappa$-symmetry of the Brink-Schwarz action for 4D massless superparticle.

As we mentioned above, $\kappa$-symmetry can be used to removed fermionic degrees of freedom. This happens exactly because of $\infty$-reducibility because the number of the components of $\theta^\alpha$, $\bar{\theta}^{\dot{\alpha}}$, and of the $\kappa$-symmetry parameter is the same. The same number of components have null-vector, null-vector for null-vector, and so on. So that, in this case, the number of the removed degrees of freedom is calculated as the infinite sum
\begin{equation}
    4-4+4- \ldots= 4 \cdot (1-1+1- \ldots) = 4 \cdot \lim_{q \to 1} (1-q+q^2-\ldots) = 4 \cdot \lim_{q \to 1} \frac{1}{1+q} =  4\cdot \dfrac{1}{2} = 2~,
\end{equation}
illustrating how only two of four fermionic degrees of freedom of massless superparticle are physically relevant. 

In contrast, the massless superparticle action in spinor moving frame formalism\footnote{The conclusions taking the Ferber-Shirafuji representation of~\eqref{eq:action_FS} are similar.}~\eqref{eq:action_M0_MF} has a manifestly \textit{irreducible} form of $\kappa$-symmetry
\begin{equation}
    \begin{array}{lclcl}
        \delta_\kappa x^\mu = i\delta_\kappa \theta \sigma^\mu \bar{\theta} - i \theta \sigma^{\mu} \delta_\kappa \bar{\theta}~, & ~ &
        \delta_{\kappa} \theta^{\alpha} = \epsilon^+ v^{{\alpha} -}~, &~& \delta_{\kappa} \bar{\theta}^{\dot{\alpha}} = \bar{\epsilon}^{+} \bar{v}^{\dot{\alpha} -}~, \\
        \delta_\kappa v^{\alpha \mp} = 0~, & ~ &   \delta_{\kappa} \rho^{\#} = 0~,
    \end{array}
    \label{eq:kappa_2}
\end{equation}
parametrised by $2$ fermionic parameters $\epsilon^{+}= \epsilon^{+}(\tau)$ and $\bar{\epsilon}^{+}= \bar{\epsilon}^{+}(\tau)$. Then each of these $\epsilon^{+}$, $\bar{\epsilon}^+$ can be used to reduce one half of the number of degrees of freedom of fermionic coordinate functions $\theta^\alpha(\tau)$ and $\bar{\theta}^{\dot{\alpha}}(\tau)$. This latter fact does not surprise us since the actions in Eqs.~\eqref{eq:action_BS} and ~\eqref{eq:action_M0_MF_SMF} are classically equivalent. In particular, it is straightforward to find the relation between parameters of $\bar{\kappa}^{\dot{\alpha}}$ of infinitely reducible $\kappa$-symmetry~\eqref{eq:kappa1} and $\epsilon^+$ irreducible $\kappa$-symmetry of the spinor moving frame formulation. Substituting \eqref{eq:p_mu} into (\ref{eq:kappa1}) we find (\ref{eq:kappa_2}) with 
\begin{equation}
    \epsilon^+ = 2\rho^\#\bar{v}^{\dot{\alpha}-} \bar{\kappa}_{\dot{\alpha}}~.
\end{equation}
This discussion demonstrates one of the advantages of spinor moving frame (or twistor-like) formalism: in its framework the $\kappa$-symmetry of superparticles and super-$p$-branes acquire an irreducible form~\cite{extraIgor_1, extraIgor_2, extraIgor_3, extraIgor_4}. This irreducibility makes it more suitable for quantization and reinforces the identification of $\kappa$-symmetry with local worldline/worldsheet/worldvolume supersymmetry of superembedding approach~\cite{origin_superembedding_1, superembedding_1, superembedding_2, superembedding_3} through the generalized action approach of~\cite{generalAction}.
 
Let us stress that this relation, as well as the transformation rules of the irreducible $\kappa$-symme\-try \eqref{eq:kappa2}, necessarily involves the constrained spinors $v^{\alpha-}$ and $\bar{v}^{\dot{\alpha}-}$. Thus the irreducible form of the $\kappa$-symmetry is a characteristic property of the spinor moving frame and other ``twistor-like'' approaches to superparticle dynamics.

In our case, the spinor moving frame variables are strictly necessary ingredients of the complete multiple D$0$-brane and mM$0$ actions which are not known in other formulations.

     \chapter[4D counterpart of multiple M-wave system]{4D counterpart of multiple M-wave system}\label{ch.4D_nAmW}
\thispagestyle{empty}

 \vspace*{-1.5cm}   
\begin{changemargin}{7.0cm}{0cm}
    \singlespacing\textcolor{cites}{    \small{  
         \begin{flushright}
         Journey before destination... It cannot be a journey if \\it doesn't have a beginning.
         \end{flushright}
   \begin{flushright}     
         {\sffamily {\textit{The Stormlight Archive III: Oathbringer}}\\{by {Brandon Sanderson}. }}
     \end{flushright}}}
    \end{changemargin}
    \vspace{12pt}

In this chapter we construct the action of $\text{D}=4$~~$\mathcal{N}=1$ non-Abelian multiwave (nAmW) system in flat superspace and prove its invariance under local worldline supersymmetry, which generalizes the so-called $\kappa$-symmetry of the massless superparticle. This $\text{D}=4$ $\mathcal{N}=1$ nAmW system, a lower dimensional counterpart of 11D multiple M$0$-brane (multiple M-wave or mM$0$) system, provides a toy model for developing the methods applicable to the study of its more complex higher-dimensional counterpart.

\section[\texorpdfstring{$\text{D}$}{D}=4 \texorpdfstring{$\mathcal{N}$}{N}=1 massless superparticle]{\boldmath\texorpdfstring{$\text{D}$}{D}=4 \boldmath\texorpdfstring{$\mathcal{N}$}{N}=1  massless superparticle}
\label{sec:4D_superparticle}
In this section we begin by coming back to the spinor moving frame formulation of the massless superparticle in $\text{D}=4$ $\mathcal{N}=1$ superspace since its action will be used to describe the center of energy sector of the complete nAmW system.  

Using~\eqref{eq:matrixRelations}, we can rewrite this action as
\begin{equation}
    S^{4\text{D}}_{0}= \int_{{\cal W}^1}\rho^{\#} v_\alpha^-\bar{v}_{\dot{\alpha}}^- \Pi^{\alpha\dot{\alpha}} = \int \text{d}\tau \rho^{\#}(\tau) v_\alpha^-(\tau) \bar{v}_{\dot{\alpha}}^-(\tau) \Pi_\tau ^{\alpha\dot{\alpha}}\;  ,
    \label{eq:S0D4=initial}
\end{equation}
where
\begin{equation}
\Pi^{\alpha\dot{\alpha}}=\text{d}\tau \Pi_\tau^{\alpha\dot{\alpha}}\; , \qquad
\Pi_\tau^{\alpha\dot{\alpha}}=\partial_\tau x^{\alpha\dot{\alpha}}(\tau)- 2i \partial_\tau \theta^{\alpha}(\tau)\bar{\theta}^{\dot{\alpha}}(\tau) +
 2i \theta^{\alpha}(\tau)\partial_\tau \bar{\theta}^{\dot{\alpha}}(\tau)\;  \qquad
 \label{Pi=dtPi}
\end{equation}
is the pull-back to the superparticle worldline of the VA 1-form
\begin{equation}
\Pi^{\alpha\dot{\alpha}}:= \Pi^\mu \tilde{\sigma}_\mu^{\dot{\alpha}\alpha}\; =\text{d}x^{\alpha\dot{\alpha}}(\tau)- 2i \text{d}\theta^{\alpha}(\tau)\bar{\theta}^{\dot{\alpha}} (\tau)+
 2i \theta^{\alpha}(\tau)\text{d}\bar{\theta}^{\dot{\alpha}}(\tau)~,
 \label{Pi=}
\end{equation}
$x^{\alpha\dot{\alpha}}(\tau)= x^{\mu}(\tau)\tilde{\sigma}_\mu^{\dot{\alpha}\alpha}$ and
$\theta^{\alpha}(\tau)=(\bar{\theta}^{\dot{\alpha}}(\tau))^* $ are bosonic vector and fermionic spinor coordinate functions of proper time $\tau$ which define embedding of the superparticle worldline ${\cal W}^1$ in $\text{D}=4$ $\mathcal{N}=1$ superspace $\Sigma^{(4|4)}$,
\begin{equation}
{\cal W}^1 \in \Sigma^{(4|4)}\; : \qquad x^{\mu}=x^{\mu}(\tau)\; ,\qquad \theta^{\alpha}=\theta^{\alpha}(\tau)\; ,\qquad  \bar{\theta}^{\dot{\alpha}}=\bar{\theta}^{\dot{\alpha}}(\tau)~.
\label{W1in}
\end{equation}
To simplify the notation, we use the same symbols for coordinate functions and coordinates (e.g. $x^\mu(\tau)$ and $x^\mu$) as well as for the pull-backs of differential forms to the worldline ${\cal W}^1$  and the forms  on target superspace $\Sigma^{(4|4)}$ (see e.g. \eqref{Pi=dtPi} and \eqref{Pi=}).

The bosonic spinor fields $v_\alpha^-$ in \eqref{eq:S0D4=initial} are constrained by \eqref{eq:v-v+=1} and define a spinor moving frame attached to the worldline.
Using~\eqref{u--=v-sv-} and~\eqref{eq:u_v_complete}, we can rewrite \eqref{eq:S0D4=initial} in the following equivalent forms
\begin{equation}
    S_0^{4\text{D}}= \int_{{\cal W}^1}\rho^{\#} \frac 1 2  u_{\alpha\dot{\alpha}}^= \Pi^{\alpha\dot{\alpha}} = \int_{{\cal W}^1}\rho^{\#}  u_{\mu}^= \Pi^{\mu} =\int_{{\cal W}^1} \rho^{\#} \text{E}^{=} \; ,
    \label{eq:S0D4==}
\end{equation}
where, at the last stage, we have introduced one of the pull-backs of the 1-forms of supervielbein adapted to the embedding of worldline in superspace, bosonic
\begin{equation}
    \text{E}^{=} =  \Pi^{\mu}u_{\mu}^= =  \frac 1 2  u_{\alpha\dot{\alpha}}^= \Pi^{\alpha\dot{\alpha}}=
 \Pi^{\alpha\dot{\alpha}}v_{\alpha}^-\bar{v}_{\dot{\alpha}}^-~,
    \label{E--=}
\end{equation}
\begin{equation}
    \text{E}^{\#} =    \Pi^{\mu}u_{\mu}^\# =  \frac 1 2  u_{\alpha\dot{\alpha}}^\# \Pi^{\alpha\dot{\alpha}} =
 \Pi^{\alpha\dot{\alpha}}v_{\alpha}^+\bar{v}_{\dot{\alpha}}^+~,
    \label{E++=}
\end{equation}
\begin{equation}
    \text{E}^{-+} = \Pi^{\mu}u_{\mu}^{-+} =  \frac 1 2  u_{\alpha\dot{\alpha}}^{-+} \Pi^{\alpha\dot{\alpha}} =
 \Pi^{\alpha\dot{\alpha}}v_{\alpha}^- \bar{v}_{\dot{\alpha}}^+ = ( \text{E}^{-+})^* \, , 
    \label{E+-=}
\end{equation}
and fermionic
\begin{equation}
     \text{E}^{\mp}= \text{d}\theta^\alpha v_\alpha^\mp \; , \qquad \bar{\text{E}}^{\mp}= \text{d}\bar{\theta}{}^{\dot\alpha} \bar{v}{}_{\dot\alpha}^\mp \; .
      \label{E-=}
\end{equation}
The first two real forms, $\text{E}^{=}$ and $\text{E}^{\#}$, will be used below to write the 4D nAmW action. 

The action~\eqref{eq:S0D4=initial} is manifestly invariant under $\text{D}=4$ ${\cal N}=1$ spacetime supersymmetry which leaves the spinor moving frame variables inert and acts on bosonic and fermionic coordinates functions by
\begin{equation}
\delta_\varepsilon x^{\alpha \dot{\alpha}} = 2i \theta^\alpha \bar{\varepsilon}^{\dot{\alpha}} - 2i \varepsilon^\alpha \bar{\theta}^{\dot{\alpha}}~, \qquad  \delta_\varepsilon \theta^\alpha = \varepsilon^\alpha~, \qquad \delta_\varepsilon \bar{\theta}^{\dot{\alpha}} = \bar{\varepsilon}^{\dot{\alpha}}~,
\label{eq:WLSUSY}
\end{equation}
with constant fermionic spinor parameter $\varepsilon^\alpha$ $=$ $(\bar{\varepsilon}^{\dot{\alpha}})^*$. It also possesses  the local fermionic  $\kappa$-symmetry
in its irreducible form (as it is described in section~\ref{sec:kappa-symmetry})
\begin{equation}
    \begin{array}{lclcl}
        \delta_\kappa x^\mu = i\delta_\kappa \theta \sigma^\mu \bar{\theta} - i \theta \sigma^{\mu} \delta_\kappa \bar{\theta}~, & ~ &
        \delta_{\kappa} \theta^{\alpha} = \kappa^+ v^{{\alpha} -}~, &~& \delta_{\kappa} \bar{\theta}^{\dot{\alpha}} = \bar{\kappa}^{+} \bar{v}^{\dot{\alpha} -}~, \\
        \delta_\kappa v_{\alpha}^{\mp} = 0~, & ~ &   \delta_{\kappa} \rho^{\#} = 0~.
    \end{array}
    \label{eq:kappa2}
\end{equation}
\subsection{Admissible variations of spinor moving frame variables and Cartan forms}
To vary the action and to clarify the structure of equations of motion, one must also vary and differentiate the moving frame and spinor moving frame variables. Although these variables obey algebraic constraints, the underlying group theoretic structure beyond them makes their differential calculus and variational problem quite simple. In fact, all differentials and variations of these fields can be expressed in terms of the SO$(1,3)$ valued Cartan forms and their variational counterparts.

As spinor frame variables are constrained by \eqref{eq:v-v+=1}, it is convenient to search for ``admissible'' variations which preserve these conditions. The same applies to the derivatives of spinor moving frame variables. The admissible derivatives preserving \eqref{eq:v-v+=1} can be expressed through the so-called Cartan forms of SO$(1,3)$ group. Since there are four complex variables in $v_{\alpha}^\pm$ and one complex condition in \eqref{eq:v-v+=1}, we can expect that there are three independent complex derivatives of spinor moving frame. These are related to three complex Cartan forms and their conjugates
\begin{equation}
    \Omega^{--}=v^{\alpha -}\text{d}v_\alpha^{-}\; , \qquad \bar{\Omega}{}^{--}=\bar{v}^{\dot{\alpha} -}\text{d}\bar{v}_{\dot{\alpha}}^{-}\; ,
    \label{eq:4D_Omega=}
\end{equation}
\begin{equation}
    \Omega^{++}=v^{\alpha +}\text{d}v_\alpha^{+}\; , \qquad \bar{\Omega}{}^{++}=\bar{v}^{\dot{\alpha} +}\text{d}\bar{v}_{\dot{\alpha}}^{+}\; ,
    \label{eq:4D_Omega++}
\end{equation}
\begin{equation}
    ~\omega^{(0)}=v^{\alpha -}\text{d}v_\alpha^{+}\; ,~~\qquad \bar{\omega}{}^{(0)}=\bar{v}^{\dot{\alpha} -}\text{d}\bar{v}_{\dot{\alpha}}^{+}\;.
    \label{eq:4D_Omega0}
\end{equation}
Indeed, using the consequences
\begin{equation}
    \delta_\alpha^{~~\beta} = v_\alpha^+ v^{\beta -} - v_\alpha^- v^{\beta +}~, \qquad \delta_{\dot{\alpha}}^{~~\dot{\beta}} = \bar{v}_{\dot{\alpha}}^+ \bar{v}^{\dot{\beta}-} - \bar{v}_{\dot{\alpha}}^- \bar{v}^{\dot{\beta}+}~, 
\end{equation}
of the constraints~\eqref{eq:v-v+=1}, we can express the derivatives of spinor frame variables i.e. $\text{d}v_\alpha^{(\beta)}$ and $\text{d}\bar{v}_{\dot{\alpha}}^{(\dot{\beta})}$ in terms of Cartan forms by
\begin{equation}
    \text{d}v_\alpha^-= -~\omega^{(0)}v_\alpha^{-}+\Omega^{--}v_\alpha^{+} \qquad \Longleftrightarrow \qquad  \text{D}v_\alpha^-:= \text{d}v_\alpha^-+\omega^{(0)}v_\alpha^{-}= \Omega^{--}v_\alpha^{+}\; ,
    \label{eq:4D_Dv-=}
\end{equation}
\begin{equation}
     ~~~~\text{d}v_\alpha^+= \omega^{(0)}v_\alpha^{+} - \Omega^{++}v_\alpha^{-} ~~~~~\qquad \Longleftrightarrow  \qquad  \text{D}v_\alpha^+:= \text{d}v_\alpha^+ -\omega^{(0)}v_\alpha^{+}= -\Omega^{++}v_\alpha^{-}\; ,
    \label{eq:4D_Dv+=}
\end{equation}
and their c.c. relations.

The expression for admissible variations
\begin{equation}
\begin{array}{lcl}
     \delta v_\alpha^-= -\iota_\delta \omega^{(0)}v_\alpha^{-}+\iota_\delta \Omega^{--}v_\alpha^{+}  & ~~  \Longleftrightarrow ~~ &  \iota_\delta \text{D}v_\alpha^-:= \delta v_\alpha^-+\iota_\delta\omega^{(0)}v_\alpha^{-}=\iota_\delta \Omega^{--}v_\alpha^{+}\; ,
\end{array} \label{eq:4D_iDv-=}
\end{equation}
\begin{equation}
    \begin{array}{lcl}
    \delta v_\alpha^+= \iota_\delta \omega^{(0)}v_\alpha^{+}-\iota_\delta \Omega^{++}v_\alpha^{-} &  ~~  \Longleftrightarrow ~~ & \iota_\delta \text{D}v_\alpha^+:= \delta v_\alpha^+ -\iota_\delta\omega^{(0)}v_\alpha^{+}= -\iota_\delta\Omega^{++}v_\alpha^{-}\; ,
\end{array}  \label{eq:4D_iDv+=}
\end{equation}
can be obtained from \eqref{eq:4D_Dv-=} and \eqref{eq:4D_Dv+=} by formal contraction with variation symbol $\iota_\delta$ as it is described in section~\ref{sec:diff_SuperForms}. Note that in the case of 1-forms this is essentially the substitution $\text{d}\mapsto \delta$.

In Eqs. \eqref{eq:4D_Dv-=} and \eqref{eq:4D_Dv+=} we also define covariant derivatives \text{D} in which the 1-form $\omega^{(0)}$ and its c.c. $\bar{\omega}^{(0)}$ play the role of a $\text{SO}(1,1) \otimes\text{U}(1)$ connection. These gauge transformations can be parametrized by $\iota_\delta \omega^{(0)}$ and its complex conjugate $\iota_\delta \bar{\omega}{}^{(0)}$, as they appear in the expressions for spinor frame variations \eqref{eq:4D_iDv-=} and \eqref{eq:4D_iDv+=}. These symmetries correspond to the gauge invariance of the massless superparticle action \eqref{eq:S0D4=initial} (or \eqref{eq:S0D4==}) provided it is supplemented by the following $\text{SO}(1,1)$ scaling of the Lagrange multiplier $\rho^{\#}$,
\begin{equation}
    \delta \rho^{\#}=  (\iota_\delta \omega^{(0)}+ \iota_\delta \bar{\omega}{}^{(0)})\rho^{\#} \;.
    \label{eq:4D_vr++=SO11}
\end{equation}
The derivatives $\text{D}$ in~\eqref{eq:4D_Dv-=} and \eqref{eq:4D_Dv+=} are covariant derivatives with respect to this gauge symmetry. Moreover, one can observe that the action \eqref{eq:S0D4==} (or \eqref{eq:S0D4=initial}) involves only $u_\mu^=$ (or $v_{_\alpha}^-$ and $\bar{v}_{\dot{\alpha}}^-$) variables. Noting that the parameters $\iota_\delta \Omega^{++}$ and $\iota_\delta \bar{\Omega}{}^{++}$, present in \eqref{eq:4D_iDv+=}, do not enter into the variation of  $u_\mu^=$ nor $v_{_\alpha}^-$, we conclude that there is an additional gauge symmetry of the system: the so-called $\mathbb{K}_2$-symmetry
\bea
{\mathbb K}_2\, :\qquad  & \delta_{\mathbb{K}_2}v_\alpha^-=0\; , \qquad \delta_{\mathbb{K}_2}v_\alpha^+=-\iota_\delta \Omega^{++} v_\alpha^- \; , & \qquad \nonumber \\
 & \delta_{\mathbb{K}_2}\bar{v}_{\dot{\alpha}}^-=0\; , \qquad \delta_{\mathbb{K}_2}\bar{v}_{\dot{\alpha}}^+=-\iota_\delta \bar{\Omega}{}^{++} \bar{v}_{\dot{\alpha}}^- \; . & \qquad \nonumber
\eea
Together the above listed gauge symmetries $[\text{SO}(1,1)\otimes \text{U}(1)]\subset \!\!\!\!\!\!\times {\mathbb K}_2$, can be used as identification relations on the set of constrained pair $\{v_\alpha^-, v_\alpha^+ \}$ which thus can be considered as a kind of homogeneous coordinate for the coset of the $\text{Spin}(1,3)= \text{SL}(2,{\mathbb C})$ group, which is isomorphic to the ${\mathbb S}^2$ sphere \cite{Delduc, Galperin_1}
\be
\{v_\alpha^\mp \} \; = \;\frac {\text{SL}(2,{\mathbb C})} {[\text{SO}(1,1)\otimes \text{U}
(1)]\subset \!\!\!\!\!\!\times {\mathbb K}_2}= {\mathbb S}^2\; .
\ee
This can be recognized as the celestial sphere of an observer in a 4-dimensional spacetime. Hence the introduction of the Lagrange multiplier $\rho^\#$, which has a St\"uckelberg-type transformation \eqref{eq:4D_vr++=SO11} under the $\text{SO}(1,1)$ gauge symmetry (i.e. it can be gauged to unity), makes the constrained bosonic spinors a kind of homogeneous coordinates on the celestial 2-sphere.

\subsection{Equations of motion and induced supergravity on the worldline}
The equations of motion derived from the action \eqref{eq:S0D4=initial} include (see Eqs. \eqref{E--=0}-\eqref{E-=0} in Appendix~\ref{sec:eom4DN10} for the complete list of equations)
\be\label{E==0}
  \text{E}^==0~, \qquad \text{E}^{+-}=\text{E}^{-+}=0\; , \qquad \text{E}^-=\bar{\text{E}}^{-}=0\; . \qquad
\ee
Thus, only the pull-backs of one bosonic and two fermionic projections of the target superspace supervielbein remain nonzero on the mass shell. These forms
\be\label{E++=sugra}
\text{E}^{\#}= \Pi^{\alpha\dot{\alpha}} v_\alpha^+ \bar{v}{}_{\dot\alpha}^+= \text{d}\tau \text{E}_\tau^{\#}\; , \qquad \text{E}^{+}= \text{d}\theta^\alpha v_\alpha^+ = \text{d}\tau \text{E}_\tau^{+}\; , \qquad\bar{\text{E}}^{+}= \text{d}\bar{\theta}{}^{\dot\alpha}\bar{v}{}_{\dot\alpha}^+ = \text{d}\tau \bar{\text{E}}_\tau^{+}\; \;  \qquad
\ee
describe 1d $\mathcal{N}=2$ supergravity induced on the worldline by its embedding into $\text{D}=4$ ${\cal N}=1$ target superspace. Indeed, under the local fermionic $\kappa$-symmetry \eqref{eq:kappa2} these 1-forms transform as
\be\label{wsSG4D}
\delta_\kappa \text{E}^\# =-4i(\text{E}^+\bar{\kappa}{}^+ - \kappa^+\bar{\text{E}}{}^+)\; , \qquad \delta_\kappa \text{E}^+=\text{D} \kappa^+\; , \qquad \delta_\kappa \bar{\text{E}}{}^+=\text{D}\bar{\kappa}{}^+\, , \qquad
\ee
and its $\tau$-components by
\be\label{wsSG4D=}
\delta_\kappa \text{E}_\tau^\# =-4i(\text{E}^+\bar{\kappa}{}^+ - \kappa^+\bar{\text{E}}{}^+)\; , \qquad \delta_\kappa \text{E}_\tau^+= \text{D}_\tau \kappa^+\; , \qquad \delta_\kappa \bar{\text{E}}{}_\tau^+= \text{D}_\tau\bar{\kappa}{}^+\, , \qquad
\ee
which are precisely the transformation rules of a $\text{1d}~~\mathcal{N}=2$ supergravity supermultiplet. We refer to this as supergravity induced by embedding, because these supergravity 1-forms are built from the embedding coordinates $x^\mu(\tau)$, $\theta^\alpha(\tau)$, $\bar{\theta}^{\dot{\alpha}}(\tau) $ and spinor moving frame variables. 

This induced supergravity provides the natural tool to couple a 1-dimensional supersymmetric matter multiplets to massless superparticle while preserving its local fermionic $\kappa$-symmetry. In particular, the action for 4D nAmW system, the counterpart of 11D mM$0$ action from \cite{Bandos_11D_mM0}, is constructed by coupling of 3D $\mathcal{N}=2$ $\text{SU}(N)$ SYM multiplet dimensionally reduced to $\text{d}=1$ to this induced 1d supergravity.  We shall present that action and exhibit its full local worldline supersymmetry in the next section.

\section[\texorpdfstring{$\text{D}$}{D}=4 \texorpdfstring{${\cal N}$}{N}=1 non-Abelian multiwave action and its worldline supersymmetry]{\boldmath\texorpdfstring{$\text{D}$}{D}=4 \texorpdfstring{${\cal N}$}{N}=1 non-Abelian multiwave action and its worldline supersymmetry}
The action of 4D non-Abelian multiwave (nAmW) system, a lower dimensional counterpart of the 11D mM$0$ action, could in principle can be obtained from this latter by dimensional reduction. In practice, however, the complicated structure of the 11D moving frame variables makes more easy to construct it directly using symmetry principles (see  \cite{Bandos_11D_mM0} for $\text{D}=11$ and  \cite{Bandos_3D_mM0} for $\text{D}=3$). On this way we arrive at
\begin{equation}
    \begin{array}{l}
         \begin{split}
              S^{4\text{D}}_{\rm nAmW} &= \int_{{\cal W}^1} \rho^{\#}\text{E}^{=} + {1\over \mu^6} \int_{{\cal W}^1} (\rho^{\#})^3
              \text{tr}\left(\bar{\widetilde{{\mathbb P}}}{\rm D} {\widetilde{\mathbb Z}} + {\widetilde{{\mathbb P}}}{\rm D} \bar{\widetilde{\mathbb Z}} - {i\over 8} {\rm D}{{\widetilde{\Psi}}}\,  \bar{{\widetilde{\Psi}}} + {i\over 8} {{\widetilde{\Psi}}} {\rm D} \bar{{\widetilde{\Psi}}}  \right) + \\
              &~\\
              & + {1\over \mu^6} \int_{{\cal W}^1} (\rho^{\#})^3 \left[ \text{E}^{\#} \widetilde{{\cal H}} + i \text{E}^{+}
              \text{tr}(\bar{{\widetilde{\Psi}}}  \bar{\widetilde{{\mathbb P}}}+  {\widetilde{\Psi}} [{\widetilde{\mathbb Z}},  \bar{\widetilde{\mathbb Z}}])+ i \bar{\text{E}}^{+}
              \text{tr}({{\widetilde{\Psi}}}  {\widetilde{{\mathbb P}}}+  \bar{{\widetilde{\Psi}}} [{\widetilde{\mathbb Z}},  \bar{\widetilde{\mathbb Z}}])\right] ,
         \end{split}
    \end{array}\label{eq:SmM0=4D}
\end{equation}
where
\begin{eqnarray}
\label{HSYM=4D} && \widetilde{{\cal H}}=   \text{tr}\left( {\widetilde{{\mathbb P}}} \bar{\widetilde{{\mathbb P}}} +  [{\widetilde{\mathbb Z}},  \bar{\widetilde{\mathbb Z}}]^2 -
{i\over 2} {\widetilde{\mathbb Z}}{ {\widetilde{\Psi}}}{ {\widetilde{\Psi}}} + {i\over 2} \bar{\widetilde{\mathbb Z}} \bar{{\widetilde{\Psi}}}  \bar{{\widetilde{\Psi}}} \right) \; .
\qquad
\end{eqnarray}
The first term of \eqref{eq:SmM0=4D} is exactly the $\text{D}=4$ massless superparticle action \eqref{eq:S0D4==}, now interpreted as the center of energy motion of the interacting multiwave system. The remaining part of the nAmW action, proportional to the dimensional parameter $1 /{\mu^{6}}$~\footnote{Notice that $\mu$ has dimension of mass in our physical system of units with  $c=\hbar =1$. Then  $[\mu^{-6}]=M^{-6}=L^6$.} contains bosonic and fermionic matrix fields. These are traceless $N\times N$ bosonic 
\begin{eqnarray}
\label{bZ=bZ0++} & {\widetilde{\mathbb Z}}= \widetilde{\mathbb Z}_{0|\#}:=\widetilde{\mathbb Z}_{0|+2}\, , \qquad &  \bar{\widetilde{\mathbb Z}}= \bar{\widetilde{\mathbb Z}}_{\#|0}:=
\bar{\widetilde{\mathbb Z}}_{+2|0}= ({\widetilde{\mathbb Z}})^\dagger \, , \qquad  \\
\label{bP=bP++3}
& \widetilde{{\mathbb P}}= \widetilde{{\mathbb P}}_{+|\#+} := \widetilde{{\mathbb P}}_{+|+3}\, ,\qquad  & \bar{\widetilde{{\mathbb P}}}= \bar{\widetilde{{\mathbb P}}}_{\#+|+} :=\bar{\widetilde{{\mathbb P}}}_{+3|+}= (\widetilde{{\mathbb P}})^\dagger\, ,\qquad  \end{eqnarray}
 and fermionic 
\begin{eqnarray}\label{bPsi=bPsi+2+1} &   { {\widetilde{\Psi}}}=
{\widetilde{\Psi}}_{\# |+}:= {\widetilde{\Psi}}_{+2 |+}\,  , \qquad & \bar{{ {\widetilde{\Psi}}}}=
\bar{{\widetilde{\Psi}}}_{+|\#}:= \bar{{\widetilde{\Psi}}}_{+|+2}=  ({\widetilde{\Psi}} )^\dagger\,   
 \end{eqnarray}
 matrix fields\footnote{The matrix field symbols in the nAmW model are covered with tildes here to distinguish them from the matrix fields of the 3D mD$0$ model, which are inert under $\text{SO}(1,1)$ but carry $\text{U}(1)$ charges (see Eqs.~\eqref{bZ0++->bZ}-\eqref{bPsi=bPsi+2+}). One might wonder why we do not apply this redefinition already at this stage, without explicitly performing dimensional reduction. The reason is that such a redefinition would introduce derivatives of the Stückelberg field directly into the action, resulting in a formulation that we find less convenient for our purposes.} all transforming under the gauge group SU$(N)$. Here the sign subindices (equivalent to the opposite sign superindices, e.g. ${\widetilde{\Psi}}_{\# |+}={\widetilde{\Psi}}^{= |-}$) indicate the transformation properties of the matrix fields under the $\text{GL}(1,{\mathbb C})= \text{SO}(1,1)\otimes \text{U}(1)$ transformations acting on the spinor frame variables $v_\alpha^\mp=v_\alpha^{(\mp |0)}$ and  $\bar{v}_{\dot\alpha}{}^\mp=\bar{v}_{\dot\alpha}{}^{(0|\mp )}$ as described in sec.~\ref{sec:4D_superparticle} above. These latter enter, besides the first term, also in the remaining part of the action~\eqref{eq:SmM0=4D} through the induced supergravity  1-forms: 1d supervielbein $\text{E}^\#$, $\text{E}^+,\bar{\text{E}}^+$ \eqref{E++=sugra} and  Cartan forms $\omega^{(0)}$, $\bar{\omega}{}^{(0)}$ \eqref{eq:4D_Omega0} which enter the covariant derivative
 \begin{equation}
       {\rm D} {\widetilde{\mathbb Z}}=  {\rm d}{\widetilde{\mathbb Z}} + 2\bar{\omega}{}^{(0)} {\widetilde{\mathbb Z}} + [\, {\mathbb A}\, , \, {\widetilde{\mathbb Z}} \,] \; ,  \qquad {\rm D} \bar{\widetilde{\mathbb Z}}=  {\rm d}\bar{\widetilde{\mathbb Z}} + 2{\omega}{}^{(0)} \bar{\widetilde{\mathbb Z}} + [\, {\mathbb A}\, , \, \bar{\widetilde{\mathbb Z}} \,] \; , 
     \label{eq:4D_DZ:=} 
\end{equation}
\begin{equation}
{\rm D}{{\widetilde{\Psi}}}={\rm d}{{\widetilde{\Psi}}}+ (2\omega^{(0)}+\bar{\omega}{}^{(0)}) {{\widetilde{\Psi}}} + [\, {\mathbb A}\, , \, {{\widetilde{\Psi}}} \,]  \;  ,  \qquad  {\rm D}
\bar{{\widetilde{\Psi}}}={\rm d}\bar{{\widetilde{\Psi}}}+ (\omega^{(0)}+2\bar{\omega}{}^{(0)}) \bar{{\widetilde{\Psi}}} + [\, {\mathbb A}\, , \, \bar{{\widetilde{\Psi}}} \,] 
     \label{eq:4D_DPsi:=}
 \end{equation}
 as connection for the  $\text{GL}(1,{\mathbb C})= \text{SO}(1,1)\otimes \text{U}(1)$ transformations. The $\text{SU}(N)$ connection 1-form $ {\mathbb A}={\rm d}\tau {\mathbb A}_\tau$, with  the traceless antihermitian $N\times N$ matrix field ${\mathbb A}_\tau$, enters into the action \eqref{eq:SmM0=4D} inside the covariant derivatives \eqref{eq:4D_DZ:=} and \eqref{eq:4D_DPsi:=}.

The dimension of the matrix matter fields are
\be
{}[ {\widetilde{\mathbb Z}}]=M =[ \bar{\widetilde{\mathbb Z}}] \; , \qquad {}[ \bar{\widetilde{\mathbb P}}]=M^2 =[ {\widetilde{\mathbb P}}] \; , \qquad [ {{\widetilde{\Psi}}}]=M^{3/2}= [ \bar{{\widetilde{\Psi}}}]\; .
\ee
The choice of such non-canonical dimension allows us to reduce the dependence on the dimensional parameter $\mu$, with dimension of mass $[\mu]=M$,  to an overall multiplier in front of the part of the action including the matrix fields.

The action \eqref{eq:SmM0=4D} is invariant under the local worldline supersymmetry transformations which act on the center of energy fields as a bit deformed version of the $\kappa$-symmetry of massless superparticle
\begin{equation}
    \begin{array}{l}
       \delta_\epsilon x^{\alpha\dot{\alpha}}=  2i \delta_\epsilon\theta^\alpha \bar{\theta}{}^{\dot{\alpha}} - 2i\theta^\alpha  \delta_\epsilon\bar{\theta}{}^{\dot{\alpha}}   + v^{\alpha +} \bar{v}{}^{\dot{\alpha}+ }
       \iota_\epsilon \text{E}^{=} \; ,  \\ 
       ~~~\delta_\epsilon\theta^\alpha =\epsilon^+ v^{\alpha -} \; , \qquad   ~~~\delta_\epsilon\bar{\theta}{}^{\dot{\alpha}}= \bar{\epsilon}^{+}  \bar{v} {}^{\dot{\alpha} -} \; , \\
       ~~\delta_\epsilon v_\alpha^\mp = 0 \; ,  ~~~~~\qquad \qquad \delta_\epsilon \bar{v}_{\dot\alpha}^\mp=0 \; , \qquad \delta_\epsilon \rho^\#=0 \; ,
    \end{array}\label{eq:4D_w-sh=susyX}
\end{equation}
where
\be\label{eq:4D_ieE--==}
\iota_\epsilon \text{E}^{=} =
  \dfrac{3i}{2\mu^6 }  (\rho^{\#})^2   \,
 {\rm tr}\left( \epsilon^{+}\bar{{\widetilde{\Psi}}} \bar{{\widetilde{{\mathbb P}}}}  +  \bar{\epsilon}{}^{+}{\widetilde{\Psi}} {\widetilde{{\mathbb P}}} -
 (\epsilon^{+}{\widetilde{\Psi}} + \bar{\epsilon}{}^{+}\bar{{\widetilde{\Psi}}} )[{\widetilde{\mathbb Z}}, \bar{\widetilde{\mathbb Z}}]
  \right)\; .
\ee
The matrix fields themselves are transformed by
\begin{equation}
    \begin{array}{lcl}
       \delta_\epsilon {\widetilde{\mathbb Z}}   = -i \epsilon^{+} \bar{{\widetilde{\Psi}}} \; ,&~&    
       ~~\delta_\epsilon {\widetilde{\mathbb Z}}   = -i \epsilon^{+} \bar{{\widetilde{\Psi}}} \; ,    
    \end{array}\label{susy-Z} 
\end{equation}
\begin{equation}
    \begin{array}{lcl}
       ~~~~\delta_\epsilon {\widetilde{{\mathbb P}}}   =   i \epsilon^{+} [{{\widetilde{\Psi}}}, {\widetilde{\mathbb Z}}]+i  \bar{\epsilon}{}^{+} [\bar{{\widetilde{\Psi}}}, {\widetilde{\mathbb Z}}]  \; ,&~&    
      ~~\delta_\epsilon \bar{\widetilde{{\mathbb P}}}   = - i \epsilon^{+} [{{\widetilde{\Psi}}}, \bar{\widetilde{\mathbb Z}}]-i  \bar{\epsilon}{}^{+} [\bar{{\widetilde{\Psi}}}, \bar{\widetilde{\mathbb Z}}]  \; ,    
    \end{array}\label{susy-P} 
\end{equation}
\begin{equation}
    \begin{array}{lcl}
        \delta_\epsilon {\widetilde{\Psi}} =   4\epsilon^{+} \bar{\widetilde{{\mathbb P}}} + 4 \bar{\epsilon}{}^{+} [{\widetilde{\mathbb Z}}, \bar{\widetilde{\mathbb Z}}] \; ,&~&    
        \delta_\epsilon \bar{{\widetilde{\Psi}}} =  4\bar{\epsilon}{}^{+} {\widetilde{{\mathbb P}}} + 4 {\epsilon}{}^{+} [{\widetilde{\mathbb Z}}, \bar{\widetilde{\mathbb Z}}] \; ,   
    \end{array}\label{susy-Psi} 
\end{equation}
\begin{equation}\label{susy-A}
     \delta_\epsilon {\mathbb A} = i\text{E}^{\#} (\epsilon^{+}  {\widetilde{\Psi}} -
 \bar{\epsilon}{}^{+}  \bar{{\widetilde{\Psi}}} )   +8i \text{E}^{+}
 \epsilon^{+}\bar{\widetilde{\mathbb Z}}  -8i \bar{\text{E}}{}^{+}
 \bar{\epsilon}{}^{+}\;    {\widetilde{\mathbb Z}}\; .
\end{equation}
In the next chapter, we perform the dimensional reduction of the 4D nAmW action to $\text{D}=3$. The resulting theory provides a candidate for the 3-dimensional counterpart of the 10D multiple D$0$-brane system; henceforth referred to as the 3D mD$0$-brane.

     \chapter[Nonlinear 3D mD0 action by dimensional reduction of the 4D nAmW system]{Nonlinear 3D mD0 action by dimensional reduction of the 4D nAmW system}\label{ch.3D_mD0}
\thispagestyle{empty}

 \vspace*{-1.5cm}   
\begin{changemargin}{4cm}{0cm}
    \singlespacing\textcolor{cites}{ \small{
         \begin{flushright}
        Always something new, always something I didn't expect,\\ and sometimes it isn't horrible.
         \end{flushright}
     \begin{flushright}     
         {\sffamily {\textit{The Wheel of Time II: The Great Hunt}}\\
         {by {Robert Jordan}. }}
     \end{flushright}}}
    \end{changemargin}
    \vspace{12pt}

In this section, we perform the dimensional reduction of $\text{D}=4$ nAmW action, which serves as a counterpart of the 11D mM$0$-brane system, down to $\text{D}=3$ and obtain in this way an action with the properties expected for 3D counterpart of 10D mD$0$ system. Furthermore, we will show that actually this procedure produces just one representative of a family of 3D actions possessing these properties. All members of the family contain in their actions some positive definite function $\mathcal{M}(\mathcal{H})$, whose meaning will be explained later. The dimensional reduction produces the action with particular $\mathcal{M}(\mathcal{H})$ given in Eq.~\eqref{cM=m+} below.

To proceed, it is convenient to first derive the action for a massive $\text{D}=3$ $\mathcal{N}=2$ superparticle by performing the dimensional reduction of the $\text{D}=4$ massless superparticle action~\eqref{eq:S0D4==}.

\section{Notation and conventions from 4D to 3D}
\label{sec:4Dto3D_conventions}
As it was already stated, $\text{D}=4$ Weyl spinor indices are denoted by dotted and undotted symbols taken from the beginning of the Greek alphabet (see section~\ref{sec:diff_SuperForms}). Symbols from the middle of the Greek alphabet are used for 4-vector indices when they are untilded, and for 3-vector indices when they carry a tilde.

Thus, the bosonic vector and fermionic spinor coordinates of $\text{D}=4$~~${\mathcal N}=1$ superspace are denoted by $z^M$ (see Eq.~\eqref{eq:superspace4D}),
while the coordinates of the $\text{D}=3$ ${\mathcal N}=2$ superspace are \begin{equation} z^{\tilde{M}}= (x^{\tilde{\mu}}, \theta^\alpha, \bar{\theta}{}^{\alpha})~ , \qquad \tilde{\mu}=0,1,2~ , \qquad \alpha=1,2~ . 
\end{equation}
 The Weyl spinor indices are raised and lowered using Levi-Civita symbols
\be
\epsilon^{\alpha\beta} = i\sigma_2= \left(\begin{matrix}0 & 1 \cr
  -1 &0 \end{matrix}\right)= -\epsilon_{\alpha\beta}=
    \epsilon^{\dot{\alpha}{\dot\beta}}= -\epsilon_{\dot{\alpha}{\dot\beta}}~.
\ee
For instance,
\be
\theta^\alpha = \epsilon^{\alpha\beta}\theta_\beta \; , \qquad \theta_\alpha = \epsilon_{\alpha\beta}\theta^\beta \; , \qquad
\ee
which similarly applies to $\text{D}=3$ spinors, and
\be
\bar{\theta}{}^{\dot\alpha} = \epsilon^{\dot{\alpha}\dot{\beta}}\bar{\theta}{}_{\dot{\beta}} \; , \qquad \bar{\theta}{}_{\dot{\alpha}} = \epsilon_{\dot{\alpha}\dot{\beta}}\bar{\theta}{}^{\dot{\beta}} \; .  \qquad
\ee
We use the following representation for the relativistic Pauli matrices (rPMs)
\begin{equation}\label{Pauli=}
    \begin{array}{l}
    \sigma_{\mu\,\alpha\dot{\beta}}  = \left\lbrace {\mathbb 1}, \sigma_1, \sigma_2, \sigma_3 \right\rbrace = \left\lbrace
    \left(\begin{matrix} 1 & 0 \cr
    0 & 1 \end{matrix}
    \right) , \left(
    \begin{matrix} 0 & 1 \cr
    1 & 0 \end{matrix}
    \right), \left(
    \begin{matrix} 0 & -i \cr
    i & \; 0 \end{matrix}
    \right), \left(
    \begin{matrix} 1 & \; 0 \cr
    0 & -1 \end{matrix}
    \right) \right\rbrace   =\\
    ~~\\
    ~~~~~~~~~~~~~= \tilde{\sigma}^{\mu\,\dot{\alpha}{\beta}}  :=
    \epsilon^{\dot{\alpha}\dot{\beta}}\epsilon^{{\beta}{\alpha}}{\sigma}_{\nu\,{\alpha}\dot{\beta}}\eta^{\nu\mu}~,
    \end{array}
\end{equation}
which obey Eqs.~\eqref{eq:sigmaRelations} and \eqref{eq:matrixRelations}.
Among the $\text{D}=4$ rPMs~\eqref{Pauli=}, only $\sigma_2$ is complex and antisymmetric. This allows us to identify the $\text{D}=3$ gamma matrices, which admit a real symmetric representation (after raising or lowering indices using the charge conjugation matrix) with the rPMs carrying indices $0$, $1$, and $3$
\bea\label{s4d=s3d}
\sigma_{\mu\,\alpha\dot{\beta}}= \left(\sigma_{\tilde{\mu}\,\alpha\dot{\beta}}, \sigma_{2\,\alpha\dot{\beta}}\right) \; , \qquad  \tilde{\sigma}_{\mu}^{\dot{\beta}\alpha}= \left(\tilde{\sigma}_{\tilde{\mu}}^{\dot{\beta}\alpha}, \tilde{\sigma}_{2}^{\dot{\beta}\alpha}\right) \, ,\\ 
\nonumber~~\\
\label{g3=s4}\gamma_{\tilde{\mu}\,\alpha{\beta}}=\sigma_{\tilde{\mu}\,\alpha\dot{\beta}}=  \left\lbrace {\mathbb 1}, \sigma_1,  \sigma_3 \right\rbrace = \left\lbrace \left(
\begin{matrix} 1 & 0 \cr
0 & 1 \end{matrix}
\right) , \left(
\begin{matrix} 0 & 1 \cr
1 & 0 \end{matrix}
\right),  \left(
\begin{matrix} 1 & \; 0 \cr
0 & -1 \end{matrix} \right)
\right\rbrace \; , \\
\nonumber~~\\
\label{tg3=ts4} \tilde{\gamma}_{\tilde{\mu}}^{{\beta}\alpha}=\tilde{\sigma}_{\tilde{\mu}}^{\dot{\beta}\alpha}= ({\mathbb 1}, -\sigma_1,  -\sigma_3) = \epsilon^{{\alpha}\gamma}\epsilon^{{\beta}\delta}\gamma_{\tilde{\mu}\,\gamma{\delta}}\; .  \qquad
\eea

\section{Massive 3D counterpart of D0-brane from massless 4D superparticle}
As a warm-up exercise, let us discuss the dimensional reduction of the $\text{D}=4$~~$\mathcal{N}=1$ massless superparticle and obtain in this manner the action for a massive $\text{D}=3$~~$\mathcal{N}=2$ superparticle, the 3D counterpart of the 10D action for a single D$0$-brane. This action will later serve as the center of mass sector of the system of multiple D$0$-branes.

\subsection{Reduction of Brink-Schwarz action}
\label{sec:BS_reduction}
Let us begin with the previously presented $\text{D}=4$ Brink-Schwarz superparticle action~\eqref{eq:action_BS}, 
\be\label{SBS=4D}
S_{\rm BS}^{4\text{D}}= \int_{\mathcal{W}^1} p_\mu \Pi^\mu +  \dfrac{e}{2} p_\mu p^\mu\; .
\ee
Let us choose $x^2$ as the direction of reduction,
\be
x^\mu = (x^{\tilde{\mu}}, x^2)\; , \qquad p_\mu = (p_{\tilde{\mu}}, p_2)\; , \qquad {\tilde{\mu}}=0,1,3 \qquad \Longleftrightarrow \qquad \mu= 0,1,2,3\; ,
\ee
which is convenient because it aligns with the representation of the $\text{D}=4$ rPMs~\eqref{Pauli=} and their relation with $\text{D}=3$ gamma matrices \eqref{g3=s4}. This corresponds to the split of the 4D VA 1-form into
\be \label{Pi2=}
\Pi^2={\rm d}x^2-i {\rm d}\theta\sigma^2\bar{\theta}+i\theta\sigma^2{\rm d}\bar{\theta}= {\rm d}x^2-\epsilon_{\alpha\beta} ( {\rm d}\theta^{\alpha} \bar{\theta}^{\beta} -\theta^{\alpha} {\rm d}\bar{\theta}^{\beta})\;
\ee
and the 3D VA 1-form
\be\label{VA=3d}
\Pi^{\tilde{\mu}}={\rm d}x^{\tilde{\mu}}-i {\rm d}\theta{\gamma}^{\tilde{\mu}}\bar{\theta} + i \theta{\gamma}^{\tilde{\mu}}{\rm d}\bar{\theta}=\frac 1 2 \Pi^{\alpha\beta} {\gamma}^{\tilde{\mu}}_{\alpha\beta} \; .
\ee
Here
\begin{equation} 
\text{d}\theta\gamma^{\tilde{\mu}} \bar{\theta}= \text{d}\theta{}^\alpha \gamma^{\tilde{\mu}}_{\alpha\beta} \bar{\theta}{}^\beta ,
\end{equation}
with symmetric 2$\times$2 matrices
\be
 \gamma^{\tilde{\mu}} _{\alpha\beta}
 = -i \gamma^{\tilde{\mu}} _\alpha{}^\sigma\epsilon_{\sigma\beta}\; , \qquad \tilde{\gamma}{}^{\tilde{\mu} \alpha\beta}
= i \epsilon^{\alpha\sigma}\gamma^{\tilde{\mu}}{}_\sigma{}^{\beta} \, \qquad
\ee
constructed from the 3D Dirac $\gamma^{\tilde{\mu}\beta}_{\alpha}$ and $\epsilon$-symbol matrices. These are imaginary in our mostly-minus metric convention and satisfy
\be
\gamma^{\tilde{\mu}}\gamma^{\tilde{\nu}} = \eta^{\tilde{\mu}\tilde{\nu}}+ i\epsilon^{\tilde{\mu}\tilde{\nu}\tilde{\sigma}}\gamma_{\tilde{\sigma}}\; , \qquad \eta^{\tilde{\mu}\tilde{\nu}}=\text{diag}(+1,-1,-1)\; .
\ee
Eq.~\eqref{VA=3d} can be represented as a symmetric spin-tensor 1-form
\be\label{VA=3d=2x2}
\Pi^{\alpha\beta}= \Pi^{\tilde{\mu}}\tilde{\gamma}_{\tilde{\mu}}^{\alpha\beta}= {\rm d}x^{\alpha\beta}- 2i {\rm d}\theta^{( \alpha}\bar{\theta}{}^{\beta )} + 2i \theta^{( \alpha}{\rm d}\bar{\theta}{}^{\beta )}\; . \qquad
\ee
The dimensional reduction can be performed by using equations of motion for the $x^2$ coordinate which reads $\text{d}p_2=0$ implying that $p_2$ is a constant,
\be\label{p2=m}
{\rm d}p_2=0 \qquad \Longrightarrow \qquad p_2=m={\rm const}\;.
\ee
Substituting the solution \eqref{p2=m} back into the action \eqref{SBS=4D}, using \eqref{Pi2=} and omitting the total derivative term, we find 
\begin{equation}
    S_{\rm dAL}^{3\text{D}}= \int_{\mathcal{W}^1} \left( p_{\tilde{\mu}} \Pi^{\tilde{\mu}} +  \dfrac{e}{2} (p_{\tilde{\mu}} p^{\tilde{\mu}}-m^2)\right)+m \int_{\mathcal{W}^1} ({\rm d}\theta^\alpha \bar{\theta}_\alpha-\theta^\alpha {\rm d}\bar{\theta}_\alpha )\; .
    \label{SdAL=}
\end{equation}
This is the $\text{D}=3$ counterpart of the De Azcárraga-Lukierski ${\mathcal N}=2$ massive superparticle action \cite{kappa_1, Azcarraga2}. This action possesses $\kappa$-symmetry, and let us note that $\kappa$-symmetry was first discovered in this massive example \cite{kappa_1}, slightly earlier than in the case of the massless superparticle \cite{kappa_2}.

From a modern perspective, \eqref{SdAL=} is the $\text{D}=3$ counterpart of super-D$0$-brane action \cite{D-branes_Eric} or, simplifying terminology, the 3D D$0$-brane action. The second term in \eqref{SdAL=} is the Wess-Zumino term of the D$0$-brane, which serves as the prototype for the Wess-Zumino term of the superstring \cite{WessZumino_term} and of higher super-$p$-branes (see \cite{p-branes_1, p-branes_2, p-branes_3, p-branes_4, D-branes_Eric, p-branes_5}, \cite{p-branes_6, p-branes_7, p-branes_8} and refs. therein).

\subsection{A suggestive ansatz for dimensional reduction of (spinor) moving frame formalism}
Our task now is to perform dimensional reduction of the spinor moving frame action for $\text{D}=4$ massless superparticle in a way that reproduces the 3D D$0$-brane action. (For a warm-up see Appendix~\ref{sec:App_3Dderivatives}, where we dimensionally reduce to the 3D massless superparticle.). To this end it is sufficient to employ the following ansatz ({\it cf.} \eqref{v=v*=})
\begin{equation}
\begin{cases}
v_\alpha^- ={\rm v}_\alpha^-  - i{\rm v}_\alpha^+  \mu^{=} \; , \qquad \cr v_\alpha^+ ={\rm v}_\alpha^+  \; , \end{cases}\qquad
\begin{cases} \bar{v}_{\dot\alpha}^- ={\rm v}_\alpha^-  + i{\rm v}_\alpha^+  \mu^{=} \; , \qquad \cr \bar{v}_{\dot\alpha}^+ ={\rm v}_\alpha^+ \; , \end{cases} 
\label{v-=v-+}
\end{equation}
where ${\rm v}_{\alpha}^{\pm}$ are real spinors satisfying
\be\label{v-v+=1=3D}
{\rm v}^{-\alpha}{\rm v}_\alpha^+=1 \; , \qquad ({\rm v}_\alpha^\pm)^*= {\rm v}_\alpha^\pm\; ,
\ee
and $\mu^{=} = \mu^{=}(\tau)$ is a real 1d scalar field.

This ansatz, although useful for the present illustrative purpose, will not be appropriate for the dimensional reduction of the nAmW system, for reasons to be explained later. Nevertheless, it provides valuable intuition regarding dimensional reduction in the spinor moving frame formalism.

Using \eqref{v-=v-+}, we can easily write that
\be\label{u--=}
u_{\alpha\dot{\beta}}^==2{\rm v}^-_{\alpha}{\rm v}^-_{\beta}+ 2{\rm v}^+_{\alpha}{\rm v}^+_{\beta} (\mu^{=} )^2+ 2i \epsilon_{\alpha\beta}\mu^{=} \quad (\; = (u_{\beta\dot\alpha}^=)^*\; )\;  . \qquad
\ee
Substituting into the original action \eqref{eq:S0D4==} splits it into parts depending only on the 3D VA 1-forms \eqref{VA=3d} and a separate element involving $\Pi^2$ \eqref{Pi2=}
\begin{equation}
S^{4\text{D}}_0 \vert_{\eqref{v-=v-+}}= \frac  1 2 \int_{\mathcal{W}^1}\rho^\# \left[{\rm v}_\alpha^-{\rm v}_\beta^-  +(\mu^=)^2 {\rm v}_\alpha^+{\rm v}_\beta^+\right]\Pi^{\alpha\beta}-
\int_{\mathcal{W}^1} \rho^\# \mu^= \left[{\rm d}x^2-\epsilon_{\alpha\beta}({\rm d}\theta^{\alpha}\bar{\theta}^{\beta}-\theta^{\alpha}{\rm d}\bar{\theta}^{\beta})\right]\; . 
\label{eq:S0D4====}
\end{equation}
Now let us use the equation of motion for $x^2$ coordinate,
\be\label{rmu=m}
{\rm d}(\rho^\#\mu^=)=0\qquad \Longrightarrow\qquad \rho^\#\mu^== m = {\rm const}\; . \ee
Substituting its solution into \eqref{eq:S0D4====} we find
\bea\label{S0D3==}
S^{4\text{D}}_0\vert_{\eqref{v-=v-+},~\eqref{rmu=m}}= \frac  1 2  \int_{\mathcal{W}^1} \rho^\# \left({\rm v}_\alpha^-{\rm v}_\beta^-  +\frac {m^2}{(\rho^\#)^2} {\rm v}_\alpha^+{\rm v}_\beta^+\right)\Pi^{\alpha\beta}+
m \int_{\mathcal{W}^1}  \epsilon_{\alpha\beta}({\rm d}\theta^{\alpha}\bar{\theta}^{\beta}-\theta^{\alpha}{\rm d}\bar{\theta}^{\beta})\; . \qquad
\eea

The term with $x^2$ has disappeared as a total derivative. We have thus obtained a 3D superparticle action, where the second term in \eqref{S0D3==} is the Wess-Zumino term of the D$0$-brane, the same as in \eqref{SdAL=}.

Now let us redefine the spinor frame variables as 
\be
{\rm v}_\alpha^2 =\sqrt{\frac {\rho^\#}{m} } {\rm v}_\alpha^- \; , \qquad {\rm v}_\alpha^1 = \sqrt{\frac m {\rho^\#} } {\rm v}_\alpha^+~.
\label{eq:v2v1}
\ee 
Then $\rho^\#$ is just removed from the first kinetic term.

The new spinors satisfy\footnote{This constraint implies that $2\times 2$ matrix $(v_{\alpha}^1, v_{\alpha}^2)$ belongs to double covering SL$(2,{\mathbb R})$ of the 3D Lorentz group SO$(1,2)$. Hence the second name of Lorentz harmonics; see below, in particular footnote~\ref{HarmFoot}, for more details and references.}
\be\label{v2v1=1}
{\rm v}^{\alpha 2}{\rm v}_\alpha^1=1\qquad \Longleftrightarrow \qquad {\rm v}^{\alpha p}{\rm v}_\alpha^q =-\epsilon^{pq}\; ,\qquad q=1,2\;
\ee
and hence form a $\text{SL}(2,{\mathbb R})$ valued matrix
\begin{equation}
{\rm v}_\alpha^q = ({\rm v}_\alpha^1,{\rm v}_\alpha^2)\;\in\; \text{SL}(2,{\mathbb R}) \; .
\label{vq=inSL}
\end{equation}
Substituting the solutions of~\eqref{eq:v2v1} for $v_\alpha^\pm$ in the action~\eqref{S0D3==} obtained by dimensional reduction of the 4D massless superparticle action \eqref{eq:S0D4==}, we arrive at
\begin{equation}
    \begin{array}{l}
        \begin{split}
            S_{\text{D}0}^{3\text{D}} &= \int_{\mathcal{W}^1} {\mathcal L}_{\text{D}0}^{3\text{D}}= \frac 1 2  m\int_{\mathcal{W}^1} {\rm v}^q_{\alpha}{\rm v}^q_{\beta}\Pi^{\alpha\beta}+
            m \int_{\mathcal{W}^1}  \epsilon_{\alpha\beta}({\rm d}\theta^{\alpha}\bar{\theta}^{\beta}-\theta^{\alpha}{\rm d}\bar{\theta}^{\beta})= \\
            &~\\
            &= \frac 1 2 m\int_{\mathcal{W}^1} {\rm u}^0_{\alpha\beta}\Pi^{\alpha\beta}+
            m \int_{\mathcal{W}^1}  \epsilon_{\alpha\beta}({\rm d}\theta^{\alpha}\bar{\theta}^{\beta}-\theta^{\alpha}{\rm d}\bar{\theta}^{\beta})~, 
        \end{split}
    \end{array} \label{SD0D3=}
\end{equation}
which is the 3D D$0$-brane action in spinor moving frame formulation \cite{10D_mD0_Igor, superD0_Igor}. 

In the second line  of \eqref{SD0D3=} we have introduced the matrix ${\rm u}^0_{\alpha\beta}={\rm v}^q_{\alpha}{\rm v}^q_{\beta}$ representing timelike normalized vector from the 3D moving frame attached to the worldline,
\bea\label{u0=}
{\rm u}^0_{\alpha\beta}={\rm u}^{0\tilde{\mu}}{\gamma}_{\tilde{\mu}\alpha\beta}={\rm v}_\alpha^1{\rm v}_\beta^1+{\rm v}_\alpha^2{\rm v}_\beta^2\; , \qquad \\  \label{u12=}
{\rm u}^1_{\alpha\beta}={\rm u}^{1\tilde{\mu}}{\gamma}_{\tilde{\mu}\alpha\beta}=2{\rm v}_{(\alpha}^1{\rm v}_{\beta)}^2\; , \qquad {\rm u}^2_{\alpha\beta}={\rm u}^{2\tilde{\mu}}{\gamma}_{\tilde{\mu}\alpha\beta}={\rm v}_\alpha^1{\rm v}_\beta^1-{\rm v}_\alpha^2{\rm v}_\beta^2~
\; . \qquad
\eea
These vectors obey the orthogonality and normalization conditions
\be
{\rm u}^0_{\tilde{\mu}}{\rm u}^{0\tilde{\mu}}=1 \; , \qquad {\rm u}^0_{\tilde{\mu}}{\rm u}^{I\tilde{\mu}}=0 \; , \qquad {\rm u}^I_{\tilde{\mu}}{\rm u}^{J\tilde{\mu}}=-\delta^{IJ} \; , \qquad I,J=1,2\;
\ee
which imply that they form $\text{SO}(1,2)$ valued 3D moving frame matrix
\be
{\rm u}_{\tilde{\mu}}{}^{\tilde{a}}=\left({\rm u}_{\tilde{\mu}}{}^{0},{\rm u}_{\tilde{\mu}}{}^{I}\right)\; \in\; \text{SO}(1,2) \qquad \text{with} \qquad \tilde{a}=0,1,2~. \ee

\subsection[Irreducible \texorpdfstring{$\kappa$}{K}-symmetry of the 3D super-D0-brane]{Irreducible \boldmath\texorpdfstring{$\kappa$}{K}-symmetry of the 3D super-D0-brane}
To analyse the local symmetries of the 3D D$0$-brane action \eqref{SD0D3=}, we compute the exterior derivative  of the Lagrangian form of the action \eqref{SD0D3=}. This reads
\bea\label{dL0D3=}
{\rm d} {\mathcal L}_{\text{D}0}^{3\text{D}} =- 2im ({\cal E}^1+i{\cal E}^2) \wedge  (\bar{{\cal E}}^1-i\bar{{\cal E}}^2) + \frac 1 2  m\Pi^{\alpha\beta}\wedge {\rm d}{\rm u}^0_{\alpha\beta}\; ,  \qquad
\eea
where ${\cal E}^q$ and $\bar{{\cal E}}^q$ are fermionic 1-forms  of the pull-back of 3D supervielbein adapted to the embedding
\be\label{cEq=}
{\cal E}^{q}= ({\cal E}^{1},{\cal E}^{2} )={\rm d}\theta^\alpha {\rm v}_\alpha^q \; , \qquad \bar{{\cal E}}{}^{q}= (\bar{{\cal E}}{}^{1},\bar{{\cal E}}{}^{2} )={\rm d}\bar{\theta}^\alpha {\rm v}_\alpha^q \; . \qquad
\ee
We do not focus now on the derivatives of the spinor moving frame variables which will be discussed later.

Below we will also need one of three bosonic 1-forms of this supervielbein
\bea\label{rmE0=}
{\rm E}^0 = \Pi^{\tilde{\mu}} {\rm u}_{\tilde{\mu}}^0= \frac 1 2 \Pi^{\alpha\beta} {\rm u}_{\alpha\beta}^0 = \frac 1 2 \Pi^{\alpha\beta}({\rm v}_\alpha^1{\rm v}_\beta^1 + {\rm v}_\alpha^2{\rm v}_\beta^2)\; , \qquad {\rm E}^I = \Pi^{\tilde{\mu}} {\rm u}_{\tilde{\mu}}^I= ( {\rm E}^1, {\rm E}^2)\; . \qquad
\eea
From \eqref{dL0D3=} (and using the formalism presented in section~\ref{sec:diff_SuperForms}), it follows that the action is invariant under the following irreducible $\kappa$-symmetry transformations
\begin{equation}
\begin{array}{l}
      ~~~~\iota_\kappa {\cal E}^1 = \kappa \; , \qquad \iota_\kappa {\cal E}^2 =  i\kappa \; , \qquad \iota_\kappa \bar{{\cal E}}^1 = \bar{\kappa} \; , \qquad \iota_\kappa \bar{{\cal E}}^2 = -i\bar{\kappa} \; ,\\
      ~\\
      \iota_\kappa \Pi^{\alpha\beta}=0\; , \qquad \delta_\kappa {\rm  v}^q_\alpha =0\qquad (\Longrightarrow \qquad \delta_\kappa {\rm  u}^0_{\alpha\beta}=0)\; ,
\end{array}
    \label{kappa=D0}
\end{equation}
or, equivalently,
\begin{equation}
\begin{array}{rcl}
    \delta_\kappa \theta^\alpha = \kappa ({\rm  v}^{\alpha 2} -i {\rm  v}^{\alpha 1}) \; , &~& \delta_\kappa \bar{\theta}{}^\alpha = \bar{\kappa } ({\rm  v}^{\alpha 2} +i {\rm  v}^{\alpha 1}) \; ,  \\
    ~\\
    \delta_\kappa x^{\alpha\beta} = 2i (\delta_\kappa \theta^{(\alpha}  \,  \bar{\theta}{}{}^{\beta )} - \theta^{(\alpha} \delta_\kappa \bar{\theta}{}^{\beta )})\; , &~& \delta_\kappa {\rm  v}^q_\alpha =0\; .
    \end{array}
\end{equation}
Let us also observe that the action \eqref{SD0D3=} is invariant under the local $\text{SO}(2)$ symmetry  rotating the spinor frame components \eqref{vq=inSL}. This acts as a local $\text{SO}(2)$ rotation of the spacelike moving frame vectors ${\rm u}_{\tilde{\mu}}^I$ and identifies the set of spinor moving frame fields ${\rm v}_\alpha^q$ as  homogeneous coordinates of the coset
\be\label{coset=3D}
\frac {\text{SL}(2,{\mathbb R})} {\text{SO}(2)} \simeq \frac {\text{SO}(1,2)}{\text{SO}(2)} \; ,
\ee
which is suitable for the description of the 3D massive particle since $\text{SO}(2)$ is the small group of 3D timelike momentum.

Another observation is that this $\text{SO}(2)$ symmetry is not manifest in the original ansatz \eqref{v-=v-+} for 4D spinor frame variables. It appears in the final action as a kind of emergent symmetry. However, if we try to use this ansatz for dimensional reduction of a more complicated nAmW system, the symmetry does not emerge automatically. This would imply that the spinor frame variables in this 3D action parametrize the $\text{SL}(2,{\mathbb R})$ group rather than coset \eqref{coset=3D} and hence carry one extra degree of freedom which actually is not wanted. Thus a refined dimensional reduction ansatz must be used.

\section{Revising the 4D M0 reduction to 3D D0. Properties of 3D spinor frame}
The dimensional reduction described above was based on the ansatz \eqref{v-=v-+} for the spinor moving frame variables, which explicitly breaks the $\text{U}(1)$ subgroup of the $\text{GL}(1,{\mathbb C})$ gauge symmetry of the 4D spinor moving frame formalism. Moreover, this ansatz restores an important $\text{SO}(2)$ gauge symmetry of the $\text{D}=3$ Lorentz harmonic approach only at the final stage, as a symmetry of the single super-D$0$-brane action \eqref{SD0D3=}. 

\subsection[\texorpdfstring{$\text{SO}(2)$}{SO(2)}=\texorpdfstring{$\text{U}(1)$}{U(1)} invariant reduction of \texorpdfstring{$\text{D}$}{D=4} spinor frame formalism]{\boldmath\texorpdfstring{$\text{SO}(2)$}{SO(2)}=\boldmath\texorpdfstring{$\text{U}(1)$}{U(1)} invariant reduction of \boldmath\texorpdfstring{$\text{D}$}{=4} spinor frame formalism}
In this section, we describe a more complicated $\text{SO}(2) \simeq \text{U}(1)$ invariant ansatz for the reduction of the $\text{D}=4$ spinor moving frame formalism to $\text{D}=3$. This ansatz is characterized by the following expressions for the reduced lightlike vectors of the 4D moving frame
\bea\label{ru--=}
\rho^\# u_{\alpha\dot\beta}^= =  {\cal M} {\rm u}^0_{\alpha\beta} + i {\cal M}\epsilon_{\alpha\beta}\; , \qquad \\
\label{ru++=}
\frac 1 {\rho^\#  }u_{\alpha\dot\beta}^\#  =  \dfrac{1}{\cal M} {\rm u}^0_{\alpha\beta} - \dfrac{i}{\cal M}\epsilon_{\alpha\beta}\; , ~\qquad
\eea
where ${\mathcal M}={\mathcal M}(\tau)$ is a new real field on the worldline. Since only these moving frame vectors appear explicitly in the nAmW action, and their reduction does not involve the vectors ${\rm u}_{\alpha\beta}^I = ({\rm u}_{\alpha\beta}^1, {\rm u}_{\alpha\beta}^2)$, we can expect that the reduced action will be invariant under $\text{SO}(2)$ rotation mixing these vectors.

The explicit form of the $\text{SO}(2)\simeq \text{U}(1)$ invariant ansatz, expressing the $\text{D}=4$ spinor moving frame variables \eqref{VinSL} in terms of 3D Lorentz harmonics ${\rm v}_\alpha^{q}=({\rm v}_\alpha^{1},{\rm v}_\alpha^{2})$ defined in \eqref{vq=inSL}, and which reproduces \eqref{ru--=} and \eqref{ru++=}, is given by
\begin{equation}
    \begin{array}{lcl}
        \sqrt{\rho^\#} v_\alpha^- =\dfrac{\sqrt{{\cal M}}}{\sqrt{2}}\, \left({\rm v}_\alpha^2 -i {\rm v}_\alpha^1 \right)\; , &~&
        \dfrac {1}{\sqrt{\rho^\#}} v_\alpha^+ =\dfrac 1{\sqrt{2}\sqrt{{\cal M}}}\, \left({\rm v}_\alpha^1 -i {\rm v}_\alpha^2 \right)\; ,\\
        \sqrt{\rho^\#} \bar{v}_{\dot\alpha}^- =
        \dfrac {\sqrt{{\cal M}}} {\sqrt{2}}\, \left({\rm v}_\alpha^2 +i {\rm v}_\alpha^1 \right)\; ,&~& \dfrac {1} {\sqrt{\rho^\#}}  \bar{v}_{\dot\alpha}^+ =  \dfrac {1}{\sqrt{2}\sqrt{{\cal M}}}\, \left({\rm v}_\alpha^1 +i {\rm v}_\alpha^2 \right)\; .
    \end{array}
    \label{rv-=v2-iv1}
\end{equation}
Notice that this implies
\be
\label{bv-=iv+}  \sqrt{\rho^\#} \bar{v}_{\dot\alpha}^- = i \dfrac {{\cal M}} {\sqrt{\rho^\#}} v_\alpha^+ \; , \qquad
 \frac {1} {\sqrt{\rho^\#}}  \bar{v}_{\dot\alpha}^+ =  i\dfrac {\sqrt{\rho^\#}} {{\cal M}}v_\alpha^-\;   \qquad
\ee
which breaks SL$(2,\mathbb{C})$ invariance of 4D action down to SL$(2,\mathbb{R})$ but preserves the complete $\text{GL}(2, \mathbb{C})$ gauge symmetry of the 4D Lorentz harmonic formalism. The $\text{SO}(1,1)$ subgroup leaves both sides of \eqref{bv-=iv+}, while $\text{U}(1) \subset \text{GL}(2, \mathbb{C})$ transformations  of both sides are nontrivial and coincident.

Using \eqref{rv-=v2-iv1}, the  (pull-backs of)  relevant  4D supervielbein forms \eqref{E--=}-\eqref{E-=} can be expressed in terms of  (pull-backs of)
3D supervielbein forms \eqref{cEq=}, \eqref{rmE0=}:
\begin{equation}
    \begin{array}{l}
       \begin{split}
           \rho^\# \text{E}^= =  {\cal M} \text{E}^0 + \frac i 2 {\cal M}\epsilon_{\alpha\beta} \Pi^{\alpha\dot{\beta}} &= {\cal M} {\rm E}^0 -{\cal M}\Pi^{2} =\\
           &=   {\cal M} {\rm E}^0 - {\cal M}({\rm d}x^2- {\rm d}\theta^\alpha \bar{\theta}_\alpha+ \theta^\alpha {\rm d}\bar{\theta}_\alpha )~,
       \end{split}
    \end{array}\label{E--=cME0+}
\end{equation}
\begin{equation}
    \begin{array}{l}
       \begin{split}
           \qquad~~~~~~ \frac 1 {\rho^\#} \text{E}^\# =  \frac 1 {{\cal M}} {\rm E}^0 - \frac i 2  \frac 1 {{\cal M}} \epsilon_{\alpha\beta} \Pi^{\alpha\dot{\beta}} &= \frac 1 {{\cal M}}  {\rm E}^0+ \frac 1 {{\cal M}} \Pi^{2}=\\
           &=  \frac 1 {{\cal M}} {\rm E}^0 +   \frac 1 {{\cal M}} ({\rm d}x^2 - {\rm d}\theta^\alpha \bar{\theta}_\alpha+ \theta^\alpha {\rm d}\bar{\theta}_\alpha )\; ,
       \end{split}
    \end{array}\label{E++=cME0-}
\end{equation}
\begin{equation}
    \begin{array}{lcl}
      ~~~~~ \dfrac 1 {\sqrt{\rho^\#}} \text{E}^+ =  \dfrac 1{\sqrt{2}\sqrt{{\cal M}}}\, \left({\cal E}^1 -i {\cal E}^2 \right)~, &~& \dfrac 1 {\sqrt{\rho^\#}} \bar{\text{E}}^+ =  \dfrac 1{\sqrt{2}\sqrt{{\cal M}}}\, \left(\bar{{\cal E}}^1 +i \bar{{\cal E}}^2 \right) \; .
    \end{array}\label{E+=cME0-}
\end{equation}
The dimensional reduction of the massless superparticle action requires to use only the first of these expressions, \eqref{E--=cME0+}. Substituting it into the action and using the equations of motion for $x^2$,
\be
{\rm d}{\cal M}=0\qquad \Longrightarrow \qquad {\cal M}=m={\rm const}\;
\ee
we arrive at the  3D super-D$0$-brane action \eqref{SD0D3=}.

\subsection{Cartan forms and covariant derivatives in 3D}
To proceed with the dimensional reduction of the nAmW action, we begin by introducing the Cartan forms of the coset $\text{SL}(2, \mathbb{R})/\text{SO}(2)$
\be
 f^{pq}:= {\rm v}^{\alpha p} {\rm d}{\rm v}_\alpha^q = f^{qp}\; , \qquad {\rm v}^{\alpha p}=\epsilon^{pq}{\rm v}^\alpha_q=\epsilon^{\alpha\beta}{\rm v}_\beta^p\; . \qquad
\ee
Due to the constraint \eqref{v2v1=1}, this matrix 1-form is symmetric, $f^{pq}=f^{qp}$ and the derivative of the 3D spinor frame variables are expressed in terms of these by
\be
{\rm d}{\rm v}_\alpha^q= {\rm v}_{\alpha p}f^{pq} \, ,  \qquad
\ee
where ${\rm v}_{\alpha p}=\epsilon_{pq}{\rm v}_{\alpha}^q$. Explicitly, this yields
\be
{\rm d}{\rm v}_\alpha^1= {\rm v}_\alpha^1f^{21}- {\rm v}_\alpha^2f^{11}\; , \qquad {\rm d}{\rm v}_\alpha^2= {\rm v}_\alpha^1f^{22}- {\rm v}_\alpha^2f^{12}\; . \qquad
\ee
It is important to note that $f^{qq}$ transforms as $\text{SO}(2)$ connection, while $f^{12}=f^{21}$ and $f^{11}-f^{22}$ 1-forms are covariant under $\text{SO}(2)$. This structure becomes transparent when using the 3D moving frame vectors defined in \eqref{u0=} and \eqref{u12=}, which allows us to identify the $\text{SO}(2)$ connection as
\be\label{u1du2}
{\rm u}^1{\rm d}{\rm u}^2=-{\rm u}^2{\rm d}{\rm u}^1= \frac 1 2 {\rm u}^{1\alpha\beta}{\rm d}{\rm u}^2_{\alpha\beta}=f^{qq}\qquad \Longleftrightarrow \qquad {\rm u}^I{\rm d}{\rm u}^J=\epsilon^{IJ}f^{qq} \;.\qquad 
\ee
Similarly, we define the $\text{SO}(2)$ covariant forms $f^I = (f^1, f^2)$, which serve as the vielbein for the coset $\text{SO}(1,2)/\text{SO}(2)$
\be\label{u0duI} f^1:=  {\rm u}^0{\rm d}{\rm u}^1 =-f^{11}+f^{22}\; ,\qquad f^2:= {\rm u}^0{\rm d}{\rm u}^2= 2f^{12}\; . \qquad
\ee
For our discussion in the next section, it is important to note that the ansatz \eqref{rv-=v2-iv1}, together with \eqref{eq:4D_Omega0}, leads to the following expressions for the 4D connections
\be\label{om0=fqq+dM}
\omega^{(0)} - \bar{\omega}^{(0)} =i{\rm v}^{\alpha q}{\rm d}{\rm v}_\alpha^q=: i f^{qq}\; , \qquad \omega^{(0)} + \bar{\omega}^{(0)} = \frac {{\rm d}\rho^\#}  {\rho^\#}\, - \frac {{\rm d}{\cal M}}  {{\cal M}}\, . \ee
The second equation implies  that the  covariant derivative of the St\"ukelberg field in the 4D nAmW system can be written in terms of the derivative of the field ${\cal M}$:
\be\label{Dr=dcM}
{\rm D}\rho^\#= {\rm d}\rho^\#- (\omega^{(0)} + \bar{\omega}^{(0)}) \rho^\#= \rho^\#  \frac {{\rm d}{\cal M}}  {{\cal M}}\, .  \qquad
\ee

\subsection[\texorpdfstring{${\cal N}$}{N}=2 3D mD0-brane action from dimensional reduction of the 4D nAmW]{\boldmath\texorpdfstring{${\cal N}$}{N}=2 3D mD0-brane action from dimensional reduction of the 4D nAmW}
\label{sec:4D_dimReduction}

Let us turn to the problem of the dimensional reduction of the $\text{D}=4$ nAmW action \eqref{eq:SmM0=4D}. This procedure simplifies if, before  substituting the above discussed ansatz \eqref{rv-=v2-iv1} for the 4D spinor frame variables, we first redefine the matrix fields by making them $\text{SO}(1,1)$ invariant. This is achieved by multiplying them by suitable powers of the $\text{D}=4$ Stückelberg field $\rho^\#$:
\begin{eqnarray}
\label{bZ0++->bZ} & \widetilde{{\mathbb Z}}= \dfrac{1}{\rho^\#} {\mathbb  Z} \, , \qquad &  \bar{\widetilde{{\mathbb Z}}}= \frac 1 {\rho^\#} \bar{\mathbb  Z} \, , \qquad  \\
\label{bP=bP++3+}
& \widetilde{{\mathbb  P}}= \dfrac{1}{(\rho^\#)^2} {\mathbb  P}\, , \qquad   & \bar{\widetilde{{\mathbb  P}}}= \frac 1 {(\rho^\#)^2} \bar{\mathbb  P} \, , \qquad    \\
\label{bPsi=bPsi+2+} &  \widetilde{\Psi}=\dfrac{1}{(\rho^\#)^{\frac 32}} {\Psi} \, , \qquad & \bar{\widetilde{ \Psi}}=\frac 1 {(\rho^\#)^{\frac 32}} \bar{{ \Psi}} \, , \qquad
 \end{eqnarray}
Using \eqref{bZ0++->bZ}-\eqref{bPsi=bPsi+2+} one can check that the new matrix fields have opposite left and right ``charges''
 \begin{eqnarray}
\label{Z=Z-+} &  {\mathbb  Z} =  {\mathbb Z}_{-|+}\, , \qquad &   \bar{\mathbb Z} = \bar{\mathbb  Z}_{+|-}\, , \qquad  \\
\label{P=P-+}
& {\mathbb P} =   {\mathbb  P}_{-|+}\, , \qquad   &  \bar{\mathbb  P} =  \bar{\mathbb  P}_{+|-}\, , \qquad    \\
\label{Psi=Psi+-} & {\Psi} =  {\Psi}_{\frac 12|-\frac 12}\, , \qquad &  \bar{{ \Psi}} =  \bar{{ \Psi}}_{-\frac 12|\frac 12}\, , \qquad
 \end{eqnarray}
confirming their $\text{SO}(1,1)$ invariance.

Such redefinition of the matrix fields clearly generates terms proportional to ${\rm d}\rho^\#$ in the action. However, after reducing the spinor moving frame sector according to the ansatz \eqref{rv-=v2-iv1}, these derivatives are replaced by those of the new scalar field ${\cal M}$, which is inert under both $\text{SO}(1,1)$ and $\text{SO}(2)$ transformations.

Indeed, using \eqref{Dr=dcM}, the covariant derivatives of the old and new matrix fields are related by
\bea\label{DZ=4D-3d}
{\rm D}\widetilde{{\mathbb Z}}=  \dfrac 1 {\rho^\#}\left({\rm D}{\mathbb Z}-  \frac {{\rm d}{\cal M}} {{\cal M}} {\mathbb Z} \right)\; , \qquad \\ \label{DbZ=4D-3d}
{\rm D}\bar{\widetilde{{\mathbb Z}}}=  \dfrac 1 {\rho^\#}\left({\rm D}\bar{{\mathbb Z}}-  \frac {{\rm d}{\cal M}} {{\cal M}} \bar{{\mathbb Z}} \right)\; , \qquad \\
\label{DPsi=4D-3d}
{\rm D}\widetilde{\Psi}=  \dfrac 1 {(\rho^\#)^{\frac 32}}\left({\rm D}\Psi- \frac 32 \frac {{\rm d}{\cal M}} {{\cal M}} \Psi \right)\; , \qquad
\\ \label{DbPsi=4D-3d}
{\rm D}\bar{\widetilde{\Psi}}=  \dfrac 1 {(\rho^\#)^{\frac 32}}\left({\rm D}\bar{\Psi}- \frac 32 \frac {{\rm d}{\cal M}} {{\cal M}} \bar{\Psi} \right)\; , ~\qquad
\eea
where (see \eqref{om0=fqq+dM})
\bea\label{DZ=3d} {\rm D}{\mathbb Z} = {\rm d}{\mathbb Z}+ if^{qq}{\mathbb Z}+ [{\mathbb A}, {\mathbb Z}] \; , ~~~ \qquad
\\ \label{DPsi=3d}  {\rm D}\Psi = {\rm d}\Psi- \frac i 2 f^{qq}\Psi+ [{\mathbb A}, \Psi]\; . \qquad \eea
Now, writing the 4D nAmW action in terms of new matrix variables and reduced spinor frame variables~\eqref{rv-=v2-iv1}, and using the  new covariant derivatives \eqref{DZ=3d}, \eqref{DPsi=3d} and the supervielbein forms \eqref{E--=cME0+}-\eqref{E+=cME0-}, we arrive at
\begin{equation}
    \begin{array}{l}
         \begin{split}
             S^{\text{4D}}_{\rm nAmW}\vert_{\eqref{rv-=v2-iv1}} &= \int_{{\cal W}^1} {\rm  E}^{0} \left({\cal M}+ \frac 1 {{\cal M}} \, \frac {{\cal H}}{\mu^6}   \right)+ \int\limits_{{\cal W}^1}  \left(- {\cal M}+ \frac 1 {{\cal M}} \, \frac {{\cal H}}{\mu^6}   \right)\, ({\rm d}x^2 - {\rm d}\theta^\alpha \bar{\theta}_\alpha+  \theta^\alpha {\rm d}\bar{\theta}_\alpha )~+\\
          &+  {1\over \mu^6} \int_{\mathcal{W}^1}
           {\rm tr}\left(\bar{\mathbb P}{\rm D} {\mathbb Z} + {\mathbb P}{\rm D} \bar{\mathbb Z} - {i\over 8} {\rm D}{ \Psi}\,  \bar{\Psi} + {i\over 8} { \Psi} {\rm D} \bar{\Psi}  \right) - {1\over \mu^6} \int_{\mathcal{W}^1} \dfrac {{\rm d}{\cal M}} {{\cal M}}
           {\rm tr}\left(\bar{\mathbb P} {\mathbb Z} + {\mathbb P} \bar{\mathbb Z}\right)  +\\
           &+ {1\over \mu^6} \int_{\mathcal{W}^1}  \frac i {\sqrt{2}\sqrt{{\cal M}}}\left( {\cal E}^{1}-i{\cal E}^{2}\right)
           {\rm tr}(\bar{\Psi}  \bar{\mathbb P}+   \Psi [{\mathbb Z},  \bar{\mathbb Z}])~+ \\  
           &+ {1\over \mu^6} \int_{\mathcal{W}^1}  \frac i {\sqrt{2}\sqrt{{\cal M}}}\left(\bar{ {\cal E}}^{1}+i\bar{{\cal E}}^{2}\right)
           {\rm tr}({\Psi}  {\mathbb P}+  \bar{\Psi} [{\mathbb Z},  \bar{\mathbb Z}])~.
         \end{split}
    \end{array}
\label{SmM0'=4D}
\end{equation}
Here, ${\cal H}$ has the same structure as $\widetilde{\cal H}$ \eqref{HSYM=4D}, but is expressed through the new $\text{SO}(1,1)$ invariant matrix fields,
\begin{eqnarray}
\label{tcH=} && {\cal H}=    {\rm tr}\left( {\mathbb P} \bar{\mathbb P} +  [{\mathbb Z},  \bar{\mathbb Z}]^2 -
{i\over 2} {\mathbb Z}{ \Psi}{ \Psi} + {i\over 2} \bar{\mathbb Z} \bar{\Psi}  \bar{\Psi} \right) \; .
\qquad
\end{eqnarray}
The complex fermionic forms ${\cal E}^q = (\bar{\cal E}^q)^*$ and real bosonic form ${\rm E}^0$ are defined in \eqref{cEq=} and \eqref{rmE0=}, respectively.

As already seen in simpler examples, the dimensional reduction requires using the equations of motion for the extra bosonic coordinate, in this case $x^2(\tau)$. This coordinate appears only once, in the second term of \eqref{SmM0'=4D}, and its equation of motion
\be\label{x2=Eq}
{\rm d} \left({\cal M}- \frac 1 {{\cal M}} \, \frac {{\cal H}}{\mu^6}   \right)=0\;
\ee
implies
\be\label{cM-cH/cM=m}
{\cal M}- \frac 1 {{\cal M}} \, \frac {{\cal H}}{\mu^6} = m= \rm const\;
\ee
where $m$ is a constant of dimension of mass. Eq. \eqref{cM-cH/cM=m} can be solved by
\be\label{cM=m+-}
{\cal M}_\pm =\frac m 2\pm \sqrt{\frac {m^2} 4+\frac {{\cal H}}{\mu^6}}\; .
\ee
Only the solution with positive sign in front of the root is physical, since it approaches $m$ when ${\cal H}\to 0$ (in particular when all matrix fields vanish). Thus we will consider
\be\label{cM=m+}
{\cal M}= {\cal M} ({{\cal H}}/\mu^6) :=\frac m 2 + \sqrt{\frac {m^2} 4+\frac {{{\cal H}}}{\mu^6}}= m  \left[ 1 + \frac {{{\cal H}}}{m^2\mu^6} + {\cal O} \left( \left(\frac {{{\cal H}}}{m^2\mu^6}\right)^2\right) \right] \; .
\ee
The last expression can be viewed as a power series expansion in $1/\mu^6$, or equivalently, as a weak field expansion in the matrix fields. Below we will refer on it  as on decomposition in coupling constant $1/{\mu^6}$ just for convenience.

Substituting the solution \eqref{cM=m+} of the $x^2$ equation of motion into the action \eqref{SmM0'=4D} we find the following action for 3D counterpart of 10D mD$0$ system:
\begin{equation}
    \begin{array}l
         \begin{split}
             S_{\rm mD0}^{\rm 3D} & = \int_{{\cal W}^1} \left[m {\rm E}^{0}+ m ({\rm d}\theta^\alpha \bar{\theta}_\alpha- \theta^\alpha {\rm d}\bar{\theta}_\alpha )\right]  + {1\over \mu^6} \int_{{\cal W}^1}  {\rm E}^{0}\frac 2 {{\cal M}} \, {{\cal H}}~+\\
             &+ {1\over \mu^6}\int_{\mathcal{W}^1}
             {\rm tr}\left(\bar{\mathbb P}{\rm D} {\mathbb Z} + {\mathbb P}{\rm D} \bar{\mathbb Z} - {i\over 8} {\rm D}{ \Psi}\,  \bar{\Psi} + {i\over 8} { \Psi} {\rm D} \bar{\Psi}  \right)~-\\
             & -  {1\over \mu^6} \int_{\mathcal{W}^1} \frac {{\rm d}{\cal M}} {{\cal M}} {\rm tr}\left(\bar{\mathbb P} {\mathbb Z}+ {\mathbb P} \bar{\mathbb Z}\right) +\\
             &+ {1\over \mu^6} \int_{\mathcal{W}^1}   \frac i {\sqrt{2}\sqrt{{\cal M}}}\left( {\cal E}^{1}-i{\cal E}^{2}\right) {\rm tr}(\bar{\Psi}  \bar{\mathbb P}+   \Psi [{\mathbb Z},  \bar{\mathbb Z}])~+ \\  
             &+ {1\over \mu^6} \int_{\mathcal{W}^1}  \frac i {\sqrt{2}\sqrt{{\cal M}}}\left(\bar{ {\cal E}}^{1}+i\bar{{\cal E}}^{2}\right)
             {\rm tr}({\Psi}  {\mathbb P}+  \bar{\Psi} [{\mathbb Z},  \bar{\mathbb Z}])~,
         \end{split}
    \end{array}
    \label{SmD0=3D}  
\end{equation}
with  ${\cal H}$ defined in \eqref{tcH=}, bosonic and fermionic 1-forms  defined in
\eqref{rmE0=} and \eqref{cEq=}, covariant derivatives defined in \eqref{DZ=3d}, \eqref{DPsi=3d}
and ${\cal M}= {\cal M} ({\cal H}/\mu^6)$ given in \eqref{cM=m+}. Actually, as we will discuss below, the action \eqref{SmD0=3D} with arbitrary (nonvanishing) function ${\cal M} ({{\cal H}}/\mu^6)$ also makes sense.

Notice that the  derivative of ${\cal M}= {\cal M} ({\cal H}/\mu^6)$ (independently of whether it is given by  \eqref{cM=m+} or considered to be arbitrary)  is proportional to $1/ {\mu^6}$ (actually to $1 / m^2\mu^6$),
\be
{\rm d}{\cal M}  = \mathcal{M}'({\cal H}/\mu^6)\text{d}\mathcal{H} = \dfrac 1 {\mu^6} \, \dfrac { {\rm d}{{\cal H}}} {2\sqrt{\dfrac {m^2} 4+\dfrac {{{\cal H}}}{\mu^6}}}\; , \qquad \mathcal{M}'(y) = \dfrac{\text{d}}{\text{d}y} \mathcal{M}(y)~,
\ee
so that at the first order in $ 1 /{\mu^6} $ the action does not contain d${{\cal H}}$ and reads
\begin{equation}
    \begin{array}l
         \begin{split}
             S_{\rm mD0}^{\rm 3D}\vert_{{\cal M}\mapsto m} &= \int_{{\cal W}^1} \left[m {\rm E}^{0}+ m ({\rm d}\theta^\alpha \bar{\theta}_\alpha- \theta^\alpha {\rm d}\bar{\theta}_\alpha )\right]  + {1\over \mu^6} \int_{{\cal W}^1}  {\rm E}^{0}\frac 2 {{m}} \, {{\cal H}}~+\\
             &+ {1\over \mu^6} \int_{\mathcal{W}^1}
             {\rm tr}\left(\bar{\mathbb P}{\rm D} {\mathbb Z} + {\mathbb P}{\rm D} \bar{\mathbb Z} - {i\over 8} {\rm D}{ \Psi}\,  \bar{\Psi} + {i\over 8} { \Psi} {\rm D} \bar{\Psi}  \right)~+\\
             &+ {1\over \mu^6} \int_{\mathcal{W}^1}   \frac i {\sqrt{2}\sqrt{{m}}}\left( {\cal E}^{1}-i{\cal E}^{2}\right) {\rm tr}(\bar{\Psi}  \bar{\mathbb P}+   \Psi [{\mathbb Z},  \bar{\mathbb Z}])~+ \\  
             &+ {1\over \mu^6} \int_{\mathcal{W}^1}  \frac i {\sqrt{2}\sqrt{{m}}}\left(\bar{ {\cal E}}^{1}+i\bar{{\cal E}}^{2}\right)
             {\rm tr}({\Psi}  {\mathbb P}+  \bar{\Psi} [{\mathbb Z},  \bar{\mathbb Z}]) ~.
         \end{split}
    \end{array}
    \label{SmD0=3D=} 
\end{equation}

\section[Worldline supersymmetry of the \texorpdfstring{$\text{D}$}{D}=3 \texorpdfstring{${\cal N}$}{N}=2 mD0-brane action]{Worldline supersymmetry of the \boldmath\texorpdfstring{$\text{D}$}{D}=3 \boldmath\texorpdfstring{${\cal N}$}{N}=2 mD0-brane action}
In this section we show that the characteristic properties of the mD$0$ system are also exhibited by a more general class of models, described by the functional \eqref{SmD0=3D} with an arbitrary positive definite function ${\cal M} = {\cal M}({\cal H}/\mu^6)$. Namely we will show that such action possesses, besides the manifest $\text{D}=3$ ${\cal N}=2$ supersymmetry, also the worldline supersymmetry generalizing the $\kappa$-symmetry \eqref{kappa=D0} of single D$0$-brane action \eqref{SD0D3=}.

\subsection{Worldline supersymmetry transformations of the center of mass variables}
Our previous experience with the 4D nAmW system and the 10D action studied in \cite{10D_mD0_Igor} suggested that, when investigating the worldline supersymmetry of the mD$0$-brane system, it is natural to assume that it acts on the center of mass variables of the mD$0$ system, i.e. on the coordinate functions and spinor frame variables,  in the same manner as the $\kappa$-symmetry of single D$0$ brane,
\begin{equation}
    \begin{array}{lr}
      \delta_\epsilon \theta^\alpha =  \dfrac 1 {\sqrt{2}} ({\rm v}^{\alpha 2}- i{\rm v}^{\alpha 1}) \epsilon ~,  & \delta_\epsilon \bar{\theta}{}^\alpha = \dfrac 1 {\sqrt{2}}  ({\rm v}^{\alpha 2}+ i{\rm v}^{\alpha 1})\, \bar{\epsilon}\; , \\
        ~& \delta_\epsilon x^{\alpha\beta}= 2i \delta_\epsilon \theta^{(\alpha}  \bar{\theta}{}^{\beta)}-  2i  \theta^{(\alpha} \delta_\epsilon \bar{\theta}{}^{\beta)}\; ,\\
        ~& \delta_\epsilon {\rm  v}^q_\alpha =0\; .
    \end{array}
    \label{kappa=mD0}
\end{equation}
In the formalism presented in section~\ref{sec:diff_SuperForms} this is tantamount to stating that
\begin{equation}
    \begin{array}{r}
\iota_\epsilon ({\cal E}^{1} + i {\cal E}^{2})=0 \; , \qquad \iota_\epsilon  (\bar{{\cal E}}{}^{1} -i \bar{{\cal E}}{}^{2})=0 \; , \qquad \iota_\epsilon {\rm E}^{0} =0\; , \qquad  \iota_\epsilon {\rm E}^{I} =0\; ,  \\
\iota_\epsilon  f^I=0\; , \qquad \iota_\epsilon  f^{qq}=0 ~~~~\qquad \Longrightarrow \qquad \delta_\epsilon {\rm v}_\alpha^\mp =0 \; , \qquad
    \end{array}
    \label{kappa==}
\end{equation}
and, consequently,
\begin{equation}
    \begin{array}{ccc}
        \iota_\epsilon  {\cal E}^{1} = \epsilon/\sqrt{2}  \qquad & \Longrightarrow & \qquad ~~\iota_{\epsilon}{\cal E}^{2} =\;  i\epsilon/\sqrt{2}  \; , \;\qquad ~\iota_\epsilon ({\cal E}^1-i {\cal E}^2)= \sqrt{2} \epsilon  \; , \\
        \iota_\epsilon  \bar{{\cal E}}{}^{1} =\bar{\epsilon } /\sqrt{2} \qquad & \Longrightarrow & ~~\qquad   \iota_\epsilon  \bar{{\cal E}}{}^{2} =-i\bar{\epsilon }/\sqrt{2} \; , \qquad  \iota_\epsilon (\bar{{\cal E}}{}^1+i \bar{{\cal E}}{}^2)= \sqrt{2} \bar{\epsilon }  \; .
    \end{array}\label{kappa=}
\end{equation}
In this form it is easier to find  the following transformation properties of the bosonic and fermionic 1-forms entering the part of the action containing the matrix fields
\bea
\label{1dSG=vE0}
\delta_{\epsilon }{\rm E}^0= -2i\frac {1}{\sqrt{2}} ({\cal E}^1-i {\cal E}^2)\bar{\epsilon } -2i\frac {1}{\sqrt{2}} (\bar{\cal E}{}^1+i \bar{\cal E}{}^2)\epsilon \; , \qquad \nonumber \\
\delta_\epsilon ({\cal E}^1-i {\cal E}^2)= \sqrt{2} {\rm D} \epsilon  \; , \qquad
\delta_\epsilon  (\bar{\cal E}{}^1+i \bar{\cal E}{}^2)=\sqrt{2} {\rm D}\bar{\epsilon } \; , \qquad
  \eea
which are the typical transformations of the $\text{d}=1$ ${\cal N}=2$ supergravity multiplet. This supergravity is induced by embedding of the worldline into $\text{D}=3$ ${\cal N}=2$ superspace  ({\it cf.} \eqref{wsSG4D} and discussion around it). Eqs. \eqref{1dSG=vE0} are useful to search for the worldline supersymmetry invariance of the part of the action containing the matrix fields.

\subsection{Worldline supersymmetry transformations of the matrix matter fields }
Now, writing the variation of the action \eqref{SmD0=3D} with \eqref{kappa=mD0}-\eqref{1dSG=vE0} and extracting from this variation the terms  proportional to D${\mathbb P}$,  D${\mathbb Z}$, D${\Psi}$ and their hermitian conjugates, we find the following equations for the basic $\kappa$-symmetry  variations of matrix ``matter'' fields
\begin{equation}
    \label{vkZ=}
\delta_\epsilon {\mathbb Z} = -\frac i {\sqrt{\cal M}}\epsilon \bar{\Psi} + \frac 1 {\mu^6} \, \frac{{\cal M}^\prime }{{\cal M}}\; \left({\mathbb Z} \delta_\epsilon {\tilde{{\cal H}}} - {\mathbb P} \Delta_\epsilon {\cal K} \right)\; ,
\end{equation}
\begin{equation}
    \delta_\epsilon \bar{{\mathbb Z}} = -\frac i {\sqrt{\cal M}}\bar{\epsilon} \Psi + \frac 1 {\mu^6} \, \frac{{\cal M}^\prime }{{\cal M}}\; \left(\bar{{\mathbb Z}}  \delta_\epsilon {\tilde{{\cal H}}} -\bar{{\mathbb P}}  \Delta_\epsilon {\cal K} \right) \;,
\end{equation}
\begin{equation}
    \begin{array}{c}
        \begin{split}
           ~~~~~~ \delta_\epsilon {\mathbb P} &= \frac i {\sqrt{\cal M}}\left(\epsilon [\Psi,{{\mathbb Z}} ]+  \bar{\epsilon}  [\bar{\Psi},{\mathbb Z} ] \right) - \frac 1 {\mu^6} \, \frac{{\cal M}^\prime }{{\cal M}}\; {\mathbb P}  \delta_\epsilon {\tilde{{\cal H}}}~+ \\
 &+ \frac 1 {\mu^6} \, \frac{2{\cal M}^\prime }{{\cal M}}\; \left([[{\mathbb Z},\bar{{\mathbb Z}}],{\mathbb Z}]+\frac i 4 \bar{\Psi}\bar{\Psi}\right) \,  \Delta_\epsilon {\cal K}\;, 
        \end{split} 
    \end{array}
\end{equation}
\begin{equation}
    \begin{array}{c}
         \begin{split}
             ~~~~~ \delta_\epsilon \bar{{\mathbb P}} &= \frac i {\sqrt{\cal M}}\left(\epsilon [\bar{{\mathbb Z}},\Psi ]+  \bar{\epsilon}  [\bar{{\mathbb Z}},\bar{\Psi} ] \right) - \frac 1 {\mu^6} \, \frac{{\cal M}^\prime }{{\cal M}}\; \bar{{\mathbb P}}  \delta_\epsilon {\tilde{{\cal H}}}-\\
             &- \frac 1 {\mu^6} \, \frac{2{\cal M}^\prime }{{\cal M}}\; \left([[{\mathbb Z},\bar{{\mathbb Z}}],\bar{{\mathbb Z}}]+\frac i 4 \Psi{\Psi}\right) \,  \Delta_\epsilon {\cal K}\;, 
         \end{split}
    \end{array}
\end{equation}
\begin{equation}
    \qquad \delta_\epsilon {\Psi} = \frac 4 {\sqrt{\cal M}}\left(\epsilon  \bar{{\mathbb P}} + \bar{\epsilon} [{\mathbb Z}, \bar{{\mathbb Z}}] ] \right) - \frac 1 {\mu^6} \, \frac{2{\cal M}^\prime }{{\cal M}}\; [\bar{\Psi},\bar{{\mathbb Z}}]   \Delta_\epsilon {\cal K}\; , \\ \label{vkbPsi=}
\end{equation}
\begin{equation}
    ~~~~~~~~  \delta_\epsilon \bar{{\Psi}} = \frac 4 {\sqrt{\cal M}}\left(\epsilon [{\mathbb Z}, \bar{{\mathbb Z}}]+  \bar{\epsilon}  {\mathbb P}] \right) + \frac 1 {\mu^6} \, \frac{2{\cal M}^\prime }{{\cal M}}\; [\Psi,{\mathbb Z} ]   \Delta_\epsilon {\cal K}\; .
\end{equation}
These expressions are treated as equations because their right-hand sides involve the variations
$\delta_{\epsilon} {{{\cal H}}}$ of $ {{{\cal H}}} $ from  \eqref{tcH=} and
\be
\Delta_\epsilon {\cal K}=\delta_\epsilon {\cal K} -\frac i{2\sqrt{{\cal M}}}\, (\epsilon\, {\nu}  +\bar{\epsilon}\bar{\nu})
\ee
where
\be
\nu = {\rm tr}\left(\bar{\Psi}\bar{{\mathbb P}} +\Psi[{\mathbb Z},\bar{{\mathbb Z}}]  \right)\; , \qquad  \bar{\nu}= {\rm tr}\left(\Psi{\mathbb P} +\bar{\Psi}[{\mathbb Z},\bar{{\mathbb Z}}]  \right)
\ee
and $\delta_{\epsilon} {\cal K}$ is the variation
 of
\be
{\cal K} ={\rm tr}(\bar{{\mathbb P}}{\mathbb Z}+{\mathbb P}\bar{{\mathbb Z}})\, .
\ee
As both of these composite variations enter with coefficients proportional to $1/\mu^6$, the equations certainly admit a solution (at least in the form of a power series in $1/\mu^6$). Moreover, there is a straightforward method for solving these equations: first, use them to derive closed algebraic expressions for $\delta_{\epsilon} {\cal H}$ and $\delta_{\epsilon} {\cal K}$; second, solve these expressions; and third, substitute the resulting solutions into the final terms of equations \eqref{vkZ=}-\eqref{vkbPsi=}.

Indeed,  using \eqref{vkZ=}-\eqref{vkbPsi=} we find
\be\label{vtcH=Eq}
 \delta_\epsilon  {{{\cal H}}} = \frac i{\sqrt{{\cal M}}}\, {\rm tr}\left[\left(\bar{\epsilon}\bar{\Psi}-\epsilon\Psi \right)\, \left([{\mathbb Z},\bar{{\mathbb P}}]+ [\bar{{\mathbb Z}},{\mathbb P}] -\frac i 4 \{\Psi\, , \bar{\Psi}\} \right)\right]   + \frac 1 {\mu^6} \, \frac {{\cal M}^\prime }{{\cal M}}\, {\frak{H} }\, \delta_\epsilon  {{{\cal H}}}
\ee
and
\be\label{vcK=Eq}
 \delta_\epsilon  {\cal K} = -\frac i{\sqrt{{\cal M}}}\, \epsilon\, {\rm tr}\left(\bar{\Psi}\bar{{\mathbb P}} -2\Psi[{\mathbb Z},\bar{{\mathbb Z}}]  \right)\, -\frac i{\sqrt{{\cal M}}}\, \bar{\epsilon}\, {\rm tr}\left(\Psi{\mathbb P} -2\bar{\Psi}[{\mathbb Z},\bar{{\mathbb Z}}]  \right)\,  + \frac 1 {\mu^6} \, \frac {{\cal M}^\prime }{{\cal M}}\, {\frak{H} }\, \Delta_\epsilon  {\cal K}\; ,
\ee
where
\be
\label{h=} {\frak{H}}= {\rm tr}\left(-2 {\mathbb P} \bar{\mathbb P} +  4 [{\mathbb Z},  \bar{\mathbb Z}]^2 -
{i\over 2} {\mathbb Z}{ \Psi}{ \Psi} + {i\over 2} \bar{\mathbb Z} \bar{\Psi}  \bar{\Psi} \right) \;
\qquad
\ee
 ({\it cf.} \eqref{tcH=}). Eqs. \eqref{vtcH=Eq} and \eqref{vcK=Eq} are closed algebraic equations for $\delta_\epsilon  {{{\cal H}}}$ and $\delta_\epsilon  {\cal K} $, respectively, which are solved by
\bea\label{vtcH=}
 \delta_\epsilon  {{{\cal H}}} = \frac i{\sqrt{{\cal M}}}\,\frac 1 {\left(1- \dfrac 1 {\mu^6} \, \dfrac {{\cal M}^\prime }{{\cal M}}\, {\frak{H} }\right)}\;   {\rm tr}\left[\left(\bar{\epsilon}\bar{\Psi}-\epsilon\Psi \right)\, \left([{\mathbb Z},\bar{{\mathbb P}}]+ [\bar{{\mathbb Z}},{\mathbb P}] -\frac i 4 \{\Psi\, , \bar{\Psi}\} \right)\right]
\eea
and
\bea\label{vcK=}
 ~~\Delta_\epsilon  {\cal K} = -\frac {3i}{\sqrt{{\cal M}}}\, \frac 1 {\left(1- \dfrac 1 {\mu^6} \, \dfrac {{\cal M}^\prime }{{\cal M}}\, {\frak{H} }\right)}\; \left[ \epsilon\, {\rm tr}\left(\bar{\Psi}\bar{{\mathbb P}} -\Psi[{\mathbb Z},\bar{{\mathbb Z}}]  \right)\,+ \bar{\epsilon}\, {\rm tr}\left(\Psi{\mathbb P} -\bar{\Psi}[{\mathbb Z},\bar{{\mathbb Z}}]  \right)\right] \; .
\eea
Thus the worldline supersymmetry transformations of the matrix matter fields are given by  \eqref{vkZ=}-\eqref{vkbPsi=}  with  \eqref{vtcH=} and \eqref{vcK=}.

\subsection{Worldline supersymmetry transformations of the non-Abelian gauge field }
Taking into account the above relations, we find that the remaining expression of the action variation contains the terms  proportional to the pull-backs of bosonic and fermionic supervielbein forms to the worldline, namely to
\be\label{E0cE0=1dSG}
{\rm E}^0\; ,\qquad ({\cal E}^1-i{\cal E}^2) \; ,\qquad  (\bar{{\cal E}}^1+i\bar{{\cal E}}^2)\;
\ee
and to the variation of SU$(N)$ gauge field 1-form $\delta {\mathbb A}$. This implies that, if the action is invariant under worldline supersymmetry, then
\be\label{vbbA=}
\delta_\epsilon {\mathbb A} =
{\rm E}^0\delta_\epsilon {\mathbb A} _0+({\cal E}^1-i{\cal E}^2)\delta_\epsilon {\mathbb A}_\eta + (\bar{{\cal E}}^1+i\bar{{\cal E}}^2)\delta_\epsilon {\mathbb A}_{\bar{\eta}}
\; .\qquad \ee
It is important to notice that $\delta_\epsilon {\mathbb A}$ enters the action variation in the trace of its product with 
\be
 {\rm tr}\left[\delta_\epsilon {\mathbb A}\, \left([{\mathbb Z},\bar{{\mathbb P}}]+ [\bar{{\mathbb Z}},{\mathbb P}] -\frac i 4 \{\Psi,\bar{\Psi}\} \right)\right]\; ,
\ee
so that the possibility of compensating all remaining terms in the action variation by choosing appropriate $\delta_\epsilon {\mathbb A}_0$, $\delta_\epsilon {\mathbb A} _\eta$ and $\delta_\epsilon {\mathbb A} _{\bar{\eta}}$ is a nontrivial check of consistency of our calculations.

For instance, this implies that the condition of vanishing the contribution proportional to ${\rm E}^0$ in the $\kappa$-symmetry variation of the action,
\be
{\rm tr}\left[\delta_\epsilon {\mathbb A}_0\, \left([{\mathbb Z},\bar{{\mathbb P}}]+ [\bar{{\mathbb Z}},{\mathbb P}] -\frac i 4 \{\Psi,\bar{\Psi}\} \right)\right] = -\frac 2 {{\cal M}}\, \left(1-\frac 1 {\mu^6} \, \frac {{\cal M}^\prime }{{\cal M}}\right)\, \delta_\epsilon  {{\cal H}}\; ,
\ee
can be solved because  $\delta_\epsilon {\cal H}$ \eqref{vtcH=} is also given by the trace of certain expression with 
\begin{equation*}
    \left([{\mathbb Z},\bar{{\mathbb P}}]+ [\bar{{\mathbb Z}},{\mathbb P}] -\frac i 4 \{\Psi ,\bar{\Psi}\} \right)~.
\end{equation*}
This allows to obtain
\bea
\delta_\epsilon {\mathbb A}_0 &=&
 - \frac  {\left(1- \dfrac 1 {\mu^6} \, \dfrac {{\cal M}^\prime }{{\cal M}}\, {\cal H}\right)} {\left(1- \dfrac 1 {\mu^6} \, \dfrac {{\cal M}^\prime }{{\cal M}}\, {\frak{H} }\right)}\;  \frac {2i}{{\cal M}\sqrt{{\cal M}}}\, \left(\bar{\epsilon}\bar{\Psi}-\epsilon\Psi \right)\; .  \qquad
 \eea
Similarly studying the terms proportional to $({\cal E}^1-i{\cal E}^2)$ and their c.c.-s, we obtain
\begin{equation}
    \begin{array}{l}
         \begin{split}
             \delta_\epsilon {\mathbb A}_\eta &= \frac {8i}{\sqrt{2}{\cal M}}\, \epsilon \bar{{\mathbb Z}}- \frac 1 {\mu^6}\, \,  \frac {{\cal M}^\prime }{2\sqrt{2}{\cal M}^2}
             \left[2i\Psi \sqrt{{\cal M}}\Delta_\epsilon {\cal K}- \frac {  3\left(\epsilon\Psi -\bar{\epsilon}\bar{\Psi}\right) {\rm tr}\left(\bar{{\mathbb P}}\bar{\Psi} - \Psi [{\mathbb Z},\bar{{\mathbb Z}}]\right)}  {\left(1- \dfrac 1 {\mu^6} \, \dfrac {{\cal M}^\prime }{{\cal M}}\, {\frak{H} }\right)} \right] =\\
             & = \frac {8i}{\sqrt{2}{\cal M}}\, \epsilon \bar{{\mathbb Z}}-   \frac 1 {\mu^6}\, \,  \frac {3{\cal M}^\prime }{2\sqrt{2}{\cal M}^2}\frac  {1} {\left(1- \dfrac 1 {\mu^6} \, \dfrac {{\cal M}^\prime }{{\cal M}}\, {\frak{H} }\right)} \times \\
             & \times
              (\bar{\epsilon}\bar{\Psi}-2{\epsilon}{\Psi}) {\rm tr}\left(\bar{{\mathbb P}}\bar{\Psi} -  \Psi [{\mathbb Z},\bar{{\mathbb Z}}]\right)  +  \bar{\epsilon}\Psi {\rm tr}\left[\left(\bar{{\mathbb P}}\bar{\Psi} - \,\Psi [{\mathbb Z},\bar{{\mathbb Z}}]\right)\right]\;
         \end{split}
    \end{array}\label{Aeta=}
\end{equation}
and
\begin{equation}
    \begin{array}{l}
         \begin{split}
             \delta_\epsilon {\mathbb A}_{\bar{\eta}} &=  - \frac {8i}{\sqrt{2}{\cal M}}\, \bar{\epsilon} {\mathbb Z}+ \frac 1 {\mu^6}\, \,  \frac {3{\cal M}^\prime }{2\sqrt{2}{\cal M}^2}
             \frac  {\epsilon\Psi -\bar{\epsilon}\bar{\Psi}} {\left(1- \dfrac 1 {\mu^6} \, \dfrac {{\cal M}^\prime }{{\cal M}}\, {\frak{H} }\right)}\;   {\rm tr}\left({\mathbb P}\Psi - \bar{\Psi} [{\mathbb Z},\bar{{\mathbb Z}}]\right) +\\
             &+ \frac 1 {\mu^6} \, \frac {i{\cal M}^\prime }{\sqrt{2}\sqrt{{\cal M}}{\cal M}}\;\bar{\Psi}\; \Delta_\epsilon {\cal K}  \, .
         \end{split}
    \end{array}
\end{equation}
The second of these equations can be further specified with the use of  \eqref{vcK=}.

\subsection{Resume on the  candidate mD0 action(s)}
Thus, we have shown that the action \eqref{SmD0=3D}, with a positive definite function ${\cal M}({\cal H}/\mu^6)$ of the matrix field Hamiltonian \eqref{tcH=}, possesses not only the target superspace $\text{D}=3$ ${\cal N}=2$ supersymmetry but also local worldline supersymmetry. This local symmetry generalizes the $\kappa$-symmetry of the single D$0$-brane action. Consequently, any of these actions can be considered as a candidate for the role of $\text{D}=3$ counterpart of the multiple D$0$-brane action.

A special representative of this family is the action where ${\cal M}({\cal H}/\mu^6)$ is given by \eqref{cM=m+}, as it arises from the dimensional reduction of the $\text{D}=4$ counterpart of the mM$0$-brane system (4D nAmW action). In general, the actions \eqref{SmD0=3D} are essentially nonlinear but the simplest representation of the family with ${\cal M}({\cal H}/\mu^6) = m = \text{const}$, which provides the counterpart to the 10D action considered as a candidate for the mD$0$ action in \cite{10D_mD0_Igor}. Our study suggests that more generic, essentially nonlinear candidates for the role of the 10D mD$0$ action should be explored, and this investigation is actually presented in the next section.

We will return to this $\text{D}=3$ counterpart of the 10D mD$0$ system in chapter~\ref{ch.quantization}, where we will construct the Hamiltonian formulation and perform the covariant quantization of the simplest\footnote{As we will see in chapter~\ref{ch.quantization}, ``simplest'' corresponds to the 3D mD$0$ representative with $\mathcal{M}(\mathcal{H}/\mu^6)=m$.} representative within this diverse zoo of 3D mD$0$ models.

      \chapter[Complete nonlinear action for supersymmetric multiple D0-brane system]{Complete nonlinear action for supersymmetric multiple D0-brane system}\label{ch.10D_mD0}
\thispagestyle{empty}

 \vspace*{-1.5cm}   
\begin{changemargin}{4cm}{0cm}
    \singlespacing\textcolor{cites}{ \small{
         \begin{flushright}
        ...tenía la rara virtud de no existir por completo\\ sino en el momento oportuno.
         \end{flushright}
     \begin{flushright}     
         {\sffamily {\textit{Cien años de soledad}}\\
         {by {Gabriel García Márquez}.}}
     \end{flushright}}}
    \end{changemargin}
    \vspace{12pt}

In this section we present a complete nonlinear action for the dynamical system of 10D multiple D$0$-branes. This action exhibits not only manifest spacetime (target superspace) supersymmetry but also worldline supersymmetry, the generalization of the local fermionic $\kappa$-symmetry characteristic of a single D$0$-brane. Moreover, as its lower dimensional 3D counterpart, this action includes an arbitrary positive definite function ${\cal M}({\cal H})$ of the SYM Hamiltonian ${\cal H}$ constructed from the matrix fields describing relative motion of mD$0$ constituents. 

The properties of this set of candidate actions will be analysed in this section, while the $11$-dimensional origin of a specific $\text{D}=10$ mD$0$ model, corresponding to a particular choice of ${\cal M}({\cal H})$, will be discussed in section~\ref{ch.11D_origin}.

\section{The 10D mD0 action and its symmetries}
The nonlinear action that we have constructed is expressed in terms of center of mass variables of mD$0$ system, which are identical to those used in the single D$0$-brane case, and matrix variables describing the relative motion of mD$0$ constituents. The field content of such systems is motivated by their low energy (and gauge fixed) description which, according to \cite{Witten_1996}, is given by U($N$) SYM theory. This generalizes the well-known fact that, after suitable gauge fixing, the low energy description of a single D$p$-brane is provided by an Abelian U$(1)$ SYM theory. Consequently, the Lorentz-covariant formulation of the mD$p$-brane system is expected to include both the fields known from the single D$p$-brane description,  which are essentially coordinate functions and Abelian gauge field, and the fields of the maximally supersymmetric SU($N$) SYM multiplet in $\text{d}=p+1$ dimensions~\cite{Witten_1996}.

Thus, focusing on the mD$0$-brane case, the set of worldline fields includes coordinate functions 
\begin{equation}
    z^M(\tau)= (x^\mu(\tau),\theta^{1 \alpha}(\tau), \theta^2_{\alpha}(\tau) )~, \qquad \mu=0,\ldots,9~, \qquad \alpha=1,\ldots,16~,
\end{equation}
which describe the embedding of the worldline ${\cal W}^1$ into the target type IIA superspace $\Sigma^{(10|32)}$ whose  bosonic and fermionic  coordinates are $z^M= (x^\mu,\theta^{\alpha 1},\theta_{\alpha}^{2})$,
\be
{\cal W}^1 \; \subset \; \Sigma^{(10|32)} \; : \qquad z^M= z^M(\tau)\; ~.
\ee
In addition, according to~\cite{Witten_1996}, the system includes matrix fields from the $\text{d}=1$ ${\cal N}=16$ SU($N$) SYM multiplet, and some auxiliary fields. The set of these latter includes the 10D spinor moving frame fields, discussed in detail below, and the matrix momentum field which allows the SYM sector to be written in a first order form.

The matrix field content of the mD$0$ system includes $9+9$ bosonic Hermitian traceless $N \times N$ matrix fields
\begin{equation}\label{matter_bosonic}
    {\mathbb X}^i(\tau)~, \quad {\mathbb P}^i(\tau)  
\end{equation}
where ${\mathbb P}^i$ serves as the conjugate momentum to the matrix field ${\mathbb X}^i$ which is a physical bosonic field of the SU$(N)$ SYM multiplet; both are labelled by the SO$(9)$ vector index $i=1,\ldots,9$. The system also includes an anti-Hermitian bosonic traceless $N \times N$ matrix valued 1-form
\begin{equation} 
{\mathbb A} = \text{d}\tau {\mathbb A}_\tau~, \label{bfA=} \end{equation}
where ${\mathbb A}_\tau(\tau)$ is a worldline gauge field valued in the Lie algebra $\mathfrak{su}(N)$. The fermionic part consists in 16 fermionic Hermitian traceless $N\times N$ matrix fields
\begin{equation}\label{matter_fermionic}
    {\boldsymbol{\Psi}}_q (\tau)~,
\end{equation}
which transform as a spinor under SO$(9)$ (i.e. as Spin$(9)$ fundamental representation) with indices $q=1,\ldots,16$. This completes the set of matrix fields which captures the dynamics of the relative motion sector of the mD$0$ system.

\subsection{10D Moving frame and spinor moving frame fields}
Finally to write the candidate for mD$0$ action we need to introduce the moving frame and spinor moving frame fields. These generalize to $\text{D}=10$ the moving frame and spinor moving frame formalism described in chapter~\ref{ch.MF_and SMF}.

The moving frame vectors are described by Lorentz group valued 10$\times$10 matrix
\be\label{harmU=10}
(u_\mu^{0}, u_\mu^{i})\in \text{SO}(1,9)~, \qquad i= 1,\ldots,9
\ee
which obey
\be\label{u02=1}
u^{\mu 0}u_\mu^{0}=1\; , \qquad u^{\mu 0} u_\mu^{i}=0\; , \qquad u^{\mu i} u_\mu^{j}=-\delta ^{ij}\; .
\ee
The spinor moving frame is described by Spin$(1,9)$ valued matrix
\begin{eqnarray}\label{harmV=10}
v_\alpha{}^q \in \text{Spin}(1,9)\;.
\end{eqnarray}
This spinor frame is related to the above described moving frame by the conditions of the Lorentz invariance of 10D sigma-matrices
\be\label{us=vsv10D} u^{(b)}_\mu \sigma^\mu_{\alpha\beta}= v_{\alpha}^q \sigma^{(b)}_{qp}v_{\beta}^p\; , \qquad
u^{(b)}_\mu \tilde{\sigma}{}_{(b)}^{qp}= v_{\alpha}^q \tilde{\sigma}{}_{\mu}^{\alpha\beta}v_{\beta}^p\, .
\ee
providing a kind of square root of the moving frame vectors in the sense of Cartan-Penrose-like relations. The indices $q, p = 1, \dots, 16$ correspond to SO$(9)$ spinor indices, reflecting the local SO$(9)$ gauge symmetry that preserves the splitting in \eqref{harmU=10},
\be\label{SO9}
u^{\mu 0}(\tau) \mapsto u^{\mu 0}(\tau) \; , \qquad u^{\mu i}(\tau) \mapsto u^{\mu j}(\tau) {\cal O}^{ji}(\tau)\; , \qquad  {\cal O}^{ji} {\cal O}^{jk}=\delta^{ik}\; .
\ee
Choosing an SO$(9)$ invariant representation of sigma-matrices
\be \sigma^{(a)}_{qp}=(\delta_{qp}, \gamma^i_{qp})=\tilde{\sigma}{}_{(a)}^{qp}\; , \qquad \gamma^i_{qp}=\gamma^i_{pq}\; , \qquad \gamma^{(i}\gamma^{j)}=\delta^{ij} \ee in which $\gamma^i_{qp}=\gamma^i_{pq}$ are $\text{d}=9$  gamma matrices, we find that \eqref{us=vsv10D} leads to
\begin{eqnarray}\label{u0s=vv}
\sigma^{{\mathbf 0}}_{\alpha\beta}:=  u_\mu^{{0}} \sigma^\mu_{\alpha\beta}=v_\alpha{}^q v_\beta{}^q \; , \qquad
~\sigma^{{\mathbf i}}_{\alpha\beta}:= u_\mu^{{i}} \sigma^\mu_{\alpha\beta}=v_\alpha{}^q \gamma^i_{qp}v_\beta{}^p \;  \qquad
\\ \nonumber {\rm and}\qquad
v_{\alpha}^q \tilde{\sigma}{}_{\mu}^{\alpha\beta}v_{\beta}^p= u_\mu^{{0}} \delta_{qp}+u_\mu^{{i}} \gamma^i_{qp}\; .  \qquad
\end{eqnarray}
Unlike their $\text{D}=4$ counterparts (see \cite{Igor_Russian} and section~\ref{sec:MF_and_SMF}), these relations~\eqref{u0s=vv}  impose strong constraints on the spinor moving frame field $v_\alpha{}^q$ reducing its initial $16\times16=256$ components to $45={\rm dim}(\text{SO}(1,9))$.
Additionally, we can define
\bea
\label{u0ts=vv}
\tilde{\sigma}^{{\mathbf 0}\alpha\beta}:= u_\mu^0 \tilde{\sigma}^{\mu \alpha\beta}=v_q{}^\alpha v_q{}^\beta \; , \qquad
\tilde{\sigma}^{{\mathbf i}\alpha\beta}:= u_\mu^i \tilde{\sigma}^{\mu\alpha\beta}=- v_q{}^\alpha  \gamma^i_{qp} v_p{}^\beta\; \qquad
\eea
where
\begin{eqnarray}\label{vs=v-1}
v_\alpha^q \tilde{\sigma}^{{\mathbf 0}\alpha\beta}=  v_q{}^\beta\; , \qquad  {\sigma}^{{\mathbf 0}}_{\alpha\beta}v_q{}^\beta  = v_\alpha{}^q\;.
\end{eqnarray}
This matrix $v_q{}^\alpha$ satisfies the relations:
\begin{eqnarray}\label{harmV-1=10D0}
v_q{}^\alpha v_\alpha{}^p=\delta_q{}^p \qquad \Longleftrightarrow \qquad
 v_\alpha{}^q v_q{}^\beta= \delta_\alpha{}^\beta
\;  \qquad
\end{eqnarray}
and hence can be identified as inverse spinor frame matrix. Below we will also use the Cartan forms
 \begin{eqnarray}\label{Omi=}
 \Omega^i= u_\mu^0 \text{d}u^{\mu i} , \qquad  \\ \label{Omij=}
  \Omega^{ij}=u_\mu^i \text{d}u^{\mu j}\;  \qquad
\end{eqnarray}
where $\Omega^i$ transforms covariantly under SO$(9)$ gauge group~\eqref{SO9}, and $\Omega^{ij}$ serves as the SO$(9)$ connection. Using \eqref{u02=1} derivatives of the moving frame can be expressed in terms of Cartan forms:
 \begin{eqnarray}\label{Du0=}
 \text{D}u_\mu^0:= \text{d}u_\mu^0= u_\mu^i \Omega^i\; , \qquad \text{D}u_\mu^{i}:=\text{d}u_\mu^{i} + u_\mu^{j}\Omega^{ji}=u_\mu^0 \Omega^{i}\; . \qquad
\end{eqnarray}
Here we have also introduced the SO(9) covariant derivatives with the composite connection given by the Cartan form $\Omega^{ji}$ \eqref{Omij=}.

The spinor moving frame derivatives are also expressed in terms of the same SO$(1,9)$ Cartan forms by
\begin{eqnarray}\label{Dv=vOm}
\text{D}v_\alpha{}^q:= \text{d}v_\alpha{}^q+ \frac 1 4 \Omega^{ij} v_\alpha{}^p\gamma^{ij}_{pq}
= \frac 1 2 \gamma^i_{qp} v_\alpha{}^p\Omega^i \; \qquad
\end{eqnarray}
which implies
\begin{eqnarray}\label{Dv-1=vOm}
\text{D}v_q^\alpha := \text{d}v_q^\alpha-  \frac 1 4 \Omega^{ij} \gamma^{ij}_{qp}v_p^\alpha
= -\frac 1 2 v_p^\alpha\gamma^i_{pq} \Omega^i \; . \qquad
\end{eqnarray}
The variation of the moving frame and spinor moving frame variables can be also expressed in terms of Cartan forms or, more precisely, in terms of contraction of these with the variational symbol:
\be
\iota_\delta \Omega^i = u^{\mu 0}\delta u_\mu^i\; , \qquad \iota_\delta \Omega^{ij} = u^{\mu i}\delta u_\mu^j\; . \qquad
\ee
The latter parametrize the SO$(9)$ transformations which will be the manifest gauge symmetry of our action. Thus, the essential variations of moving frame vectors and spinor moving frame matrices are given by
\bea
&& \delta u_\mu^0= u_\mu^i \iota_\delta  \Omega^i\; , \qquad \delta u_\mu^{i}=u_\mu^0 \iota_\delta  \Omega^{i} \label{eq:delta_u0}
\; , \qquad  \\
&& \delta v_\alpha{}^q
= \frac 1 2 \gamma^i_{qp} v_\alpha{}^p\iota_\delta  \Omega^i\; , \qquad \delta v_q^\alpha  =-\frac 1 2 \gamma^i_{qp}v_p^\alpha \iota_\delta  \Omega^i~. \label{eq:delta_vq_vq}
\eea
The moving frame formalism allows to define the Lorentz invariant 1-forms on the worldline by contracting the pull-back of the VA 1-form of the type IIA superspace
\be\label{Pi=VA}
\Pi^\mu = \text{d}x^\mu -i\text{d}\theta^1\sigma^\mu\theta^1 -i\text{d}\theta^2\tilde{\sigma}^\mu\theta^2 \;  \ee
with the moving frame vectors:
\be\label{E0:=}
\text{E}^0= \Pi^\mu u_\mu^0 \; , \qquad \text{E}^i= \Pi^\mu u_\mu^i \; . \qquad
\ee
Similarly, spinor moving frame formalism allows to define on the worldline the following Lorentz invariant fermionic 1-forms:
\be\label{Eq1:=}
\text{E}^{1q}= \text{d}\theta^{\alpha 1} v_\alpha^{\; q}\, , \qquad \text{E}^{2}_{q}= \text{d}\theta_{\alpha}^{ 2} v_q^\alpha\; . \qquad
\ee

\subsection{Candidate mD0 action(s)}
\label{sec:10D_mD0_candidates}
The action for $N$ nearly coincident D$0$-brane system has the form
\begin{equation}
    \begin{array}{l}
        \begin{split}
             S_{\text{mD0}} &= m \int_{\mathcal{W}^1} \text{E}^{0} -im  \int_{\mathcal{W}^1} (\text{d}\theta^1\theta^2-\theta^1 \text{d}\theta^2) - \frac 1 {\mu^6} \int_{\mathcal{W}^1} \frac {\text{d} {\cal M}}{ {\cal M} }
             {\rm tr} ({\mathbb P}^i{\mathbb X}^i)~  + \\
             &~\\
             &+  \frac 1 {\mu^6} \int_{\mathcal{W}^1}   \left[\text{tr}\left({\mathbb P}^i \text{D} {\mathbb X}^i + 4i {\boldsymbol{ \Psi}}_q \text{D}
             {\boldsymbol{ \Psi}}_q  \right) +  \frac 2 {\cal M} \text{E}^{0}\, {\cal H}\right]~+\\
             &~\\
             &+  \frac 1 {\mu^6}  \int_{\mathcal{W}^1}
             \frac 1 {\sqrt{2{\cal M}}}(\text{E}{}^{1q}-\text{E}{}^{2}_{q}) {\rm tr}\left[-4i (\gamma^i {\boldsymbol{ \Psi}})_q  {\mathbb P}^i  + {1\over 2} (\gamma^{ij} {\boldsymbol{ \Psi}})_q  [{\mathbb X}^i, {\mathbb X}^j]  \right] ~,
        \end{split}
    \end{array} \label{eq:10D_SmD0=} 
\end{equation}
where  $m$ and $\mu$ are constants of dimension of mass. In the above equation, $\text{E}^{0}$ is the projection \eqref{E0:=} of the pull-back of VA 1-form \eqref{Pi=VA} to the vector $u_\mu^0$ of the moving frame \eqref{harmU=10}, $\text{E}{}^{1q}$ and $\text{E}{}^{2}_{q}$ are the contractions \eqref{Eq1:=} of the pull-backs of the fermionic forms. The combination of bosonic ${\mathbb P}^i$ and ${\mathbb X}^i$ and fermionic ${\boldsymbol{ \Psi}}_q$ matrix fields
\begin{eqnarray}
\label{HmM0==} {\cal H} = {1\over 2} \text{tr}\left( {\mathbb P}^i {\mathbb P}^i \right)   - {1\over 64}
\text{tr}\left[ {\mathbb X}^i ,{\mathbb X}^j \right]^2  - 2\,  \text{tr}\left({\mathbb X}^i\,{\boldsymbol{ \Psi}}\gamma^i {\boldsymbol{ \Psi}}\right)~, 
  \end{eqnarray}
which has the form of 1d $\mathcal{N}=16$ SU$(N)$ SYM Hamiltonian; abusing a bit the terminology, we will call it below the relative motion Hamiltonian.

Actually the two first terms in \eqref{eq:10D_SmD0=} formally coincide with the action of single D$0$-brane, i.e. massive $\text{D}=10$ type IIA superparticle in its spinor moving frame formulation \cite{10D_mD0_Igor, superD0_Igor} (see Appendix~\ref{sec:Appen:10D_kappa} for more details). In this case $m$ plays the role of the superparticle mass. In contrast, the constant $\mu$ characterizes the interaction of the center of mass and relative motion sector as well as the self-interaction of this latter. Notice that to simplify and to make more transparent the dependence of the action on this parameter we have chosen non-canonical dimensions for the matrix matter fields \eqref{matter_bosonic} and \eqref{matter_fermionic}.

In particular, with this choice of dimensions of matrix fields, the relative motion Hamiltonian ${\cal H}$ \eqref{HmM0==} is $\mu$-independent. However its dimension becomes (mass$^6$) so that ${\cal H}/\mu^6$ is dimensionless.

The calligraphic ${\cal M}$ in \eqref{eq:10D_SmD0=} is a {\it positive definite function} of this dimensionless combination of the relative motion Hamiltonian and coupling constant,
\be
{\cal M}= {\cal M}({\cal H}/\mu^6)\; ,
\ee
for which, the consistency requires the function ${\cal M}={\cal M}( {\cal H}/ \mu^6)$  to be positive and we will assume this below,
\be\label{cM>0}
{\cal M}( {\cal H}/ \mu^6) > 0\; .
\ee
In section~\ref{ch.11D_origin}, we will see how a particular case of the action \eqref{eq:10D_SmD0=} with
\be\label{cap5:cM=m+}
{\cal M} = \frac m 2 +  \sqrt{\frac {m^2} {4}+\frac {\cal H} {\mu^6}}\;
\ee
can be obtained by dimensional reduction of the 11D multiple M-wave (or M$0$-branes) system from \cite{Bandos_11D_mM0, Meliveo}, which is similar to dimensional reduction of its $\text{D}=4$ counterpart described in chapter~\ref{ch.3D_mD0}. Another representative of the family \eqref{eq:10D_SmD0=} with ${\cal M} =m$ was studied in \cite{10D_mD0_Igor} where it was noticed that it cannot be obtained by dimensional reduction from 11D mM$0$ action.

The covariant derivatives in the second line of \eqref{eq:10D_SmD0=}
\begin{equation}
    \text{D}{\mathbb X}^i =   \text{d}{\mathbb X}^i   - \Omega^{ij} {\mathbb
     X}^j+ [{\mathbb A},    {\mathbb X}^i] \; ,
\label{DXi=_derivatives}
\end{equation}
\begin{equation}
     \text{D}\Psi_q  = \text{d}{\boldsymbol{ \Psi}}_q
     -{1\over 4} \Omega^{ij} \gamma^{ij}_{qp} {\boldsymbol{ \Psi}}_p+ [{\mathbb A},
     {\boldsymbol{ \Psi}}_q ] \; 
     \label{DPsi=_derivatives} 
\end{equation}
contain, beside the SU$(N)$ gauge field \eqref{bfA=}, also the composite SO$(9)$ connection. The formal Ricci identities for such a covariant derivatives\footnote{These can be calculated on the extension of  the worldline to some space of two or more dimensions. Such an extension can be realized by considering the forms depending on both differentials and variations of the variables, i.e. on $\text{d}x^\mu$ and $\delta x^\mu$, etc.}  read
\begin{equation}\label{DD=}
    \begin{array}{l}
         ~~\text{DD}{\mathbb X}^i= \Omega^i\wedge \Omega^j \, {\mathbb X}^j + [{\mathbb F}, {\mathbb X}^i]~,  \\
        \text{DD}{\boldsymbol{ \Psi}}_q = \dfrac 1 4 \, \Omega^i\wedge \Omega^j\, (\gamma^{ij}{\boldsymbol{ \Psi}})_q  + [{\mathbb F}, {\boldsymbol{ \Psi}}_q]~,
    \end{array}
\end{equation}
where  $\mathbb{F} = \text{d}\mathbb{A} - \mathbb{A} \wedge \mathbb{A}$ is the formal 2-form field strength of the 1-form gauge field $\mathbb{A}$. When deriving \eqref{DD=} we have used the Maurer-Cartan equations
\be\label{MC=Eq}
\text{D} \Omega^i= \text{d}\Omega^i- \Omega^j\wedge \Omega^{ji} = 0 \; , \qquad \text{d}\Omega^{ij}+\Omega^{ik}\wedge \Omega^{kj} =-\Omega^i\wedge \Omega^j\;
\ee
which can be found by taking formal exterior derivatives of \eqref{Du0=}.

\section{Local worldline supersymmetry}
The action \eqref{eq:10D_SmD0=} is manifestly invariant under 10D type IIA superPoincar\'e transformations, including spacetime (actually target superspace) supersymmetry with constant fermionic parameters $\epsilon^{\alpha 1}$ and $\epsilon_\alpha^2$ acting nontrivially only on the center of mass variables,
\bea\label{susy=IIA}
 \delta_\epsilon \theta^{1 \alpha}= \epsilon^{\alpha 1} \; , \qquad \delta_\epsilon \theta_{\alpha}^{2}=\epsilon_\alpha^2 \; ,  \qquad \delta_\epsilon v_\alpha^q=0\;~~, \\ \delta_\epsilon x^\mu =i\theta^1 \sigma^\mu \epsilon^1+i\theta^2 \tilde{\sigma}{}^\mu \epsilon^2\; . \nonumber
\eea
It is also invariant  under the SU($N$) gauge symmetry acting on the matrix matter fields by its adjoint representation, provided the $\mathfrak{su}(N)$ valued 1-form
${\mathbb A}$ transforms as SU$(N)$ connection, as well as under the SO$(9)$ symmetry acting by vector representation on index $i$ of $u_\mu^i$, ${\mathbb X}^i$, ${\mathbb P}^i$ and by its spinor representation on index $q$ of ${\boldsymbol{ \Psi}}_q$ and $v_\alpha^q$. 

Furthermore the action is invariant under local fermionic worldline supersymmetry which is parametrized by fermionic function $\kappa^q(\tau)$ carrying  spinor index of SO$(9)$. It acts on the center of mass variables exactly in the same manner as irreducible $\kappa$-symmetry of single D$0$-brane in its spinor moving frame formulation \cite{10D_mD0_Igor, superD0_Igor} (see Appendix~\ref{sec:Appen:10D_kappa} for details),
\bea\label{kappa=}
&& \delta_\kappa \theta^{1 \alpha}=  {\kappa^q} v_q^\alpha/{\sqrt{2}} \; , \qquad \delta_\kappa \theta_{\alpha}^{2}= -  {\kappa^q} v_\alpha{}^q/{\sqrt{2}} \; , \qquad  \nonumber \\ && ~~\delta_\kappa x^\mu =i\delta_\kappa\theta^1 \sigma^\mu \theta^1+i\delta_\kappa\theta^2 \tilde{\sigma}{}^\mu \theta^2  \; , \qquad    \\
 && ~~\delta_\kappa v_\alpha^q=0\qquad \Longrightarrow \qquad  \delta_\kappa u_{{\mu}}^{0} =0= \delta_\kappa u_{{\mu}}^{i}\; , \nonumber 
 \eea
hence notation $\kappa^q(\tau)$. On the matrix fields sector, the action of worldline supersymmetry includes nonlinear terms some of which are proportional to the derivative of the function ${\cal M}({\cal H}/{\mu^6})$ with respect to its argument (hence to additional power of ${1}/{\mu^6}$),
\be
\delta {\cal M} ({\cal H}/{\mu^6}) = \dfrac {1}{{\mu^6}} {\cal M}^\prime ({\cal H}/{\mu^6}) \,  \delta {\cal H} , \quad \text{with }
{\cal M}^\prime  (y)= \dfrac {\text{d}}{\text{d} y} {\cal M} (y)\; . \qquad
\ee
The worldline supersymmetry transformations of the matrix matter fields are (see Appendix~\ref{sec:Append:10D_mD0} for their derivation)
\begin{equation}
    \delta_\kappa {\mathbb X}^i  = \frac {4i}{\sqrt{{\cal M}}}\, \kappa\gamma^i{\boldsymbol{ \Psi}}   + \frac 1 {\mu^6} \,  \frac {{\cal M}^\prime}{{\cal M}} \, \delta_\kappa {\cal H}\;   {\mathbb X}^i -
\frac 1 {\mu^6} \,  \frac {{\cal M}^\prime}{{\cal M}} \, \Delta_\kappa {\cal K}\,  {\mathbb P}^i \; , \label{susy=X}
\end{equation}
\begin{equation}
    \begin{array}{c}
         \begin{split}
             \delta_\kappa {\mathbb P}^i &=  - \frac {1}{\sqrt{{\cal M}}}\, [\kappa\gamma^{ij}{\boldsymbol{ \Psi}},  {\mathbb X}^j]
             -\frac 1 {\mu^6} \,  \frac {{\cal M}^\prime}{{\cal M}} \, \delta_\kappa {\cal H}  {\mathbb P}^i +\\
             &~\\
             &+\frac 1 {\mu^6} \,  \frac {{\cal M}^\prime}{{\cal M}} \Delta_\kappa {\cal K}
            \left( \frac 1 {16} [[ {\mathbb X}^i, {\mathbb X}^j], {\mathbb X}^j]-\gamma^i_{pq} \{{\boldsymbol{ \Psi}}_p, {\boldsymbol{ \Psi}}_q\} \right) ,\; 
         \end{split}
    \end{array} \label{susy=P} 
\end{equation}
\begin{equation}
    \delta_\kappa {\boldsymbol{ \Psi}}_q=-\frac 1 {2\sqrt{{\cal M}}}\, (\kappa\gamma^{i})_q {\mathbb P}^i  - \frac {i}{16\sqrt{{\cal M}}}\, (\kappa\gamma^{ij})_q [ {\mathbb X}^i, {\mathbb X}^j]
 - \frac {i}  {4\mu^6} \,  \frac {{\cal M}^\prime}{{\cal M}} \, \Delta_\kappa {\cal K}\, [(\gamma^{i}{\boldsymbol{ \Psi}})_q , {\mathbb X}^i]\; . \label{susy=Psi}
\end{equation}
Here
\bea \label{kappaH=}
\delta_\kappa {\cal H} =  \frac 1{ \sqrt{{\cal M}}}\, \frac {{\rm tr} \left(\kappa^q{\boldsymbol{ \Psi}}_q\left( [{\mathbb X}^i, {\mathbb P}^i] -4i\{{\boldsymbol{ \Psi}}_{q}, {\boldsymbol{ \Psi}}_{q}\}\right) \right) } {1+\dfrac 1 {\mu^6} \,  \dfrac {{\cal M}^\prime}{{\cal M}} \, {\frak H} } \; \qquad
\eea
with
\be\label{frakH=} {\frak H}:= \text{tr}\left( {\mathbb P}^i {\mathbb P}^i \right)   + {1\over 16}
\text{tr}\left[ {\mathbb X}^i ,{\mathbb X}^j \right]^2 + 2\,  \text{tr}\left({\mathbb X}^i\, \mathbf{\Psi}\gamma^i {\boldsymbol{ \Psi}}\right)\;
\ee
is the worldline supersymmetry variation of the relative motion Hamiltonian \eqref{HmM0==} and
\bea \label{kappaK=}
\Delta_\kappa {\cal K} = \frac 1{ 2\sqrt{{\cal M}}}\, \frac { {\rm tr}  \left(4i (\kappa\gamma^i {\boldsymbol{ \Psi}}) {\mathbb P}^i + {5\over 2}
(\kappa\gamma^{ij} {\boldsymbol{ \Psi}})  [{\mathbb X}^i, {\mathbb X}^j]  \right)}{1+\dfrac 1 {\mu^6} \,  \dfrac {{\cal M}^\prime}{{\cal M}} \, {\frak H} } . \;
\eea
This latter is related to the worldline supersymmetry variation of ${\cal K}= {\rm tr} ({\mathbb X}^i\,{\mathbb P}^i)$
by
\be
\label{DkappaK=}
\Delta_\kappa {\cal K}=\delta_\kappa ( {\rm tr} ({\mathbb X}^i\,{\mathbb P}^i))+ \frac 1 {2\sqrt{{\cal M}}}i\kappa^q\nu_q
\ee
where
\bea\label{inu=}
i\nu_q := {\rm tr} \left(-4i (\gamma^i {\boldsymbol{ \Psi}})_q  {\mathbb P}^i + {1\over 2}
(\gamma^{ij} {\boldsymbol{ \Psi}})_q  [{\mathbb X}^i, {\mathbb X}^j]  \right)\; . \qquad
\eea
In terms of the above blocks the worldline supersymmetry variation of the SU$(N)$ connection 1-form (gauge field)  can be written as (see Appendix~\ref{sec:Append:10D_mD0} for its derivation)
\begin{equation}
    \begin{array}{l}
    \begin{split}
     \delta_\kappa {\mathbb A} &= - \frac 2 {{\cal M}\sqrt{{\cal M}}}\, \text{E}^0\,  (\kappa^q{\boldsymbol{ \Psi}}_q) \frac {\left(1- \dfrac 1 {\mu^6} \, \dfrac {{\cal M}^\prime}{{\cal M}}{\cal H}\right)}{\left(1+ \dfrac 1 {\mu^6} \, \dfrac {{\cal M}^\prime}{{\cal M}}\, {\frak H}\right)} + \frac 1 {\sqrt{2}{\cal M}} \, (\text{E}^{1q}-\text{E}_q^2)(\gamma^i\kappa)_q \,{\mathbb X}^i - \\
     &-  (\text{E}^{1q}-\text{E}_q^2)  \frac 1 {\mu^6} \, \frac {{\cal M}^\prime}{\sqrt{2}{\cal M}^2} \,  \frac  1 {\left(1+ \dfrac 1 {\mu^6} \, \dfrac {{\cal M}^\prime}{{\cal M}}\, {\frak H}\right)}  \kappa^p\, {{\boldsymbol{ \Psi}}_{(q}\,{\rm tr}\left[4i (\gamma^i {\boldsymbol{ \Psi}})_{p)} {\mathbb P}^i +\frac 5 2
    (\gamma^{ij} {\boldsymbol{ \Psi}})_{p)}  [\mathbb{X}^i, {\mathbb X}^j]  \right] }\,  .
    \end{split}
    \end{array}\label{susy=A}
\end{equation}

\section{Equations of motion}
In this section we write the complete set of equations of motion for the multiple D$0$-brane system which follow from our action \eqref{eq:10D_SmD0=}.

\subsection{Equations for the center of mass variables}
Beginning from the center of mass variables, we regroup  the variation with respect to the coordinate  functions, $\delta z^M(\tau)=( \delta x^\mu(\tau), \delta \theta^{\alpha 1}(\tau), \delta \theta^2_{\alpha}(\tau))$, into
\be 
\iota_\delta \text{E}^i = \delta z^M \text{E}_M^a (z) u_a^i\; , \qquad \iota_\delta \text{E}^0 = \delta z^M \text{E}_M^a (z) u_a^0\; \qquad
\ee
and
\be
\iota_\delta (\text{E}^{q1} - \text{E}_q^2) \; , \qquad \iota_\delta (\text{E}^{q1} + \text{E}_q^2) \; , \qquad
\ee
where
\be
\iota_\delta \text{E}^{q1} = \delta \theta^{\alpha 1}v_\alpha{}^q \; , \qquad
\iota_\delta \text{E}_q^{2} = \delta \theta^2_{\alpha}v_q{}^\alpha\; . \qquad
\ee
This choice simplifies the form of the equations of motion for the coordinate functions (originally defined as $\frac{\delta S_{\text{mD}0}}{\delta z^M}=0$) and splits their set in a Lorentz-covariant manner into
\bea\label{Omi()=0}
\Omega^i \left(m + \frac{2}{\mu^6}\frac{\mathcal{H}}{\mathcal{M}} \right) = 0\; , \qquad \\ \label{E1+E2=Omi}
\left(m + \frac{1}{\mu^6} \frac{\mathcal{H}}{\mathcal{M}} \right)\left(\text{E}^{1q} + \text{E}^2_{q} \right) = \frac{-i}{4 \sqrt{2\mathcal{M}} \mu^6}\gamma^i_{qp}i\nu_p\Omega^i\; ,
\eea
 and
\bea\label{dH()=0}
\frac{2}{\mathcal{M}} \left(1 - \frac{1}{\mu^6}\frac{\mathcal{M'}}{\mathcal{M}} \mathcal{H} \right)\text{d}\mathcal{H}=0\; , \qquad \\
\label{eq:Dnu}
\frac{1}{\sqrt{2\mathcal{M}}}i\text{D}\nu_q - \frac{1}{\mu^6} \frac{1}{2\sqrt{2 \mathcal{M}}} \frac{\mathcal{M'}}{\mathcal{M}} i \nu_q \text{d} \mathcal{H} - \frac{2i}{\mathcal{M}} \left(\text{E}^{1q} - \text{E}^2_{q} \right) \mathcal{H} = 0\; ,
\eea
where
\be\label{inu=}
i\nu_q:= {\rm tr} \left(-4i (\gamma^i {\boldsymbol{ \Psi}})_q  {\mathbb P}^i + {1\over 2}
(\gamma^{ij} {\boldsymbol{ \Psi}})_q  [{\mathbb X}^i, {\mathbb X}^j]  \right)\; . \qquad
\ee
As we will show below, Eqs. \eqref{dH()=0} and \eqref{eq:Dnu} are dependent, meaning they are satisfied identically when the other equations of motion are taken into account. This reflects the Noether identities associated with the worldline reparametrization invariance and the worldline supersymmetry ($\kappa$-symmetry) whose ``local parameters'' can be identified with
$\iota_\delta \text{E}^0$ and $\iota_\delta (\text{E}^{q1}-\text{E}_q^2)$, respectively. With this in mind, we begin by analysing \eqref{Omi()=0} and \eqref{E1+E2=Omi},  along with the equations for the spinor frame variables. We then proceed to the matrix equations, which will allow us to confirm the aforementioned dependence of \eqref{dH()=0} and \eqref{eq:Dnu}).

Given that the function ${\cal M}(\mathcal{H} / \mu^6)$ appearing in our action is positive definite (as stated in \eqref{cM>0}), Eq. \eqref{Omi()=0} immediately implies
\be\label{Omi=0}
\Omega^i = 0 \, . \ee
Using this result, the fermionic equation \eqref{E1+E2=Omi} simplifies to 
\be\label{E1+E2=0}
\text{E}^{1q} + \text{E}^2_{q}  =0  \; .
\ee
Finally, the essential variations of the moving frame and spinor moving frame variables are given by~\eqref{eq:delta_u0}, \eqref{eq:delta_vq_vq}
\be
\delta u_\mu^0= u_\mu^i \iota_\delta \Omega^i\, , \qquad  \delta v_\alpha{}^q=  \frac 1 2 \, \iota_\delta \Omega^i\, \gamma^i_{qp}  v_\alpha{}^p\, , \qquad \delta v_q{}^\alpha=-  \frac 1 2 \, \iota_\delta \Omega^i\, v_p{}^\alpha \gamma^i_{pq}  \, , \qquad
\ee
these leads to the bosonic vector equation
\begin{equation}\label{Ei()=ff+Om}
\text{E}^i \left(m + \frac{2}{\mu^6} \frac{\mathcal{H}}{\mathcal{M}} \right) - \frac{1}{2 \mu^6  \sqrt{2 \mathcal{M}}}\left(\text{E}^{1q} + \text{E}^2_{q} \right) \gamma^{i}_{qp}i\nu_p - \frac 2 {\mu^6 } \, \text{tr} \left( \mathbb{P}^{\left[i \right.} \mathbb{X}^{\left. j \right]} + i {\boldsymbol{ \Psi}} \gamma^{ij} {\boldsymbol{ \Psi}} \right)\, \Omega^j = 0\; ,
\end{equation}
where $i\nu_q$ is defined in  \eqref{inu=}. Substituting \eqref{Omi=0} and the fermionic equations \eqref{E1+E2=0} into this, we find that \eqref{Ei()=ff+Om}  reduces to the simple condition
\begin{equation}\label{Ei=0}
 \text{E}^i  = 0\; .
\end{equation}
Thus the complete set of equations governing the center of mass sector is given by
Eqs. \eqref{Omi=0}, \eqref{E1+E2=0} and \eqref{Ei=0}. This coincides precisely with the set of equations of motion for a single D$0$-brane in its spinor moving frame formulation \cite{superD0_Igor}.

\subsection{Equations of the relative motion of the constituents of mD0 system}
The equations of motion for the matrix fields describing the relative motion of the mD$0$ constituents read
\begin{equation}
    \begin{array}{l}
         \begin{split}
             \text{D}\mathbb{X}^i =  -\frac{2}{\mathcal{M}} \left(1 - \frac{1}{\mu^6}\frac{\mathcal{M'}}{\mathcal{M}}\mathcal{H} \right)&\text{E}^0 \mathbb{P}^i + \frac{1}{\mu^6}\frac{\mathcal{M'}}{\mathcal{M}} \left( \mathbb{X}^i  \text{d}\mathcal{H} - \mathbb{P}^i  \text{d}\mathcal{K}\right)~+ \qquad   \\
             &\\
             &+ \frac{1}{\sqrt{2\mathcal{M}}} \left(\text{E}^{1q} - \text{E}^2_{q} \right) \left(4i \left(\gamma^i {\boldsymbol{ \Psi}} \right)_q - \frac{1}{2\mu^6}\frac{\mathcal{M'}}{\mathcal{M}} i \nu_q \mathbb{P}^i \right)\; , 
         \end{split}
    \end{array} \label{DXi=}
\end{equation}
\begin{equation}
    \begin{array}{l}
         \begin{split}
               \text{D} \mathbb{P}^i &=  \frac{2}{\mathcal{M}} \left[ \left(1- \frac{1}{\mu^6} \frac{\mathcal{M'}}{\mathcal{M}}\mathcal{H} \right)\text{E}^0 + \frac{1}{\mu^6}\frac{\mathcal{M'}}{4 \sqrt{2 \mathcal{M}}} \left(\text{E}^{1q} - \text{E}^2_{q} \right) i\nu_q \right] \times \\
               &~\\
               &\times \left(\frac{1}{16} \left[[\mathbb{X}^i, \mathbb{X}^j],\mathbb{X}^j  \right] - \gamma^i_{pr} \left\lbrace{\boldsymbol{ \Psi}}_p, {\boldsymbol{ \Psi}}_r \right\rbrace  \right) - \frac{1}{\sqrt{2 \mathcal{M}}}\left(\text{E}^{1q}- \text{E}^2_{q}\right) [\left(\gamma^{ij}  {\boldsymbol{ \Psi}} \right)_q, \mathbb{X}^j]
               +\\
               &~\\
               &+\frac{1}{\mu^6} \frac{\mathcal{M'}}{\mathcal{M}}\text{d}\mathcal{K} \left(\frac{1}{16} \left[[\mathbb{X}^i, \mathbb{X}^j],\mathbb{X}^j  \right] - \gamma^i_{pr} \left\lbrace {\boldsymbol{ \Psi}}_p,  {\boldsymbol{ \Psi}}_r \right\rbrace  \right) - \frac{1}{\mu^6}\frac{\mathcal{M'}}{\mathcal{M}} \mathbb{P}^i \text{d}\mathcal{H}\; ,
         \end{split}
    \end{array} \label{DPi=}
\end{equation}
\begin{equation}
    \begin{array}{l}
         \begin{split}
              ~~\text{D}{\boldsymbol{ \Psi}}_q &=  - \frac{1}{2 \sqrt{2\mathcal{M}}}\left(\text{E}^{1p}- \text{E}^2_{p} \right) \left(\gamma^i_{pq}\mathbb{P}^i + \frac{i}{8}\gamma^{ij}_{pq}[\mathbb{X}^i,\mathbb{X}^j] \right)~-\\
             &~\\
             & - \frac{i}{2\mathcal{M}} \left[ \left(1 - \frac{1}{\mu^6} \frac{\mathcal{M'}}{\mathcal{M}}\mathcal{H}\right)\text{E}^0 +  \frac{\mathcal{M'}}{2\mu^6} \text{d}\mathcal{K}   + \frac{1}{4 \mu^6}\frac{\mathcal{M'}}{\sqrt{2\mathcal{M}}}\left(\text{E}^{1p}- \text{E}^2_{p} \right) i\nu_p  \right] [(\gamma^i {\boldsymbol{ \Psi}})_q, \mathbb{X}^i] 
         \end{split}
    \end{array} \label{DPsi=}
\end{equation}
and
\begin{equation}\label{Gauss=}
[\mathbb{X}^i, \mathbb{P}^i] = 4i \left\lbrace {\boldsymbol{ \Psi}}_q, {\boldsymbol{ \Psi}}_q \right\rbrace~.
\end{equation}
In  Eqs. \eqref{DXi=}-\eqref{DPsi=}
 ${\cal H}$ is given in \eqref{HmM0==},
\be\label{cK=}
 {\cal K}= {\rm tr} ({\mathbb X}^i\,{\mathbb P}^i)\;
\ee
and $i\nu_q$ is defined in \eqref{inu=}.
Eq. \eqref{Gauss=} appears as a result of variation with respect to the  worldline gauge field $\mathbb{A} = \text{d}\tau \mathbb{A}_\tau$ and is the (non-Abelian version of the)  Gauss law of 10D SYM  reduced to 1d.

Using Eqs. \eqref{DXi=}-\eqref{DPsi=} and \eqref{Gauss=}  one can check (by straightforward although a bit involving calculations) that
\be\label{dH=0}
\text{d}{\cal H}=0
\ee
and
\be\label{Dnu=EH}
i\text{D}\nu_q= \frac {2\sqrt{2}}{\sqrt{{\cal M}}}(\text{E}^{1q}-\text{E}_q^2){\cal H}\; ,
\ee
which implies that  Eqs. \eqref{dH()=0} and \eqref{eq:Dnu} are satisfied identically (as we have announced and discussed just after their derivation).

Taking into account \eqref{dH=0} we find that the equations for bosonic  matrix matter fields, \eqref{DXi=} and \eqref{DPi=} simplify a bit:
\begin{equation}
    \begin{array}{l}
         \begin{split}
             ~~~~ ~~~~~~~~~~\text{D}\mathbb{X}^i &= -\frac{2}{\mathcal{M}} \left(1 - \frac{1}{\mu^6}\frac{\mathcal{M'}}{\mathcal{M}}\mathcal{H} \right)\text{E}^0 \mathbb{P}^i - \frac{1}{\mu^6}\frac{\mathcal{M'}}{\mathcal{M}} \mathbb{P}^i  \text{d}\mathcal{K}~+ \qquad \\
             &~\\ 
             & ~~~~~~~~~~ ~~~~~~~~~ ~~~~~~~~~~~ ~~~~~~~~~~+\frac{1}{\sqrt{2\mathcal{M}}} \left(\text{E}^{1q} - \text{E}^2_{q} \right) \left(4i \left(\gamma^i {\boldsymbol{ \Psi}} \right)_q - \frac{1}{2\mu^6}\frac{\mathcal{M'}}{\mathcal{M}} i \nu_q \mathbb{P}^i \right)\; ,
         \end{split}
    \end{array}\label{DXi==}
\end{equation}
\begin{equation}
    \begin{array}{l}
         \begin{split}
             \text{D} \mathbb{P}^i &=  \frac{2}{\mathcal{M}} \left[ \left(1- \frac{1}{\mu^6} \frac{\mathcal{M'}}{\mathcal{M}}\mathcal{H} \right)\text{E}^0 + \frac{1}{\mu^6}\frac{\mathcal{M'}}{4 \sqrt{2 \mathcal{M}}}  \left(\text{E}^{1q} -\text{E}^2_{q} \right) i\nu_q \right] \times  \qquad \qquad \\
             &~\\
             & ~~~~~~~~~~~~~~~~~~~~~~~~~~~~~~~~~~~~~~~~~~~~~~~~~~~~~~~~  \times  \left(\frac{1}{16} \left[[\mathbb{X}^i, \mathbb{X}^j],\mathbb{X}^j  \right] - \gamma^i_{pr} \left\lbrace{\boldsymbol{ \Psi}}_p, {\boldsymbol{ \Psi}}_r \right\rbrace  \right)~.
         \end{split}
    \end{array}\label{DPi==}
\end{equation}
This form of equations is not yet final since it involves $\text{d}\mathcal{K}={\rm tr} \left(
\text{D} \mathbb{X}^i \, \mathbb{P}^i+\mathbb{X}^i\text{D} \mathbb{P}^i
\right) $. Calculating the r.h.s. of this expression with the use of \eqref{DXi==} and \eqref{DPi==}, we find
\begin{equation}
    \begin{array}{l}
         \begin{split}
             \text{d}\mathcal{K} &=  - \frac 2 {{\cal M}}  \frac {\mathfrak{H} }{ \left(1 + \frac{1}{\mu^6} \frac{\mathcal{M'}}{\mathcal{M}}\mathfrak{H} \right)} \left(1- \frac{1}{\mu^6} \frac{\mathcal{M'}}{\mathcal{M}}\mathcal{H} \right)  \text{E}^0  + \frac {(\text{E}^{1q}-\text{E}^2_q)}{\sqrt{2{\cal M}}} \times \qquad \qquad \\
             &~\\
             & ~~~~~~ ~~~~~~ ~~~~~~~ ~~~~~~~~~~~~~~~\times \frac { {\rm tr} \left(4i (\gamma^i {\boldsymbol{ \Psi}})_q  {\mathbb P}^i + (\gamma^{ij} {\boldsymbol{ \Psi}})_q  [{\mathbb X}^i, {\mathbb X}^j]\right) -  \frac{1}{2\mu^6} \frac{\mathcal{M'}}{\mathcal{M}}\mathfrak{H}\, i\nu_q  } { \left(1 + \frac{1}{\mu^6} \frac{\mathcal{M'}}{\mathcal{M}}\mathfrak{H} \right)}\,  
         \end{split}
    \end{array}\label{dcK==}
\end{equation}
with \be\label{frakH=} {\frak H}:= \text{tr}\left( {\mathbb P}^i {\mathbb P}^i \right)   + {1\over 16}
\text{tr}\left[ {\mathbb X}^i ,{\mathbb X}^j \right]^2 + 2\,  \text{tr}\left({\mathbb X}^i\, {\boldsymbol{ \Psi}}\gamma^i {\boldsymbol{ \Psi}}\right)\; .
\ee
Thus the final form of the equations for the matrix matter (or relative motion) fields is given by
\eqref{DXi==}, \eqref{DPi==} and \eqref{DPsi=} with $\text{d}\mathcal{K}$ substituted by the r.h.s. of \eqref{dcK==}. While the resulting expressions may appear frighteningly complicated, we will show that they can be significantly simplified by an appropriate choice of gauge fixing for the symmetries of the system. Before proceeding with the gauge fixing, however, we must first address the issue of the mass of the mD$0$ system.

\subsection{Mass of mD0 system and its center of mass motion}
It is instructive to compute the canonical momentum for the center of mass coordinate function of mD$0$ system
\be\label{pmu=}
p_\mu = \left(m + \frac{2}{\mu^6} \frac{\mathcal{H}}{\mathcal{M}}\,\right) u_\mu^0=: \frak{M} u_\mu^0  \, .
\ee
The mass $ \frak{M}$ of the mD$0$ is defined by the square of this 10-momentum, $p_\mu p^\mu =  \mathfrak{M}^2$,  and is thus given by
\be\label{Mass=}
\mathfrak{M} = m + \frac{2}{\mu^6} \frac{\mathcal{H}}{\mathcal{M}}\,\, .
\ee
Note that this mass depends essentially on the choice of the positive definite function ${\mathcal{M}}(\mathcal{H}/\mu^6)$ in the action \eqref{eq:10D_SmD0=}.
However, as a consequence of \eqref{dH=0}, this mass is conserved,
\be\label{dfM=0}
\text{d}\frak{M}=0\; .
\ee
It is important that the mass depends on the relative motion of the mD$0$ constituents. As a result, in this supersymmetric model, the relative motion influences the center of mass motion in the same way it does in the purely bosonic dynamical system from \cite{Myers}.

The constant parameter $m$ in the action \eqref{eq:10D_SmD0=} corresponds to the mass of the center of mass motion of mD$0$ when the relative motion sector is in its ground state, i.e. when ${\cal H}=0$. As we will see below, supersymmetric configurations of mD$0$ systems  satisfy this condition.
Therefore, any excitation of the relative motion inevitably increases the mass of the center of mass of the mD$0$ system (due to the positivity condition \eqref{cM>0}) and, as we will also show, necessarily breaks supersymmetry.

Now observe that, by definition of Cartan forms, $\text{d}u_\mu^0= u_\mu^i\Omega^i$. Hence, Eq.~\eqref{Omi=0} implies that on-shell we have
\be \label{du0=0}
\text{d}u_\mu^0=0\; .
\ee
Together with \eqref{dfM=0}, this implies that the momentum \eqref{pmu=} is conserved
\be \text{d}p_\mu=0~. \ee

\subsection[Gauge fixing of the local SO(9) and  SU\texorpdfstring{($N$)}{(N)} symmetries]{Gauge fixing of the local SO(9) and  SU\boldmath\texorpdfstring{($N$)}{(N)} symmetries}
\label{sec:convenientGF}
For future use, let us also notice that, on the surface defined by  Eq. \eqref{Omi=0}, the derivatives of the orthogonal moving frame vectors $u_\mu^i$ decompose entirely along these vectors,
\be
\text{d}u_\mu^i= \Omega^{ij} u_\mu^j \; .
\ee
Furthermore, since $\Omega^{ij}$ transforms as a connection under local gauge \text{SO}$(9)$ symmetry of the mD$0$ action \eqref{eq:10D_SmD0=}, and since any 1-dimensional connection can be gauged away, we can fix the gauge
\be\label{Omij=0}
\Omega^{ij}= 0 \ee
in which the 9 spacelike vectors of the moving frame become constant
\be\label{dui=0}
\text{d}u_\mu^i= 0  \; .
\ee
Similarly, the local $\text{SU}(N)$  gauge symmetry can be used to set the $\mathfrak{su}(N)$ valued gauge field to zero,
\be
{\mathbb A}= 0\; .
\ee
 In this gauge, the covariant derivative $\text{D}=\text{d}\tau \text{D}_\tau$ reduces to a (proper-)time derivative,
\be
\text{D}{\mathbb X}^i= \text{d}\tau \frac{\text{d}} {\text{d}\tau} {\mathbb X}^i=  \text{d}\tau  \dot{{\mathbb X}}{}^i\; , \qquad  \text{D}{\mathbb P}^i= \text{d}\tau \frac{\text{d}}{\text{d}\tau} {\mathbb P}^i=  \text{d}\tau  \dot{{\mathbb P}}{}^i\; , \qquad
\text{D}{\boldsymbol{ \Psi}}_q= \text{d}\tau \frac{\text{d}}{\text{d}\tau} {\boldsymbol{ \Psi}}_q=  \text{d}\tau  \dot{\boldsymbol{ \Psi}}_q\; .  \qquad
\ee
Notice that, as far as moving frame and spinor moving frame variables are now (proper-)time independent, we obtain
\be\label{E0=gauge}
\text{E}^0=\text{d}\tau \text{E}_\tau^0= \text{d}x^{{\mathbf 0}} -i \text{d}\theta^{1q}\,\theta^{1q}-i \text{d}\theta_q^{2} \, \theta_q^{2} \; , \qquad \text{E}^{q1}= \text{d}\theta^{1q}\; , \qquad \text{E}_q^{2}= \text{d}\theta_q^{2}\; , \qquad
\ee
and
\be\label{Ei=gauge}
\text{E}^i=\text{d}\tau \text{E}_\tau^i= \text{d}x^{{\mathbf i}} -i \text{d}\theta^{1q}\gamma^i_{qp}\theta^{1p}+i \text{d}\theta_q^{2} \gamma^i_{qp} \theta_p^{2} \; , \qquad
\ee
where
\be\label{x0=xu0}
x^{{\mathbf 0}}=x^\mu u_\mu^0\; , \qquad x^{{\mathbf i}}=x^\mu u_\mu^i\; , \qquad
\theta^{1q}=\theta^{\alpha 1} v_\alpha{}^q \; , \qquad  \theta^2_{q}=\theta^2_\alpha v^\alpha_q \;   \qquad
\ee
 are the supersymmetric generalizations of co-moving coordinates for the center of mass of the mD$0$-brane.

\subsection[Gauge fixing of \texorpdfstring{$\kappa$}{k}-symmetry and reparametrization symmetry]{Gauge fixing of \boldmath\texorpdfstring{$\kappa$}{k}-symmetry and reparametrization symmetry}
\label{secGFk}
Using the consequences \eqref{E0=gauge} and the gauge fixing condition \eqref{Omij=0}, Eq.~\eqref{E1+E2=0} simplifies to
\be\label{dth1=-dth2}
 \text{d}\theta^{1q}= - \text{d}\theta_q^{2}\; , \qquad
\ee
This equation is invariant under global supersymmetry and local $\kappa$-symmetry, under which (in the coordinate basis \eqref{x0=xu0}) we have
\begin{equation}
    \begin{array}{lcl}
        \delta  \theta^{1q} =\epsilon^{1q}+ \kappa^q/\sqrt{2} \; , \qquad \quad \delta \theta_q^{2} =  \epsilon_q^{2}- \kappa^q/\sqrt{2}~, \\
        ~\\
        \delta x^{{\mathbf 0}}=i\text{d}\theta^{1q}\, (\epsilon^{1q}- \kappa^q/\sqrt{2})+i\theta_q^{2} \, ( \epsilon_q^{2}+ \kappa^q/\sqrt{2}) ~.
    \end{array}
\end{equation}
This implies
\be
\delta \text{E}^0= -2i ( \text{d}\theta^{1q} -\text{d} \theta_q^{2}) \kappa^q/\sqrt{2}\; .
\ee
Clearly, we can now use the worldline supersymmetry to fix one (but not both) of the fermionic coordinate functions to zero (as the $\kappa$-symmetry of single D$0$-brane can be). Let us choose
\be\label{th2=0}
\theta_q^2=0 \; .
\ee
This gauge is preserved by a combination of global supersymmetry and local worldline supersymmetry satisfying
\be\label{e2=k}
\epsilon^2_q = \kappa^q/\sqrt{2}\; .
\ee
This condition also implies that the local worldline supersymmetry parameter becomes a constant spinor at this stage. In this gauge~\eqref{th2=0}, Eq.~\eqref{dth1=-dth2} simplifies to
\be\label{dth1=0}
 \text{d}\theta^{1q}= 0\; , \qquad
\ee
which further reduces $\text{E}^{0}$and $\text{E}^i$ to: \begin{equation} 
\text{E}^0 = \text{d}x^{\mathbf{0}}~ , \qquad \text{E}^i = \text{d}x^{\mathbf{i}}~.
\end{equation}
This gauge choice remains supersymmetric because now
\be
\delta x^{{\mathbf 0}}= i (\epsilon^{1q}+\epsilon^2_q)\theta^{1q}\; , \qquad \delta x^{{\mathbf i}}= i (\epsilon^{1q}+\epsilon^2_q)\gamma^i_{qp}\theta^{1p}\;
\ee
and both r.h.s.-s are constants due to \eqref{dth1=0}, $\text{d}v_\alpha^q=0$ and $\text{d}\epsilon^{\alpha 1,2}=0$.

Having fixed $\kappa$-symmetry, we may now fix the reparametrization symmetry by setting
\be\label{dt=dx0}
\text{d}\tau= \text{d}x^{{\mathbf 0}}=\text{E}^0 \qquad \Longrightarrow \qquad \text{E}_\tau^0= \dot{x}^{{\mathbf 0}}=1\;
\ee
still preserving the supersymmetry. However, for convenience, we will not fix $\tau$-reparametrization symmetry at this stage and instead preserve its explicit form for subsequent analysis.

\subsection{Gauge fixed form of the field equations}
Thus, the above gauge fixing and the field equations for  the center of mass variables imply that
\bea\label{du=0=dv} \text{d}u^{0}_\mu=0\; , \qquad \text{d}u^{i}_\mu=0\; , \qquad  \text{d}v_\alpha^q=0\; ,  \\ \label{Ei=0b}
 \qquad \text{E}^i =\text{d}x^{{\mathbf i}} = 0\; , \\
 \text{E}^{1q}=0\; , ~~\text{E}^2_q=0\; ,~
\eea
and that $\text{E}^0=\text{d}\tau \dot{x}^{\mathbf 0}$. With this in mind, the equations for the matrix fields reduce to
\bea \label{DXi=gf}
&&~~\dot{\mathbb{X}}^i = -\frac{2}{\mathcal{M}} \left(1 - \frac{1}{\mu^6}\frac{\mathcal{M'}}{\mathcal{M}}\mathcal{H} \right)\dot{x}^{\mathbf 0} \mathbb{P}^i - \frac{1}{\mu^6}\frac{\mathcal{M'}}{\mathcal{M}} \mathbb{P}^i  \dot{\mathcal{K}}\; , ~~~~~~~~~~~~
\\ \label{DPi=gf}
&&~~\dot{\mathbb{P}}^i =\left[ \frac{2}{\mathcal{M}}  \left(1- \frac{1}{\mu^6} \frac{\mathcal{M'}}{\mathcal{M}}\mathcal{H} \right)\dot{x}^{\mathbf 0} + \frac{1}{\mu^6} \frac{\mathcal{M'}}{\mathcal{M}}\dot{\mathcal{K}} \right] \left(\frac{1}{16} \left[[\mathbb{X}^i, \mathbb{X}^j],\mathbb{X}^j  \right] - \gamma^i_{pr} \left\lbrace{\boldsymbol{ \Psi}}_p, {\boldsymbol{ \Psi}}_r \right\rbrace  \right)~, ~~~~~~~~~~~~
 \\
\label{DPsi=gf} 
&&\dot{\boldsymbol{ \Psi}}_q = - \frac{i}{2\mathcal{M}} \left[ \left(1 - \frac{1}{\mu^6} \frac{\mathcal{M'}}{\mathcal{M}}\mathcal{H}\right)\dot{x}^{\mathbf 0} +  \frac{\mathcal{M'}}{2\mu^6} \dot{\mathcal{K}}    \right] [(\gamma^i {\boldsymbol{ \Psi}})_q, \mathbb{X}^i]  \; , ~~~~~~~~~~~~
\eea
where
\be\label{dcK=gf}
\dot{\mathcal{K}}= - \frac 2 {{\cal M}}  \frac {\mathfrak{H} }{ \left(1 + \dfrac{1}{\mu^6} \dfrac{\mathcal{M'}}{\mathcal{M}}\mathfrak{H} \right)}\,  \left(1- \dfrac{1}{\mu^6} \dfrac{\mathcal{M'}}{\mathcal{M}}\mathcal{H} \right) \, \dot{x}^{\mathbf 0}  \qquad
\ee
with ${\frak H}$ defined in \eqref{frakH=}. Substituting \eqref{dcK=gf}, we find that the final gauge fixed form of Eqs. \eqref{DXi=gf}-\eqref{DPsi=gf} is
\bea\label{DXi==gf}
&& ~~\dot{\mathbb{X}}{}^i = -\frac{2}{\mathcal{M}} \, \frac {\left(1 - \frac{1}{\mu^6}\frac{\mathcal{M'}}{\mathcal{M}}\mathcal{H} \right)} { \left(1 + \frac{1}{\mu^6} \frac{\mathcal{M'}}{\mathcal{M}}\mathfrak{H} \right)} \dot{x}^{\mathbf{0}}\, \mathbb{P}^i \; , \qquad
\\ \label{DPi==gf}
&& ~~\dot{ \mathbb{P}}{}^i = \frac{2}{\mathcal{M}}\frac {\left(1 - \frac{1}{\mu^6}\frac{\mathcal{M'}}{\mathcal{M}}\mathcal{H} \right)} { \left(1 + \frac{1}{\mu^6} \frac{\mathcal{M'}}{\mathcal{M}}\mathfrak{H} \right)} \dot{x}^{\mathbf{0}} \,\left(\frac{1}{16} \left[[\mathbb{X}^i, \mathbb{X}^j],\mathbb{X}^j  \right] - \gamma^i_{pr} \left\lbrace{\boldsymbol{ \Psi}}_p, {\boldsymbol{ \Psi}}_r \right\rbrace  \right) \; ,\qquad \\
\label{DPsi==gf} 
&& \dot{\boldsymbol{ \Psi}}_q = - \frac{i}{2\mathcal{M}} \, \frac {\left(1 - \frac{1}{\mu^6}\frac{\mathcal{M'}}{\mathcal{M}}\mathcal{H} \right)} { \left(1 + \frac{1}{\mu^6} \frac{\mathcal{M'}}{\mathcal{M}}\mathfrak{H} \right)}\, \dot{x}^{\mathbf 0}  [(\gamma^i {\boldsymbol{ \Psi}})_q, \mathbb{X}^i]  \; .
\eea
Notice that, although one might fix the gauge $\dot{x}^{\mathbf 0}=1$ \eqref{dt=dx0}, we prefer to keep it unfixed to emphasize the invariance of our equations of motion under $\tau$-reparametrization, a feature that will be important in the discussion below.

\subsection{Relation of the relative motion equations of mD0 with  SYM equations}
Formally, we can define the new time variable by
\be\label{dt=}
\text{d}t = \text{d}x^{{\mathbf 0}} \,  \frac{2}{\mathcal{M}}\frac {\left(1 - \frac{1}{\mu^6}\frac{\mathcal{M'}}{\mathcal{M}}\mathcal{H} \right)} { \left(1 + \frac{1}{\mu^6} \frac{\mathcal{M'}}{\mathcal{M}}\mathfrak{H} \right)} \qquad \Longleftrightarrow \qquad
\frac {\text{d}t(\tau)} {\text{d}\tau} = \dot{x}^{{\mathbf 0}} \,  \frac{2}{\mathcal{M}}\frac {\left(1 - \frac{1}{\mu^6}\frac{\mathcal{M'}}{\mathcal{M}}\mathcal{H} \right)} { \left(1 + \frac{1}{\mu^6} \frac{\mathcal{M'}}{\mathcal{M}}\mathfrak{H} \right)} \; ,
\ee
and rewrite the above equations using the derivative $\frac{\text{d}}{\text{d}t}$ arriving at
\bea\label{DXi=SYM}
&& ~~\frac{\text{d}}{\text{d}t} {\mathbb{X}}{}^i = -\, \mathbb{P}^i \; , \qquad
\\ \label{DPi=SYM}
&&~~\frac{\text{d}}{\text{d}t} { \mathbb{P}}{}^i = \frac{1}{16} \left[[\mathbb{X}^i, \mathbb{X}^j],\mathbb{X}^j  \right] - \gamma^i_{pr} \left\lbrace {\boldsymbol{ \Psi}}_p, {\boldsymbol{ \Psi}}_r \right\rbrace \; ,\qquad \\
\label{DPsi=SYM} 
&& \frac{\text{d}}{\text{d}t} {\boldsymbol{ \Psi}}_q = - \frac{i}{4} \,   [(\gamma^i {\boldsymbol{ \Psi}})_q, \mathbb{X}^i]  \;
\eea
which coincide with the equations of motion of the 1d reduction of 10D SYM theory in the gauge ${\mathbb A}=0$.

However, this procedure does not correspond to a $\tau$-reparametrization (i.e. a 1d general coordinate transformation of the proper time). The simplest way to see this is to appreciate that the definition of \eqref{dt=}  for the ``new time'' is actually invariant under the $\tau$-reparametrization and therefore cannot serve as a gauge fixing condition for this symmetry\footnote{In fact, since equations \eqref{DXi=gf}-\eqref{DPsi=gf} are manifestly reparametrization invariant, they cannot be transformed into \eqref{DXi=SYM}-\eqref{DPsi=SYM} via a $\tau$-reparametrization.}.

The other more physical observation is that changing the form of the positive definite function ${\cal M}({\cal H})$ affects the physical properties of the system. In particular, it changes the mass ${\frak M}$ of the mD$0$ system ({\it{cf.}} \eqref{Mass=}), which depends on the choice of $\mathcal{M}(\mathcal{H})$. Hence, such a change cannot be achieved by transformation of a gauge
symmetry of the system.

Thus what we have found is not a kind of gauge equivalence, but rather an interesting correspondence between the gauge fixed form of the relative motion equations of our mD$0$ system and those of maximally supersymmetric ${\cal N}=16$  1d SU($N$) SYM theory. Note, however, that the formal definition \eqref{dt=} of the new time variable involves the matrix fields ${\mathbb X}^i$, ${\mathbb P}^i$, ${\boldsymbol{ \Psi}}_q$ and thus, in general, differs for different solutions of the field equations. Indeed, while  ${\cal H}$ (and hence ${\cal M}({\cal H})$) is constant on-shell, the same is not true for ${\frak H}={\frak H}({\mathbb X}^i, {\mathbb P}^i, {\boldsymbol{ \Psi}}_q)$.
However, for special solutions with $\mathcal{H} = 0$ (implying also $\mathfrak{H} = 0$), the variable $t$ becomes proportional to $x^{\mathbf{0}}$ via a constant rescaling.  In this case, the correspondence between the mD$0$ relative motion solutions and SU($N$) SYM equations is one-to-one.

As we will see in a moment,  this situation arises for bosonic supersymmetric solutions of the mD$0$ equations.

\section{Supersymmetric bosonic solutions}
\label{sec:SUSYbosSol}
As we have discussed in sec.~\ref{secGFk}, on the mass shell, one of two center of mass fermions can be gauged away using worldline supersymmetry ($\kappa$-symmetry), as shown in \eqref{th2=0}, and then the remaining fermionic coordinate function becomes constant \eqref{dth1=0}. By setting this constant to zero as well,  we arrive at a purely bosonic center of mass configuration with
\be\label{th1=0}
\theta^{\alpha 1}=0 \; , \qquad \theta_\alpha^2=0 \; .
\ee
This configuration still preserves half of the target (super)space supersymmetry, with the two supersymmetry spinor parameters related to a single $\text{SO}(9)$ spinor parameter of the worldline supersymmetry via constant spinor frame variables (fixed in the gauge \eqref{du=0=dv}) by
\be\label{e1=k}
\epsilon^{\alpha 1} = -\kappa^qv_q^\alpha /\sqrt{2}\; , \qquad \epsilon_\alpha^2 = \kappa^qv_\alpha^q/\sqrt{2}\;.
\ee
Then, after fixing the reparametrization invariance by setting $x^{{\mathbf 0}}(\tau)=\tau$,  the general solution of the bosonic center of mass equations takes the form
\be\label{x=x0+pt}
x^\mu (\tau)  = x^\mu_0 +\frac {p^\mu }{\frak{M}}\tau \; ,
\ee
where $p^\mu$ and $\frak{M}$ are defined in \eqref{pmu=} and \eqref{Mass=}. These are constants due to \eqref{dH=0} which is the Noether identity associated with reparametrization symmetry. Let us stress that any of these bosonic solutions describing the  center of mass motion preserves $1/2$ of the type IIA spacetime supersymmetry, which can be preserved or broken by the relative motion of the mD$0$ constituents.

Originally, the matrix fields that encode the relative motion of the mD$0$ constituents are inert under spacetime supersymmetry. However, after gauge fixing the worldline supersymmetry via Eq.~\eqref{th2=0}, this is not longer the case. The worldline supersymmetry parameter, which acts on the matrix fields, becomes identified with the second spacetime supersymmetry parameter through Eq.\eqref{e2=k}. Furthermore, setting the remaining fermionic coordinate function to zero, \eqref{th1=0},  similarly identifies it with the first supersymmetry parameter \eqref{e1=k}. As a result, spacetime supersymmetry now acts nontrivially on the matrix fields, coinciding with worldline supersymmetry, now parametrised by a constant fermionic spinor $\kappa^q$.

Setting to zero the fermionic matrix field
\be\label{Psi=0} {\boldsymbol{ \Psi}}_q=0
\ee
we can still preserve the worldline supersymmetry (completely or partially), and hence also a part of the spacetime supersymmetry if the following Killing spinor equation (see \eqref{susy=Psi})
\be
\label{susy=Psi=b}\delta_\kappa {\boldsymbol{ \Psi}}_q=-\frac 1 {2\sqrt{{\cal M}}}\, (\kappa\gamma^{i})_q {\mathbb P}^i  - \frac {i}{16\sqrt{{\cal M}}}\, (\kappa\gamma^{ij})_q [ {\mathbb X}^i, {\mathbb X}^j]=0
\;
\ee
admits a nontrivial solution. Notice that the contribution of the positive definite function ${\cal M}({\cal H})$ can be factored out, thus reducing the equation to
\be
\label{susy=Psi=bb}  \kappa^p {\mathbb L}_{pq}:= \kappa^p \left(\gamma^{i}_{pq} {\mathbb P}^i  - \frac {i}{8}\, (\gamma^{ij})_{pq} [ {\mathbb X}^i, {\mathbb X}^j]\right)=0\;
\;
\ee
which formally coincides with the BPS condition in the case of (the first order formulation of) the maximally supersymmetric SU($N$) SYM theory\footnote{The term ``formally'' highlights that  the relation between ${\mathbb P}^i$ and $\dot{{\mathbb X}}^i$ in the generic mD$0$ action involves the function ${\cal M}({\cal H})$ and, for ${\cal M}^\prime \not= 0$, additional dependence on $\mathbb{P}^i$ and $\mathbb{X}^i$ through the function $\frak{H}$ defined in Eq.\eqref{frakH=}; see also Eq.\eqref{DPi==gf}.}. It also formally coincides with the equation defining the supersymmetry which can be  preserved by purely bosonic solutions of 11D mM$0$ equations \cite{Bandos_11D_mM0}, and thus our discussion below parallels that analysis in \cite{Bandos_11D_mM0}.

Preservation of the 1/2 of the global supersymmetry implies, in the light of \eqref{e1=k}, that Eq. \eqref{susy=Psi=bb} does not impose any restriction on the parameter $\kappa^q$. This, in turn, requires that the matrix ${\mathbb L}_{pq}$ vanishes, i.e.  $\gamma^{i}_{pq} {\mathbb P}^i  - \frac {i}{8}\, (\gamma^{ij})_{pq} [ {\mathbb X}^i, {\mathbb X}^j]=0$. Then
\be\label{vacuum}
 {\mathbb P}^i=0\; , \qquad [ {\mathbb X}^i, {\mathbb X}^j]=0 \, ,
\ee
which characterizes the vacuum of the relative motion of the mD$0$ constituents. In this case ${\cal H} =0$ and, indeed \eqref{vacuum} follows directly from this condition.

To examine whether more parts of spacetime supersymmetry can be preserved we analyze Eq.~\eqref{susy=Psi=bb}, which is actually a set of 16 traceless SU($N$) matrix equations. To reduce it to a more manageable form, we multiply it by the matrix 
\begin{equation}
    \tilde{{\mathbb L}}_{qr}:= \left(\gamma^{i}_{qr} {\mathbb P}^i  + \frac {i}{8}\, (\gamma^{ij})_{qr} [ {\mathbb X}^i, {\mathbb X}^j]\right)~,
\end{equation}
which differs by relative sign from the matrix ${{\mathbb L}}_{qr}$ in  \eqref{susy=Psi=bb}, and trace the result with respect to the $\text{SU}(N)$ indices. In such a way we arrive at
\be
\label{kpL2tL2=0}  \kappa^p {\rm tr} ({\mathbb L}\tilde{\mathbb L})_{pq}:=
\kappa^q \left({\rm tr}( {\mathbb P}^i{\mathbb P}^i)  - \frac {1}{32}\, {\rm tr}([ {\mathbb X}^i, {\mathbb X}^j]^2)\right) + \frac {i}{4}\, \kappa^p  (\gamma^{j})_{pq}{\rm tr}( {\mathbb P}^i [ {\mathbb X}^i, {\mathbb X}^j])=0\; ,
\ee
where we have used
\begin{equation}
    \gamma^{ijkl} {\rm tr}\left([\mathbb{X}^i,\mathbb{X}^j][\mathbb{X}^k, \mathbb{X}^l] \right)=\gamma^{ijkl}~ {\rm tr}\left( \mathbb{X}^i [\mathbb{X}^j, [\mathbb{X}^k, \mathbb{X}^l]] \right)=0~,
\end{equation}
which follows from the Jacobi identity $\left[\mathbb{X}^{[j}, \left[\mathbb{X}^k, \mathbb{X}^{l]} \right] \right]\equiv 0$. Now one can recognize in the multiplier of the first term the  bosonic limit of the relative motion Hamiltonian   $\mathcal{H}$  \eqref{HmM0==} multiplied by $2$ and also appreciate that the coefficient for the second term vanishes as a consequence of the (bosonic limit of the) Gauss constraint \eqref{Gauss=}, ${\rm tr}( {\mathbb P}^i [ {\mathbb X}^i, {\mathbb X}^j])={\rm tr}( [{\mathbb P}^i , {\mathbb X}^i] {\mathbb X}^j)=0$. Thus, if we use this constraint, the final form of the consistency condition \eqref{kpL2tL2=0}  for the supersymmetry preservation by mD$0$ configuration simplifies to
\begin{equation}\label{kc=0}
\kappa^q \mathcal{H} = 0~. 
\end{equation}
Thus the BPS condition for a purely bosonic supersymmetric configuration of the mD$0$ system is just the vanishing of the (bosonic limit of the) relative motion Hamiltonian,
\begin{eqnarray}
\label{cHb=0} {\cal H} = {1\over 2} \text{tr}\left( {\mathbb P}^i {\mathbb P}^i \right)   - {1\over 64}
\text{tr}\left[ {\mathbb X}^i ,{\mathbb X}^j \right]^2  =0\; .  \quad
  \end{eqnarray}
As we have already noticed, the general solution is given by the vacuum \eqref{vacuum}, meaning that only $1/2$ of the spacetime supersymmetry can be preserved by the mD$0$ system.

This result allows us to conclude that any supersymmetric solution of the maximally supersymmetric SU($N$) SYM theory yields a corresponding supersymmetric solution of the mD$0$ equations for any positive definite function ${\cal M}({\cal H})$ since the relative motion mD$0$ equations then differ from the SYM equations by rescaling of the time variable by the constant factor ${\cal M}(0)/2$. These solutions form a family characterized by distinct properties of the center of mass motion $x^\mu_0$ and $p^\mu = \frak{M}u^{0\mu}$ in \eqref{x=x0+pt}. Thus, as far as the supersymmetric configurations of the relative motion sector in the mD$0$ model is concerned, these are in one-to-one correspondence with those solutions of the SU($N$) SYM equations.

The correspondence of the relative motion equations of the mD$0$ brane constituents and the maximally supersymmetric $\text{d}=1$ ${\cal N}=16$ SU($N$) SYM equations can be also used to search for non-supersymmetric solutions of the mD0 equations. See Appendix~\ref{sec:App_nonSUSY} for an example.

     \chapter{Dimensional reduction of the 11D mM0 action and its relation to mD0 system} \label{ch.11D_origin}
\thispagestyle{empty}

\vspace*{-1.5cm}   
\begin{changemargin}{7cm}{0cm}
    \singlespacing\textcolor{cites}{ \small{
         \begin{flushright}
         No, not Northward. Upward.
         \end{flushright}
     \begin{flushright}     
         {\sffamily {\textit{Flatland: A Romance of Many Dimensions}}\\
         {by {Edwin A. Abbott}. }}
     \end{flushright}}}
    \end{changemargin}
    \vspace{12pt}
    
In this chapter we demonstrate that a specific representative of the family of the candidate mD$0$ actions \eqref{eq:10D_SmD0=}, namely the one with ${\cal M}({\cal H}/\mu^6)$ as given in Eq.\eqref{cM=m+}, can be derived by dimensional reduction of the 11D mM$0$ action proposed in \cite{Bandos_11D_mM0}.

\section{11D mM0 action and its symmetries}
The mM$0$ brane system, which is conjectured to arise as the decompactification limit (M-theory lifting) of mD$0$ system, contains the same matrix fields as the $\text{d}=1$ ${\cal N}=16$ SYM on its worldline. However, as we will see, the set of variables describing the center of energy sector differs from that of the mD$0$ system. 

\subsection{Center of energy variables for mM0}
The center of energy degrees of freedom include the coordinate functions
\begin{equation}
z^{\underline{M}}(\tau)=  (X^{\underline{\mu}}(\tau), \Theta^{\underline{\alpha}}(\tau))\; , \qquad {\underline{\mu}}=0,1,\ldots,9,10\; , \qquad
{\underline{\alpha}}=1,\ldots,32\;
\end{equation}
which describe the embedding of the worldline into the 11D superspace $\Sigma^{(11|32)}$ with coordinates $z^{\underline{M}}=  (X^{\underline{\mu}}, \Theta^{\underline{\alpha}})$. These are supplemented by spinor moving frame fields (described below) and a Lagrange multiplier
$\rho^\# (\tau)$ whose role will also be clarified below.

The 11D VA 1-form is given by
\be\label{Pi=11}
\Pi^{\underline{\mu}}= \text{d}X^{\underline{\mu}} - i \text{d}\Theta \Gamma^{\underline{\mu}}\Theta = (\Pi^{\mu}, \Pi^{*})~,
\ee
where $\Gamma^{\underline{\mu}}_{\underline{\alpha}\underline{\beta}}$ are real symmetric  $32\times 32$ matrices 
\begin{equation}
    \Gamma^{\underline{\mu}}_{\underline{\alpha}\underline{\beta}}= \Gamma^{\underline{\mu}} {}_{\underline{\alpha}}{}^{\underline{\gamma}}C_{\underline{\gamma}\underline{\beta}} = \Gamma^{\underline{\mu}}_{\underline{\beta}\underline{\alpha}}~
\end{equation}
constructed from the 11D Dirac matrices $\Gamma^{\underline{\mu}}{}_{\underline{\alpha}}{}^{\underline{\gamma}} = - (\Gamma^{\underline{\mu}}{}_{\underline{\alpha}}{}^{\underline{\gamma}})^*$ obeying the Clifford algebra
\begin{equation}\label{Cliff}
\Gamma^a\Gamma^b+\Gamma^b\Gamma^a = 2\eta^{ab}{\mathbb 1}_{32\times 32}
\end{equation} 
and the charge conjugation matrix $C$, which is antisymmetric and imaginary in our mostly minus signature:  $C_{\underline{\gamma}\underline{\beta}}= - C_{\underline{\beta}\underline{\gamma}}=- (C_{\underline{\gamma}\underline{\beta}})^*$. We will also need the symmetric matrices  with upper indices
\begin{equation}
    \tilde{\Gamma}^{{\underline{\mu}}  \; \underline{\alpha}\underline{\beta}}= \tilde{\Gamma}^{{\underline{\mu}}  \; \underline{\beta}\underline{\alpha}}= C^{\underline{\alpha}\underline{\gamma}} {\Gamma}^{\underline{\mu}}{}_{\underline{\gamma}}{}^{\underline{\beta}}
\end{equation}
 which can be used to write the Clifford algebra \eqref{Cliff} in the form
 \begin{equation}
     \Gamma^{(\underline{\mu}}{}_{\underline{\alpha}\underline{\delta}} \tilde{\Gamma}^{\underline{\nu}) \underline{\delta}\underline{\beta}}=\eta^{\underline{\mu}\underline{\nu}} \delta_{\underline{\alpha}}^{\underline{\beta}}~.
 \end{equation}
We use the following SO$(1,9)$ invariant decomposition of 11D fermionic Majorana spinor coordinates on two 10D Majorana-Weyl (with opposite chiralities)
\be
\Theta^{\underline{\alpha}}= \left(\begin{matrix} \theta^{1\alpha} \cr
\theta^2_{\alpha}
\end{matrix}\right)\; . \qquad
\ee
Then, with the appropriate $\text{SO}(1,9)$ invariant representation of the 11D gamma matrices (presented in Appendix~\ref{sec:App_Gamma}), the 11D $\mathcal{N}=1$ VA 1-form  decomposes, as indicated already in \eqref{Pi=11}, yielding the 10D $\mathcal{N}=2$ VA 1-form \eqref{Pi=VA}
\be\label{Pi=10}
\Pi^{{\mu}}= \text{d}X^{{\mu}} - i\text{d}\theta^1 \sigma^{{\mu}}\theta^1 - i\text{d}\theta^2 \tilde{\sigma}^{{\mu}}\theta^2
\ee
and the scalar (with respect to SO$(1,9)$) 1-form
\be\label{Pi*=}
\Pi^{*}= \text{d}X^{*} + i\text{d}\theta^1 \theta^2 + i \text{d}\theta^2 \theta^1\; .
\ee
The 11D spinor moving frame variables (or 11D Lorentz harmonics \cite{Galperin_2}; see
\cite{Bandos_11D_mM0, Meliveo} and refs. therein) are  defined as rectangular blocks of Spin$(1,10)$ valued matrix
\begin{eqnarray}\label{harmV=11D}
V_{\underline{\alpha}}^{(\underline{\beta})}= \left(\begin{matrix} v_{\underline{\alpha} {q}}^{\; +} ,~~  v_{\underline{\alpha} q}^{\; -}
  \end{matrix}\right) \in \text{Spin}(1,10)\;  \qquad
\end{eqnarray}
which serves as a square root of the $\text{SO}(1,10)$ valued moving frame matrix
\begin{eqnarray}\label{uaib=}
U_{\underline{\mu}}^{(\underline{a})}= \left( \frac 1 2 \left(U_{\underline{\mu} }^\#+ U_{\underline{\mu} }^=\right) , U_{\underline{\mu} }^i\, , \frac 1 2 \left(U_{\underline{\mu} }^\# -U_{\underline{\mu}}^=\right)\right)  \in \text{SO}(1, 10)\;  \qquad
\end{eqnarray}
constructed from two lightlike vectors $U^=$ and $U^\#$ normalized in their contraction, together with 9 orthonormal spacelike vectors $U^i$ orthogonal to that two,
\begin{eqnarray}\label{uu=0}
&& U_{\underline{\mu}}^{=}U^{{\underline{\mu}}=}=0 \; , \qquad
  U_{\underline{\mu}}^{\#}U^{{\underline{\mu}}\#}=0 \; , \qquad  U_{\underline{\mu}}^{=}U^{\underline{\mu}\#}=2 \; , \qquad \\ \label{uui=0} &&~ U_{\underline{\mu}}^{i}U^{{\underline{\mu}}=}=0 \; , \qquad ~U_{\underline{\mu}}^{i}U^{{\underline{\mu}}\#}=0 \; , \qquad  ~~U_{\underline{\mu}}^{i}U^{{\underline{\mu}}j}=-\delta^{ij} \; . \qquad
\end{eqnarray}
The moving frame vectors also obey $U_{\underline{\mu}}^{(\underline{c})}\eta_{(\underline{c})(\underline{d})}U_{\underline{\nu}}^{(\underline{d})}=\eta_{{\underline{\mu}}{\underline{\nu}}}$, which can be written in the form of
\begin{eqnarray}\label{I=UU}
\delta_{\underline{\mu}}{}^{\underline{\nu}}= {1\over 2}U_{\underline{\mu}}^{=}U^{{\underline{\nu}}\#}+ {1\over 2}U_{\underline{\mu}}^{\#}U^{{\underline{\nu}}=}-  U_{\underline{\mu}}^{i}U^{{\underline{\nu}}i} \; . \qquad
\end{eqnarray}
The square root relations discussed above encode the Lorentz invariance of the 11D Dirac matrices. They take the form,
\begin{eqnarray}\label{VGVt=G11} V\Gamma_{\underline{\mu}} V^T =  U_{\underline{\mu}}^{({\underline{a}})} {\Gamma}_{(\underline{a})}\, , \qquad V^T \tilde{\Gamma}^{({\underline{a}})}  V = \tilde{\Gamma}^{{\underline{\mu}}} U_{\underline{\mu}}^{({\underline{a}})}\;
 \, . \qquad \end{eqnarray}
The spinor moving frame variables also satisfy the constraint
\be \label{VCVt=C11}
 VCV^T=C \;,  \qquad
\ee
which reflects the Lorentz invariance of the charge conjugation matrix. From this, it follows that the blocks of the inverse spinor moving frame matrix obeying
\begin{eqnarray}\label{v-qv+p=11}
&
v^{-{\underline{\alpha}}}_{{q}}   v_{{\underline{\alpha}} {p}}^{\; +}=\delta_{{q}{p}}
 \; ,  \qquad & v^{-{\underline{\alpha}}}_{{q}}   v_{{\underline{\alpha}} q}^{\; -}=0 \;  , \qquad
 \nonumber  \\
 & v^{+{\underline{\alpha}}}_{{q}}  v_{{\underline{\alpha}} {p}}^{\; +}=0
 \;  , \qquad & v^{+{\underline{\alpha}}}_{{q}} v_{{\underline{\alpha}} {p}}^{\; -} = \delta_{qp} \;   \qquad
\end{eqnarray}
are constructed from the spinor frame variables of~\eqref{harmV=11D} as
 \begin{eqnarray}
\label{V-1=CV-A} v_{q}^{+ {\underline{\alpha}}}  =  i C^{{\underline{\alpha}}{\underline{\beta}}}v_{{\underline{\beta}} q}^{\; +  }   \, ,
\qquad  v_{q}^{- {\underline{\alpha}}}  =  -i C^{{\underline{\alpha}}{\underline{\beta}}}v_{{\underline{\beta}} q}^{\; -  }   \,.  \qquad
 \end{eqnarray}
Using an appropriate (SO$(1,1)\otimes \text{SO}(9)$ invariant) representation of the 11D gamma matrices (see Appendix \ref{sec:App_Gamma}), Eqs.~\eqref{VGVt=G11} split into
\begin{eqnarray}\label{u==v-v-=11D}
 &
  ~~v^-_{{q}} \tilde{\Gamma}_{{\underline{\mu}}}v^-_{{p}}=U_{\underline{\mu}}^= \delta_{{q}{p}}   \; , & \qquad  U_{\underline{\mu}}^= \Gamma^{\underline{\mu}}_{\underline{\alpha}\underline{\beta}}= 2v_{\underline{\alpha} q}{}^- v_{\underline{\beta} q}{}^-  \; ,  \qquad \\
\label{v+v+=u++}
&  \;  v_{{q}}^+ \tilde{\Gamma}_{{\underline{\mu}}} v_{{p}}^+ =U_{{\underline{\mu}}}^{\# } \delta_{{q}{p}}\; , & \qquad~ U_{ {\underline{\mu}}}^{\# } {\Gamma}^{ {\underline{\mu}}}_{ {{\underline{\alpha}}} {{\underline{\beta}}}} =2 v_{{{\underline{\alpha}}}{q}}{}^{+}v_{{{\underline{\beta}}}{q}}{}^{+} \; , \qquad \\
 \label{uIs=v+v-=11D}
&~ v_{{q}}^- \tilde {\Gamma}_{{\underline{\mu}}} v_{{p}}^+=U_{ {\underline{\mu}}}^{i} \gamma^i_{q{p}}\; , &\qquad
 ~~U_{{\underline{\mu}}}^{i} {\Gamma}^{{\underline{\mu}}}_{{\underline{\alpha}}{\underline{\beta}}} =2 v_{( {{\underline{\alpha}}}|{q} }{}^- \gamma^i_{qp}v_{|{{\underline{\beta}}}){p}}{}^{+} \; , \quad  \end{eqnarray}
where $\gamma^i_{qp}=\gamma^i_{p{q}}$ are the SO$(9)$ Dirac matrices obeying
\be \gamma^{i}_{qr}\gamma^{j}_{rp} +\gamma^{j}_{qr}\gamma^{i}_{rp} =2\delta^{ij}\delta_{qp}\; . 
\ee
The derivatives of the moving frame and spinor moving frame variables can be expressed in terms of Cartan forms
\bea\label{Omij=11D}\underline{\Omega}^{ij}:=  U^{\underline{\mu}i}\text{d} U_{\underline{\mu}}^j\, ,  \qquad \\ \label{Om0=11D} \underline{\Omega}^{(0)}:=  U^{\underline{\mu}=}\text{d} U_{\underline{\mu}}^{\#}\, ,  \qquad  \\ \label{Om--j=11D} \underline{\Omega}^{\#j}:=  U^{\underline{\mu}\#}\text{d} U_{\underline{\mu}}^j\, , \qquad \underline{\Omega}^{=j}:=  U^{\underline{\mu}=}\text{d} U_{\underline{\mu}}^j\,  \qquad
\eea (see e.g. \cite{Igor_amplitudes_1, Igor_amplitudes_2} and refs. therein for more details). The form $\underline{\Omega}^{ij}$ transforms as a connection under the SO$(9)$ symmetry which acts on both the 9-vector indices $i,j$ and the 16-component spinor indices $q,p$ of the moving frame variables and the matrix fields. This is the 11D counterpart of the Cartan form  $\Omega^{ij}$ from the 10D SO$(1,9)/$SO$(9)$ spinor moving frame (Lorentz harmonics) formalism.

The Cartan form $\underline{\Omega}^{(0)}$ transforms as a connection under the SO$(1,1)$ group, under which the moving frame, spinor moving frame variables, and the Lagrange multiplier $\rho^\#$ scale according to the weights indicated by their sign indices. In particular, since  $\#=++$ we have 
\bea\label{SO11}
U^\#_{\underline{\mu}} \mapsto e^{2\beta } U^\#_{\underline{\mu}}\; , \qquad U^=_{\underline{\mu}} \mapsto e^{-2\beta } U^=_{\underline{\mu}}\; , \qquad U^i_{\underline{\mu}} \mapsto  U^i_{\underline{\mu}}\; , \qquad \nonumber  \\
\underline{\Omega}^{(0)}  \mapsto   \underline{\Omega}^{(0)} +d\beta   \; , \qquad \underline{\Omega}^{ij}  \mapsto   \underline{\Omega}^{ij} \; , \qquad  \nonumber \\
\rho^\#\mapsto e^{2\beta }\rho^\#\;  \qquad
\eea
under SO$(1,1)$ group. The remaining Cartan forms, $\underline{\Omega}^{\#j}$ and $\underline{\Omega}^{=j}$, transform covariantly under both SO$(9)$ and SO$(1,1)$ which are gauge symmetries of the mM$0$ action.

\subsection{Matrix fields describing the relative motion of mM0 constituents}
The matrix fields describing the relative motion of the mM0 constituents are exactly the same as those appearing in the mD$0$ system. These include bosonic traceless $N \times N$ matrices ${\mathbb X}^i$ and ${\mathbb P}^i$, carrying the SO(9) vector indices, fermionic traceless $N \times N$ matrices ${\boldsymbol{ \Psi}}_q$, carrying SO(9) spinor index $q=1,\ldots,16$; and a bosonic traceless $N \times N$ matrix 1-form ${\mathbb A}= \text{d}\tau {\mathbb A}_\tau$ corresponding to the SU($N$) gauge connection on the worldline.

\subsection{The mM0 action}
The action for the 11D mM0 system, originally proposed in \cite{Bandos_11D_mM0}, can be written in the form
\begin{equation}
    \begin{array}{l}
         \begin{split}
             S_{\text{mM}0} &= \int_{\mathcal{W}^1} \rho^{\#}\, \underline{\text{E}}^{=} +  \frac 1 {\mu^6} \int_{\mathcal{W}^1}  \left(  \text{tr}\left({\mathbb P}^i \text{D} {\mathbb X}^i + 4i \mathbf{ \Psi}_q \text{D}
             {\boldsymbol{ \Psi}}_q  \right) + \underline{\text{E}}^{\#} \,\frac 1 {\rho^{\#}}\, {\cal H} \right)+ \\
             &+ \frac 1 {\mu^6}  \int_{\mathcal{W}^1}
            \underline{\text{E}}{}^{+q} \,\frac 1 {\sqrt{\rho^{\#}}}\, \text{tr}\left(-4i (\gamma^i {\boldsymbol{ \Psi}})_q  {\mathbb P}^i + {1\over 2}
            (\gamma^{ij} {\boldsymbol{ \Psi}})_q  [{\mathbb X}^i, {\mathbb X}^j]  \right) -  \frac 1 {\mu^6} \int_{\mathcal{W}^1} \frac {\text{D}\rho^\# }{\rho^\# }
            \text{tr} ({\mathbb P}^i{\mathbb X}^i)~,
         \end{split}
    \end{array}\label{SmM0=}
\end{equation}
in which the matrix fields are redefined (with respect to those in \cite{Bandos_11D_mM0}) to be inert under the SO$(1,1)$ symmetry acting on the spinor moving frame variables and Lagrange multiplier $\rho^\#$, as shown in \eqref{SO11}.

The covariant derivatives of the matrix fields appearing in \eqref{SmM0=} are given by
 \begin{equation}\label{DXi=11D}
      \text{D}{\mathbb X}^i  := \text{d}{\mathbb X}^i   - \underline{\Omega}^{ij} {\mathbb
X}^j+ [{\mathbb A},    {\mathbb X}^i] \; ,
 \end{equation}
 \begin{equation} \label{DPsi=11D} 
     ~~~~~~~~~~~~\text{D}{\boldsymbol{ \Psi}}_q  := \text{d} {\boldsymbol{ \Psi}}_q
   -{1\over 4} \underline{\Omega}^{ij} \gamma^{ij}_{qp} {\boldsymbol{ \Psi}}_p+ [{\mathbb A},
 {\boldsymbol{ \Psi}}_q ] \;
 \end{equation}
and include, besides the SU($N$) connection 1-form ${\mathbb A}$, also the composite SO$(9)$ connection  $\underline{\Omega}^{ij}$ \eqref{Omij=11D}. In addition, the derivative of the Lagrange multiplier $\rho^{\#}$
\be
\text{D}\rho^\#= \text{d}\rho^\#- 2\underline{\Omega}^{(0)} \rho^\#\;
\ee
includes the SO$(1,1)$ connection \eqref{Om0=11D}. Let us recall that these composite connections are Cartan forms arising from the nonvanishing components of the derivatives of the moving frame and spinor moving frame variables.

The moving frame and spinor moving frame variables also enter the action explicitly through the projections
\be\label{E++=}
\underline{\text{E}}^\#=\Pi^{\underline{\mu}} U_{\underline{\mu}}^\# \; , \qquad \underline{\text{E}}^==\Pi^{\underline{\mu}} U_{\underline{\mu}}^= \; , \qquad
\ee
 where $\Pi^{\underline{\mu}}$ is the pull-back of the 11D bosonic supervielbein 1-form (see Eq.~\eqref{Pi=11}), and
\be\label{E+q=}
\underline{\text{E}}^{+q} = \text{d}\Theta^{\underline{\alpha}}\, v^{\; +}_{\underline{\alpha} q}\,
\ee
is the pull-back of the fermionic supervielbein form.

The ``relative motion Hamiltonian'' in \eqref{SmM0=}
\begin{equation}
\label{HmM0=} {\cal H} = {1\over 2} \text{tr}\left( {\mathbb P}^i {\mathbb P}^i \right) - {1\over 64}
\text{tr}\left[ {\mathbb X}^i ,{\mathbb X}^j \right]^2  - 2\,  \text{tr}\left({\mathbb X}^i\, {\boldsymbol{ \Psi}}\gamma^i{\boldsymbol{ \Psi}}\right)  \qquad
  \end{equation}
coincides with that of the mD$0$ system (\textit{cf}. Eq. \eqref{HmM0==}).

The action \eqref{SmM0=} is manifestly invariant under the 11D target space supersymmetry and, as shown in \cite{Bandos_11D_mM0}, possesses an additional 16 parametric local worldline supersymmetry, generalizing the $\kappa$-symmetry  of single M$0$-brane action formulated in the spinor moving frame approach~\cite{BRST_M0, Igor_BRST_2}.

The properties of this mM$0$ system described by the action \eqref{SmM0=} were further studies in \cite{Meliveo}. The problem of its dimensional reduction to $\text{D}=10$ was addressed in \cite{10D_mD0_Igor}, but not resolved there. The reason was that a suitable convenient choice of the basic matrix fields was not found in \cite{10D_mD0_Igor}. This issue was recently overcome, first for the simpler case of $\text{D}=4$ counterpart of mM$0$ system described in chapter~\ref{ch.4D_nAmW}. Using this insight in~\cite{Unai_3} we have successfully rewritten the mM$0$ action from \cite{Bandos_11D_mM0} in terms of more convenient matrix variables, as presented in Eq.~\eqref{SmM0=}. The dimensional reduction of this reformulated mM$0$ action to 10D is the subject of the next subsection.

\section{Dimensional reduction of mM0 action to D=10}
\label{sec:dimRed}
As we have already noticed, with the suitable representation for 11D gamma matrices and charge conjugation matrix (see Appendix \ref{sec:App_Gamma}, in particular Eqs.~\eqref{G11=s10}, \eqref{tG11=s10}, and \eqref{C11=10}), the 11D VA 1-form can be decomposed as in \eqref{Pi=11} into the 10D VA 1-form \eqref{Pi=10} and SO$(1,9)$ invariant 1-form \eqref{Pi*=}.

\subsection{SO(1,9) invariant form oh the 11D spinor moving frame and basic 1-forms}
The next step is to solve the  constraints defining the above described 11D  spinor moving frame variables in terms of the 10D Spin$(1,9)/\text{Spin}(9)$ spinor ones \eqref{harmV=10}.
The convenient form of the solution is
\begin{eqnarray}\label{hV11=hv10}
\frac 1 {\sqrt{\rho^\#}}  v_{{\underline{\alpha}} q}^{\; +}
=\frac 1 {\sqrt{2}}  \frac 1 {\sqrt{{\cal M}}}\left(\begin{matrix} v_{{\alpha}}^{\; q} \cr  -v_q{}^\alpha
  \end{matrix}\right)\; , \qquad  \sqrt{\rho^\#} \,v_{{\underline{\alpha}} q}^{\; -}
=\frac 1 {\sqrt{2}}\, \sqrt{{\cal M}}\,  \left(\begin{matrix} v_{{\alpha}}^{\; q} \cr v_q{}^\alpha
  \end{matrix}\right)\;  \qquad
\end{eqnarray}
which implies the complementary relations
 \begin{eqnarray}\label{hV-1=hv10}
 \frac 1 {\sqrt{\rho^\#}}  v^{+\underline{\alpha}}_q
=\frac 1 {\sqrt{2}} \frac 1 {\sqrt{{\cal M}}}  \left(\begin{matrix}  v_q{}^\alpha \cr v_{{\alpha}}^{\; q}
  \end{matrix}\right)\; , \qquad   \sqrt{\rho^\#} \, v^{-\underline{\alpha}}_q
=\frac 1 {\sqrt{2}} \, \sqrt{{\cal M}}\, \left(\begin{matrix}  v_q{}^\alpha \cr -v_{{\alpha}}^{\; q}
  \end{matrix}\right)\; .   \qquad
\end{eqnarray}
These expressions involve both the Lagrange multiplier $\rho^\#$ from the mM$0$ action and an arbitrary function ${\cal M}(\tau)$ of the proper time $\tau$ parametrizing the center of energy  worldline of the mM$0$ system.

Importantly, this is not merely an ansatz, but the general solution of the constraints \eqref{u==v-v-=11D}-\eqref{uIs=v+v-=11D} and \eqref{VCVt=C11}. This can be confirmed by counting the number of degrees of freedom, modulo natural gauge symmetries, on both sides of the relation
\eqref{hV11=hv10} which gives $1+9=1+9$. Indeed, both sides include one scalar ($\rho^\#$ or ${\cal M}$) and spinor frame variables parametrizing cosets isomorphic to ${\mathbb S}^9$ sphere: 
\begin{equation}
    \dfrac {\text{SO}(1,10)}{(\text{SO}(1,1)\otimes \text{SO}(9)) \subset\!\!\!\!\times K_9}\simeq {\mathbb S}^9 \qquad \text{and} \qquad \frac {\text{SO}(1,9)}{\text{SO}(9)}\simeq {\mathbb S}^9
\end{equation}
respectively (see e.g. \cite{Igor_amplitudes_1, Igor_amplitudes_2} and refs. therein for more details). Notice also that the l.h.s. of \eqref{hV11=hv10} preserves the characteristic $\text{SO}(1,1)$ gauge symmetry \eqref{SO11} of the 11D spinor moving frame formalism (which was taken into account in the above counting of the degrees of freedom). From the 11D perspective, the solution \eqref{hV11=hv10} explicitly breaks the full Lorentz symmetry $\text{SO}(1,10)$ down to its $\text{SO}(1,9)$ subgroup, which becomes Lorentz symmetry of the reduced 10D theory.

Substituting \eqref{hV11=hv10} into (the pull-backs of) 11D fermionic 1-form entering the mM0 action \eqref{SmM0=}, we obtain
\be  \underline{\text{E}}{}^{+q} = {\sqrt{\rho^{\#}}}\, \frac 1 {\sqrt{2{\cal M}}} (\text{E}^{1q} - \text{E}_q^2) \; , \ee
where
\be\label{E1q=} \text{E}^{1q}=\text{d}\theta^{1\alpha} v_{\alpha}^q \; , \qquad \text{E}^2_{q}=\text{d}\theta^2_{\alpha} v_q^{\alpha}\;  \qquad \ee
can be naturally identified with the pull-backs of the 10D type IIA supervielbein forms \eqref{Eq1:=}.

Now, substituting \eqref{hV11=hv10} into Eqs. \eqref{u==v-v-=11D} and \eqref{v+v+=u++}
and using the suitable representation for 11D gamma matrices (see Appendix \ref{sec:App_Gamma}) as well as \eqref{u0s=vv} we find that the 11D moving frame vectors are related to its 10D counterparts by
\begin{equation}\label{U==}
U^=_{\mu}= \frac {{\cal M}} {\rho^\#} u^0_{\mu}\; , \qquad U^=_{*}= - \frac {{\cal M}} {\rho^\#}\; ,
\end{equation}
\begin{equation}\label{U++=}
    U^\#_{\mu} =  \frac {\rho^\#}{{\cal M}} \, u^0_{\mu} \; , \qquad U^\#_{*}  = \frac {\rho^\#}{{\cal M}}\;
\end{equation}
so that
\be
\rho^\#\underline{\text{E}}^{=}= {\cal M}({\rm E}^0 -\Pi^*) \; , \qquad \frac {\underline{\text{E}}^{\#}}{\rho^\#}= \frac 1 {{\cal M}}({\rm E}^0 +\Pi^*) \; .  \qquad
\ee
Moreover, from \eqref{uIs=v+v-=11D} we find
\be\label{Ui=ui} 
U_{\underline{\mu}}^i = \delta_{\underline{\mu}}{}^\nu u_\nu^i\;
\ee
which can be used to identify the composite $\text{SO}(9)$ connection of the  mM$0$ system with its 10D counterpart \eqref{Omij=},
\be\label{Omij11=10}\underline{\Omega}^{ij}:=  U^{\underline{\mu}i}\text{d} U_{\underline{\mu}}^j = u^{{\mu}i}\text{d} u_{{\mu}}^j=: {\Omega}^{ij}\; .
\ee
Thus the covariant derivative of the matrix fields in the mM$0$ action (Eqs. \eqref{DXi=11D} and  \eqref{DPsi=11D}) coincide with \eqref{DXi=} and \eqref{DPsi=} used in the mD$0$ action \eqref{eq:10D_SmD0=}.

As far as $\text{SO}(1,1)$ connection is concerned, we find from \eqref{U==} and \eqref{U++=}
\be \underline{\Omega}{}^{(0)}= \frac 1 4 U^{=\underline{\mu}}\text{d}U_{\underline{\mu}}^\# = \frac 1 2 \left( \frac {\text{d}\rho^{\#}}{\rho^{\#}}- \frac {\text{d}{\cal M}}{{\cal M}}\right)\; ,   \ee
which gives
\be
\text{D}\rho^\# = \text{d}\rho^\#- 2 \rho^\# \underline{\Omega}^{(0)} = \rho^\# \frac {\text{d}{\cal M}}{{\cal M}}\; .
\ee
Thus, after using in \eqref{SmM0=} the relations \eqref{hV11=hv10} and their consequences, $\rho^\#$ disappears from the 11D action and it is fully replaced by ${\cal M}$.

\subsection{Dimensional reduction of the mM0 action}
With the solution \eqref{hV11=hv10} and the splitting \eqref{Pi=11}, the mM0 action \eqref{SmM0=}  acquires the form
\begin{equation}
    \begin{array}{l}
         \begin{split}
             S_{\text{mM}0}\vert_{\eqref{hV11=hv10}} &= \int_{\mathcal{W}^1} {\cal M} (\text{E}^{0}-\Pi^*)  +  \frac 1 {\mu^6} \int_{\mathcal{W}^1} \frac 1 {\cal M} (\text{E}^{0}+\Pi^*)\, {\cal H} +\\
             &+ \frac 1 {\mu^6} \int_{\mathcal{W}^1}    \text{tr}\left({\mathbb P}^i \text{D} {\mathbb X}^i + 4i {\boldsymbol{ \Psi}}_q \text{D}
             {\boldsymbol{ \Psi}}_q  \right)-  \frac 1 {\mu^6} \int_{\mathcal{W}^1} \frac {\text{d} {\cal M}}{ {\cal M} }
             \text{tr} ({\mathbb P}^i{\mathbb X}^i) +\\
             & +   \frac 1 {\mu^6}  \int_{\mathcal{W}^1}
             \frac 1 {2\sqrt{{\cal M}}}(\text{E}{}^{1q}-\text{E}{}^{2}_{q}) \, \text{tr}\left(-4i (\gamma^i {\boldsymbol{ \Psi}})_q  {\mathbb P}^i + {1\over 2}
             (\gamma^{ij} {\boldsymbol{ \Psi}})_q  [{\mathbb X}^i, {\mathbb X}^j]  \right)~, 
         \end{split}
    \end{array} \label{SmM0=V11-v10}  
\end{equation}
where the covariant derivatives are given by \eqref{DXi=} and \eqref{DPsi=}, and the relative motion Hamiltonian ${\cal H}$ has the form of \eqref{HmM0==}. Here,
\be
\text{E}^0=\Pi^\mu u_\mu^0~.
\ee
with $\Pi^\mu$ from \eqref{Pi=10} (see  \eqref{E0:=}), $\Pi^*$ is defined in \eqref{Pi*=} and, according to \eqref{E1q=},
\be\label{E1q-2=} \text{E}^{1q}-\text{E}^2_{q} = \text{d}\theta^{1\alpha} v_{\alpha}^q - \text{d}\theta^2_{\alpha} v_q^{\alpha}\; .  \qquad \ee
To complete the dimensional reduction, we derive the equation of motion,  from the action \eqref{SmM0=V11-v10}, for the 11th bosonic coordinate field $X^*$,
\be\label{cMEq=}
\text{d}\left({\cal M} - \frac 1 {{\cal M}}\, \frac {\cal H} {\mu^6}\right)=0\;,
\ee
and substituting its solution back into the functional \eqref{SmM0=V11-v10}. This solution can be equivalently written as
\be\label{cMEq=m}
{\cal M} - \frac 1 {{\cal M}}\, \frac {\cal H} {\mu^6}=m \;
\ee
for some integration constant $m$. Its solution (with nonvanishing limit when ${\cal H}\mapsto 0$) is
\be\label{cM=m+s}
{\cal M} = \frac m 2 +  \sqrt{\frac {m^2} {4}+\frac {\cal H} {\mu^6}}\; .
\ee
Substituting this result back into the action \eqref{SmM0=V11-v10} we find the 10D mD$0$ action \eqref{eq:10D_SmD0=} with a particular function ${\cal M}= {\cal M} \left( {{\cal H}} /{\mu^6} \right)$ given by the above expression \eqref{cM=m+s}.

Thus, among the family of models described by Eq.~\eqref{eq:10D_SmD0=}, this specific case is singled out as having a clear origin in 11D M-theory.

     \chapter{Towards field theory of multiple D0-branes: quantization of simplest 3D mD0 prototype} \label{ch.quantization}
\thispagestyle{empty}

 \vspace*{-1.5cm}   
\begin{changemargin}{4cm}{0cm}
    \singlespacing\textcolor{cites}{ \small{
         \begin{flushright}
         They were close to the end of the beginning...
         \end{flushright}
     \begin{flushright}     
         {\sffamily {\textit{The Dark Tower I: The Gunslinger }}\\
         {by {Stephen King}.}}
     \end{flushright}}}
    \end{changemargin}
    \vspace{12pt}

In this thesis we have constructed and presented a completely supersymmetric nonlinear action that exhibits the properties expected of a multiple D$0$-brane system. Its quantization is anticipated to yield an interesting supersymmetric field theory formulated in the superspace extended by additional matrix coordinates, which can provide important insights into the study of String/M-Theory. 

As a first step toward this goal, in this chapter we develop the Hamiltonian formulation and perform a covariant quantization of the simplest 3-dimensional counterpart (see chapter~\ref{ch.3D_mD0}) of the 10-dimensional multiple D$0$-brane model (see chapter~\ref{ch.10D_mD0}). Through this process, we derive a supersymmetric equations of the field theory of this simplest 3D counterpart of mD$0$ system and discuss some of its solutions.

\section{Simplest 3D counterpart of the 10D mD0 action}
In this section we recall the fields used to describe the 3D mD$0$ counterpart. However, as we will discuss in section~\ref{sec:3DcomplexSMF}, we adopt a different representation for the 3D spinor moving frame variables compared to the one used in chapter~\ref{ch.3D_mD0}.

\subsection{Physical and auxiliary fields of the 3D mD0 system}
The dynamical variables of 3D mD$0$ system, can be naturally decomposed into the center of mass variables and the matrix fields describing the relative motion of the mD$0$ constituents. This latter set contains the fields of $\text{D}=3$ SU($N$) SYM model dimensionally reduced to $\text{d}=1$. These are traceless $N \times N$  bosonic matrix fields
\begin{equation}
    {\mathbb Z}=|| {\mathbb Z}_i{}^j(\tau)  ||  \qquad \text{and} \qquad \bar{{\mathbb Z}}= ||\bar{{\mathbb Z}}_i{}^j(\tau) || ~,
\end{equation}
 which are related by Hermitian conjugation
\be\label{Z=bZ*}
{{\mathbb Z}} = (\bar{{\mathbb Z}})^\dagger \qquad \Longleftrightarrow \qquad {{\mathbb Z}}_i{}^j = (  \bar{{\mathbb Z}}_j{}^i)^* \; , \qquad {\rm tr}\, {{\mathbb Z}} = {{\mathbb Z}}_i{}^i=0\; ,  \qquad i,j=1,\ldots,N\; ,
\ee
and two Hermitian conjugate  traceless $N \times N$ fermionic matrix fields\footnote{Here we have replaced the notation for the matrix fermionic fields from $\Psi \mapsto {\boldsymbol{\Psi}}$ compared to chapter~\ref{ch.3D_mD0}. This distinction allows us to differentiate between the fermionic matrix fields ${\boldsymbol{\Psi}}$ and the fermionic coordinates $\Psi$ that will appear in the quantization of the mD$0$ system in section~\ref{sec:quantization}.}
\begin{equation}
    {\boldsymbol{ \Psi}}=||{{\boldsymbol{ \Psi}}}_i{}^j(\tau) || \qquad \text{and} \qquad \bar{{\boldsymbol{ \Psi}}}=|| \bar{{\boldsymbol{ \Psi}}}_i{}^j (\tau) ||~,
\end{equation}
satisfying
\be\label{Psi=bPsi*}
{{\boldsymbol{ \Psi}}}=\bar{{\boldsymbol{ \Psi}}}{}^\dagger \qquad \Longleftrightarrow \qquad {{\boldsymbol{ \Psi}}}_i{}^j =( \bar{{\boldsymbol{ \Psi}}}_j{}^i )^*\; , \qquad {\rm tr}\,  {{\boldsymbol{ \Psi}}} \equiv  {{\boldsymbol{ \Psi}}}_i{}^i=0~,
\ee
as well as anti-Hermitian 1-form
\be
{\mathbb A}=\text{d}\tau {\mathbb A}_\tau= \text{d}\tau || {\mathbb A}_\tau{}_i{}^j (\tau)|| =- {\mathbb A}^\dagger \qquad \Longleftrightarrow \qquad   {\mathbb A}_\tau{}_i{}^j=- ( {\mathbb A}_\tau{}_j{}^i)^*\; .
\ee
All these fields depend  on the proper time $\tau$ which parametrizes the center of mass worldline ${\cal W}^1$ of the 3D mD$0$ system. This is described by the same variables as that used to describe a single D$0$-brane in the spinor moving frame formulation. The center of mass degrees of freedom are described by 3-vector bosonic and two complex conjugate fermionic spinor coordinate functions
\be
z^M(\tau) = (x^\mu(\tau), \theta^\alpha (\tau), \bar{\theta}{}^\alpha (\tau)), \quad   \quad  \bar{\theta}{}^\alpha (\tau)=(\theta^\alpha (\tau))^*\; , \quad    \mu=0,1,2\; , \quad \alpha=1,2 \; , 
\ee
which parametrize the worldline ${\cal W}^1$ as a trajectory in flat $\text{D}=3$ ${\cal N}=2$ superspace $\Sigma^{(3|4)}$ with coordinates
$z^M = (x^\mu, \theta^\alpha , \bar{\theta}{}^\alpha)$,
\bea
{\cal W}^1 \; \subset \; \Sigma^{(3|4)}\; : \qquad z^M=z^M(\tau) \; .
\eea
To write the action we also need to introduce some auxiliary fields, in particular the spinor moving frame variables, which will be described below. However, as a preliminary step, it is useful to revisit the de Azcárraga-Lukierski action, formulated purely in terms of the coordinate functions introduced above. 

\subsection{3D version of the de Azcárraga-Lukierski action for a single D0-brane}
When single D$0$-brane is considered, its action can be written only in terms of the above described coordinate functions. We present it here as a reference point.

The standard action for single D$0$-brane \cite{D-branes_Eric} is given by 10D counterpart of the $\text{D}=4$ action originally constructed by de Azc\'arraga and Lukierski \cite{kappa_1, Azcarraga2}. Here we discuss its 3D counterpart given by
\be\label{SD0=A+L}
S_{\text{AL}}=- m\int d\tau \left(\sqrt{E_\tau^a E_{\tau}^b\eta_{ab}}+\dot{\theta}^\alpha \bar{\theta}_\alpha- \theta^\alpha \dot{\bar{\theta}}_\alpha \right)\,
\ee
where the dot denotes a derivative with respect to proper time $\tau$, and
\be\label{Eta=}
E_\tau^a =\dot{x}{}^a - i \dot{\theta}\gamma^a \bar{\theta} + i {\theta}\gamma^a \dot{\bar{\theta}}\; , \qquad
\ee
is the pull-back (coefficient of $\text{d}\tau$) of the VA 1-form,
\be\label{Ea=}
E^a=  \text{d}{x}{}^a - i \text{d}{\theta}\gamma^a \bar{\theta} + i {\theta}\gamma^a \text{d}{\bar{\theta}}=: E^a(z)\; , \qquad E^a(z(\tau))=
\text{d}\tau E_\tau^a \; , \qquad a=0,1,2\; . \qquad
\ee
(See sections~\ref{sec:4Dto3D_conventions} and~\ref{sec:BS_reduction} for the notation).

As key property of the action \eqref{SD0=A+L} is its invariance under local fermionic $\kappa$-symmetry
\bea
&& \delta_\kappa x^a =  i \delta_\kappa{\theta}\gamma^a \bar{\theta} - i {\theta}\gamma^a \delta_\kappa{\bar{\theta}}\; , \qquad \\ 
&& \delta_\kappa{\theta}{}^\alpha=\kappa_\beta \left(\epsilon -i\tilde{\gamma}_a E_\tau^a /\sqrt{E_\tau^bE_{\tau b}}\right)^{\beta\alpha} \; , \qquad \delta_\kappa\bar{\theta}{}^\alpha=\bar{\kappa}_\beta \left(\epsilon +i\tilde{\gamma}_a E_\tau^a /\sqrt{E_\tau^bE_{\tau b}}\right)^{\beta\alpha} \; , \nonumber
\eea
which, as we have emphasized on several occasions throughout this thesis, implies that the ground state of the system preserves $1/2$ of the spacetime supersymmetry \cite{Bergshoeff_Tomas}, making it a BPS state, the 3D counterpart of the D$0$-brane BPS state in String Theory.

Thus, moving frame and spinor moving frame variables are not obligatory for describing a single D$0$-brane: in this case, the spinor moving frame formulation (as presented in chapter~\ref{ch.3D_mD0}) is classically equivalent to the formulation provided by the de Azcárraga-Lukierski action described above. However, the action for the 3D mD$0$ system is known only in its form involving spinor moving frame variables which we are going to describe now.

\subsection{3D complex spinor moving frame variables (Lorentz harmonics)}
\label{sec:3DcomplexSMF}
In chapter~\ref{ch.3D_mD0} we have used the real spinor moving frame variables $(v_{\alpha}^1, v_{\alpha}^2)$ subject to the constraint $v^{\alpha 2}v_{\alpha}^1=1$. However, for the purpose of this chapter, we found it much more convenient to describe the spinor frame variables using complex spinors
\be\label{w=} w_\alpha= \frac 1 {\sqrt{2}} (v_{\alpha}^1-iv_{\alpha}^2)\qquad \text{  and its c.c.} \qquad
\bar{w}_\alpha= \frac 1 { \sqrt{2}} (v_{\alpha}^1+iv_{\alpha}^2)\ee
which obey the relation
\be \label{bww=i}
\bar{w}{}^\alpha w_\alpha=i\,.
\ee 
This allows us to write the unity decomposition and to decompose the unit antisymmetric tensors as
\be
\delta_\alpha{}^\beta = \, i \bar{w}{}_\alpha w^\beta- i w_\alpha  \bar{w}{}^\beta\; , \qquad \epsilon_{\alpha\beta} = -i \bar{w}{}_\alpha w_\beta+ i w_\alpha  \bar{w}{}_\beta\; .
\ee
Taking these constraints into account, we find that the derivatives of these complex spinors (providing a 3D version of the 4D Newman-Penrose diad~\cite{Newman}) are expressed by
\be\label{Dw=}
\text{D}w_\alpha :=\text{d}w_\alpha + ia w_\alpha =  if \bar{w}_\alpha \; , \qquad \text{D} \bar{w}_\alpha :=\text{d} \bar{w}_\alpha - ia  \bar{w}_\alpha =- i \bar{f} {w}_\alpha  \; , \qquad
\ee
\noindent in terms of Cartan forms\footnote{These relate to the real Cartan forms $f^{pq}=v^{\alpha p}\text{d}v_{\alpha}^q$  used in chapter~\ref{ch.3D_mD0} by
\be
f=\frac 1 2 (f^{11}-f^{22}) -if^{12} \; , \qquad \bar{f}=\frac 1 2 (f^{11}-f^{22}) +if^{12}  \; , \qquad a=\frac 12 (f^{11}+f^{22})=\frac 12 f^{qq} \; .
\ee}
\be\label{f=wdw}
f= w^\alpha \text{d}w_\alpha\; , \qquad \bar{f} = \bar{w}^\alpha \text{d}\bar{w}_\alpha\; , \qquad a = w^\alpha \text{d}\bar{w}_\alpha= \bar{w}^\alpha \text{d}w_\alpha\; . \qquad
\ee
The first two of these provide a covariant basis of the cotangent space to the coset SU$(1,1)/\text{U}(1)= \text{SL}(2,{\mathbb R})/\text{SO}(2)$, while the third, $a$, transforms as a U$(1)$ connection. This allowed to introduce the U$(1)$ covariant derivatives $\text{D}$ in \eqref{Dw=}. It is straightforward to check that these Cartan forms satisfy the Maurer-Cartan structure equations
\be
\text{D}f=\text{d}f+2i f\wedge a =0\; , \qquad \text{D}\bar{f}= \text{d}\bar{f}-2i \bar{f}\wedge a =0\; , \qquad \text{d}a=-if\wedge \bar{f} \; . \
\ee
The moving frame vectors (a 3D version of the complex 4D lightlike Newman-Penrose tetrad \cite{Newman}) are constructed  as bilinears of the complex spinor $w$ and $\bar{w}$
\be\label{u=wgw}
u_a^{(0)}=w\gamma_a\bar{w}\; , \qquad u_a=w\gamma_aw\; , \qquad \bar{u}_a=\bar{w}\gamma_a\bar{w}\; . \qquad
\ee
Due to the properties of $\text{D}=3$ gamma matrices, these vectors obey
\begin{equation}
    \begin{array}{lclcl}
      u_a^{(0)}u^{a(0)}=1\; ,   &   u_a^{(0)}u^{a}=0\; , & u_a^{(0)}\bar{u}{}^{a}=0~,  \\
    \end{array} \label{u0u0=1}
\end{equation}
\begin{equation}
    \begin{array}{lclcl}
      u_au^{a}=0\; ,   & \bar{u}_{a}\bar{u}{}^{a}=0~, & u_a\bar{u}{}^{a}=-2 ~.  \\
    \end{array} \label{uu=0}
\end{equation}
Furthermore, they satisfy
\be\label{ug=ww}
u_a^{(0)}\gamma^a_{\alpha\beta}=2 w_{(\alpha}\bar{w}_{\beta )}\; , \qquad u_a\gamma^a_{\alpha\beta}=2 w_{\alpha}{w}_{\beta}\; , \qquad \bar{u}_a\gamma^a_{\alpha\beta}=2 \bar{w}_{\alpha}\bar{w}_{\beta}\; . \qquad
\ee
Eqs.~\eqref{u0u0=1} and \eqref{uu=0} imply that the real $3\times3$  matrix
\be\label{uinO} 
u_a^{(b)}=\left(u_a^{(0)}, \frac 1 2 (u_a+\bar{u}_a), \frac 1 {2i} (u_a-\bar{u}_a)\right)\;\in \;  \text{O}(1,2) \;  \ 
\ee 
belongs to O$(1,2)$ group. Moreover, their construction from spinor bilinears further implies\footnote{To see this we can take the constant ``vacuum'' value of the spinors $v_\alpha^q=\delta_\alpha^q$ in \eqref{w=} and find that the vectors are  $u_a^{(0)}=\delta_a^0$, $u_a=-\delta_a^0+i\delta_a^1$, $\bar{u}_a=-\delta_a^0-i\delta_a^1$ and obey \eqref{euuu=-2i} with our representation for 3D gamma matrices (see section~\ref{sec:4Dto3D_conventions}).}
\be\label{euuu=-2i}
\epsilon^{abc}u_a^{(0)}u_b\bar{u}_c =-2i\qquad \Longleftrightarrow \qquad u_{[b}\bar{u}_{c]} =-i\epsilon_{abc}u^{a(0)}\; , \qquad
\ee
which ensures this matrix to be SO$(1,2)$ valued (not just O$(1,2)$ valued as guaranteed by \eqref{u0u0=1}, \eqref{uu=0}),
\be\label{uinSO} u_a^{(b)}=\left(u_a^{(0)}, \frac 1 2 (u_a+\bar{u}_a), \frac 1 {2i} (u_a-\bar{u}_a)\right)\;\in \;  \text{SO}(1,2) \; , \qquad \ee
meaning that this set of vectors defines an oriented Lorentz frame. Thus \eqref{ug=ww}, or equivalently \eqref{u=wgw}, establish how the moving frame is constructed from the complex spinor moving frame.

The derivatives of the moving frame vectors are expressed through the Cartan forms~\eqref{ug=ww}
\be\label{du0}
\text{d}u^{(0)}_a=if\bar{u}_a - i\bar{f}u_a \; , \qquad \text{D}u_a= \text{d}{u}_a+ 2ia {u}_a=2ifu^{(0)}_a  \; , \qquad \text{D}\bar{u}_a=\text{d}\bar{u}_a- 2ia \bar{u}_a=-2i\bar{f}u^{(0)}_a   \; . \qquad
\ee
Consequently, Cartan forms~\eqref{f=wdw} can be equivalently written as
\be\label{f=udu}
f=\frac i 2 {u}{}^a \text{d}u^{(0)}_a \; , \qquad \bar{f}=-\frac i 2 \bar{u}{}^a \text{d}u^{(0)}_a  \; , \qquad a=-\frac i 4 \bar{u}{}^a \text{d}u_a  =\frac i 4 u^a \text{d}\bar{u}_a  \; . \qquad
\ee
When differentiating a function $\mathfrak{h}(\bar{w}, w)$ of the spinor frame variables only, the constraint~\eqref{bww=i} implies that its exterior derivative decomposes on the Cartan forms~\eqref{f=wdw}
\be
\text{d}= \text{d}\bar{w}_\alpha \frac {\partial}  {\partial \bar{w}_\alpha} + \text{d}{w}_\alpha \frac {\partial }  {\partial  w_\alpha}= if \bar{{\mathbb D}}  - i\bar{f} {\mathbb D} + ia {\mathbb D}^{(0)}\; ,
\ee
where we have used Eqs.~\eqref{Dw=} at the second stage and the covariant derivatives are defined as
\bea\label{bbD:=}
{\mathbb D} = w_\alpha \frac {\partial} {\partial \bar{w}_\alpha}  \; , \qquad  \bar{{\mathbb D}}= \bar{w}_\alpha\frac {\partial} {\partial {w}_\alpha}  \; , \qquad {\mathbb D}^{(0)}= \bar{w}_\alpha \frac {\partial} {\partial \bar{w}_\alpha} - {w}_\alpha \frac {\partial} {\partial {w}_\alpha}  \; . 
\eea
These covariant derivatives have the characteristic property to ``annihilate'' the constraint \eqref{bww=i} 
\be {\mathbb D} ( \bar{w}{}^{\alpha} w_\alpha ) =0  \; , \qquad  \bar{{\mathbb D}} (\bar{w}{}^{\alpha} w_\alpha )=0 \; , \qquad {\mathbb D}^{(0)} (\bar{w}{}^{\alpha} w_\alpha )=0  \; \qquad \ee
and obey the $\mathfrak{su}(1,1)\simeq \mathfrak{su}(2,{\mathbb R})$ algebra
\be
 {}[{\mathbb D} \, , \, \bar{{\mathbb D}}]= {\mathbb D}^{(0)} \; , \qquad  [ {\mathbb D}^{(0)}\, , \, {\mathbb D}]=- 2{\mathbb D}   \; , \qquad  [{\mathbb D} \, , \, \bar{{\mathbb D}}]= 2 \bar{{\mathbb D}}\; . \qquad
\ee
Finally, using the spinor and moving frame vectors, we can split the pull-back of the $\text{D}=3$ ${\cal N}=2$ superspace supervielbein forms
\be\label{Ea==}
E^a=  \delta^a_\mu \text{d}{x}{}^\mu - i \text{d}{\theta}\gamma^a \bar{\theta} + i {\theta}\gamma^a \text{d}{\bar{\theta}}\; , \qquad E^\alpha = \text{d}\theta^\alpha \; , \qquad \bar{E}{}^\alpha =\text{d}\bar{\theta}{}^\alpha \qquad
\ee
into a set of three bosonic and four fermionic Lorentz invariant 1-forms
\begin{equation}\label{E0=}
    {\rm E}^{(0)}=E^au_a^{(0)}\; , \qquad \qquad {\rm E}=E^au_a\; , \qquad \bar{{\rm E}}=E{}^a\bar{u}_a\; ,
\end{equation}
\begin{equation}\label{Ew=}
   \qquad  \qquad  \qquad     \qquad  \qquad   ~~~~~~\text{E}^w= \text{d}\theta^\alpha w_\alpha \; , \qquad \text{E}^{\bar{w}}= \text{d}\theta^\alpha {\bar{w}}_\alpha \; , 
\end{equation}
\begin{equation} \label{bEw=}
    \qquad  \qquad  \qquad     \qquad  \qquad   ~~~~~~ \bar{\text{E}}{}^w= \text{d}\bar{\theta}{}^\alpha w_\alpha \; , \qquad \bar{\text{E}}{}^{\bar{w}}= \text{d}\bar{\theta}{}^\alpha {\bar{w}}_\alpha\; . 
\end{equation}

\subsection{Lagrangian of the simplest 3D counterpart of the mD0 system}
The Lagrangian for the 3D counterpart of the 10D mD$0$ action contains two constants with dimension of mass,, $m$ and $\mu$, as well as an arbitrary positive definite function  \be\label{cM=cMcH}
{\cal M}={\cal M}({\cal H}/\mu^6)\quad
\ee
defined in terms of the bosonic composite constructed from matrix fields\begin{eqnarray}
\label{cH=} && {\cal H}=  {\rm tr}\left( {\mathbb P} \bar{\mathbb P} +  [{\mathbb Z},  \bar{\mathbb Z}]^2 -
{i\over 2} {\mathbb Z}{\boldsymbol{ \Psi}}{\boldsymbol{ \Psi}} + {i\over 2} \bar{\mathbb Z} \bar{\boldsymbol{ \Psi}}  \bar{\boldsymbol{ \Psi}} \right) \;
.\qquad
\end{eqnarray}
This expression can be recognized as the Hamiltonian of 3D SU($N$) SYM model reduced to $\text{d}=1$. For convenience, let us also introduce a pair of conjugate fermionic currents
\begin{equation}
{\nu} := \text{tr}( {\boldsymbol{ \Psi}} \mathbb{P} + \bar{\boldsymbol{ \Psi}}[\mathbb{Z}, \bar{\mathbb{Z}}])~, \qquad {\bar{\nu}} := \text{tr}(\bar{\boldsymbol{ \Psi}} \bar{\mathbb{P}} + {\boldsymbol{ \Psi}}[\mathbb{Z}, \bar{\mathbb{Z}}])~,\qquad
\end{equation}
which can be identified as the SYM supercurrents. The Lagrangian corresponding to the action~\eqref{SmD0=3D} can then be written as
\begin{equation}
    \begin{array}{l}
         \begin{split}
             &\mathcal{L}_\tau[m,\mu, \mathcal{M}({\cal H})] =  m {\rm E}_\tau^{0}+ m  (\dot{\theta}^\alpha \bar{\theta}_\alpha- \theta^\alpha \dot{\bar{\theta}}_\alpha ) + {1\over \mu^6} {\rm E}_\tau^{0}\frac 2 {{\cal M}}  {{\cal H}}  + {1\over \mu^6}
             \frac {\dot{{\cal M}}} {{\cal M}}
             {\rm tr}\left(\bar{\mathbb P} {\mathbb Z}+ {\mathbb P} \bar{\mathbb Z}\right) +~~~~~~~~~~~~~~~~~ \\
             & + {1\over \mu^6}
             {\rm tr}\left(\bar{\mathbb P}{\rm D}_\tau {\mathbb Z} + {\mathbb P}{\rm D}_\tau \bar{\mathbb Z} - {i\over 8} {\rm D}_\tau{ {\boldsymbol{ \Psi}}}\,  \bar{\boldsymbol{ \Psi}} + {i\over 8} {\boldsymbol{ \Psi}} {\rm D}_\tau \bar{\boldsymbol{ \Psi}} \right) + {1\over \mu^6}  \frac i {\sqrt{{\cal M}}}\left( \dot{\theta}{}^\alpha\, w_\alpha \bar{\nu} +  \dot{\bar{\theta}}{}^{\alpha}\bar{w}_{\alpha}\, \nu \right)
         \end{split}
    \end{array}\label{eq:3DmD0_L}
\end{equation}
Here ${\rm D}_\tau $ denotes covariant derivatives which include not only the $\mathfrak{su}(N)$ valued connection ${\mathbb A}=\text{d}\tau {\mathbb A}_\tau$, but also a composite U$(1)$ defined by the Cartan form $a=\text{d}\tau a_\tau$ \eqref{f=wdw} (see also \eqref{f=udu})
\begin{equation}
    \begin{array}{lcl}
        \text{D}_\tau {\mathbb Z}= \dot{{\mathbb Z}}+2ia_\tau  {\mathbb Z} +[{\mathbb A}_\tau ,{\mathbb Z}] \; , &~&  \text{D}_\tau {\boldsymbol{ \Psi}}= \dot{{\boldsymbol{ \Psi}}}-i a_\tau  {\boldsymbol{ \Psi}} +[{\mathbb A}_\tau ,{\boldsymbol{ \Psi}}]\; ,\\
          \text{D}_\tau \bar{{\mathbb Z}}= \dot{\bar{{\mathbb Z}}}-2ia_\tau  \bar{{\mathbb Z}} +[{\mathbb A}_\tau ,\bar{{\mathbb Z}}] \; , &~& \text{D}_\tau \bar{{\boldsymbol{ \Psi}}}= \dot{ \bar{{\boldsymbol{ \Psi}}}}+i a_\tau   \bar{{\boldsymbol{ \Psi}}} +[{\mathbb A}_\tau , \bar{{\boldsymbol{ \Psi}}}]\; .
    \end{array}\label{DtZ:=a}
\end{equation}
Two special members within this family of Lagrangians are defined by function \eqref{cM=cMcH} being constant
\be\label{cM=m}
{\cal M} = m={\text{const}}\; ,
\ee
and by specific function
\be\label{cap7:cM=m+}
{\cal M} = \frac m 2 +  \sqrt{\frac {m^2} {4}+\frac {\cal H} {\mu^6}}\; .
\ee
The latter was derived in section~\ref{sec:4D_dimReduction} by dimensional reduction of the 4D counterpart of the mM$0$ action~\cite{Bandos_11D_mM0}, while the former is the simplest representative of the family which can be recognized as 3D counterpart of the 10D action  discussed previously in~\cite{10D_mD0_Igor}. 

In this chapter, we focus on the simplest model with constant ${\cal M}=m$, with the Lagrangian
\begin{equation}
    \begin{array}{l}
         \begin{split}
             \mathcal{L}_\tau[m,\mu] &=  m {\rm E}_\tau^{0}+ m  (\dot{\theta}^\alpha \bar{\theta}_\alpha- \theta^\alpha \dot{\bar{\theta}}_\alpha ) + {1\over \mu^6} {\rm E}_\tau^{0}\frac 2 {m}  {{\cal H}}  +  {1\over \mu^6}  \frac i {\sqrt{m}}\left( \dot{\theta}{}^\alpha\, w_\alpha \bar{\nu} +  \dot{\bar{\theta}}{}^{\alpha}\bar{w}_{\alpha}\, \nu \right)+  \\
             & + {1\over \mu^6}
             {\rm tr}\left(\bar{\mathbb P}{\rm D}_\tau {\mathbb Z} + {\mathbb P}{\rm D}_\tau \bar{\mathbb Z} - {i\over 8} {\rm D}_\tau{ {\boldsymbol{ \Psi}}}\,  \bar{\boldsymbol{ \Psi}} + {i\over 8} {\boldsymbol{ \Psi}} {\rm D}_\tau \bar{\boldsymbol{ \Psi}} \right) ~.
         \end{split}
    \end{array}\label{eq:3DmD0_L_mConstw}
\end{equation}
We will develop Hamiltonian formalism for this model and perform its quantization, thereby deriving the equation of the field theory of simplest 3D counterpart of the mD$0$-brane.

\section{Hamiltonian formalism of the simplest 3D prototype of mD0 brane model}
In this section we develop Hamiltonian formalism for our simplest 3-dimensional counterpart of the mD$0$ system.

\subsection{Canonical momenta and Poisson brackets}
In general, the canonical momenta $p_A$ are defined as derivatives of the Lagrangian $L(q^A,\dot{q}{}^A)$ with respect to generalized velocities $\dot{q}^A=\frac{\text{d}}{\text{d}\tau}q^A$ associated with the coordinates $q^A$,
\be\label{pA:=} 
p_A=  \dfrac{\partial {L}}{\partial \dot{q}^A}~ .
\ee
They are canonically conjugate to these coordinates which implies the following structure of the Poisson brackets algebra
\begin{equation*}
        [p_A, q^B\}_{\text{PB}}= -(-)^{\varepsilon (A)\varepsilon (B)}[q^B, p_A\}_{\text{PB}}=-\delta_A^B \; ,
\end{equation*}
\begin{equation}\label{PB}
    [q^A, q^B\}_{\text{PB}}= -(-)^{\varepsilon (A)\varepsilon (B)}[q^B, q^A\}_{\text{PB}}=0 \; , ~~ [p_A, p_B\}_{\text{PB}}= -(-)^{\varepsilon (A)\varepsilon (B)}[p_B, p_A\}_{\text{PB}}=0 \; ,
\end{equation}
where $\varepsilon (A)$ is Grassmann (fermionic) parity of $p_A$ and $q^A$ equal to $0$ for bosons and $1$ for fermions\footnote{The Poisson brackets are essentially defined  by relations \eqref{PB} and the Leibniz rules $[fg,h\}_{\text{PB}}=f [g,h\}_{\text{PB}}+ (-)^{\varepsilon (f)\varepsilon (h) } [f,h\}_{\text{PB}}g$, where $f$, $g$ and $h$ are arbitrary functions of  $q^A$ and  $p_A$. An explicit expression is given by 
\begin{equation*}
    [f, g\}_{\text{PB}}= (-)^{\varepsilon (A)\varepsilon (f) } \left( \frac {\partial f}  {\partial q^A } \, \frac {\partial g}  {\partial p_A } - (-)^{\varepsilon (A)}  \frac {\partial f}  {\partial p_A } \, \frac {\partial g}  {\partial q^A }\right)~.
\end{equation*}}.

The canonical Hamiltonian $H_0(p_A, q^A)$ is then defined by Legendre transform of the Lagrangian\footnote{We simplify the definition of the Legendre transform here since this is sufficient for our purposes. For a more precise treatment and examples of systems where this distinction is crucial see e.g. section 3 of \cite{ModMax} and references therein.}
\be\label{H0=dqp-L}
H_0 (p_A,q^A)=\dot{q}{}^Ap_A - L(q^A,\dot{q}{}^A)\;
\ee
and the evolution of the dynamical system is described by the Hamiltonian equations
\be\label{evol=H0}
\dot{f}(q,p)= [{f}(q,p)\, ,\, H_0]_{\text{PB}}\; \qquad \Longleftrightarrow \qquad  \begin{cases} \dot{q}^A=  \dfrac {\partial H_0}  {\partial  p_A} \; , \cr \cr  \dot{p}_A= - \dfrac {\partial H_0}  {\partial  q^A}\; .
 \end{cases}
\ee
In our case, we first denote the momenta conjugate to the bosonic and fermionic coordinate functions by
\begin{equation}
\begin{array}{ccccc}\label{Pa:=}
p_a = \dfrac{\partial \mathcal{L}_\tau}{\partial \dot{x}^a}~,&~& \bar{\Pi}_\alpha = \dfrac{\partial \mathcal{L}_\tau}{\partial \dot{\theta}^\alpha}~,&~& {\Pi}_\alpha = \dfrac{\partial \mathcal{L}_\tau}{\partial \dot{\bar{\theta}}^\alpha}
\end{array}
\end{equation}
so that their nonvanishing Poisson brackets are
\begin{equation}
    \begin{array}{lll}
        ~& ~~~~~[p_a, x^b]_{\text{PB}} =-[x^b, p_a ]_{\text{PB}} = -\delta_a^b~, & \\
        ~\\
         ~ &~ \{\bar{\Pi}_\alpha, \theta^\beta\}_{\text{PB}} =\{\theta^\beta, \bar{\Pi}_\alpha\}_{\text{PB}} = -\delta_\alpha^\beta~, \qquad & \{{\Pi}_\alpha, \bar{\theta}^\beta\}_{\text{PB}} =\{ \bar{\theta}^\beta, {\Pi}_\alpha\}_{\text{PB}} = -\delta_\alpha^\beta~.
    \end{array}
\end{equation}

\subsection{Covariant momenta and Poisson/Dirac brackets in the spinor moving frame sector}
Similarly, we can introduce the canonical momenta conjugate to spinor frame variables
\be
\bar{P}_\alpha =\frac {\partial \mathcal{L}_\tau} {\partial {\dot{w}}^\alpha} \; , \qquad P_\alpha =\frac {\partial \mathcal{L}_\tau} {\partial {\dot{\bar{w}}}{}^\alpha}
\ee
which have nonvanishing Poisson brackets
\be
{}[\bar{P}{}^\alpha, w_\beta]_{\text{PB}}={- \delta^\alpha{}_\beta{}}\; ,  \qquad {}[P^\alpha, \bar{w}_\beta]_{\text{PB}}={- \delta^\alpha{}_\beta{}}\; .   \qquad
\ee
When working with these canonical momenta, we have to consider the condition \eqref{bww=i} as a second class constraint (in terminology of Dirac \cite{Dirac}; see below for further discussion). It turns out that this constraint\footnote{Here and below, $\approx$ denotes weak equality in the sense of \cite{Dirac}, i.e. relations between coordinates and momenta which can be used only after all Poisson brackets are calculated.} is conjugate to
\be\label{wP+cc=0}
w_\alpha\bar{P}{}^\alpha+ \bar{w}_\alpha{P}{}^\alpha \approx 0 \; ,
\ee
 which also appears in our model. Then we could resolve this pair of second class constraints by passing from Poisson brackets to the corresponding Dirac brackets (see \cite{3D_quantization_Bandos} and refs. therein), after which both \eqref{bww=i} and \eqref{wP+cc=0} can be considered as satisfied in the strong sense.
 
Alternatively, this procedure can be overcame by introducing the so-called covariant momenta
\be\label{frak-d=}
{\mathfrak d}= w_\alpha P^\alpha  \; , \qquad  \bar{{\mathfrak d}}= \bar{w}_\alpha\bar{P}{}^\alpha \; , \qquad {\mathfrak d}^{(0)} =\bar{w}_\alpha P^\alpha - w_\alpha \bar{P}{}^\alpha  \; , \qquad
\ee
which have the Poisson brackets with  spinor frame variables
\bea\label{fdw=PB}
&  {} [{\mathfrak d}, w_\alpha ]_{\text{PB}}= 0\; ,  \qquad &[{\mathfrak d}, \bar{w}_\alpha ]_{\text{PB}}=  - {w}_\alpha  \; , \qquad \nonumber \\ 
 &  [\bar{{\mathfrak d}}, w_\alpha ]_{\text{PB}}= - \bar{w}_\alpha \; ,  \qquad &[\bar{{\mathfrak d}}, \bar{w}_\alpha ]_{\text{PB}}=  0\; , \qquad   \\ & {} [{\mathfrak d}^{(0)}, w_\alpha ]_{\text{PB}}=  w_\alpha \; ,   \qquad &[{\mathfrak d}^{(0)}, \bar{w}_\alpha ]_{\text{PB}}=  - \bar{w}_\alpha \; . \qquad \nonumber
\eea
It is straightforward to verify that these covariant momenta have vanishing Poisson brackets with the constraints \eqref{bww=i}. Therefore, by working only with the covariant momenta, we can consider the constraint \eqref{bww=i} as satisfied in the strong sense.

Notice that \eqref{fdw=PB} imply the following Poisson brackets of the covariant momenta with moving frame vectors
\bea
  {} [{\mathfrak d},u_a^{(0)} ]_{\text{PB}}=-u_a \; ,  \qquad &[{\mathfrak d}, u_a ]_{\text{PB}}= 0  \; , \qquad & [{\mathfrak d}, \bar{u}_a ]_{\text{PB}}=-2u_a^{(0)}   \; , \qquad   \nonumber  \\ {} [\bar{{\mathfrak d}},u_a^{(0)} ]_{\text{PB}}=-\bar{u}_a  \; ,  \qquad &[\bar{{\mathfrak d}}, u_a ]_{\text{PB}}=-2 u_a^{(0)}  \; , \qquad & [\bar{{\mathfrak d}}, \bar{u}_a ]_{\text{PB}}=0   \; , \qquad
      ~\\
      {} [{\mathfrak d}^{(0)}, u_a^{(0)} ]_{\text{PB}}=0\; , \;  \qquad & [{\mathfrak d}^{(0)}, u_a ]_{\text{PB}}=2u_a\; ,  \qquad  & [{\mathfrak d}^{(0)}, \bar{u}_a ]_{\text{PB}}=-2 \bar{u}_a\; .  \nonumber
\eea
The covariant momenta thus generate the $\mathfrak{so}(1,2) \simeq \mathfrak{su}(1,1)$ algebra under Poisson brackets
\bea
 & {} [{\mathfrak d}^{(0)},{\mathfrak d} ]_{\text{PB}}=2{\mathfrak d} \; , \;  \qquad  [{\mathfrak d}^{(0)}, \bar{{\mathfrak d}} ]_{\text{PB}}=-2\bar{{\mathfrak d}}\; , & \qquad \nonumber \\ 
 ~\\
 & {} [{\mathfrak d},\bar{{\mathfrak d}} ]_{\text{PB}}={\mathfrak d}^{(0)} \; .  & \qquad  \nonumber
\eea
To express the canonical Hamiltonian \eqref{H0=dqp-L} in terms of the Lagrangian and covariant momenta, we start from the standard relation
\begin{equation}
    \text{d}\tau H_0= \text{d}w_\alpha \bar{P}^\alpha + \text{d}\bar{w}_\alpha {P}^\alpha +\ldots-\mathcal{L}~,
\end{equation}
substitute for the derivatives of $w_\alpha $ and  $\bar{w}_\alpha$ their expressions in terms of Cartan forms, Eq.~\eqref{Dw=} (valid when constraint \eqref{bww=i} is satisfied), and use \eqref{frak-d=} arriving at
\be\label{H0=frak-d}
\text{d}\tau H_0=  ia {\mathfrak d}^{(0)}+ if\bar{{\mathfrak d}}-i\bar{f} {\mathfrak d}+\ldots-\mathcal{L}\; ~~ \Longleftrightarrow ~~ H_0=  ia_\tau  {\mathfrak d}^{(0)}+ if_\tau \bar{{\mathfrak d}}-i\bar{f}_\tau  {\mathfrak d}+\ldots-\text{d}\tau \mathcal{L}_\tau\; . \qquad
\ee
This relation makes it manifest that the covariant momenta correspond to derivatives of the Lagrangian with respect to the Cartan forms
\be\label{fd=dL-df}
 {\mathfrak d}^{(0)}=- i \frac {\partial \mathcal{L}_\tau}{\partial a_\tau} \; , \qquad {\mathfrak d}= i \frac {\partial \mathcal{L}_\tau}{\partial \bar{f}_\tau} \; , \qquad   \bar{{\mathfrak d}}=-i \frac {\partial \mathcal{L}_\tau}{\partial {f}_\tau}\; . \qquad
\ee

\subsection{Simplifying  Poisson/Dirac brackets in the sector of matrix fields }
The canonical momenta of the bosonic matrix fields are
\bea
 \dfrac{\partial \mathcal{L}}{\partial \dot{\mathbb{Z}}_j^i}\approx\dfrac{1}{\mu^6} \bar{\mathbb{P}}^j_i ~, \qquad \dfrac{\partial \mathcal{L}}{\partial \dot{\bar{\mathbb{P}}}{}^j_i} \approx 0 \; , \qquad \nonumber \\  \nonumber \\  \dfrac{\partial \mathcal{L}}{\partial \dot{\bar{\mathbb{Z}}}_j^i}\approx \dfrac{1}{\mu^6} \mathbb{P}^j_i ~,
\qquad \dfrac{\partial \mathcal{L}}{\partial \dot{{\mathbb{P}}}{}^j_i} \approx 0 \; . \qquad  \eea
These relations correspond to resolved pairs of second class constraints, allowing us to exclude the variables $\mathbb{P}^j_i$, $ \bar{\mathbb{P}}^j_i$ and their momenta from the consideration.

However, a more convenient equivalent procedure is to remove the momenta conjugate to $ \mathbb{P}^j_i$ and $ \bar{\mathbb{P}}^j_i$ and instead replace the momenta conjugate to $\mathbb{Z}= \mathbb{Z}_i^j$ and $\bar{\mathbb{Z}}= \bar{\mathbb{Z}}_i^j$
by $ \bar{\mathbb{P}}^j_i$ and $ \mathbb{P}^j_i$. Then, the nonvanishing Poisson brackets for these new phase space variables (in fact, Dirac brackets) are\footnote{Note that the last terms on the r.h.s.'s of Eqs. \eqref{bbPbbZ=PB}, $- \dfrac{1}{N} \delta^j_i \delta^l_k$, ensure their consistency with the tracelessness condition of matrices \eqref{Z=bZ*}.}
\begin{equation}\label{bbPbbZ=PB}
\begin{array}{ccl}
[\bar{\mathbb{P}}^j_i, \mathbb{Z}_k{}^l ]_{\text{PB}} = -\mu^6 \left( \delta^l_i \delta_k^j - \dfrac{1}{N} \delta^j_i \delta^l_k\right)~, &~\qquad & [\mathbb{P}^j_i, \bar{\mathbb{Z}}_k{}^l]_{\text{PB}} = -\mu^6 \left( \delta^l_i \delta_k^j - \dfrac{1}{N} \delta^j_i \delta^l_k\right)~.
\end{array}
\end{equation}
A similar approach is standard when constructing Hamiltonian mechanics for fermionic fields with canonical kinetic term, where the conjugate momentum of fermionic field is identified with the field itself. In our case, the fermionic matrix fields also possess such a kinetic term.

The canonical momenta for the fermionic matrix field ${\boldsymbol{ \Psi}} = {\boldsymbol{ \Psi}}_i^j$ and its Hermitian conjugate $\bar{\boldsymbol{ \Psi}} = \bar{\boldsymbol{ \Psi}}_i^j$ are given by
\begin{equation} \label{PPsi=bPsi}
\begin{array}{ccc}
\Pi_{\bar{\boldsymbol{ \Psi}}} = \dfrac{\partial \mathcal{L}}{\partial \dot{\bar{\boldsymbol{ \Psi}}}} \approx  - \dfrac{i}{8 \mu^6} {\boldsymbol{ \Psi}}~,&\qquad  &\bar{\Pi}_{\boldsymbol{ \Psi}} = \dfrac{\partial \mathcal{L}}{\partial \dot{\boldsymbol{ \Psi}}} \approx - \dfrac{i}{8 \mu^6} \bar{\boldsymbol{ \Psi}}~.
\end{array}
\end{equation}
The canonical Poisson brackets with nonvanishing r.h.s. are
\begin{equation}
\left\lbrace {\boldsymbol{ \Pi}}_i{}^j, \bar{\boldsymbol{ \Psi}}_k{}^l \right\rbrace_{\text{PB}} =\left\lbrace {\boldsymbol{ \Psi}}_i{}^j, \bar{\boldsymbol{ \Pi}}_k{}^l \right\rbrace_{\text{PB}} =  - ( \delta^l_i \delta_k^j - \dfrac{1}{N} \delta^j_i \delta^l_k )~
\end{equation}
where, for shortness, we write ${\boldsymbol{ \Pi}}_i{}^j:=(\Pi_{\bar{\boldsymbol{ \Psi}}})_i{}^j$ and $\bar{\boldsymbol{ \Pi}}_i{}^j:=(\bar{\Pi}_{\bar{\boldsymbol{\Psi}}})_i{}^j$.

From these, we see that \eqref{PPsi=bPsi} are second class constraints. Since they are not in explicitly resolved form, to treat them as strong equalities, we have to replace Poisson brackets by Dirac brackets
\begin{equation}
\begin{split}
[\ldots,\ldots\rbrace_{\text{D}} = [\ldots,\ldots \rbrace_{\text{PB}} &- 4i\mu^6 [\ldots,(\Pi_{\bar{\boldsymbol{ \Psi}}} + \frac{i}{8\mu^6} {\boldsymbol{ \Psi}})_i^j \rbrace_{\text{PB}}[(\bar{\Pi}_{{\boldsymbol{ \Psi}}} + \frac{i}{8\mu^6} \bar{\boldsymbol{ \Psi}})_j^i,\ldots \rbrace_{\text{PB}}-\\ \\
&- 4i\mu^6 [\ldots,(\bar{\Pi}_{{\boldsymbol{ \Psi}}}  \frac{i}{8\mu^6} \bar{\boldsymbol{ \Psi}})_i^j \rbrace_{\text{PB}}[(\Pi_{\bar{\boldsymbol{ \Psi}}} + \frac{i}{8\mu^6} {\boldsymbol{ \Psi}})_j^i,\ldots \rbrace_{\text{PB}}~.
\end{split}
\end{equation}
With these Dirac brackets, the complex fermionic matrix field and its Hermitian conjugate become canonically conjugate
\begin{equation}\label{PsibPsi=DB}
\left\lbrace {\boldsymbol{ \Psi}}_i{}^j, \bar{\boldsymbol{ \Psi}}_k{}^l \right\rbrace_{\text{D}} = -4i\mu^6 ( \delta^l_i \delta_k^j - \dfrac{1}{N} \delta^j_i \delta^l_k )~, \qquad \left\lbrace {\boldsymbol{ \Psi}}_i{}^j, {\boldsymbol{ \Psi}}_k{}^l \right\rbrace_{\text{D}} = 0 \; , \qquad \left\lbrace \bar{\boldsymbol{ \Psi}}_i{}^j, \bar{\boldsymbol{ \Psi}}_k{}^l \right\rbrace_{\text{D}} = 0\; . \qquad
\end{equation}
Below for simplicity, and in line with the common practice of imposing conditions of the form \eqref{PsibPsi=DB} as canonical, we will call these Dirac brackets Poisson brackets and write \eqref{PsibPsi=DB} as
\begin{equation}\label{PsibPsi=PB}
\left\lbrace {\boldsymbol{ \Psi}}_i{}^j, \bar{\boldsymbol{ \Psi}}_k{}^l \right\rbrace_{\text{PB}} = -4i\mu^6 ( \delta^l_i \delta_k^j - \dfrac{1}{N} \delta^j_i \delta^l_k )~, \qquad \left\lbrace {\boldsymbol{ \Psi}}_i{}^j, {\boldsymbol{ \Psi}}_k{}^l \right\rbrace_{\text{PB}} = 0 \; , \qquad \left\lbrace \bar{\boldsymbol{ \Psi}}_i{}^j, \bar{\boldsymbol{ \Psi}}_k{}^l \right\rbrace_{\text{PB}} = 0\; . \qquad
\end{equation}
This is more convenient also because we have already used implicitly the passage to Dirac brackets in the spinor moving frame sector of our model.

\subsection{Primary constraints}
With our Lagrangian \eqref{eq:3DmD0_L_mConstw}, we compute the canonical momenta  \eqref{pA:=}  and covariant momenta \eqref{fd=dL-df} for the fields in both the center of mass sector and the gauge field. This leads us to the following set of \textit{primary constraints} of the system
\begin{equation}
\Phi_a := p_a - mu_a^{(0)} - \dfrac{2}{\mu^6}\dfrac{\mathcal{H}}{m} u_a^{(0)} \approx 0~,
\label{eq:Phi_a}
\end{equation}
\begin{equation}
d_\alpha :=  \bar{\Pi}_\alpha + i p_a (\gamma^a \bar{\theta})_\alpha - m \bar{\theta}_\alpha - \dfrac{i}{\mu^6 \sqrt{ m}} w_\alpha\, \bar{\nu} \approx 0~, 
\label{eq:Phif} 
\end{equation}
\begin{equation}
\bar{d}_\alpha := {\Pi}_\alpha + i p_a (\gamma^a \theta)_\alpha + m \theta_\alpha - \dfrac{i}{\mu^6 \sqrt{m}} \bar{w}_\alpha\, \nu \approx 0~, 
\label{eq:barPhif}
\end{equation}
\begin{equation}\label{fd=0}
     {\mathfrak d}\approx 0~,
\end{equation}
\begin{equation}\label{bfd=0}
     \bar{{\mathfrak d}} \approx 0 \; ,
\end{equation}
\begin{equation}\label{U=0}
    U:= {\mathfrak d}^{(0)}-  \frac 2{\mu^6} \, {\cal B}\approx 0~,
\end{equation}
through
\be\label{cB:=} {\cal B}:= {\rm tr}\left(\bar{\mathbb{P}}\mathbb{Z}-\mathbb{P}\bar{\mathbb{Z}}+\frac i 8 {\boldsymbol{ \Psi}}\bar{{\boldsymbol{ \Psi}}}\right)~,
\ee
and
\begin{equation}
\mathbb{P}_{\mathbb{A}}:= \frac {\partial L}{\partial \dot{\mathbb{A}}_\tau} \approx 0~.
\label{eq:PA}
\end{equation}
The last relation is the only primary constraint imposed just on the matrix variables. However, the matrix fields also contribute to the constraints derived from the definition of momenta of the center of mass variables through ${\cal H}$ of \eqref{cH=} in \eqref{eq:Phi_a}, via
\begin{equation}
 \bar{\nu}=\text{tr}(\bar{\boldsymbol{ \Psi}} \bar{\mathbb{P}} + {\boldsymbol{ \Psi}} [\mathbb{Z}, \bar{\mathbb{Z}}])\; \qquad {\text{and}} \qquad \nu=\text{tr}({\boldsymbol{ \Psi}} \mathbb{P} + \bar{\boldsymbol{ \Psi}} [\mathbb{Z}, \bar{\mathbb{Z}}]) \;
\label{nu=}
\end{equation}
in \eqref{eq:Phif}  and \eqref{eq:barPhif}. 

\subsection{Canonical Hamiltonian and secondary constraints}
The canonical Hamiltonian of our system is defined by the Legendre transform of the Lagrangian, which we write in terms of SO$(1,2)/\text{SO}(2)$ Cartan forms and covariant momenta (see equation \eqref{H0=frak-d}) as
\begin{equation}\label{H0:=}
\begin{split}
\text{d}\tau H_0 &= \text{d}x^{a}p_{a}  + \text{d}\theta^\alpha  \bar{\Pi}_\alpha +\text{d} \bar{\theta}^\alpha {\Pi}_\alpha  + ia \tilde{\mathfrak{d}}^{(0)} + i f \bar{{\mathfrak{d}}} - i \bar{f} {\mathfrak{d}} +\\
&+ \frac{1}{\mu^6} \text{tr}(\text{d}\mathbb{Z}\bar{\mathbb{P}}) + \frac{1}{\mu^6} \text{tr}(\text{d}\bar{\mathbb{Z}}\mathbb{P}) -	\frac{i}{8 \mu^6} \text{tr}(\text{d}{\boldsymbol{ \Psi}} \bar{{\boldsymbol{ \Psi}}}) -\frac{i}{8 \mu^6} \text{tr}(\text{d}\bar{{\boldsymbol{ \Psi}}} {\boldsymbol{ \Psi}}) + \text{tr}(\text{d}\mathbb{A} \mathbb{P}_{\mathbb{A}}) - \mathcal{L}_{\text{mD}0} ~.
\end{split}
\end{equation}
For our Lagrangian  \eqref{eq:3DmD0_L_mConstw},  after  extracting the primary constraints,  \eqref{H0:=} reduces to
\begin{eqnarray}\label{H0=trAG}
H_0 = E_\tau^a\Phi_a+\dot{\theta}{}^\alpha d_\alpha +\dot{\bar{\theta}}{}^\alpha \bar{d}_\alpha
+ ia_\tau U + i f_\tau \bar{\tilde{\mathfrak{d}}} - i \bar{f}_\tau \tilde{\mathfrak{d}} +
 \text{tr}(\text{d}\mathbb{A} \mathbb{P}_{\mathbb{A}}) + {\rm tr} ({\mathbb A}_\tau {\mathbb G})  \approx  {\rm tr}  ({\mathbb A}_\tau {\mathbb G})~,
\end{eqnarray}
where $\mathbb{G}$ is a $N\times N$ anti-Hermitian traceless matrix
\begin{equation}\label{bbG:=}
\mathbb{G}:= \dfrac{1}{\mu^6} \left( [\bar{\mathbb{Z}}, \mathbb{P}] + [\mathbb{Z}, \bar{\mathbb{P}}] - \frac{i}{4} \{{\boldsymbol{ \Psi}}, \bar{\boldsymbol{ \Psi}} \} \right) \; .
\end{equation}
The last expression, valid in the weak sense, can be used to check whether the preservation of the primary constraints under evolution, as defined by Hamiltonian equations \eqref{evol=H0}, produces secondary constraints. Indeed, this happens: requiring the preservation of the primary constraint \eqref{eq:PA},  $\dot{\mathbb{P}}_{\mathbb{A}} = [\mathbb{P}_{\mathbb{A}}, H_0] \approx 0$, leads to the \textit{secondary constraint}
\begin{equation}
\mathbb{G}= \dfrac{1}{\mu^6} \left( [\bar{\mathbb{Z}}, \mathbb{P}] + [\mathbb{Z}, \bar{\mathbb{P}}] - \frac{i}{4} \{{\boldsymbol{ \Psi}}, \bar{\boldsymbol{ \Psi}} \} \right) \approx 0\; .
\label{eq:Gauss}
\end{equation}
This reveals the role of the 1d gauge field ${\mathbb A}_\tau$ as a Lagrange multiplier for the constraint \eqref{eq:Gauss},  which we will call the {\it Gauss law}, keeping in mind its counterpart in higher dimensional supersymmetric gauge theories. Furthermore, since it can be shown that constraint \eqref{eq:PA} is the first class and generates a gauge symmetry corresponding to an arbitrary shift of ${\mathbb A}_\tau$, we will simplify the presentation by omitting the conjugate variables  ${\mathbb A}_\tau$ and ${\mathbb P}_{\mathbb A}$ from further consideration\footnote{More precisely, one can understand this reduction of the phase space as fixing the gauge ${\mathbb A}_\tau =0$ using the symmetry generated by the constraint ${\mathbb P}_{\mathbb A}$ and then treating the constraint  \eqref{eq:PA} as strong equality.}.

As a result, the Gauss law \eqref{eq:Gauss} implies the canonical Hamiltonian vanishes weakly
\begin{equation}\label{H0=0}
    H_0\approx 0\; .
\end{equation}

\subsection{First class constraints and gauge symmetries}
The vanishing of canonical Hamiltonian \eqref{H0=0} implies that the total Hamiltonian of our system, as introduced by Dirac in \cite{Dirac}, is a linear combination of the constraints,
\begin{eqnarray}\label{H=LMs}
H = b^a \Phi_a + {\kappa}^\alpha d_\alpha + \bar{ {\kappa}}^\alpha \bar{d}_\alpha + ik\bar{{\mathfrak d}} - i\bar{k} {\mathfrak d} + ik^{(0)} \left({\mathfrak d}{}^{(0)} -   \frac 2{\mu^6} \, {\cal B}\right) + \text{tr}(\mathbb{Y}\mathbb{G})~,
\end{eqnarray}
where the coefficients in this expression, functions of proper time $\tau$ known as Lagrange multipliers\footnote{Notice the reappearance of the pure gauge 1d gauge field $ {\mathbb A}_\tau $, which we removed from consideration using the gauge symmetry generated by \eqref{eq:PA}, now reincarnated as the Lagrange multiplier ${\mathbb Y}$ for the Gauss constraints \eqref{eq:Gauss}.}, are restricted by requiring preservation of all the constraints under the evolution \cite{Dirac},
\be\label{evol=H}
\frac{\text{d}}{\text{d}\tau} (\text{constraints})= [\text{constraints}\,,\, H]_{\text{PB}}\approx 0 \; \qquad
\ee
 ({\it cf.} \eqref{evol=H0})\footnote{Generically this procedure might result in appearance of further secondary constraints, but in this case the system instead produces equations for the multipliers.}.  The solution of Eqs.~\eqref{evol=H} expresses the original Lagrange multipliers in terms of fewer variables acting as (true) Lagrange multipliers for the first class constraints that generate, on the Poisson brackets, gauge symmetries of the dynamical system under consideration.

To specify the system of equations for Lagrange multipliers, it is useful first to determine the algebra of constraints under Poisson brackets. Let us start from the Gauss law constraint  \eqref{bbG:=} which is (an anti-Hermitian traceless) matrix ${\mathbb G}= {\mathbb G}_i{}^j$ different matrix elements of which may have nonvanishing brackets. Actually these form a representation of the $\mathfrak{su}(N)$ algebra. To write it in a compact and comprehensive form, it is convenient  to introduce ``reference'' traceless matrices $\mathbb{Y}$ and $\mathbb{Y}^\prime$ and compute the brackets of the traces of ${\mathbb G}$ with these matrices. In such a way we find the $\mathfrak{su}(N)$ algebra in the form
\begin{equation} \label{GG=G}
[\text{tr}(\mathbb{Y}\mathbb{G}), \text{tr}(\mathbb{Y}^\prime \mathbb{G})]_{\text{PB}} = \text{tr}\left([\mathbb{Y}, \mathbb{Y}^\prime ]\mathbb{G}\right)~.
\end{equation}
Furthermore, ${\mathbb G}$ has vanishing Poisson brackets with all other constraints, which reflecting their SU$(N)$ invariance (see Appendix \ref{App=PB} for technical details). Thus, ${\mathbb G}$ is a first class constraint generating the SU$(N)$ gauge symmetry of the model.

The remaining constraint algebra, given by Eqs.~\eqref{eq:Phi_a}-\eqref{eq:PA}, is  characterised by the following nonvanishing Poisson brackets (see Appendix \ref{App=PB}):
\begin{equation}
    \begin{array}{ccc}
       [\mathfrak{d},\mathfrak{d}^{(0)} - \frac{2}{\mu^6} \mathcal{B}]_{\text{PB}}= -2 \mathfrak{d}~,   &~& [\mathfrak{d},\Phi_a]{}_{\text{PB}}= \left(m + \frac{2}{\mu^6}\frac{\mathcal{H}}{m} \right) u_a~, 
    \end{array}
\end{equation}
\begin{equation}
    [\mathfrak{d},\bar{d}_\alpha]_{\text{PB}} = \frac{i}{\mu^6 \sqrt{m}} w_\alpha \nu~, 
\end{equation}
\begin{equation}
    [\mathfrak{d},\bar{\mathfrak{d}}]_{\text{PB}} = \mathfrak{d}^{(0)}~,
\end{equation}
\begin{equation}
    \begin{array}{ccc}
       [\bar{\mathfrak{d}},\mathfrak{d}^{(0)} - \frac{2}{\mu^6} \mathcal{B}]_{\text{PB}} = 2 \bar{\mathfrak{d}}~, &~&[\bar{\mathfrak{d}},\Phi_a]_{\text{PB}} = \left(m + \frac{2}{\mu^6}\frac{\mathcal{H}}{m} \right) \bar{u}_a~, 
    \end{array}
\end{equation}
\begin{equation}
    [\bar{\mathfrak{d}},d_\alpha]_{\text{PB}} = \frac{i}{\mu^6 \sqrt{m}} \bar{w}_\alpha \bar{\nu}~,
\end{equation}
\begin{equation}
    [\Phi_a ,d_\alpha]_{\rm{PB}}= \; \frac {2i}{m\sqrt{m}}\, u^{(0)}_a w_\alpha \,  {\rm tr}({\boldsymbol{ \Psi}}{\mathbb G}) \; ,
\end{equation}
\begin{equation}
    [\Phi_a  ,\bar{d}_\alpha]_{\rm{PB}}= -\frac {2i}{m\sqrt{m}}\, u^{(0)}_a \bar{w}_\alpha\, {\rm tr}(\bar{{\boldsymbol{ \Psi}}}{\mathbb G}) \, ,
\end{equation}
and
\begin{equation}
    \{d_\alpha ,d_\beta \}_{\rm{PB}} = -\frac {8i}{m}\, w_\alpha w_\beta\,  {\rm tr}(\bar{{\mathbb Z}}{\mathbb G}) \; ,
\end{equation}
\begin{equation}
    \{\bar{d}_\alpha ,\bar{d}_\beta \}_{\rm{PB}}= \; \; \frac {8i}{m}\, \bar{w}_\alpha\bar{w}_\beta\, {\rm tr}({\mathbb Z}{\mathbb G}) \, ,
\end{equation}
\begin{equation}
    \begin{array}{c}
         \begin{split}
             \{d_\alpha ,\bar{d}_\beta \}_{\rm{PB}} &= -2ip_a\gamma^a_{\alpha\beta}+2m\epsilon_{\alpha\beta}+\frac {4i}{\mu^6}\, w_\alpha \bar{w}_\beta \, \frac {{\cal H}}{m}  = \\
             & = -2i\Phi_a\gamma^a_{\alpha\beta} -4i  w_\beta\bar{w}_\alpha  \,  \left(m +\frac {1}{\mu^6}\,\frac {{\cal H}}{m}\right)\; .
         \end{split}
    \end{array}
\end{equation}
Using this algebra, the requirement of constraint preservation under evolution, described by Eq.~\eqref{evol=H} with Hamiltonian \eqref{H=LMs}, implies that the Lagrange multipliers must satisfy a system of equations solved by
\bea 
k=0\; ,~~~~ \qquad  \bar{k}=0\; ,  && \\
b^au_a=0\; , \qquad b^a\bar{u}_a=0   && \Longrightarrow  \qquad b^a= b u^{(0)a}\; ,  \\
\bar{\kappa}^\beta w_\beta=0   && \Longrightarrow \qquad  \bar{\kappa}^\beta=\bar{\kappa}w^\beta \; ,   \\
\kappa^\beta \bar{w}_\beta=0  && \Longrightarrow \qquad \kappa^\beta =\kappa  \bar{w}^\beta\; .
\eea
Thus, the total Hamiltonian reads
\begin{eqnarray}\label{H=1st}
H = b \, u^{a(0)} \Phi_a + {\kappa}  \, \bar{w}^\alpha d_\alpha + \bar{ {\kappa}}\, w^\alpha \bar{d}_\alpha+ ik^{(0)} \left({\mathfrak d}{}^{(0)} -   \frac 2{\mu^6} \, {\cal B}\right) + \text{tr}(\mathbb{Y}\mathbb{G})~.
\end{eqnarray}
The arbitrary Lagrange multipliers $b$, $\kappa$, $\bar{ {\kappa}}$, $k^{(0)}$ and $\mathbb{Y}_j{}^i$ (with  $\mathbb{Y}_i{}^i=0$) reflect the  gauge symmetries of our model: reparametrization, local worldline supersymmetry ($\kappa$-symmetry), U$(1)$ and SU$(N)$ symmetries. In \eqref{H=1st} they are multiplied by the constraints generating these symmetries,
\begin{equation}\label{Phi0=0}
\Phi^{(0)} := p_a u^{a(0)} - \left( m + \frac {2}{\mu^6}\,\frac {{\cal H}}{m}\right)\approx 0~,
\end{equation}
\begin{equation}\label{bwd=0}
\bar{w}^\alpha d_\alpha := \bar{w}^\alpha (  \bar{\Pi}_\alpha + i p_a (\gamma^a \bar{\theta})_\alpha -m \bar{\theta}_\alpha) + \frac{1}{\mu^6}  \frac{1}{\sqrt{m}}\bar{{\nu}}  \approx 0~,
\end{equation}
\begin{equation}\label{wbd=0}
w^\alpha \bar{d}_\alpha := w^\alpha ({\Pi}_\alpha + i p_a (\gamma^a \theta)_\alpha + m\theta_\alpha ) - \frac{1}{\mu^6}  \frac{1}{\sqrt{m}} {\nu} \approx 0~,
\end{equation}
\begin{equation}\label{U0=0}
U^{(0)}:= \mathfrak{d}^{(0)} -  \frac 2 {\mu^6} {\cal B} \approx 0~, \qquad
\end{equation}
with ${\cal B}$ defined in  \eqref{cB:=}, and secondary constraint \eqref{eq:Gauss}.

\subsection{Second class constraints}
The remaining constraints are the second class ones. They can be collected in conjugate  pairs
\begin{equation}
    \mathfrak{d} \approx 0~, \qquad \bar{\mathfrak{d}} \approx 0~, \label{eqs:frak_d} 
\end{equation}
\begin{equation}
    \Phi := u^a \Phi_a = u^a p_a \approx 0~, \qquad \bar{\Phi} := \bar{u}^a \Phi_a = \bar{u}^a p_a \approx 0~, \label{eqs:Phi}  
\end{equation}
\begin{equation}
    w^{\alpha} d_{\alpha} = w^\alpha ( \bar{\Pi}_{\alpha} + i p_a(\gamma^a \bar{\theta})_\alpha -m \bar{\theta}_\alpha) \approx 0~, \qquad \bar{w}^{\alpha} \bar{d}_{\alpha} = \bar{w}^\alpha({\Pi}_\alpha + ip_a(\gamma^a \theta )_\alpha + m\theta_\alpha) \approx 0~, \label{eqs:wd} 
\end{equation}
as can be seen from their Poisson brackets, particularly
\begin{eqnarray}
  \label{fdbP=}  [\mathfrak{d}, \bar{\Phi}]_{\text{PB}} = - 2 \Phi^{(0)} - 2 \left( m + \frac{2}{\mu^6}\frac{\mathcal{H}}{m} \right) \approx - 2 \left( m + \frac{2}{\mu^6}\frac{\mathcal{H}}{m} \right)~,  ~~~~~\\  
  \nonumber \\ \label{bfdP=} 
 [\bar{\mathfrak{d}}, \Phi]_{\text{PB}} = - 2 \Phi^{(0)} - 2 \left( m + \frac{2}{\mu^6}\frac{\mathcal{H}}{m} \right) \approx - 2 \left( m + \frac{2}{\mu^6}\frac{\mathcal{H}}{m} \right)~, ~~~~~ \\   \label{wdbwbd=} 
 ~ \nonumber \\
\{w^\alpha d_\alpha, \bar{w}^\alpha \bar{d}_\alpha\}_{\text{PB}}=- 2i \Phi^{(0)} - 4i \left(m + \frac{1}{\mu^6} \frac{\mathcal{H}}{m} \right) \approx - 4i \left(m + \frac{1}{\mu^6} \frac{\mathcal{H}}{m} \right)~. 
\end{eqnarray}

\subsection{The algebra of first and second class constraints}
Notice that there are also other nonvanishing brackets among the second class constraints:
\begin{eqnarray} \label{fdfbd=}
{}[ \mathfrak{d}, \bar{\mathfrak{d}} ]_{\text{PB}} = \mathfrak{d}^{(0)} =  \left( \mathfrak{d}^{(0)} - \frac{2}{\mu^6} \mathcal{B} \right) +\frac{2}{\mu^6} \mathcal{B}  \approx \frac{2}{\mu^6}\mathcal{B}~, \qquad \\ \nonumber \\  \label{fdbwbd=}[\mathfrak{d}, \bar{w}^\alpha \bar{d}_\alpha]_{\text{PB}} = -w^\alpha \bar{d}_\alpha - \frac{1}{\mu^6 \sqrt{m}}\nu \approx - \frac{1}{\mu^6 \sqrt{m}}\nu~,\qquad \\  \nonumber \\   \label{bfdwd=}
 [\bar{\mathfrak{d}},  w^\alpha d_\alpha]_{\text{PB}}= -\bar{w}^\alpha {d}_\alpha+ \frac{1}{\mu^6 \sqrt{m}}\bar{\nu} \approx \;  \frac{1}{\mu^6 \sqrt{m}}\bar{\nu}~.\qquad~
\end{eqnarray}
The algebra of the first and second class constraints is summarized in Table~\ref{table:mD0-centr} (below), where we omit the first class constraints ${\mathbb G}_i{}^j$ \eqref{bbG:=} since they only have nonvanishing brackets with themselves, \eqref{GG=G}. We also introduce the notation
\be\label{M(H)}
M({\cal H}):= m+\frac {1}{\mu^6}\frac {{\cal H}}{m} \; .
\ee
This should not be confused with ${\cal M}({\cal H})$ which was discussed in the first sections of this chapter as well as in chapters~\ref{ch.3D_mD0}-\ref{ch.11D_origin}.

The appearance of the positively definite additive contributions $M({\cal H})$ or $M(2{\cal H})$ in the r.h.s. of the Poisson brackets clearly indicates the pairing of fermionic and bosonic second class constraints.

The algebra of constraints is sufficiently complicated to hamper the use of BRST quantization with standard ``abelization'' scheme for second class constraints \cite{Egorian, Batalin}. The technical problem comes from the non-canonical form of the bosonic second class constraints algebra in which, besides the nonvanishing brackets \eqref{fdbP=},  \eqref{bfdP=} and \eqref{wdbwbd=}, the other brackets \eqref{fdfbd=}, \eqref{fdbwbd=} and  \eqref{bfdwd=} also do not vanish in the weak sense.

\begin{table}[h!]
\resizebox{\textwidth}{!}{\begin{tabular}{c||cccc||cccccc}
 $[\ldots,\ldots \}_{\text{PB}}$
&   ${U}{}^{(0)}$ & $\Phi^{(0)}$& $\bar{w}d$ & $w\bar{d}$ & $\mathfrak{d}$ & $\bar{\mathfrak{d}}$ & $\Phi$ &
$\bar{\Phi}$ &  $wd$ & $\bar{w}\bar{d}$ \\
 \hline \hline
 \\
$U^{(0)}$& 0 & 0 & $-\bar{w}d$ & $w\bar{d}$ &  2$\mathfrak{d}$ & -2$\bar{\mathfrak{d}}$ & 2$\Phi$ &
$-2\bar{\Phi}$ &  $wd$ & $-\bar{w}\bar{d}$
\\
$\Phi^{(0)}$ & 0 & 0 & $-\frac {2{\rm tr} ({\boldsymbol{\Psi}} {\mathbb G})}{m\sqrt{m}}$ & $-\frac {2{\rm tr} (\bar{{\boldsymbol{\Psi}}} {\mathbb G})}{m\sqrt{m}}$ & $\Phi$ & $\bar{\Phi}$ & 0 & 0 & 0 & 0 \\
$\bar{w}d$  & $\bar{w}d$  & $\frac {2{\rm tr} ({\boldsymbol{ \Psi}} {\mathbb G})}{m\sqrt{m}}$ & $\frac {8i{\rm tr} (\bar{Z}{\mathbb G})}{m}$ &  $-2i\Phi^{(0)}$ & $wd$ & 0 & 0 & 0 & 0 & $-2i\bar{\Phi}$ \\
$w\bar{d}$  & $-w\bar{d}$  & $\frac {2{\rm tr} (\bar{{\boldsymbol{ \Psi}}} {\mathbb G})}{m\sqrt{m}}$ &  $-2i\Phi^{(0)}$  & $-\frac {8i{\rm tr} ({Z}{\mathbb G})}{m}$  & 0 & $\bar{w}\bar{d}$  & 0 & 0 & $-2i\Phi$ & 0 \\
 \hline \hline
 \\
${\mathfrak{d}}$ &  $-2{\mathfrak{d}}$ & $-\Phi $ & $-wd$ & 0 & 0 & \fbox{$\begin{matrix}U^{(0)}+\cr +\frac {2{\cal B}}{\mu^6}\end{matrix}$ }& 0 & \fbox{$\begin{matrix}-2\Phi^{(0)}-\cr -2M(2{\cal H})\end{matrix}$} & 0 & \fbox{$\begin{matrix}-w\bar{d}-\cr -\frac {\nu}{\mu^6\sqrt{m}}\end{matrix}$}\\
$\bar{\mathfrak{d}}$  &  $2\bar{\mathfrak{d}}$ & $-\bar{\Phi}$ & 0 &$-\bar{w}\bar{d}$ & \fbox{$\begin{matrix}-U^{(0)}-\cr -\frac {2{\cal B}}{\mu^6}\end{matrix}$ }& 0 & \fbox{$\begin{matrix}-2\Phi^{(0)}-\cr -2M(2{\cal H})\end{matrix}$} & 0 & \fbox{$\begin{matrix}-\bar{w}d+\cr +\frac {\bar{\nu}}{\mu^6\sqrt{m}}\end{matrix}$}  & 0 \\
 $\Phi$ & $-2\Phi$&  0 & 0 & 0 & 0 & \fbox{$\begin{matrix}2\Phi^{(0)}+\cr +2M(2{\cal H})\end{matrix}$}  & 0 & 0  & 0 & 0 \\
 $\bar{\Phi}$ &  $2\bar{\Phi}$  & 0 &0 & 0  & \fbox{$\begin{matrix}2\Phi^{(0)}+\cr +2M(2{\cal H})\end{matrix}$} & 0 & 0 & 0 & 0 & 0 \\
  $wd$ & -$wd$ & 0 & 0 & $-2i\Phi $ & 0 & \fbox{$\begin{matrix}\bar{w}d-\cr -\frac {\bar{\nu}}{\mu^6\sqrt{m}}\end{matrix}$}  & 0 & 0 & 0 & \fbox{$\begin{matrix}-2i\Phi^{(0)}-\cr -4iM({\cal H})\end{matrix}$}\\
   $ \bar{w}\bar{d}$ &  $ \bar{w}\bar{d}$  & 0&  $-2i\bar{\Phi} $ & 0 &  \fbox{$\begin{matrix}w\bar{d}+\cr +\frac {{\nu}}{\mu^6\sqrt{m}}\end{matrix}$} & 0  & 0 & 0 & \fbox{$\begin{matrix}-2i\Phi^{(0)}-\cr -4iM({\cal H})\end{matrix}$} & 0\\
 \hline \hline
\end{tabular}}
\caption{Algebra of the first and second class constraints on the Poisson brackets  of the simplest 3D mD0 system. $M({\cal H}):= m+\frac {1}{\mu^6}\frac {{\cal H}}{m}$ is defined in \eqref{M(H)}.
}
\label{table:mD0-centr}
\end{table}
Moreover, when attempting to apply the Gupta-Bleuler quantization scheme in its canonical forms, we found that for this algebra of constraints the corresponding system of equations (in supercoordinate representation) admits only the trivial solution. We demonstrate this explicitly for the simplest case of vanishing all the matrix and fermionic fields in Appendix~\ref{GB-failed}. This is the only example known to the authors where the Gupta-Bleuler simplified method of quantization of dynamical system fails.

Hence, the only remaining option, capable of overcoming these difficulties, is to apply the Dirac bracket procedure in the bosonic sector of the system. Notably, this procedure can be significantly simplified (in fact, reduced to explicit resolution of the bosonic second class constraints) once we pass to the so-called \textit{analytic basis} in the center of mass sector of the configuration superspace. We will carry out this change of coordinate basis in the next section.

\section{Hamiltonian mechanics in the analytical basis}
In order to simplify the constraint structure and enable an explicit resolution of the pairs bosonic second class constraints, in this section we reformulate the Hamiltonian mechanics in the analytical (Lorentz harmonic) basis. 

\subsection{Constraints in the analytical basis}
Let us begin by noticing that the center of mass sector of the configurational space of our dynamical system can be identified with the so-called Lorentz harmonic superspace, an extended enlarged superspace with coordinates\footnote{\label{HarmFoot} Here, ``extended'' refers to ${\cal N}=2$ supersymmetry while ``enlarged'' reflects the presence of additional bosonic spinor (spinor moving frame or Lorentz harmonic) coordinates. The terminology goes back to \cite{Igor_Russian}, which began the line of seminal papers on Harmonic superspaces  \cite{Galperin2, Galperin3, lightcone, harmonic}. See Appendix \ref{App=D0} for further discussion.}
\be\label{c-basis}
\Sigma^{(3+3|4)} = \{ (x^a,\theta^\alpha , \bar{\theta}^\alpha , w_\alpha , \bar{w}_\alpha )\} =: \{ (z^M , w_\alpha , \bar{w}_\alpha )\} \; , \qquad \bar{w}^\alpha{w}_\alpha =i\; .
\ee
The complete configuration superspace also includes $2(N^2-1)$ bosonic and $(N^2-1)$  fermionic matrix coordinates, and its coordinate basis can be chosen as
 \begin{equation}\label{c-basisALL}
\Sigma^{(4+2N^2|3+N^2)}  =  \{(x^a,\theta^\alpha , \bar{\theta}^\alpha , w_\alpha , \bar{w}_\alpha
; {\mathbb Z}, \bar{{\mathbb Z}}, {\boldsymbol{ \Psi}}  )\} =: \{ (z^M , w_\alpha , \bar{w}_\alpha; {\mathbb Z}, \bar{{\mathbb Z}}, {\boldsymbol{ \Psi}}  )\}~. 
\end{equation}
This implies that $ \bar{{\boldsymbol{\Psi}}}$ is considered to be momentum conjugate to ${{\boldsymbol{ \Psi}}}$.

Clearly, other coordinate bases can be chosen. In our case, it is convenient {\it to perform a change of coordinate in the center of mass sector} \eqref{c-basis} of the whole configurational superspace, choosing in it the analytic coordinate basis
\be
 \Sigma^{(3+3|4)} = \{ ({\rm x}{}^{(0)},{\rm x}_A,\bar{{\rm x}}_A\; ,  \theta^w , {\theta}^{\bar{w}} , \bar{\theta}^{{w}} ,\bar{\theta}^{\bar{w}}  ; w_\alpha , \bar{w}_\alpha  )\} =:\{ (z^{(M)}_{An}, \; w_\alpha , \bar{w}_\alpha\; , {\mathbb Z}, \bar{{\mathbb Z}}, {\boldsymbol{ \Psi}},  \bar{{\boldsymbol{ \Psi}}}   )\} \; , \qquad
\ee
with
\begin{equation}\label{x0=}
    {\rm x}{}^{(0)}:=x^au_a^{(0)}\; ,
\end{equation}
\begin{equation}\label{xA=}
    \begin{array}{lcr}
       {\rm x}_A : = {\rm x}\, -2i \theta^w \bar{\theta}^{{w}}\;  ,& ~ &  {\rm x}:= x^au_a\; ,
    \end{array}
\end{equation}
\begin{equation}\label{thw=}
    \begin{array}{lcr}
      \theta^w := \theta^\alpha w_\alpha~,& ~ &   \theta^{\bar{w}}:= \theta^\alpha \bar{w}_\alpha~,
    \end{array}
\end{equation}
\begin{equation}\label{bxA=}
    \begin{array}{lcr}
       \bar{\text{x}}_A := \bar{{\rm x}} +2i\theta^{\bar{w}} \bar{\theta}^{\bar{w}} \;,  & ~ &  \bar{{\rm x}}:= x^a \bar{u}_a\; ,
    \end{array}
\end{equation}
\begin{equation}\label{bthw=}
    \begin{array}{lcr}
      \bar{\theta}^w := \bar{\theta}^\alpha w_\alpha~,& ~ &  \bar{\theta}^{\bar{w}}:= \bar{\theta}^\alpha \bar{w}_\alpha~,
    \end{array}
\end{equation}
(see Appendix \ref{App=D0}, particularly Eqs.~\eqref{xA=D0}-\eqref{bthw=D0}, {\it cf.}  \cite{Igor_Russian, lightcone, harmonic}). In this analytical basis, the Lagrangian form ${\cal L}=\text{d}\tau \mathcal{L}_\tau$ with $\mathcal{L}_\tau$ from \eqref{eq:3DmD0_L_mConstw} reads 
\begin{equation}
    \begin{array}{l}
         \begin{split}
             \mathcal{L} &= \text{d}{\rm x}^{(0)} \left(m + \frac{2}{\mu^6} \frac{\mathcal{H}}{m} \right)  -  ifm\; \left[\bar{{\rm x}}_A \, \left(1+ \frac 2 {\mu^6} \frac {\cal H}{m^2}\right) -  \frac {4i} {\mu^6} \frac {\cal H}{m^2} \,{\theta}^{\bar{w}} \bar{\theta}^{\bar{w}} + \frac i {\mu^6 } \frac 1{m\sqrt{m}} {\theta}^{\bar{w}}\bar{\nu}\right] + \\
             &~\\
             &+ i \bar{f}m  \left[ {\rm x}_A\, \left(1+ \frac 2 {\mu^6} \frac {\cal H}{m^2}\right)+ \frac {4i} {\mu^6} \frac {\cal H}{m^2} \,\theta^w \bar{\theta}^{{w}}+ \frac i {\mu^6 } \frac 1{m\sqrt{m}} \bar{\theta}^w\nu \right]  -\\
             &~\\
             &-2i \, \left( \text{d}\theta^{\bar{w}} -ia\theta^{\bar{w}}  \right) \, \bar{\theta}^w\,  \left(m + \frac{1}{\mu^6} \frac{\mathcal{H}}{m} \right)  -2i \, \left(\text{d}\bar{\theta}^w+ia\bar{\theta}^w  \right)\, \theta^{\bar{w}} \, \left(m + \frac{1}{\mu^6} \frac{\mathcal{H}}{m} \right) -\\
             &~\\
             &-2i \, \left( \text{d}\theta^{{w}} +ia\theta^{{w}}  \right) \, \left( \frac{1}{\mu^6} \frac{\mathcal{H}}{m}\, \bar{\theta}^{\bar{w}} -\frac{1}{\mu^6}  \frac{1}{2\sqrt{m}}\bar{\nu} \right)
             -2i \, \left(\text{d}\bar{\theta}^{\bar{w}}-ia\bar{\theta}^{\bar{w}} \right)\, \left( \frac{1}{\mu^6} \frac{\mathcal{H}}{m}\, {\theta}^{w} -\frac{1}{\mu^6}  \frac{1}{2\sqrt{m}}\nu \right)+\\
             &~\\
             &+ {1\over \mu^6}{\rm tr}\left(\bar{\mathbb P}{\rm D}_\tau {\mathbb Z} + {\mathbb P}{\rm D}_\tau \bar{\mathbb Z} -{i\over 8} {\rm D}_\tau{{\boldsymbol{ \Psi}}}\,  \bar{{\boldsymbol{ \Psi}}} + {i\over 8} {{\boldsymbol{ \Psi}}} {\rm D}_\tau \bar{{\boldsymbol{ \Psi}}}  \right)~.
         \end{split}
    \end{array}\label{cL=mD0=An}
\end{equation}
It is convenient to introduce the notation
\be\label{tcH:=}
\tilde{{\cal H}}= \frac{1}{\mu^6} \frac{\mathcal{H}}{m}\; , \qquad
\tilde{\nu}= \frac{1}{\mu^6}  \frac{1}{2\sqrt{m}}\nu = (\bar{\tilde{\nu}})^*  \qquad
\ee
which allows the Lagrangian to be written more compactly
\begin{equation}
    \begin{array}{l}
        \begin{split}
            \mathcal{L} &= \text{d}{\rm x}^{(0)}\, (m+2\tilde{\mathcal{H}})  -  i f \left[\bar{{\rm x}}_A (m+2\tilde{\mathcal{H}}) -  {4i} \tilde{\cal H} \,{\theta}^{\bar{w}} \bar{\theta}^{\bar{w}} + 2i {\theta}^{\bar{w}}\bar{\tilde{\nu}}\right] + \\
            &~\\
            & +i \bar{f} \left[ {{\rm x}}_A \, (m+2\tilde{\mathcal{H}}) +  {4i} \tilde{\cal H} \,{\theta}^{{w}} \bar{\theta}^{{w}} + 2i \bar{\theta}^w\tilde{\nu} \;\right] -2i \left( \text{d}\theta^{\bar{w}} -ia\theta^{\bar{w}}  \right) \, \bar{\theta}^w \, (m+\tilde{\mathcal{H}}) -\\
            &~\\
            &- 2i \left(\text{d}\bar{\theta}^w+ia\bar{\theta}^w  \right)\, \theta^{\bar{w}}  \, (m+\tilde{\mathcal{H}}) - 2i \, \left( \text{d}\theta^{{w}} +ia\theta^{{w}}  \right) \, \left( \tilde{\mathcal{H}}\, \bar{\theta}^{\bar{w}} -\bar{\tilde{\nu}} \right) -\\
            &~\\
            &-2i  \left(\text{d}\bar{\theta}^{\bar{w}}-ia\bar{\theta}^{\bar{w}} \right)\, \left(\tilde{\mathcal{H}}\, {\theta}^{w} -\tilde{\nu} \right) + {1\over \mu^6}{\rm tr}\left(\bar{\mathbb P}{\rm D}_\tau {\mathbb Z} + {\mathbb P}{\rm D}_\tau \bar{\mathbb Z}-{i\over 8} {\rm D}_\tau{{\boldsymbol{ \Psi}}}\,  \bar{{\boldsymbol{ \Psi}}} + {i\over 8} {{\boldsymbol{ \Psi}}} {\rm D}_\tau \bar{{\boldsymbol{ \Psi}}}  \right)~.
        \end{split} 
    \end{array} \label{cL=mD0=An=}
\end{equation}
Then the canonical Hamiltonian is defined by
\begin{equation}
    \begin{array}{l}
        \begin{split}
        \text{d}\tau H_0 &= \text{d}{\rm x}^{(0)}p^{(0)} + \text{d}{\rm x}_A \bar{p} + \text{d}\bar{\rm x}_A p + \text{d}\theta^w \Pi^\theta_w + \text{d}\theta^{\bar{w}} \Pi^\theta_{\bar{w}} + \text{d} \bar{\theta}^w \bar{\Pi}^{\bar{\theta}}_w + \text{d} \bar{\theta}^{\bar{w}} \bar{\Pi}^{\bar{\theta}}_{\bar{w}} + \\
        &~\\
        &+ ia \tilde{\mathfrak{d}}^{(0)} + i f \bar{\tilde{\mathfrak{d}}} - i \bar{f} \tilde{\mathfrak{d}} + \frac{1}{\mu^6} \text{tr}(\text{d}\mathbb{Z}\bar{\mathbb{P}})+ \frac{1}{\mu^6} \text{tr}(\text{d}\bar{\mathbb{Z}}\mathbb{P}) -	\frac{i}{8 \mu^6} \text{tr}(\text{d}{\boldsymbol{ \Psi}} \bar{{\boldsymbol{ \Psi}}}) -\\
        &~\\
        &-\frac{i}{8 \mu^6} \text{tr}(\text{d}\bar{{\boldsymbol{ \Psi}}} {\boldsymbol{ \Psi}}) + \text{tr}(\text{d}\mathbb{A} \mathbb{P}_{\mathbb{A}}) - \mathcal{L}_{\text{mD}0}~.
        \end{split}
    \end{array} \label{H0=analyt}
\end{equation}
In \eqref{H0=analyt} the momenta for the bosonic coordinate functions are related to their central basis counterparts by
\begin{equation}\label{up=-2p}
\begin{array}{ccccc}
p^{(0)}:= u^{a(0)}p_a~,&~&p := -\dfrac{1}{2}u^a p_a~,&~&\bar{p} := -\dfrac{1}{2}\bar{u}^a p_a
\end{array}
\end{equation}
and have the nonvanishing Poisson brackets
\begin{equation}
\begin{array}{ccccc}
\left[p^{(0)}, {\rm x}^{(0)} \right]_{\text{PB}} = -1~,&~&\left[p, \bar{{\rm x}}_A \right]_{\text{PB}} = -1~,&~&\left[\bar{p}, {\rm x}_A \right]_{\text{PB}} = -1~.
\end{array}
\end{equation}
The momenta conjugate to the fermionic coordinate functions of the analytical basis obeying
\begin{equation}
\begin{array}{ccccc}
{}\{{\Pi}^\theta_w , \theta^w\}_{\text{PB}} = -1~,&~& \{ {\Pi}^\theta_{\bar{w}} , \theta^{\bar{w}}\}_{\text{PB}} = -1~,\\
~\\
\{ \bar{\Pi}^{\bar{\theta}}_w , \bar{\theta}^w\}_{\text{PB}} = -1~, &~& \{ \bar{\Pi}^{\bar{\theta}}_{\bar{w}} , \bar{\theta}^{\bar{w}}\}_{\text{PB}} = -1~
\end{array}
\end{equation}
are related to their central basis cousins by the somewhat more complicated relations
\begin{equation}\label{Pi-PiAn}
\begin{array}{ccc}
\Pi^\theta_{w}:= -i\bar{w}^\alpha \Pi_\alpha+2i \bar{\theta}^w\bar{p}~,&~& \Pi^\theta_{\bar{w}}:= iw^\alpha \Pi_\alpha - 2i \bar{\theta}^{\bar{w}}p~,\\
~\\
\bar{\Pi}^{\bar{\theta}}_{w}:= -i\bar{w}^\alpha \bar{\Pi}_\alpha -2i \theta^w \bar{p} ~,&~& \bar{\Pi}^{\bar{\theta}}_{\bar{w}}:= iw^\alpha \bar{\Pi}_\alpha+ 2i \theta^{\bar{w}}p~.
\end{array}
\end{equation}
The covariant momenta in the analytical basis, denoted by tilde to distinguish then from their central basis counterparts, satisfy
\begin{equation}
\begin{array}{lll}
{}[\tilde{\mathfrak{d}}^{(0)} , {z}_{An}^{(M)}]_{\text{PB}} =0 \; , \qquad {}[\tilde{\mathfrak{d}}, {z}_{An}^{(M)}]_{\text{PB}} =0 \; , \qquad  {}[\bar{\tilde{\mathfrak{d}}}, {z}_{An}^{(M)}]_{\text{PB}} =0 \; , \qquad
\end{array}
\end{equation}
while
\begin{equation}
\begin{array}{lll}
{}[{\mathfrak{d}}^{(0)} , z^{M}]_{\text{PB}} =0 \; , \qquad {}[{\mathfrak{d}},  z^{M}]_{\text{PB}} =0 \; , \qquad  {}[\bar{{\mathfrak{d}}}, z^{M}]_{\text{PB}} =0 \; . \qquad
\end{array}
\end{equation}
Eq.~\eqref{H0=analyt} also implicitly encodes the definitions of the canonical momenta as derivatives of the Lagrangian with respect to velocities, \eqref{pA:=}, as well as the covariant momenta as derivatives with respect to Cartan forms, \eqref{fd=dL-df}. Calculating these for our Lagrangian \eqref{cL=mD0=An=}, we obtain the primary constraints ({\it cf.} \eqref{eq:Phi_a}-\eqref{eq:PA})
\begin{eqnarray}\label{p0=An}
\Phi^{(0)} := p^{(0)} - (m + 2\tilde{\mathcal{H}}) \approx 0~,   \qquad
 \\ \nonumber \\  \label{pA=0} - \frac 1 2\  \Phi := p \approx 0~,    \qquad \\ \nonumber \\  \label{bpA=0} - \frac 1 2\bar{\Phi} :=  \bar{p} \approx 0~,    \qquad \\ \nonumber \\ \label{fd=An}
 \tilde{{\mathfrak d}}+ {{\rm x}}_A \, (m+2\tilde{{\mathcal{H}}}) +  {4i} \tilde{{\cal H}} \,{\theta}^{{w}} \bar{\theta}^{{w}} + 2i \bar{\theta}^w\tilde{\nu}  \approx 0 \; , \qquad  \\ \nonumber
 \\  \label{bfd=An}  \bar{\tilde{{\mathfrak d}}} + \bar{{\rm x}}_A \, (m+2\tilde{\mathcal{H}}) -  {4i} \tilde{\cal H} \,{\theta}^{\bar{w}} \bar{\theta}^{\bar{w}} + 2i {\theta}^{\bar{w}}\bar{\tilde{\nu}}\approx 0 \; , \qquad 
\end{eqnarray}
\begin{eqnarray} \label{U=An} \tilde{U}:= \tilde{{\mathfrak d}}^{(0)}-4i \theta^{\bar{w}}  \bar{\theta}^w  \, (m+\tilde{\mathcal{H}})+ 4i \theta^{{w}}  \bar{\theta}^{\bar{w}}\,\tilde{\mathcal{H}}\,   -2i \theta^{{w}} \bar{\tilde{\nu}} +2i \bar{\theta}^{\bar{w}} \tilde{{\nu}}  -   2\tilde{{\cal B}}\approx 0 \; , \qquad \\ \nonumber \\   \label{tB=} \text{where} \qquad \tilde{{\cal B}} := \frac 1 {\mu^6} {\cal B}:= \frac 1 {\mu^6}{\rm tr}\left(\bar{{\mathbb P}}{\mathbb Z}-{\mathbb P}\bar{{\mathbb Z}}+\frac i 8 {{\boldsymbol{ \Psi}}}\bar{{{\boldsymbol{ \Psi}}}}\right) ,   \qquad
\end{eqnarray}

\begin{eqnarray}\label{dw=}
 d_w:= \Pi^\theta_w - 2i \left( \bar{\tilde{\nu}}- \bar{\theta}^{\bar{w}} \tilde{{\mathcal{H}}} \right) \approx 0~, \qquad  \\ \nonumber \\  \label{dbw=}  \qquad d_{\bar{w}}:= \Pi^\theta_{\bar{w}}+ 2i\left(m+    \tilde{{\mathcal{H}}} \right)\bar{\theta}^w \approx 0~, \qquad  \\ \nonumber  \\  \label{bdw=}  \bar{d}_w := \bar{\Pi}^{\bar{\theta}}_w + 2i\left(m+    \tilde{{\mathcal{H}}} \right) \theta^{\bar{w}}\approx 0~, \qquad   \\ \nonumber
 \\ \label{bdbw=} \bar{d}_{\bar{w}} := \bar{\Pi}^{\bar{\theta}}_{\bar{w}}  - 2i \left( \tilde{{\nu}}- {\theta}^{{w}} \tilde{{\mathcal{H}}} \right) \approx 0~ \qquad
\end{eqnarray}
and \eqref{eq:PA}
\begin{equation}
\mathbb{P}_{\mathbb{A}}:= \frac {\partial L}{\partial \dot{\mathbb{A}}_\tau} \approx 0~.
\label{eq:PA=An}
\end{equation}
Exactly as in the central basis analysis, the requirement of preservation of this last constraint leads to the secondary constraint \eqref{eq:Gauss},
\begin{equation}
\mathbb{G}= \dfrac{1}{\mu^6} \left( [\bar{\mathbb{Z}}, \mathbb{P}] + [\mathbb{Z}, \bar{\mathbb{P}}] - \frac{i}{4} \{{\boldsymbol{ \Psi}}, \bar{{\boldsymbol{ \Psi}}} \} \right) \approx 0
\label{eq:Gauss=An}
\end{equation}
and, as a result, the canonical Hamiltonian vanishes in the weak sense
\begin{equation}
  \label{H0=0=An}
  H_0\approx 0\; .
\end{equation}
That is, the total Hamiltonian is a linear combination of constraints or, more precisely, of the first class constraints.

The set of first class constraints
includes Eqs.~\eqref{p0=An}, \eqref{dw=}, \eqref{bdbw=},
\begin{eqnarray}
\label{Phi0=0I}
\Phi^{(0)}\approx 0\; , \qquad d_w\approx 0\; , \qquad \bar{d}_{\bar{w}}\approx 0\; , \qquad
\end{eqnarray}
and the sum of \eqref{U=An} with a linear combination of the second class constraints \eqref{pA=0},  \eqref{bpA=0}, \eqref{dbw=} and \eqref{bdw=},
\begin{equation}
    \begin{array}{l}
         \begin{split}
             \tilde{\tilde{U}}^{(0)}& := \tilde{U}^{(0)} - 2 \text{x}_A \bar{p} + 2 \bar{\text{x}}_A p - \bar{\theta}^w \bar{d}_w + \theta^{\bar{w}}d_{\bar{w}} =\\
             &~\\
             &=  \tilde{{\mathfrak d}}^{(0)} - 2 \text{x}_A \bar{p} + 2 \bar{\text{x}}_A p - \bar{\theta}^w \bar{\Pi}^{\bar{\theta}}_w + \theta^{\bar{w}} \Pi^{\theta}_{\bar{w}} +4i \theta^{w}  \bar{\theta}^{\bar{w}} \tilde{\mathcal{H}} - 2i \theta^w \bar{\tilde{\nu}} + 2i \bar{\theta}^{\bar{w}} \tilde{\nu} - 2 \tilde{{\cal B}} \approx 0~.
         \end{split}
    \end{array}\label{ttU=}
\end{equation}
It also includes~\eqref{eq:PA=An} (reflecting the gauge nature of the 1d gauge field) and the Gauss constraint \eqref{eq:Gauss=An}. The remaining constraints are second class.

In what follows, we will not need the complete Poisson bracket algebra of the constraints. Instead of calculating it directly, we can use their relations to the central basis constraints, together with the knowledge of the algebra of these latter. In the case of first class constraint \eqref{Phi0=0} and second class constraints \eqref{eqs:Phi}, this relation is just a coincidence,
\begin{equation}
\Phi^{(0)} := p_a u^{(0)} - (m + 2 \tilde{\cal H}) = p^{(0)} -( m+2 \tilde{{\cal H}})~,
\end{equation}
\begin{equation}
\Phi := \Phi_a u^a = - 2p~,
\end{equation}
\begin{equation}
\bar{\Phi} := \Phi_a \bar{u}^a = -2 \bar{p}~.
\end{equation}
The same is true for  the Gauss constraints \eqref{eq:Gauss=An} and for \eqref{eq:PA=An}. However, for the remaining constraints, the relations are nontrivial: 
\begin{equation}
    \begin{array}{l}
         \begin{split}
             U^{(0)}:= \mathfrak{d}^{(0)} - 2 \tilde{\mathcal{B}} &= \tilde{\mathfrak{d}}^{(0)} - 2 \text{x}_A \bar{p} + 2 \bar{\text{x}}_Ap - \theta^w \Pi^\theta_w - \bar{\theta}^w \bar{\Pi}^{\bar{\theta}}_{w} +  \theta^{\bar{w}} \Pi^\theta_{\bar{w}} + \bar{\theta}^{\bar{w}} \bar{\Pi}^{\bar{\theta}}_{\bar{w}} - 2 \tilde{\mathcal{B}} =\\
             &~\\
             &= \tilde{\tilde{U}}{}^{(0)}- \theta^w d_w + \bar{\theta}^{\bar{w}} \bar{d}_{\bar{w}}~,
         \end{split}
    \end{array}
\end{equation}
\begin{equation}
\bar{w}^\alpha d_\alpha := \bar{w}^\alpha (  \bar{\Pi}_\alpha + i p_a (\gamma^a \bar{\theta})_\alpha - m \bar{\theta}_\alpha) + 2\bar{\tilde{\nu}}  =
i d_w- \Phi^{(0)} \bar{\theta}^{\bar{w}}~,
\end{equation}
\begin{equation}
w^\alpha \bar{d}_\alpha := w^\alpha ({\Pi}_\alpha + i p_a (\gamma^a \theta)_\alpha + m\theta_\alpha ) - 2\tilde{\nu}  =- i \bar{d}_{\bar{w}}  + \Phi^{(0)} \theta^w~,
\end{equation}
\begin{equation}
w^\alpha d_\alpha := w^\alpha(  \bar{\Pi}_\alpha + i p_a (\gamma^a \bar{\theta})_\alpha - m \bar{\theta}_\alpha) =  -i d_{\bar{w}} + 4 p \bar{\theta}^{\bar{w}}  + \Phi^{(0)} \bar{\theta}^w~,
\end{equation}
\begin{equation}
\bar{w}^\alpha \bar{d}_\alpha := \bar{w}^\alpha({\Pi}_\alpha + i p_a (\gamma^a \theta)_\alpha + m\theta_\alpha )  =i \bar{d}_w - 4p \theta^w - \Phi^{(0)} \theta^{\bar{w}}~,
\end{equation}
\begin{equation}
\mathfrak{d} = \tilde{\mathfrak{d}} + 2(\text{x}^{(0)} + i \theta^{\bar{w}}\bar{\theta}^w + i \theta^w \bar{\theta}^{\bar{w}})p + (\text{x}_A + 2i \theta^w \bar{\theta}^w)p^{(0)} + \bar{\theta}^w \bar{\Pi}^{\bar{\theta}}_{\bar{w}} + \theta^w \Pi^{\theta}_{\bar{w}}~,
\end{equation}
\begin{equation}
\bar{\mathfrak{d}} = \bar{\tilde{\mathfrak{d}}} + 2(\text{x}^{(0)} - i \theta^{\bar{w}}\bar{\theta}^w - i \theta^w \bar{\theta}^{\bar{w}})\bar{p} + (\bar{\text{x}}_A - 2i \theta^{\bar{w}} \bar{\theta}^{\bar{w}})p^{(0)} + \theta^{\bar{w}} \Pi^\theta_w + \bar{\theta}^{\bar{w}} \bar{\Pi}^{\bar{\theta}}_{{w}}~.
\end{equation}
Using  these expressions, the algebra of constraints  in the analytical basis can be read from Table \ref{table:mD0-centr} and  Eqs.~\eqref{fdbP=}-\eqref{bfdwd=}. Below we will need only part of this algebra, summarized (essentially) in Table~\ref{table:mD0=An}.

\subsection{Resolving bosonic second class constraints}
Actually, as we have already stated, we do not require the complete algebra of the constraints in detail. Instead, we can observe that the bosonic second class constraints in the analytic basis, Eqs.~\eqref{bpA=0}, \eqref{fd=An} and  \eqref{pA=0}, \eqref{bfd=An}, are explicitly resolved (in Dirac's terminology~\cite{Dirac}). This means that
in these pairs of constraints, one element (namely  \eqref{fd=An} and \eqref{bfd=An}) can be rewritten to express the coordinates ${\rm x}_A$ and $\bar{\rm x}_A$ in terms of other variables, while the conjugate elements (\eqref{bpA=0} and \eqref{pA=0}) imply the vanishing of the momenta conjugate to these dependent coordinates. In such cases, the procedure of changing the Poisson brackets by Dirac brackets can be replaced by just setting momenta to zero in the strong sense
\be
\bar{p}=0\; , \qquad p=0\; ,
\ee
and substituting ${\rm x}_A$ and $\bar{\rm x}_A$ in all places when these appear by their expressions obtained from solving \eqref{fd=An} and \eqref{bfd=An},
\bea  \label{xA=An}
 {{\rm x}}_A = -\frac{ \tilde{{\mathfrak d}}+  {4i} \tilde{{\cal H}} \,{\theta}^{{w}} \bar{\theta}^{{w}} +2i \bar{\theta}^w\tilde{\nu}}{m+2\tilde{{\mathcal{H}}}}   \; , \qquad   \bar{{\rm x}}_A = -\frac{\bar{\tilde{{\mathfrak d}}} - {4i} \tilde{\cal H} \,{\theta}^{\bar{w}} \bar{\theta}^{\bar{w}} + 2i {\theta}^{\bar{w}}\bar{\tilde{\nu}}} {m+2\tilde{{\mathcal{H}}}}   \; . \qquad
\eea
Fortunately, in our case, ${\rm x}_A$ and $\bar{{\rm x}}_A$ do not contribute to other constraints. Therefore, the above prescription is tantamount to reducing the configuration space by omitting the directions corresponding to these coordinates. Equivalently, one can omit \eqref{fd=An} and \eqref{bfd=An} the set of constraints, thereby converting \eqref{bpA=0} and \eqref{pA=0} into first class constraints. Then, ${{\rm x}}_A$ and $\bar{{\rm x}}_A$ can be gauged away using the symmetries generated by \eqref{bpA=0} and \eqref{pA=0}, thus arriving at the same result as described above.

For the fermionic second class constraints, since the canonically conjugate constraints are also complex conjugate, they can be quantized via the Gupta-Bleuler method. At the classical level, this is equivalent to omitting one of the conjugate constraints (say, $d_{\bar{w}}$), thereby effectively promoting its conjugate $\bar{d}_w$ to a first class constraint.

{\color{red}{
\begin{table}[h!]
\resizebox{\textwidth}{!}{\begin{tabular}{c||c|c|c|c|c}
 $[\ldots,\ldots \}_{\text{PB}}$
& $\Phi^{(0)}$&  $\tilde{\tilde{U}}^{(0)}$  & $d_w$ & $\bar{d}_{\bar{w}}$ &
 $ \bar{d}_w$ \\
 \hline \hline
 ~\\
$\Phi^{(0)}$& 0 & $ \boxed{\begin{matrix} 2i \left( \theta^w \text{tr}(\tilde{{\boldsymbol{ \Psi}}} \tilde{{\mathbb{G}}}) \right. +\cr  \left. +\bar{\theta}^{\bar{w}} \text{tr}(\bar{\tilde{{\boldsymbol{ \Psi}}}}\tilde{{\mathbb{G}}})\right)\end{matrix}}$ & $2i\text{tr}(\tilde{{\boldsymbol{ \Psi}}} \tilde{{\mathbb{G}}})$ & $-2i \text{tr}(\bar{\tilde{{\boldsymbol{ \Psi}}}} \tilde{{\mathbb{G}}})$   & 0 \\
~\\
$\tilde{\tilde{U}}^{(0)}$ & $\boxed{\begin{matrix}-2i \left(\theta^w \text{tr}(\tilde{{\boldsymbol{ \Psi}}} \tilde{{\mathbb{G}}}) \right. +\cr  \left.+\bar{\theta}^{\bar{w}} \text{tr}(\bar{\tilde{{\boldsymbol{ \Psi}}}} \tilde{{\mathbb{G}}})\right)\end{matrix}}$ & 0 &  $\boxed{\begin{matrix}2\theta^w \left(3 \bar{\theta}^{\bar{w}}\text{tr}(\tilde{{\boldsymbol{ \Psi}}} \tilde{{\mathbb{G}}}) \right. -\cr -\left. 4i \text{tr}(\bar{\mathbb{Z}} \tilde{{\mathbb{G}}}) \right)\end{matrix}}$ & $\boxed{\begin{matrix}2 \bar{\theta}^{\bar{w}}\left(3\theta^w \text{tr}(\bar {\tilde{{\boldsymbol{ \Psi}}}} \tilde{{\mathbb{G}}}) \right. -\cr -\left.
4i \text{tr}(\mathbb{Z} \tilde{{\mathbb{G}}}) \right)\end{matrix}}$ &  $\boxed{\begin{matrix} -\bar{d}_w -\cr -2 \theta^{\bar{w}} \bar{\theta}^{\bar{w}} \text{tr}(\bar{\tilde{{\boldsymbol{ \Psi}}}} \tilde{{\mathbb{G}}}) + \cr +2 \theta^w\theta^{\bar{w}}  \text{tr}(\tilde{{{\boldsymbol{ \Psi}}}} \tilde{{\mathbb{G}}}) \end{matrix}}$  \\
~\\
 $d_w$ &  $-2i \text{tr}\tilde{{\boldsymbol{ \Psi}}} \tilde{{\mathbb{G}}})$ & $\boxed{\begin{matrix}-2\theta^w \left(3 \bar{\theta}^{\bar{w}}\text{tr}(\tilde{{\boldsymbol{ \Psi}}} \tilde{{\mathbb{G}}}) \right. -\cr -\left. 4i \text{tr}(\bar{\mathbb{Z}} \tilde{{\mathbb{G}}}) \right)\end{matrix}}$ & $\boxed{\begin{matrix} 4 \bar{\theta}^{\bar{w}} \text{tr}(\tilde{{\boldsymbol{ \Psi}}}\tilde{{\mathbb{G}}}) -\cr - 8i \text{tr}(\bar{\mathbb{Z}}\tilde{{\mathbb{G}}})\end{matrix}}$ & $\boxed{\begin{matrix}2 \left(\theta^w \text{tr}(\tilde{{\boldsymbol{ \Psi}}} \tilde{{\mathbb{G}}}) \right. +\cr  + \left.\bar{\theta}^{\bar{w}} \text{tr}(\bar{\tilde{{\boldsymbol{ \Psi}}}} \tilde{{\mathbb{G}}})\right)\end{matrix}}$ &  $2 \theta^{\bar{w}}  \text{tr}(\tilde{{\boldsymbol{ \Psi}}} \tilde{{\mathbb{G}}})$ \\
 ~\\
 $\bar{d}_{\bar{w}}$ & $2i \text{tr}(\bar{\tilde{{\boldsymbol{ \Psi}}}} \tilde{{\mathbb{G}}})$ & $\boxed{\begin{matrix}-2\bar{\theta}^{\bar{w}}\left(3\theta^w \text{tr}(\bar {\tilde{{\boldsymbol{ \Psi}}}} \tilde{{\mathbb{G}}}) \right. -\cr -\left. 4i \text{tr}(\mathbb{Z} \tilde{{\mathbb{G}}}) \right)\end{matrix}}$  & $\boxed{\begin{matrix}2 \left(\theta^w \text{tr}(\tilde{{\boldsymbol{ \Psi}}} \tilde{{\mathbb{G}}}) \right. +\cr  \left.+\bar{\theta}^{\bar{w}} \text{tr}(\bar{\tilde{{\boldsymbol{ \Psi}}}} \tilde{{\mathbb{G}}}) \right)\end{matrix}}$ & $\boxed{\begin{matrix}-4  \theta^{w} \text{tr}(\bar{\tilde{{\boldsymbol{ \Psi}}}}\tilde{{\mathbb{G}}})+ \cr +8i \text{tr}({\mathbb{Z}}\tilde{{\mathbb{G}}})\end{matrix}}$ &  $-2 \theta^{\bar{w}}  \text{tr}(\bar{\tilde{{\boldsymbol{ \Psi}}}} \tilde{{\mathbb{G}}})$  \\
 ~\\
 $ \bar{d}_w$ &  0 &   $\boxed{\begin{matrix} \bar{d}_w +\cr +2 \theta^{\bar{w}} \bar{\theta}^{\bar{w}} \text{tr}(\bar{\tilde{{\boldsymbol{ \Psi}}}} \tilde{{\mathbb{G}}}) - \cr -2 \theta^w\theta^{\bar{w}}  \text{tr}(\tilde{{{\boldsymbol{ \Psi}}}} \tilde{{\mathbb{G}}}) \end{matrix}}$ & $2 \theta^{\bar{w}}  \text{tr}(\tilde{{\boldsymbol{ \Psi}}} \tilde{{\mathbb{G}}})$ & $-2 \theta^{\bar{w}}  \text{tr}(\bar{\tilde{{\boldsymbol{ \Psi}}}} \tilde{{\mathbb{G}}})$ & 0  \\
 ~\\
 \hline \hline
\end{tabular}}
\caption{Closed algebra of the ``effective first class constraints'' of the mD0 system. For compactness, we have omitted the lines and columns corresponding to the Gauss constraint \eqref{eq:Gauss=An} (which have only one nonvanishing element on their crossing) and to the first class  constraint  \eqref{eq:PA=An} (which are rows and columns of zeros), and we have also rescaled the Gauss constraint and the fermionic matrix fields as follows: $\tilde{{\mathbb G}}:= \frac 1 m {\mathbb G}\,$, $ \tilde{{\boldsymbol{ \Psi}}} := \frac 1 {\sqrt{m}} \; {\boldsymbol{ \Psi}}\,$,  $ \bar{\tilde{{\boldsymbol{ \Psi}}}} := \frac 1 {\sqrt{m}} \; \bar{{\boldsymbol{ \Psi}}}$. }
\label{table:mD0=An}
\end{table}
}}

The essential part of the algebra of the resulting effective first class constraints is collected in Table~\ref{table:mD0=An}. We have omitted rows and columns corresponding to the constraint \eqref{eq:PA=An}, which has vanishing brackets with all others, and the Gauss constraint matrix \eqref{eq:Gauss=An}, whose only nonvanishing bracket is with itself, \eqref{GG=G}. To simplify expressions on the r.h.s., we introduce in Table~\ref{table:mD0=An} rescaled expressions for Gauss constraint and for the fermionic matrix fields:
\be\label{tG=G/m}
\tilde{{\mathbb G}}:= \frac 1 m {\mathbb G}\, \qquad  \tilde{{\boldsymbol{ \Psi}}} := \frac 1 {\sqrt{m}} \; {\boldsymbol{ \Psi}}\, \qquad  \bar{\tilde{{\boldsymbol{ \Psi}}}} := \frac 1 {\sqrt{m}} \; \bar{{\boldsymbol{ \Psi}}}\; . \qquad
\ee
Notice that all the terms in the r.h.s. of the brackets resumed in Table \ref{table:mD0=An} are proportional to the Gauss law constraint and thus vanish when this is taken into account

\section{Quantization of multiple D0-brane system and its field theory equations}
\label{sec:quantization}
Quantization implies replacing the phase space variables with operators, whose commutators (and anticommutators) are determined by the Poisson brackets of their classical counterparts, according to Dirac's prescription
\begin{equation}\label{Dirac=quant}
[\ldots,\ldots\}_{\text{PB}}\quad \mapsto \quad \frac 1 i\, [\ldots,\ldots\}~.
\end{equation}
In the supercoordinate representation, the bosonic and fermionic configuration space variables maintain their $\text{(quasi-)}$classical description, i.e. they are ordinary numbers ($c$-number) or elements of a Grassmann algebra ($a$-number), respectively. Their momenta are represented by differential operators such that the Dirac prescription~\eqref{Dirac=quant} holds.

In particular, for the bosonic and fermionic center of mass coordinate functions and their momenta, quantization proceeds by replacing them with the corresponding coordinates and differential operators as
\begin{equation}
\begin{array}{rcccl}
\hat{\text{x}}^{(0)} = \text{x}^{(0)}~,&~& \hat{\text{x}}_A = \text{x}_{A}~,&~&\hat{\bar{\text{x}}}_A = \bar{\text{x}}_A~,\\
~\\
\hat{p}^{(0)} = -i \partial_{\text{x}^{(0)}} ~,&~& \hat{\bar{p}}_A = -i \partial_{\text{x}_A}~,&~&\hat{p}_A = -i \partial_{\bar{\text{x}}_A}~,
\end{array}
\label{eq:bosOperators}
\end{equation}
and
\begin{equation}
\begin{array}{rcrclcl}
\hat{\theta}^w = \theta^w~,&~&\hat{\theta}^{\bar{w}} = \theta^{\bar{w}}~,&~& \hat{\bar{\theta}}^w = \bar{\theta}^w~,&~&\hat{\bar{\theta}}^{\bar{w}} = \bar{\theta}^{\bar{w}}~,\\
~\\
\hat{\Pi}^{\theta}_w = -i \partial_{\theta^w} ~,&~& \hat{\Pi}^{\theta}_{\bar{w}}= -i \partial_{\theta^{\bar{w}}}~,&~&\hat{\bar{\Pi}}^{\bar{\theta}}_w = -i \partial_{\bar{\theta}^w}~,&~&\hat{\bar{\Pi}}^{\bar{w}}_{\bar{\theta}} = -i \partial_{\bar{\theta}^{\bar{w}}}~,
\end{array}
\label{eq:ferOperators}
\end{equation}
where the notation
\begin{equation}
\begin{array}{ccrcrcr}
~&~&\partial_{\text{x}^{(0)}} = \dfrac{\partial}{\partial \text{x}^{(0)}}~,&~&\partial_{\text{x}_A} = \dfrac{\partial}{\partial \text{x}_A}~,&~&\partial_{\bar{\text{x}}_A} = \dfrac{\partial}{\partial \bar{\text{x}}_A}~,\\
~\\
\partial_{\theta^w} = \dfrac{\partial}{\partial \theta^w}~,&~&\partial_{\theta^{\bar{w}}} = \dfrac{\partial}{\partial \theta^{\bar{w}}}~,&~&\partial_{\bar{\theta}^w} = \dfrac{\partial}{\partial \bar{\theta}^w}~,&~&\partial_{\bar{\theta}^{\bar{w}}} = \dfrac{\partial}{\partial \bar{\theta}^{\bar{w}}}~
\end{array}
\end{equation}
has been used. For the complex spinors frame variables, the covariant momenta are expressed in terms of covariant derivatives
\be
\mathbb{D}^{(0)} =  \bar{w}_\alpha \dfrac{\partial}{\partial \bar{w}_\alpha} - {w}_\alpha \dfrac{\partial}{\partial {w}_\alpha}~,\qquad \mathbb{D} =w_\alpha \dfrac{\partial}{\partial \bar{w}_\alpha}~,\qquad \bar{\mathbb{D}} =  \bar{w}_\alpha\dfrac{\partial}{\partial {w}_\alpha}~,
\ee
so that
\begin{equation}
\begin{array}{lcrcr}
~&~&\hat{w}_\alpha = w_\alpha~,&~&\hat{\bar{w}}_\alpha = \bar{w}_\alpha~,\\
~\\
\hat{\mathfrak{d}}{}^{(0)}=-i\mathbb{D}^{(0)} ~,&~&\hat{\mathfrak{d}}=-i \mathbb{D} ~,&~&\hat{\bar{\mathfrak{d}}}= -i\bar{\mathbb{D}} ~.
\end{array}
\end{equation}
For the bosonic matrix fields, quantization in coordinate representation is standard
\begin{equation}\label{PZ=ddZ}
\begin{array}{rccl}
\hat{\mathbb{Z}}_i{}^j = \mathbb{Z}_i{}^j ~,&~& \hat{\bar{\mathbb{Z}}}_i{}^j  = \bar{\mathbb{Z}}_i{}^j ~,\\
~\\
\hat{\bar{\mathbb{P}}}_i{}^j  = -i \mu^6 \dfrac{\partial}{\partial \mathbb{Z}_j{}^i}~,&~&\hat{\mathbb{P}}_i{}^j  = -i \mu^6 \dfrac{\partial}{\partial \bar{\mathbb{Z}}_j{}^i}~,
\end{array}
\end{equation}
while for the fermionic matrix variables obeying
\begin{equation}\label{hPsihbPsi=}
\left\lbrace \hat{{\boldsymbol{ \Psi}}}_i{}^j, \hat{\bar{{\boldsymbol{ \Psi}}}}_k{}^l \right\rbrace = \; 4\mu^6 ( \delta^l_i \delta_k^j - \dfrac{1}{N} \delta^j_i \delta^l_k )~, \qquad \left\lbrace \hat{{\boldsymbol{ \Psi}}}^j_i, \hat{{{\boldsymbol{ \Psi}}}}_k^l \right\rbrace = 0 \; , \qquad \left\lbrace \hat{\bar{{\boldsymbol{ \Psi}}}}^j_i, \hat{\bar{{\boldsymbol{ \Psi}}}}_k^l \right\rbrace = 0\;  \qquad
\end{equation}
one employs the holomorphic representation (analogous to that for the Heisenberg algebra), in which the creation operator is represented by differentiation with respect to classical counterpart of the annihilation operator,
\begin{equation}\label{bPsi=ddPsi} \hat{{\boldsymbol{ \Psi}}}^j_i = {\boldsymbol{ \Psi}}^j_i~,\qquad \hat{\bar{{\boldsymbol{ \Psi}}}}^j_i = \; 4 \mu^6 \dfrac{\partial}{\partial {\boldsymbol{ \Psi}}^i_j}~.
\end{equation}
Then, the quantum counterparts of the SYM supercurrents and Hamiltonian are represented by the following differential operators
\begin{equation}\label{hnu=}
\hat{\tilde{\nu}}=\frac{1}{\mu^6}  \frac{1}{2\sqrt{m}} \hat{\nu}=  \frac{1}{2\sqrt{m}} \text{tr}\left(-i {\boldsymbol{ \Psi}} \dfrac{\partial}{\partial \overline{\mathbb{Z}}} +4 [\mathbb{Z}, \bar{\mathbb{Z}}]  \dfrac{\partial}{\partial {\boldsymbol{ \Psi}}}\right)~, \qquad \end{equation}
\begin{equation} \label{hbnu=}
\hat{\bar{\tilde{\nu}}}=\frac{1}{\mu^6}  \frac{1}{2\sqrt{m}} \hat{\bar{\nu}}= \frac{1}{2\sqrt{m}} \text{tr}\left(-4i  {\mu^6} \dfrac{\partial}{\partial \mathbb{Z}}  \dfrac{\partial}{\partial {\boldsymbol{ \Psi}}}+  \frac{1}{\mu^6} {\boldsymbol{ \Psi}} [\mathbb{Z}, \bar{\mathbb{Z}}]\right)~,  \qquad
\end{equation}
\begin{equation}\label{hcH=}
\hat{\tilde{{\cal H}}}= \frac{1}{\mu^6} \frac{\hat{\mathcal{H}}}{m}= \frac{1}{m} {\rm tr}\left( -\mu^6 \dfrac{\partial}{\partial \mathbb{Z}}\dfrac{\partial}{\partial \bar{\mathbb{Z}}} +8i\mu^6\bar{\mathbb Z} \dfrac{\partial}{\partial {\boldsymbol{ \Psi}}}\dfrac{\partial}{\partial {\boldsymbol{ \Psi}}}  +  \frac{1}{\mu^6} [{\mathbb Z},  \bar{\mathbb Z}]^2 -
{i\over 2}  \frac{1}{\mu^6} {\mathbb Z}{{\boldsymbol{ \Psi}}}{{\boldsymbol{ \Psi}}} \right) ~,
\end{equation}
and the U$(1)$ generator acting on the matrix variables is
\begin{equation}\label{hcB=}
\hat{\tilde{\mathcal{B}}}= \frac{1}{\mu^6} \hat{\mathcal{B}}= i \text{tr} \left(
\bar{\mathbb{Z}}\dfrac{\partial}{\partial \bar{\mathbb{Z}}}- \mathbb{Z} \dfrac{\partial}{\partial \mathbb{Z}}+\frac{1}{2} {\boldsymbol{ \Psi}} \dfrac{\partial}{\partial {\boldsymbol{ \Psi}}}\right)~.
\end{equation}
The state vector is represented by a function dependent on the configuration space coordinates\footnote{To be precise, we should also say that $ \Xi  $ depends on the 1d gauge field ${\mathbb A}_\tau$, but then we should subject it to the quantum constraint \eqref{eq:PA} which reads
$\dfrac{\partial}{\partial \dot{\mathbb{A}}_\tau}\Xi =0 $ and implies just independence on  ${\mathbb A}_\tau$. We allowed ourselves to ``straighten'' the presentation by omitting this stage.},
 \begin{equation}\label{Xi=}
 \Xi = \Xi\left({\rm x}^{(0)}, \bar{w}, w; \theta^w , \theta^{\bar{w}} , \bar{\theta}{}^w , \bar{\theta}{}^{\bar{w}}; {\mathbb{Z}}, \bar{\mathbb{Z}};
 {\boldsymbol{ \Psi}} \right)~.
\end{equation}
The physical states are represented by the state vectors obeying equations obtained as quantum version of the effective first class constraints which read
\begin{equation}\label{id-xx0}
\left[ -i \partial_{\text{x}^{(0)}} - \left(m + 2 \hat{\tilde{\mathcal{H}}} \right) \right]\Xi = 0~,
\end{equation}
\begin{equation}\label{D0Xi-=0}
i\hat{\tilde{\tilde{U}}}^{(0)}\Xi= \left( \mathbb{D}^{(0)} - \bar{\theta}^w \partial_{\bar{\theta}^w} + \theta^{\bar{w}} \partial_{\theta^{\bar{w}}} - 4 \theta^{w} \bar{\theta}^{\bar{w}} \hat{\tilde{\mathcal{H}}} + 2 \theta^w \hat{\bar{\tilde{\nu}}} - 2 \bar{\theta}^{\bar{w}} \hat{\tilde{\nu}} - 2i \hat{\tilde{\mathcal{B}}} - q \right)\Xi = 0~,
\end{equation}
\begin{equation}\label{d-dthw}
\left[ \partial_{\theta^w}+ 2 \left(\hat{\bar{\tilde{\nu}}} - \bar{\theta}^{\bar{w}} \hat{\tilde{\mathcal{H}}} \right) \right]\Xi = 0~,
\end{equation}
\begin{equation}\label{d-dbthbw}
\left[ \partial_{\bar{\theta}^{\bar{w}}} + 2 \left(\hat{\tilde{\nu}} - \theta^w \hat{\tilde{\mathcal{H}}} \right) \right]\Xi = 0~,
\end{equation}
\begin{equation}\label{d-dbthw}
\left[\partial_{\bar{\theta}^w} -2 \left( m + \hat{\tilde{\mathcal{H}}} \right) \theta^{\bar{w}} \right]\Xi = 0~,
\end{equation}
where $\hat{\tilde{\nu}}$, $\hat{\bar{\tilde{\nu}}}$ and $ \hat{\tilde{\mathcal{H}}} $ are given in \eqref{hnu=}, \eqref{hbnu=} and \eqref{hcH=}, and the quantum SU$(N)$ Gauss law constraint
\begin{equation}\label{hGXi=0}
i \hat{\mathbb{G}}^j_i \Xi =  \left(\bar{\mathbb{Z}}^k_i \dfrac{\partial}{\partial \bar{\mathbb{Z}}^k_j} - \bar{\mathbb{Z}}^j_k \dfrac{\partial}{\partial  \bar{\mathbb{Z}}_k^i}  +  \mathbb{Z}^k_i \dfrac{\partial}{\partial \mathbb{Z}^k_j} - \mathbb{Z}^j_k \dfrac{\partial}{\partial  \mathbb{Z}_k^i} + {\boldsymbol{ \Psi}}^k_i \frac{\partial}{\partial 	{\boldsymbol{ \Psi}}_j^k} - {\boldsymbol{ \Psi}}^j_k \dfrac{\partial}{\partial {\boldsymbol{ \Psi}}_k^i}  \right)\Xi = 0~,
\end{equation}
Notice that, when calculating the quantum generator of SU$(N)$ transformations ${\mathbb{G}}$ in \eqref{hGXi=0} we have to use the $\text{``qp'' ordering}$ for the commutators and anticommutators of matrices operators. Specially, we have
\begin{equation}
\begin{array}{l}
[{\bar{\mathbb{Z}}}, {\mathbb{P}}]{}_i{}^j \mapsto  - i \mu^6 \left(\bar{\mathbb{Z}}_i{}^k \dfrac{\partial}{\partial \bar{\mathbb{Z}}_j{}^k} - \bar{\mathbb{Z}}_k{}^j \dfrac{\partial}{\partial  \bar{\mathbb{Z}}_k^i} \right)~, \qquad  [{\mathbb{Z}}, {\bar{\mathbb{P}}}]{}_i{}^j \mapsto  - i \mu^6 \left(\mathbb{Z}_i{}^k \dfrac{\partial}{\partial \mathbb{Z}_j{}^k} - \mathbb{Z}_k{}^j\dfrac{\partial}{\partial  \mathbb{Z}_k{}^i} \right)~,\\
~\\
~~~~~~~~~~~~~~~~~~~~~~~~~~~~~~~~~~~~~~~~~~~~~~~~~~\{{{\boldsymbol{ \Psi}}},{\bar{{\boldsymbol{ \Psi}}}}  \}{}_i{}^j \mapsto 4 \mu^6 \left( {\boldsymbol{ \Psi}}_i^{\,k} \dfrac{\partial}{\partial 	{\boldsymbol{ \Psi}}_j^{\,k}} - {\boldsymbol{ \Psi}}_k^{\, j}  \dfrac{\partial}{\partial {\boldsymbol{ \Psi}}_k^{\, i}} \right)~.
\end{array}
\end{equation}
Without this ordering, the resulting expression for ${\mathbb{G}}$ would fail to be traceless as it should be. Thus, Eqs.~\eqref{id-xx0}, \eqref{D0Xi-=0}, \eqref{d-dthw}, \eqref{d-dbthbw}, \eqref{d-dbthw} and \eqref{hGXi=0} define the (super)field theory corresponding to the simplest 3D counterpart of multiple D$0$-brane system in the configuration superspace \eqref{Xi=}.

It is convenient to allow the state vector to depend also on additional complex bosonic coordinates ${\rm x}_A$ and its c.c. $\bar{\rm x}_A$,
 \begin{equation}\label{Xi==}
 \Xi = \Xi\left({\rm x}^{(0)}, {\rm x}_A, \bar{\rm x}_A, \bar{w}, w; \theta^w , \theta^{\bar{w}} , \bar{\theta}{}^w , \bar{\theta}{}^{\bar{w}}; {\mathbb{Z}}, \bar{\mathbb{Z}};
 {\boldsymbol{ \Psi}} \right)~,
\end{equation}
but we impose the additional constraint removing this dependence:
\begin{equation}\label{d-dxAXi=0}
 \dfrac{\partial}{\partial {\rm x}_A}\Xi =0  \; ,  \qquad \dfrac{\partial}{\partial \bar{\rm x}_A}\Xi =0   \; .  
\end{equation}
This formulation is equivalent to the original one, but it makes explicit that the state vector can be viewed as dependent on the central basis coordinates,
\begin{equation}\label{Xi=centr}
 \Xi = \tilde{\Xi}\left(x^a, \bar{w}, w; \theta^\alpha, \bar{\theta}{}^\alpha;  {\mathbb{Z}}, \bar{\mathbb{Z}};
 {\boldsymbol{ \Psi}} \right)~.
\end{equation}
This perspective is useful for interpreting the resulting (super)field theory equations and write this in a more conventional form, at least in some particular cases such as $N=1$ case discussed in the next subsection. In particular, in this simplest case for $N=1$, the equations can be reformulated in terms of derivatives\footnote{Here, we have introduced the notation
\begin{equation}
    \partial_\alpha = \dfrac{\partial}{\partial  \theta^\alpha}~, \qquad  \bar{\partial}_\alpha = \dfrac{\partial}{\partial  \bar{\theta}^\alpha~.}
\end{equation}} and covariant derivatives in the central basis
\bea\label{ddxa=an}
\partial_a= u_a^{(0)} \partial_{{\rm x}^{(0)} } + u_a \partial_{{\rm x}_A} + \bar{u}_a \partial_{\bar{{\rm x}}_A}\; , \qquad  \\ 
\label{Dal=an}
\text{D}_\alpha = \partial_\alpha + i(\gamma^a\bar{\theta})_\alpha \partial_a =
w_\alpha \left(\partial_{\theta^w } +i\bar{\theta}{}^{\bar{w}}  \partial_{{\rm x}^{(0)} }  \right) + \bar{w}_\alpha \left(\partial_{\theta^{\bar{w}} } +4i\bar{\theta}{}^{\bar{w}} \partial_{\bar{\rm x}_A} +i\bar{\theta}{}^{{w}}\partial_{{\rm x}^{(0)}}  \right)\; ,  \\ 
\label{bDal=an}
\bar{\text{D}}_\alpha = \bar{\partial}_\alpha + i(\gamma^a{\theta})_\alpha \partial_a=
w_\alpha \left(\bar{\partial}_{\bar{\theta}{}^w} +4i{\theta}{}^{w} \partial_{{\rm x}_A} +i{\theta}{}^{\bar{w}}\partial_{{\rm x}^{(0)}} \right) +  \bar{w}_\alpha \left(\bar{\partial}_{\bar{\theta}}{}^{\bar{w}} +i{\theta}{}^{w}  \partial_{{\rm x}^{(0)}}  \right)\; ,
\eea
and the spinor frame variables can be integrated out,  thus arriving to a more conventional description of field theory in terms of usual (central basis) spacetime coordinates.

But before considering the $N=1$ case, let us note that, in the general case of arbitrary $N>1$, it is possible to change the variables in the relative motion sector by decomposing the matrix fields
\begin{equation}\label{Z=ZITI}
\begin{array}{lcccr}
\hat{\mathbb{Z}}_i^j = \sqrt{2} \hat{{Z}}^I T_{Ii}^{~j}~,&~& \hat{\bar{\mathbb{Z}}}_i^j = \sqrt{2} \hat{\bar{{Z}}}^I T_{Ii}^{~j}~,&~&
\hat{{\boldsymbol{ \Psi}}}_i^j = \sqrt{2} \hat{{\Psi}}^I T_{Ii}^{~j}~,\\
~\\
\hat{\bar{\mathbb{P}}}_i^j = \sqrt{2} \hat{\bar{{P}}}^I T_{Ii}^{~j}~,~&~& \hat{\mathbb{P}}_i^j = \sqrt{2} \hat{{P}}^I T_{Ii}^{~j}~,&~& \hat{\bar{{\boldsymbol{ \Psi}}}}_i^j =  \sqrt{2} \hat{\bar{{\Psi}}}^I T_{Ii}^{~j}~,
\end{array}
\end{equation}
on the SU$(N)$ generators $T_{Ii}^{~j}$, with $I,J,K=1,\ldots N^2-1\;$, which satisfy
\begin{equation}
    \begin{array}{c}
        [T_I,  T_J]=if_{IJK}T_K \; , \qquad f_{IJK}=f_{[IJK]} \; , \qquad f_{IJK}f_{I'JK}=N\delta_{II'}\; ,\\
        ~\\
         {\rm tr} (T_I T_J)= \dfrac 1 2 \delta_{IJ}\; , \qquad T_I T_J = \dfrac 1 2 \left(\dfrac 1 N \delta_{IJ} {\mathbb{1}}+\left(d_{IJK}+if_{IJK}\right)T_K \right)\; , \qquad d_{IJK}=d_{(IJK)}\; ,\\
         ~\\
         T_{I}{}_i{}^j  T_{I}{}_k{}^l =\dfrac 1 2 \left(\delta_i{}^l  \delta{}_k{}^j- \frac 1 N\delta{}_i{}^j  \delta{}_k{}^l\right)\qquad \Longrightarrow \qquad (T_{I}T_{I})_i{}^j= C_F\delta{}_i{}^j \; , \qquad C_F=\dfrac {N^2-1}{N}\;
    \end{array}
\end{equation}
and some other relations which can be found, e.g., in \cite{Haber}. In the simplest case of SU$(2)$ (for $N=2$),
\be
\text{SU}(2):\qquad T_I{}_i{}^j= \frac 1 2 \sigma_I{}_i{}^j\; , \qquad  f_{IJK}=\epsilon_{IJK}\; , \qquad  d_{IJK}=0\; , \qquad I,J,K=1,2,3.
\ee
The coefficients in \eqref{Z=ZITI} are chosen in such a way that the nontrivial commutation relations of the
basic operators have (almost) standard form
\begin{equation}
\begin{array}{rcccl}
[\hat{{P}}^I, \hat{\bar{Z}}^J] = - i\mu^6 \delta^{IJ}~,&~&[\hat{\bar{P}}^I, \hat{Z}^J] = - i\mu^6 \delta^{IJ}~,&~&\{\hat{{\Psi}}{}^I, \hat{\bar{{\Psi}}}{}^J\} = \; 4\mu^6 \delta^{IJ}~.
\end{array}
\end{equation}
Thus, in coordinate representation, the operators are realized by
\begin{equation}
\begin{array}{lcrcclcl}
\hat{{Z}}^I ={Z}^I~,&~& \hat{\bar{{P}}}^I = -i \mu^6 \dfrac{\partial}{\partial {Z}^I}~, &~&~& \hat{\bar{{Z}}}^I = \bar{{Z}}^I~,&~&\hat{{P}}^I = -i \mu^6 \dfrac{\partial}{\partial \bar{{Z}}^I}~, \qquad\\
\\
~&~& \hat{{\Psi}}^I = {\Psi}^I~,&~&~&\hat{\bar{{\Psi}}}^I = \;4 \mu^6 \dfrac{\partial}{\partial {\Psi}^I}~,&~&~
\end{array}
\end{equation}
where ${Z}^I= ({Z}^1, {Z}^2,\ldots , {Z}^{N^2-1})$ are complex bosonic ($c$-number) coordinates and $ {\Psi}^I= (\Psi^1, \Psi^2,\ldots ,\Psi^{N^2-1})$ are fermionic ($a$-number) coordinates,
\be
 {\Psi}^I {\Psi}^J=- {\Psi}^J {\Psi}^I\; .
\ee
In terms of these relative motion coordinates,   the quantum generator of SU$(N)$ transformations \eqref{hGXi=0} reads
\begin{equation}\label{hGXi==0}
\hat{\mathbb{G}}^j_i  =: 2\hat{{G}}^IT^{~j}_{Ii}= 2 f_{IJK}\left(\bar{{Z}}^I \dfrac{\partial}{\partial \bar{{Z}}^J}  +  {Z}^I \dfrac{\partial}{\partial {Z}^J}  +  {\Psi}^I \frac{\partial}{\partial 	{\Psi}^J}  \right)T_{Ki}{}^{j}~,
\end{equation}
while the part of the U$(1)$ generator acting on the matrix field \eqref{hcB=} is
\begin{equation}\label{hcB==}
\hat{\tilde{\mathcal{B}}}= \frac{1}{\mu^6} \hat{\mathcal{B}}= i  \left(
\bar{{Z}}^I\dfrac{\partial}{\partial \bar{{Z}}^I}- {Z}^I \dfrac{\partial}{\partial {Z}^I}+ \frac 1 2 {\Psi}^I \dfrac{\partial}{\partial {\Psi}^I}\right)~.
\end{equation}
Similarly, the SYM supercharges \eqref{hnu=}, \eqref{hbnu=} take the form
\begin{equation}\label{tnu=}
\hat{\tilde{\nu}}=\frac{1}{\mu^6}  \frac{1}{2\sqrt{m}} \hat{\nu}=  - \frac{i}{2\sqrt{m}}\left( {\Psi}^I \dfrac{\partial}{\partial \bar{{Z}}^I} - 4\sqrt{2} f_{IJK}  {Z}^I \bar{{Z}}^J  \dfrac{\partial}{\partial {\Psi}^K}\right)~, \qquad \end{equation}
\begin{equation}\label{tbnu=}
\hat{\bar{\tilde{\nu}}}=\frac{1}{\mu^6}  \frac{1}{2\sqrt{m}} \hat{\bar{\nu}}= -\frac{i}{2\sqrt{m}}
\left(4  {\mu^6} \dfrac{\partial}{\partial {Z}^I}  \dfrac{\partial}{\partial {\Psi}^I}-  \frac{\sqrt{2}}{\mu^6} f_{IJK}  {Z}^I \bar{{Z}}^J {\Psi}^K\right)~,  \qquad
\end{equation}
and the SYM Hamiltonian \eqref{hcH=}
\begin{equation}\label{htcH=}
    \begin{array}{l}
     \begin{split}
         \hat{\tilde{{\cal H}}} &= \frac{1}{m} \left( -\mu^6 \dfrac{\partial}{\partial {Z}^I}\dfrac{\partial}{\partial \bar{{Z}}^I}
        - 4\sqrt{2} \mu^6 \bar{Z}^I f_{IJK} \dfrac{\partial}{\partial {\Psi}^J}\dfrac{\partial}{\partial {\Psi}^K}  \right. -\\
         &~\\
         & \left. -\dfrac{2}{\mu^6} f_{IJM} f_{KLM} {Z}^I \bar{{Z}}^J {Z}^K \bar{\mathbb{Z}}^L+ {\sqrt{2} \over 4\mu^6} f_{IJK}  {Z}^I{\Psi}^J {\Psi}^K \right) ~.
     \end{split}
    \end{array}
\end{equation}
These expressions can be used in the equations \eqref{id-xx0}-\eqref{d-dbthw} imposed on the state vector superfield (``wavefunction'')
 \begin{equation}\label{Xi===}
 \Xi = \Xi\left({\rm x}^{(0)}, {\rm x}_A, \bar{{\rm x}}_A, \bar{w}, w; \theta^w , \theta^{\bar{w}} , \bar{\theta}{}^w , \bar{\theta}{}^{\bar{w}}; {{Z}}{}^I, \bar{{Z}}{}^I;
{\Psi}^I \right)~,
\end{equation}
which also satisfies \eqref{d-dxAXi=0}. This is manifestly a superfield on superspace with $5+N^2$ bosonic and $3+N^2$ fermionic directions (specifically, $9$ bosonic and $7$ fermionic directions in the simplest case $N=2$).

\section{Some properties of 3D mD0 field theory equations}
\subsection[\texorpdfstring{$N$}{N}=1: Field theory of single D0-brane]{\boldmath\texorpdfstring{$N$}{N}=1: Field theory of single D0-brane}
If we set the number $N$ of D$0$-branes to unity, $N=1$, the matrix fields disappear so that the state vector becomes superfield in a center of mass superspace. This superspace is the usual superspace enlarged by spinor frame variables (Lorentz harmonic superspace in terminology of \cite{Igor_Russian})
\begin{equation}\label{Xi=cen1N}
 \Xi_0:= \Xi \vert_{N=1}= \tilde{\Xi}_0\left(x^a; \theta^\alpha, \bar{\theta}{}^\alpha ; \bar{w}, w\right)= \Xi_0\left({\rm x}^{(0)}, {\rm x}_A, \bar{{\rm x}}_A, \bar{w}, w; \theta^w , \theta^{\bar{w}} , \bar{\theta}{}^w , \bar{\theta}{}^{\bar{w}}\right)~.
\end{equation}
The above system of equations for this states vector then reduces to
\begin{equation}\label{d-dthw1N}
\partial_{\theta^w}\Xi_0 = 0~,
\qquad
\partial_{\bar{\theta}^{\bar{w}}}\Xi_0 = 0~,
\end{equation}
\begin{equation}\label{d-dbthw1N}
\left[\partial_{\bar{\theta}^w} -2  m  \theta^{\bar{w}} \right]\Xi_0 = 0~,
\end{equation}
\begin{equation}\label{d-dxA=0}
\partial_{ {\rm x}_A}\Xi_0 =0  \; ,  \qquad \partial_{ \bar{\rm x}_A}\Xi_0 =0  \; ,  \qquad
\end{equation}
\begin{equation}\label{id-xx01N}
\left( -i \partial_{\text{x}^{(0)}} - m  \right)\Xi_0 = 0~,
\end{equation}
\begin{equation}\label{D0Xi-=01N}
\left( \mathbb{D}^{(0)} - \bar{\theta}^w \partial_{\bar{\theta}^w} + \theta^{\bar{w}} \partial_{\theta^{\bar{w}}} - q \right)\Xi_0 = 0~.
\end{equation}
This system of equations describes the free (super)field theory of single 3D counterpart of D$0$ brane, i.e. the massive ${\cal N}=2$ superparticle in $\text{D}=3$. In Appendix \ref{App=D0}, we show how this system is obtained directly by quantizing the massive superparticle in its spinor moving frame formulation. Of course, the quantization of such a simple system in the standard formulation described by the de Azc\'arraga-Lukierski action \eqref{SD0=A+L} is much simpler; however, in our case, the more complicated quantization makes sense as it describes $N=1$ limit of our simplest counterpart for multiple D$0$-brane system, which is known presently in its spinor moving frame formulation only.

Eqs.~\eqref{d-dthw1N} and \eqref{d-dxA=0} imply that the state vector superfield $\Xi$ is independent of ${\rm x}_A, \bar{{\rm x}}_A; \theta^w, \bar{\theta}^{\bar{w}}$ coordinates,
\begin{equation}\label{Xi0=}
\Xi_0=  \Xi_0\left({\rm x}^{(0)},  \bar{w}, w;  \theta^{\bar{w}} , \bar{\theta}{}^w \right)\; .
\end{equation}
Eqs. \eqref{d-dbthw1N} and
\eqref{id-xx01N} are easily solved by
\be
\Xi_0 = e^{im{\rm x}^{(0)} - 2m {\theta}^{\bar{w}}\bar{\theta}^{{w}}} \, \chi ( {\theta}^{\bar{w}}, \bar{w}_\alpha,  w_\alpha) \; \equiv  e^{im{\rm x}^{(0)} } \, (1- 2 m{\theta}^{\bar{w}}\bar{\theta}^{{w}})\, \chi ( {\theta}^{\bar{w}}, \bar{w}_\alpha,  w_\alpha) \;  \qquad
\ee
and then \eqref{D0Xi-=01N} implies
\be\label{D0chi=}
\left({\mathbb D}^{(0)} + {\theta}^{\bar{w}}\partial_{{\theta}^{\bar{w}}}-q\right) \chi ( {\theta}^{\bar{w}}, \bar{w}_\alpha,  w_\alpha)
=0 \; .  \qquad
\ee
Decomposing $\chi$ in the powers of (fermionic and hence nilpotent) ${\theta}^{\bar{w}}$,
\be
\chi ( {\theta}^{\bar{w}}, \bar{w}_\alpha,  w_\alpha)= \phi (\bar{w}_\alpha,  w_\alpha) + i {\theta}^{\bar{w}}\, \xi ( \bar{w}_\alpha,  w_\alpha)
\ee
we find from \eqref{D0chi=} that the components of these superfields are bosonic and fermionic functions of complex normalized bosonic spinors (functions on SU$(1,1)\approx \text{SL}(2,{\mathbb R})$) with definite U$(1)$ charges $q$ and $(q-1)$,
\bea\label{D0phi=}
\phi =\phi^q ( \bar{w}_\alpha,  w_\alpha) \; , \qquad \left({\mathbb D}^{(0)} -q\right) \phi^q ( \bar{w}_\alpha,  w_\alpha)
=0 \; .  \qquad \\ \nonumber \\ \label{D0xi=} \xi =\xi^{(q-1)}(\bar{w}_\alpha,  w_\alpha)  \; , \qquad \left({\mathbb D}^{(0)} -(q-1)\right) \xi^{(q-1)}(\bar{w}_\alpha,  w_\alpha)
=0 \; .  \qquad
\eea
These definite U$(1)$ charges imply that the fields are functions on the coset  SU$(1,1)/\text{U}(1)$\footnote{More precisely, they are sections of a fiber bundle $\pi: \text{SU}(2) \mapsto \text{SU}(2)/\text{U}(1)$ with structure group U$(1)$, which can be identified with Hopf fibration of ${\mathbb S}^3=\text{SU}(2)$ over ${\mathbb S}^2=\text{SU}(2)/\text{U}(1)$ with fiber ${\mathbb S}^1=\text{U}(1)$. However, such mathematical subtleties are beyond the scope of this thesis.}. As far as $\theta^{\bar{w}}$ carries charge one with respect to U$(1)$ symmetry generated by the operator in \eqref{D0chi=}, the latter equation implies that the superfield $\chi$ itself has definite charge $q$ with respect to this U$(1)$ symmetry,
\be
\chi=\chi^q(\bar{w}_\alpha,  w_\alpha; {\theta}^{\bar{w}} ) \; .
\ee
Eqs. \eqref{D0phi=} and \eqref{D0xi=} imply the following homogeneity properties of the bosonic and fermionic components under U$(1)$ gauge transformations
\begin{equation}
    \begin{array}{c}
        \phi^q ( e^{-i\vartheta}\bar{w}_\alpha, e^{i\vartheta} w_\alpha)=e^{-iq\vartheta}\phi^q ( \bar{w}_\alpha,  w_\alpha)\; , \\
        ~\\
       \quad  ~~~~\xi^{q-1}( e^{-i\vartheta}\bar{w}_\alpha, e^{i\vartheta} w_\alpha)=e^{-i(q-1)\vartheta}\xi^{q-1}(\bar{w}_\alpha,  w_\alpha) \; .
    \end{array}
\end{equation}

\subsubsection{Spacetime interpretation I}
To clarify the physical meaning of these functions, let us observe that, as a result of the constraints ( Eqs.~\eqref{bww=i} and \eqref{p0=An}-\eqref{bpA=0}), which imply $p_a= mu_a^{(0)}$ and
\begin{equation}
    p_a\gamma^a_{\alpha\beta}=mu_a^{(0)}\gamma^a_{\alpha\beta}=2m\bar{w}_{(\alpha} w_{\beta )}~,
\end{equation}
 we can write
\be
\bar{w}_{\alpha} w_{\beta }=\frac 1 {2m}p_a\gamma^a_{\alpha\beta}+\frac i {2}\epsilon_{\alpha\beta}\; , \qquad p_ap^a=m^2\; .
\ee
As a result, the dependence on the product $\bar{w}_{\alpha} w_{\beta }$ can be identified with a dependence on the on-shell momentum of a massive particle. For integer $q\geq 1$ we can then write the following solution of Eqs.~\eqref{D0phi=} and \eqref{D0xi=}
\begin{equation}\label{phiq=+}
    \begin{array}{lcr}
       ~~~~~ \phi^q(\bar{w},w)=\bar{w}^{\alpha_1} \ldots \bar{w}^{\alpha_q} \phi_{\alpha_1\ldots \alpha_q}(p)\vert_{p^2=m^2}\; ,\\
        ~&~& \qquad q \geq 1\; .\\
        \xi^{q-1}(\bar{w},w)=\bar{w}^{\alpha_1} \ldots \bar{w}^{\alpha_{q-1}} \xi_{\alpha_1\ldots \alpha_{q-1}}(p)\vert_{p^2=m^2}\; .
    \end{array}
\end{equation}
According to the general spin-statistics theorem\footnote{Notice that in 3D, the set of possible statistics is not exhausted by the Bose-Einstein and Fermi-Dirac options and includes also anyons  \cite{Wilczek_1, Wilczek_2, Isengo}, such as quartions \cite{Volkov_quartions}, which can be collected in exotic  supermultiplets \cite{Sorokin_2}. We will not address these possibilities in the present thesis.}, for even $q$ the field $\phi^q(\bar{w},w)$ should be bosonic and $\xi^{q-1}(\bar{w},w)$ fermionic, and vice versa for odd $q$.

For $q\leq 0$, a similar solution is
\begin{equation}\label{phiq=-}
    \begin{array}{lcr}
       ~~~~~ \phi^q(\bar{w},w)={w}^{\alpha_1} \ldots {w}^{\alpha_{-q}} \phi_{\alpha_1\ldots \alpha_{-q}}(p)\vert_{p^2=m^2}\; ,\\
        ~&~& \qquad  q \leq 0\; .\\
        \xi^{q-1}(\bar{w},w)={w}^{\alpha_1} \ldots {w}^{\alpha_{-q-1}} \xi_{\alpha_1\ldots \alpha_{-q-1}}(p)\vert_{p^2=m^2}\; .
    \end{array}
\end{equation}
Finally, for $q=1$ we have
\begin{equation}\label{phiq=0}
\phi^{(1)}(\bar{w},w)=\bar{w}^{\alpha} \phi_{\alpha}(p)\vert_{p^2=m^2}\; , \qquad
\xi^{(0)}(\bar{w},w)= \xi(p)\vert_{p^2=m^2}\; .
\end{equation}
with fermionic $ \phi_{\alpha}(p)$ and bosonic $\xi(p)$.

However, since it is not manifestly clear how/whether these analytic solutions can be normalized, we will next describe the solution in terms of a different type of functions.

\subsubsection{Spacetime interpretation II}
To pass to a more standard description in terms of central basis coordinates, let us summarize the preceding discussion: the state vector superfield for a single massive 3D ${\cal N}=2$  superparticle takes the form
\be\label{Xiq=phiqexp}
\Xi_0=\Xi_0^q= e^{im{\rm x}^{(0)}}\left(\phi^q(\bar{w}_\alpha,  w_\alpha) +i {\theta}^{\bar{w}}\xi^{q-1}(\bar{w}_\alpha,  w_\alpha) - 2m{\theta}^{\bar{w}}\bar{\theta}^{{w}}{\phi}{}^q(\bar{w}_\alpha,  w_\alpha)\right)\; ,
\ee
where $\phi^q(\bar{w}_\alpha,  w_\alpha)$ and $\xi^{q-1}(\bar{w}_\alpha,  w_\alpha)$ satisfy \eqref{D0phi=} and \eqref{D0xi=}. Using \eqref{x0=}, \eqref{thw=}, \eqref{bthw=} (or \eqref{ddxa=an}-\eqref{bDal=an}) and  acting with a covariant derivative in central basis on this superfield, we obtain
\be\label{(bD-imth)Xiq=0}
(\bar{\text{D}}_\alpha +im\theta_\alpha)\Xi_0^q=0\; ,
\ee
which can be obtained from a (generalized) dimensional reduction of the chirality constraint in $\text{D}=4$, \footnote{The fermionic operator on the l.h.s. of \eqref{(bD-imth)Xiq=0} is the result of the  dimensional reduction of the 4D fermionic covariant derivative $\bar{\text{D}}_{\dot{\alpha}}=\bar{\partial}_{\dot{\alpha}}+i(\theta\sigma^{\underline{a}})_{\dot{\alpha}}{\partial}_{\underline{a}}$ with ${\partial}_{x^\perp}:={\partial}_{x^2}=im$.}. We also have
\be\label{(d+imu0)Xiq=0}
(\partial_a-imu_a^{(0)}) \Xi_0^q=0
\ee
which implies the Klein-Gordon equation
\be\label{(d*d+m2)Xiq=0}
(\partial_a\partial^a +m^2) \Xi_0^q=0 \; . \ee
Consider the leading component of the superfield \eqref{Xiq=phiqexp}, given by $e^{im{\rm x}^{(0)}}\phi^q(\bar{w}_\alpha,  w_\alpha)$. Roughly speaking (that is, following the structure of \eqref{phiq=+}), for the case $q=0$ this can be written as
\be \Xi_0^{(0)}\vert_{_{\theta =0}} = e^{im{\rm x}^{(0)}}\phi^{(0)}(\bar{w}_\alpha,  w_\alpha)=  e^{ip_a {x}^a} \phi(p)\vert_{p^2=m^2}  \; , \qquad \ee
where we used \eqref{p0=An}-\eqref{bpA=0}. In this form it becomes evident that the state vector depends on both momentum and coordinates. To obtain a conventional state vector (a ``wavefunction'') in coordinate representation, we should integrate over the on-shell momentum.

In the standard textbook notation this reads
\be
\phi(x)=\left. \int \frac {d^2\vec{p}} {(2\pi)^2 p^0} \left(e^{ip_a {x}^a} \phi(p)+ e^{-ip_a {x}^a} \phi^*(p)\right)\right|_{p^0=\sqrt{\vec{p}^2+m^2}}\; .
\ee
Our twistor-like representation describes only one of two terms in the integrand of the above equation, since the sign of $p^0$ is fixed by \eqref{Phi0=0}. This relation also allows us to determine the expression for integrating measure in terms of moving frame and spinor moving frame variables. Indeed, it implies (see \eqref{du0})
\be\label{dpa=}
\text{d}p_a=im f\bar{u}_a -im\bar{f}u_a \; , \qquad
\ee
and, using \eqref{euuu=-2i},
\be
\frac 1 2 \epsilon^{abc}\text{d}p_b\wedge \text{d}p_c=-2i u^{a(0)} \bar{f}\wedge f \; .
\ee
Therefore, the natural correspondence is
\be
{\text{d}^2\vec{p}} \; \longleftrightarrow \;  \bar{f}\wedge f
\ee
and the standard coordinate representation of the state vector field with $q=0$ is
\be\label{phix=0}
\phi (x)= \int \bar{f}\wedge f  \, e^{imx^au_a^{(0)}} \phi^{(0)}(\bar{w},w)\; , \qquad q=0 \; , \qquad
\ee
where (let us recall)
\be
 u_a^{(0)}=\bar{w}\gamma_aw\; , \qquad f=w^\alpha \text{d}w_\alpha =(\bar{f})^*\; , \qquad \bar{w}^\alpha w_\alpha=i\; .
\ee
This construction can be easily generalized to the case of nonzero $q$ giving
\be\label{phix=+q}
\phi_{\alpha_1\ldots \alpha_q} (x)= \int \bar{f}\wedge f  \, w_{\alpha_1}\ldots  w_{\alpha_q} e^{im x^au_a^{(0)}} \phi^{q}(\bar{w},w)\; , \qquad q> 0 \; , \qquad
\ee
and
\be\label{phix=-q}
\phi_{\alpha_1\ldots \alpha_{-q}} (x)= \int \bar{f}\wedge f  \, \bar{w}_{\alpha_1}\ldots  \bar{w}_{\alpha_{-q}} e^{im x^au_a^{(0)}} \phi^{q}(\bar{w},w)\; , \qquad q< 0 \; . \qquad
\ee
It is important to emphasize that when using expressions involving $\phi^q(\bar{w},w)$, we have to be careful about the class of functions we work with. In particular, we should not substitute Eqs.~\eqref{phiq=+}-\eqref{phiq=0} for $\phi^{q}(\bar{w},w)$ since the corresponding integrals would diverge. We will not work out the precise mathematical formulation of the problem here (see \cite{Igor_Russian} and \cite{Zima_1, Zima_2} for a more rigorous treatment of $\text{D}=4$ case), but rather state that Eqs. \eqref{phix=+q} and \eqref{phix=-q} describe the solutions of the wave equation, assuming  $\phi^{q}(\bar{w},w)$ are chosen such that the integrals converge.

Similarly, the standard superfield representation of the state vector can be written as
\bea\label{XiZ=q}
\Xi_{\alpha_1\ldots \alpha_{|q|}} (x,\theta,\bar{\theta})= \begin{cases} \bigintss \bar{f}\wedge f  \, w_{\alpha_1}\ldots  w_{\alpha_q} \; e^{im x^au_a^{(0)}-2m\theta{\bar{w}}\,\bar{\theta}w} \; \chi^{q}(\bar{w},w, \theta \bar{w})\; , \qquad q\geq 0 \; , \qquad \cr \cr   \bigintss \bar{f}\wedge f  \, \bar{w}_{\alpha_1}\ldots  \bar{w}_{\alpha_{-q}} e^{im x^au_a^{(0)}-2m\theta{\bar{w}}\,\bar{\theta}^w}\; \chi^{q}(\bar{w},w, \theta{\bar{w}})\; , \qquad q< 0 \; , \qquad \end{cases} 
\eea
where $\theta\bar{w}= \theta^\alpha \bar{w}_\alpha =(\bar{\theta}w)^*$ and $u_a^{(0)}= \bar{w}\gamma_aw$. These superfields obey 
\bea\label{(bD-imth)Xi=0}
(\bar{\text{D}}_\alpha +im\theta_\alpha)\;\Xi_{\alpha_1\ldots \alpha_{|q|}} (x,\theta,\bar{\theta})=0\; , \qquad
\eea
\begin{equation}
    (\text{D}^\alpha \text{D}_\alpha - 2im \bar{\theta}^\alpha\text{D}_\alpha -m^2\bar{\theta}^\alpha \bar{\theta}_\alpha)\;\Xi_{\alpha_1\ldots \alpha_{|q|}} (x,\theta,\bar{\theta})=0\; 
    \label{eq:DalphaDalpha_covariant}
\end{equation}
and Eq.~\eqref{(d*d+m2)Xiq=0}
\be\label{(d*d+m2)Xi=0}
(\partial_a\partial^a +m^2)\; \Xi_{\alpha_1\ldots \alpha_{|q|}} (x,\theta,\bar{\theta}) =0 \; .
\ee
Moreover, one can easily check that for $q\not= 0$ the superfield \eqref{XiZ=q} obeys the following Dirac equation
\be\label{DiracXi=mXi}
\partial^{\alpha\alpha_1} \Xi_{\alpha_1\alpha_2\ldots \alpha_{|q|}}=-m\,\frac q {|q|}\Xi^{\alpha}{}_{\alpha_2\ldots \alpha_{|q|}}  \,  
\ee
and, for $q < 0$,
\begin{equation}
    (\text{D}^\alpha -im\bar{\theta}^\alpha) \;\Xi_{\alpha \alpha_1\ldots \alpha_{-q}} = 0~.
    \label{eq:Dalpha_covariant}
\end{equation}
For $q>0$ we do not have equation of such type, indicating that the corresponding supermultiplet is reducible.

In the case $q<0$, Eq.~\eqref{DiracXi=mXi} can be derived using $\bar{\text{D}}_\alpha$ to~\eqref{eq:Dalpha_covariant}, while  Eq.~\eqref{(d*d+m2)Xi=0} can be obtained in a similar way from~\eqref{DiracXi=mXi} or by applying $\bar{\text{D}}^\alpha\bar{\text{D}}_\alpha$ to \eqref{eq:DalphaDalpha_covariant}. 

The irreducible multiplet for $q=0$ contains massive complex scalar and spinor fields. Its equations can be obtained via dimensional reduction from those of massless scalar multiplet in $\text{D}=4$. Eqs.~\eqref{(bD-imth)Xi=0}-\eqref{eq:Dalpha_covariant} describe a massive vector multiplet $(\frac{1}{2},1)$ for $q=-1$, a massive supermultiplet with $3\text{d}$ gravitino $(1,\frac{3}{2})$ for $q-2$, a massive supergravity multiplet $(\frac{3}{2},2)$ for $q=-3$, and higher spin supermultiplets for $q\leq -4$.

Thus the quantum state vector of $\text{D}=3$ ${\cal N}=2$ superparticle can be identified with superfield in the standard $\text{D}=3$~${\cal N}=2$ superspace obeying Eqs.~\eqref{(bD-imth)Xi=0}-\eqref{DiracXi=mXi}. This description is equivalent to that in terms of a function defined on the analytic basis Lorentz harmonic superspace \eqref{Xi=cen1N} obeying \eqref{d-dthw1N}-\eqref{D0Xi-=01N}.

Unfortunately, in the general case of $N>1$, the first type of description is not available (or, perhaps more precisely, is not practical). This is because the bosonic and fermionic matrix coordinates of the 3D mD$0$ configurational superspace are not inert under the U$(1)$ symmetry. Therefore, we have to work with the description of the second type, in terms of superfield depending on the coordinates of the analytical basis of center of mass superspace \eqref{Xi==}, satisfying the set of equations
\eqref{id-xx0}-\eqref{hGXi=0}.

\subsection[Dependence on Grassmann center of mass coordinates for generic case \texorpdfstring{$N>$}{N>}1.]{Dependence on Grassmann center of mass coordinates for generic case \boldmath\texorpdfstring{$N>$}{N>}1.}
The set of fermionic equations for the state vector \eqref{d-dthw}-\eqref{d-dbthw} can be solved in the general case of $N > 1$. To this end, let us first observe that the commutators and anticommutators of the SYM currents simplify significantly when acting on the state vector obeying Eq. \eqref{hGXi=0},
\begin{eqnarray}\label{hnuhbnu=-H}
&& \hat{\tilde{\nu}}^2 \,\Xi =  0\; , \qquad  \hat{\bar{\tilde{\nu}}}^2 \,\Xi =  0\; , \qquad  \{\hat{\tilde{\nu}} , \hat{\bar{\tilde{\nu}}}  \}\,\Xi   =\hat{\tilde{\mathcal{H}}}\,\Xi~, \\ 
&& {}[\hat{\tilde{\nu}}, \hat{\tilde{\mathcal{H}}}]\, \Xi = 0 \; , \qquad {}[\hat{\bar{\tilde{\nu}}}, \hat{\tilde{\mathcal{H}}}]\, \Xi = 0 \; ,
\end{eqnarray}
which, in particular, leads to 
\begin{equation}
    e^{-2\theta^w\hat{\bar{\tilde{\nu}}}}\hat{{\tilde{\nu}}}= \hat{{\tilde{\nu}}}e^{-2\theta^w\hat{\bar{\tilde{\nu}}}} - 2\theta^w\hat{\tilde{\mathcal{H}}}~.
\end{equation} 
It is then straightforward to verify that Eqs.~\eqref{d-dthw}-\eqref{d-dbthw} admit the (formal) solution
\begin{equation}
    \begin{array}{l}
         \begin{split}
             \Xi &= e^{-2\theta^w\hat{\bar{\tilde{\nu}}}}e^{-2\bar{\theta}{}^{\bar{w}}\hat{{\tilde{\nu}}}}  e^{+2\theta^w\bar{\theta}{}^{\bar{w}}\hat{\tilde{\mathcal{H}}}}e^{-2\theta^{\bar{w}}\bar{\theta}{}^{{w}}(m+\hat{\tilde{\mathcal{H}}})}\;\left({\mathfrak{B}} + \theta^{\bar{w}} {\mathfrak{F}}\right) = \\
             &~\\
             &= \left(1- 2\theta^w\hat{\bar{\tilde{\nu}}}- 2\bar{\theta}{}^{\bar{w}}\hat{{\tilde{\nu}}}   +2\theta^w\bar{\theta}{}^{\bar{w}}(\hat{\tilde{\mathcal{H}}}-2\hat{\bar{\tilde{\nu}}}\hat{{\tilde{\nu}}} )-2\theta^{\bar{w}}\bar{\theta}{}^{{w}}(m+\hat{\tilde{\mathcal{H}}})+ \right. \\
             &~\\
             & +4\theta^w\theta^{\bar{w}}\bar{\theta}{}^{{w}}\hat{\bar{\tilde{\nu}}}(m+\hat{\tilde{\mathcal{H}}})+4\theta^{\bar{w}}\bar{\theta}{}^{{w}}\bar{\theta}{}^{\bar{w}}\hat{{\tilde{\nu}}}(m+\hat{\tilde{\mathcal{H}}})-\\
             &~\\
             & \left. -4 \theta^w \bar{\theta}^{\bar{w}} \theta^{\bar{w}} \bar{\theta}^w (\hat{\tilde{\mathcal{H}}} - 2 \hat{\bar{\tilde{\nu}}}\hat{\tilde{\nu}})( m + \hat{\tilde{\mathcal{H}}}) \right)\; \left({\mathfrak{B}} + \theta^{\bar{w}} {\mathfrak{F}}\right) ~,
         \end{split}
    \end{array}
\end{equation}
where
\be
{\mathfrak{B}}={\mathfrak{B}}\left(\text{x}^{(0)}, \bar{w}, w; {\mathbb{Z}}, \bar{\mathbb{Z}};{\boldsymbol{ \Psi}} \right) \; , \qquad
{\mathfrak{F}}={\mathfrak{F}}\left(\text{x}^{(0)}, \bar{w}, w;  {\mathbb{Z}}, \bar{\mathbb{Z}};{\boldsymbol{ \Psi}} \right)
\ee
are superfields of opposite statistics (one bosonic, one fermionic), depending on the spinor moving frame variables and the matrix coordinates of our enlarged superspace. They should obey
\be
\hat{{\mathbb G}}_j{}^i  {\mathfrak{B}}= 0 \; , \qquad \hat{{\mathbb G}}_j{}^i  {\mathfrak{F}}= 0 \;  \qquad
\ee
with $\hat{\mathbb{G}}_j{}^i$ defined in \eqref{hGXi=0} (or \eqref{hGXi==0}), implying that $\mathfrak{B}$ and $\mathfrak{F}$ are constructed from SU$(N)$ invariant combinations of the matrix fields. Moreover, they obey
\be\label{U0fB=qfB}
 \left( \mathbb{D}^{(0)} - 2i \hat{\tilde{\mathcal{B}}} - q \right) {\mathfrak{B}}= 0  \; , \qquad  \left( \mathbb{D}^{(0)} - 2i \hat{\tilde{\mathcal{B}}} - (q-1) \right) {\mathfrak{F}}= 0\; , \qquad
\ee
which fix their U$(1)$ charges under transformations acting on the spinor frame and matrix fields
\be\label{fB=fBq}
 {\mathfrak{B}}=  {\mathfrak{B}}^{(q)} \left(\text{x}^{(0)}, \bar{w}, w; {\mathbb{Z}}, \bar{\mathbb{Z}};{\boldsymbol{ \Psi}} \right)  \; , \qquad  {\mathfrak{F}}=  {\mathfrak{F}}^{(q-1)}\left(\text{x}^{(0)},  \bar{w}, w; {\mathbb{Z}}, \bar{\mathbb{Z}};{\boldsymbol{ \Psi}} \right)   \; . \qquad
\ee
Additionally, as a result of \eqref{id-xx0}, these functions obey the Schrödinger equation with a shifted SYM Hamiltonian, where $\text{x}^{(0)}$ coordinate playing the role of time,
\begin{equation}\label{id-xx0fB=}
\left( i \partial_{\text{x}^{(0)}} + m + 2 \hat{\tilde{\mathcal{H}}} \right) {\mathfrak{B}}^{(q)}  = 0~, \qquad \left( i \partial_{\text{x}^{(0)}} + m + 2 \hat{\tilde{\mathcal{H}}} \right) {\mathfrak{F}}^{(q-1)}  = 0~.
\end{equation}
To derive \eqref{U0fB=qfB} from \eqref{D0Xi-=0}, we use the relations
\begin{equation}\label{D0exp-=}
    \begin{array}{l}
         \begin{split}
             i\hat{\tilde{\tilde{U}}}^{(0)} &e^{-2\theta^w\hat{\bar{\tilde{\nu}}}} = \left( \mathbb{D}^{(0)} - \bar{\theta}^w \partial_{\bar{\theta}^w} + \theta^{\bar{w}} \partial_{\theta^{\bar{w}}} -4 \theta^{w} \bar{\theta}^{\bar{w}} \hat{\tilde{\mathcal{H}}} + 2 \theta^w \hat{\bar{\tilde{\nu}}} - 2 \bar{\theta}^{\bar{w}} \hat{\tilde{\nu}} - \right.\\
             &~\\
             &\left.  - 2i \hat{\tilde{\mathcal{B}}} - q \right) e^{-2\theta^w\hat{\bar{\tilde{\nu}}}} = e^{-2\theta^w\hat{\bar{\tilde{\nu}}}}  \left( \mathbb{D}^{(0)} - \bar{\theta}^w \partial_{\bar{\theta}^w} + \theta^{\bar{w}} \partial_{\theta^{\bar{w}}}  + 2 \bar{\theta}^{\bar{w}} \hat{\tilde{\nu}} - 2i \hat{\tilde{\mathcal{B}}} - q \right)  \; ,
         \end{split}
    \end{array}
\end{equation}
\begin{equation}
    \begin{array}{l}
    \begin{split}
        &\left( \mathbb{D}^{(0)} - \bar{\theta}^w \partial_{\bar{\theta}^w} + \theta^{\bar{w}} \partial_{\theta^{\bar{w}}}  - 2 \bar{\theta}^{\bar{w}} \hat{\tilde{\nu}} - 2i \hat{\tilde{\mathcal{B}}} - q \right) e^{-2\bar{\theta}{}^{\bar{w}}\hat{{\tilde{\nu}}}}  = \\
        &~\\
        &=  e^{-2\bar{\theta}{}^{\bar{w}}\hat{{\tilde{\nu}}}}  \left( \mathbb{D}^{(0)} - \bar{\theta}^w \partial_{\bar{\theta}^w} + \theta^{\bar{w}} \partial_{\theta^{\bar{w}}}  - 2i \hat{\tilde{\mathcal{B}}} - q \right)\; ,
    \end{split}
    \end{array}
\end{equation} 
which follows from the commutation relations of the U$(1)$ generator $\mathcal{B}$ with the SYM supercurrents
\be[\hat{{\tilde{\nu}}} , \mathcal{B}]\, = \frac i 2 \mu^6 \hat{{\tilde{\nu}}},\; \qquad [\hat{\bar{\tilde{\nu}}}, \mathcal{B}]\, =  -\frac i 2 \mu^6 \hat{\bar{\tilde{\nu}}}~.\; \qquad
\ee
Naturally, Eq. \eqref{id-xx0fB=} admits the formal solution\footnote{This solution is formal because it involves exponent of operator like $e^{-it \frac{\hat{\mathcal{H}}}{\hslash}}$ which can be used to write a formal solution of an arbitrary Schrödinger equation $i \hslash \dfrac{\partial}{\partial t} \psi = \hat{\mathcal{H}}\psi$.}
\begin{equation}\label{fB=fBq=}
    \begin{array}{c}
        {\mathfrak{B}}^{(q)} \left(\text{x}^{(0)}, \bar{w}, w; {\mathbb{Z}}, \bar{\mathbb{Z}};{\boldsymbol{ \Psi}} \right) =e^{i\text{x}^{(0)}(m+2\hat{\tilde{\mathcal{H}}})} \check{{\mathfrak{B}}}{}^{(q)} \left(\bar{w}, w; {\mathbb{Z}}, \bar{\mathbb{Z}};{\boldsymbol{ \Psi}} \right)  \; , \\
        ~\\
        {\mathfrak{F}}^{(q-1)}\left(\text{x}^{(0)}, \bar{w}, w; {\mathbb{Z}}, \bar{\mathbb{Z}};{\boldsymbol{ \Psi}} \right)=e^{i\text{x}^{(0)}(m+
2\hat{\tilde{\mathcal{H}}})}  \check{{\mathfrak{F}}}{} ^{(q-1)}\left( \bar{w}, w; {\mathbb{Z}}, \bar{\mathbb{Z}};{\boldsymbol{ \Psi}} \right)  \; ,
    \end{array}
\end{equation}
 but its usefulness is restricted. Below we will discuss some more explicit particular solutions of our equations. 

\subsection{Solutions with BPS-type configurations of the SYM sector}
Let us consider the generic case $N>1$ but impose on the state vector superfield the following set of conditions: \bea\label{nuXi=0}
\hat{\tilde{\nu}}\, \Xi =0\; , \qquad \\
\label{bnuXi=0} \hat{\bar{\tilde{\nu}}} \,\Xi =0\; , \qquad   \\ \label{cHXi=0} \hat{\tilde{\mathcal{H}}} \,\Xi =0\; ,  \qquad
\eea
where $ \hat{{\tilde{\nu}}}$,  $ \hat{\bar{\tilde{\nu}}}$, $ \hat{\tilde{\mathcal{H}}}$ are defined in Eqs. \eqref{hnu=}, \eqref{hbnu=}, and \eqref{hcH=}, respectively. Under these conditions, Eqs. \eqref{id-xx0}-\eqref{d-dbthw} reduce to Eqs. \eqref{d-dthw1N}-\eqref{D0Xi-=01N}, now imposed on the state vector superfield depending on matrix coordinates, as given in \eqref{Xi==}. Furthermore, in this case, the center of mass sector of the equations of motion decouples from the equations governing the relative motion. The latter are now specified by Eqs. \eqref{nuXi=0}-\eqref{cHXi=0} and \eqref{hGXi=0}.

Therefore, we can search for a solution in the form with factorized center of mass motion, i.e. in the form of a sum of products
\be
\Xi^q=\sum\limits{q^\prime} {\Xi}_0^{q-q^\prime}(\text{x}^{(0)}, \bar{w},w;\, \theta^{\bar{w}}, \bar{\theta}{}^w )\; \underline{\Xi}{}^{q^\prime}  (Z^I,\bar{Z}^I,\Psi^I)~,
\ee
where ${\Xi}_0^{q-q^\prime}$ depends on the center of mass variables and obeys Eqs. \eqref{d-dthw1N}, \eqref{d-dbthw1N}–\eqref{id-xx01N}, as well as
\begin{equation}\label{D0Xi-=01Nq'}
\left( \mathbb{D}^{(0)} - \bar{\theta}^w \partial_{\bar{\theta}^w} + \theta^{\bar{w}} \partial_{\theta^{\bar{w}}} - (q-q^\prime) \right){\Xi}_0^{q-q^\prime} = 0~,
\end{equation}
and  $\underline{\Xi}{}^{q^\prime}  (Z^I,\bar{Z}^I,\Psi^I)$ depends on the relative motion variables and obeys Eqs. \eqref{nuXi=0}–\eqref{cHXi=0}, \eqref{hGXi=0}, together with
\begin{equation}\label{cBXi=q'}
\left(
 \hat{\tilde{\mathcal{B}}}-\frac i 2 \, q^\prime \right)\,\underline{\Xi}{}^{q^\prime}  (Z^I,\bar{Z}^I,\Psi^I)=0 \; ,
\end{equation}
with the operators defined in Eqs.~\eqref{hGXi==0}-\eqref{tbnu=}.

The set of Eqs. \eqref{tnu=}-\eqref{htcH=} and \eqref{hGXi=0} (with \eqref{hGXi==0}) coincides with the equations for the supersymmetric vacua of the 1d reduction of the 3D SYM model in the Schrödinger representation (see \cite{Nair} for a study of 3D Yang-Mills (YM) theory in this representation). Related analyses of the ground state of YM, SYM, and the BFSS Matrix model were carried out in \cite{Frohlich, Lin, Hoppe}, and references therein.

The solutions for ${\Xi}_0^{q-q^\prime}$ correspond to the solutions of equations for a single D$0$-brane, which we have already discussed in the previous (sub)section. Here, we focus on Eqs. \eqref{htcH=}, \eqref{tnu=}, \eqref{tbnu=}, and \eqref{hGXi=0} for the simplest case of $N=2$.

\subsubsection[\texorpdfstring{$N$}{N}=2 case: field theory  of simplest nontrivial 3D counterpart of mD$0$ system]{\boldmath\texorpdfstring{$N$}{N}=2 case: field theory  of simplest nontrivial 3D counterpart of mD0 system}
Let us search for a solution of mD$0$ equations that obeys the additional conditions \eqref{nuXi=0}
in the case of $N=2$, corresponding to a system of two D$0$-branes and strings ending both on the same and on different D$0$-branes. In this case the decomposition of the superfield $\underline{\Xi}{}^{q^\prime}  (Z^I,\bar{Z}^I,\Psi^I)$ in the fermionic coordinates   $ \Psi^I = (\Psi^1 ,  \Psi^2 , \Psi^3)$ reads
\be\label{Xq'=}
\underline{\Xi}{}^{q^\prime}  (Z^I,\bar{Z}^I,\Psi^I)=\mathfrak{f}{}^{q^\prime}  (Z,\bar{Z})+ \Psi^I \mathfrak{f}{}_I^{q^\prime -1} (Z,\bar{Z})+\frac 1 2  \Psi^I  \Psi^J\mathfrak{f}{}_{IJ}{}^{q^\prime -2} (Z,\bar{Z}) +\Psi^{\wedge 3}\mathfrak{f}{}{}^{q^\prime -3}  (Z,\bar{Z})
 \ee
where
\be
\Psi^{\wedge 3}:= \frac 1 {3!}\epsilon_{IJK} \Psi^I  \Psi^J \Psi^K = \Psi^1  \Psi^2 \Psi^3
\ee
is the basic monomial of maximal degree constructed from three fermionic variables.

We now substitute this decomposition into Eqs.~\eqref{cBXi=q'}, \eqref{cHXi=0} , \eqref{nuXi=0}, \eqref{bnuXi=0} and  \eqref{hGXi=0} with \eqref{hGXi==0}. First of all,  Eq.~\eqref{cBXi=q'} clearly implies that the component functions in the decomposition \eqref{cBXi=q'} are eigenfunctions of the operator
\begin{equation}\label{hcB0==}
-2i\hat{\tilde{\mathcal{B}}}_0 = \left(
2\bar{{Z}}^I\dfrac{\partial}{\partial \bar{{Z}}^I}-2 {Z}^I \dfrac{\partial}{\partial {Z}^I}\right)~,
\end{equation}
with eigenvalues (having the meaning of U$(1)$ charges) indicated by the superscripts on these component fields.

Next, note that the anticommutators of the operators in \eqref{nuXi=0} and \eqref{bnuXi=0} can be expressed in terms of the operators in \eqref{cHXi=0} and \eqref{hGXi=0}. Therefore, it is sufficient to solve former two equations.

Substituting \eqref{Xq'=} into Eq.~\eqref{nuXi=0} (and re-denoting $q'$ by $q$), we find that this superfield equation leads to the following set of four equations for the component functions
\bea\label{Psi0}
\epsilon_{IJK} Z^J\bar{Z}^K {\mathfrak{f}}_I^{(q-1)}=0\; , \qquad
\\ \nonumber \\ \label{Psi1}
\bar{\partial}_{I} {\mathfrak{f}}^{(q)}=4\sqrt{2}
\epsilon_{JKL} Z^J\bar{Z}^K  {\mathfrak{f}}^{(q-2)}_{LI}\; , \qquad
\\ \nonumber \\ \label{Psi2}
{}  \bar{\partial}_{[I} {\mathfrak{f}}_{J]}{}^{(q-1)}= 4\sqrt{2}
 Z^{[I}\bar{Z}^{J]}  {\mathfrak{f}}^{(q-3)} \; , \qquad
\\ \nonumber \\  \label{Psi3}
{}  \bar{\partial}_{[I} {\mathfrak{f}}_{JK]}{}^{(q-2)}=0  \; , \qquad
\eea
where (and below)
\be
\label{bd=ddbZ}
\bar{\partial}_I := \frac \partial {\partial \bar{Z}{}^{I}}\; ,  \qquad {\partial}_I := \frac \partial {\partial {Z}{}^{I}}\; . \qquad
\ee
By Poincar\'e lemma, Eq. \eqref{Psi3} is solved by
\be\label{fIJ=dIfJ}
 {\mathfrak{f}}^{(q-2)}_{IJ}=
{}  \bar{\partial}_{[I} {\mathfrak{f}}_{J]}{}^{(q)} \;.
\ee
So that, Eq.~\eqref{Psi1} takes the form
\begin{equation}\label{Psi1==}
    \begin{array}{l}
         \begin{split}
             \bar{\partial}_{I} {\mathfrak{f}}^{(q)}& =4\sqrt{2}\epsilon_{JKL} Z^J\bar{Z}^K   \bar{\partial}_{[L} {\mathfrak{f}}_{I]}{}^{(q)} ~ \Longleftrightarrow~\\
             &~\\
             &\Longleftrightarrow~ \bar{\partial}_{I} \left( {\mathfrak{f}}^{(q)}+2\sqrt{2}
\epsilon_{JKL} Z^J\bar{Z}^K   {\mathfrak{f}}_{L}{}^{(q)}\right) =2\sqrt{2}
\epsilon_{JKL} Z^J\bar{Z}^K   \bar{\partial}_{L} {\mathfrak{f}}_{I}{}^{(q)} \; . 
         \end{split}
    \end{array}
\end{equation}
Similarly, one can find that Eq. \eqref{bnuXi=0} for superfield \eqref{Xq'=} with \eqref{fIJ=dIfJ} implies the following three equations (since the highest component of Eq.~\eqref{bnuXi=0} vanishes identically)
\bea \label{Psi0=}
\mu^6\, \partial_{I}  {\mathfrak{f}}^{(q-1)}_{I} =0 \; , \qquad
\\ \nonumber \\ \label{Psi1=}
\mu^6\, \partial_{J}  \bar{\partial}_{[I} {\mathfrak{f}}^{(q)}_{J]}=- \frac 1 {2\sqrt{2} \mu^{6}}\epsilon_{IJK}Z^J \bar{Z}{}^{K}f^{(q)}\; ,
\\
 \nonumber \\
 \label{Psi2=}  \mu^6\,  \partial_{I}  {\mathfrak{f}}^{(q-3)} = - \frac 1 {\sqrt{2} \mu^{6}} Z^{[I}\bar{Z}{}^{J]} {\mathfrak{f}}^{(q-1)}_{J}\; .  \qquad
\eea
In Appendix \ref{AppBornOppenhemer} we present the Born-Oppenheimer-like method to search for asymptotic solution to this system of equations, in the line of \cite{Frohlich} (see also \cite{Lin}).

Thus, in this chapter we have carried out the covariant quantization of the simplest 3-dimen\-sional counterpart of the 10D mD$0$ system. Through this process, we derived the equations of the 3D mD$0$ field theory, which is defined on the superspace extended by additional matrix bosonic and fermionic coordinates. We have also studied some particular solutions of these equations and have discussed their properties.

These results represent an initial step toward the development of the (super)field theory of 10D mD$0$-brane system which, as we expect, will shed a new light on structure and properties of String Theory


     \fancyhead[LE]{\fancyplain{}{\textcolor{chapters}{\textbf{\thepage}}}}
\fancyhead[RO]{\fancyplain{}{\textcolor{chapters}{\textbf{\thepage}}}}
\cfoot{\fancyplain{\sffamily\textcolor{chapters}{{\thepage}}}{}}
\chapter*{Conclusions and Outlook}
\addcontentsline{toc}{chapter}{Conclusions and Outlook}
\thispagestyle{empty}
\vspace*{-1.5cm}

 {   \singlespacing\textcolor{cites}{    \small{
\begin{changemargin}{6cm}{0cm}
\singlespacing\textcolor{cites}{ \small{
         \begin{flushright}Ho sfilato via la mia vita miei desideri. Se tu potessi\\ 
        risalire il mio cammino, li troveresti uno dopo l'altro,\\
        incantati, immobili, fermati lì per sempre a segnare\\
        la rotta di questo viaggio strano che a nessuno\\
        mai ho raccontato se non a te.
         \end{flushright}
     \begin{flushright}     
         {\sffamily {\textit{Novecento: Un monologo}}\\
         {by {Alessandro Baricco}. }}
     \end{flushright}}}
    \end{changemargin}}}
 }   
    \vspace{12pt}

While a complete supersymmetric description of the $N$ nearly coincident D$p$-branes system is still far from being achieved, this thesis takes some steps toward such a description, both in classical and quantum domains, beginning with the system of multiple D$0$-branes and laying the groundwork to guide possible future progress in this direction.

Motivated by the result presented in~\cite{D-branes_Eric}, where it was shown that the dimensional reduction of the 11D single M$0$-brane reproduces the action for a 10D single D$0$-brane, it was natural to expect that the action for the 10D mD$0$-brane (mD$0$) should arise as the dimensional reduction of some 11D action; namely, a non-Abelian generalization of the single M$0$-brane system. This idea was pursued by the supervisor of this thesis, who constructed such a non-Abelian 11D action in flat superspace~\cite{Bandos_11D_mM0} by extending to the M$0$-brane action in its spinor moving frame formulation (which is described in chapter~\ref{ch.MF_and SMF}) with an additional sector of matrix fields and interacting terms involving center of energy variables.

However, a previous attempt to derive such 10D action for mD$0$ system from this 11D mM$0$ system was not successful~\cite{10D_mD0_Igor}; instead a candidate action for such 10D mD$0$ system was constructed in~\cite{10D_mD0_Igor} by coupling of 1d reduction of the 10D SU$(N)$ SYM to the supergravity induced by the embedding of the center of mass worldline into the 10D type IIA superspace. 

This unsuccessful attempt of reduction of mM$0$ action (along other arguments previously discussed in chapter~\ref{ch.intro}) served as the starting point for the investigation presented in this thesis. Basing on the idea that the 10D mD$0$ action should have an 11D origin, and wishing first to approach the problem on a simpler example, in chapter~\ref{ch.4D_nAmW} we have constructed the action~\eqref{eq:SmM0=4D}, describing the 4D nAmW counterpart of the 11D mM$0$ system in flat superspace. This toy model, which was explicitly constructed from first principles, is doubly supersymmetric (that is, it is invariant under both target superspace supersymmetry and local worldline supersymmetry), a characteristic property expected for $p$-branes and multiple $p$-brane systems in String/M-theory. As such, it provides a simplified but concrete framework for approaching the problem of the 11D origin of the 10D mD$0$ action.

In chapter~\ref{ch.3D_mD0}, the dimensional reduction of the 4D nAmW action~\eqref{eq:SmM0=4D} was performed and thus a 3D counterpart of mD$0$ action was obtained. But moreover, the study there showed (following~\cite{Unai_1}) the existence of a set of $\text{D}= 3$ $\mathcal{N}=2$ candidate actions~\eqref{SmD0=3D} for describing the lower dimensional counterpart of the 10D mD$0$ system. All the member of this set exhibit the properties expected for such a 3D counterpart: it includes all bosonic and fermionic fields that one anticipates from considering the very low energy gauge fixed description given by the U$(N)$ SYM action~\cite{Witten_1996} and the known actions for single D$p$-branes~\cite{p-branes_1, p-branes_2, p-branes_3, p-branes_4, D-branes_Eric, p-branes_5}. Moreover, it is doubly supersymmetric, i.e. invariant under both $\text{D}= 3$ $\mathcal{N}=2$ target superspace supersymmetry and under local worldline supersymmetry, generalizing the $\kappa$-sym\-me\-try of the single D$0$-brane (in the irreducible form characteristic for the spinor moving frame formulation~\cite{superD0_Igor}). The generic form of the action of this set includes an arbitrary positive definite function $\mathcal{M}(\mathcal{H}/\mu^6)$ of the SYM Hamiltonian $\mathcal{H}$, constructed from the bosonic and fer\-mionic matrix fields describing the relative motion of the constituents of the 3D mD$0$ system (we roughly call it ``relative motion Hamiltonian''). A key feature is that the system remains doubly supersymmetric regardless of the specific choice of this positive definite function  $\mathcal{M}(\mathcal{H}/\mu^6)$. A representative of this set of actions~\eqref{SmD0=3D}, which can be obtained directly via the dimensional reduction of the 4D nAmW action~\eqref{eq:SmM0=4D}, corresponds to a particular choice of $\mathcal{M}(\mathcal{H}/\mu^6)$ given in \eqref{cM=m+}.  

Following the line developed in chapter~\ref{ch.3D_mD0}, we have obtained in chapter~\ref{ch.10D_mD0} a (set of) nonlinear action(s) for the dynamical system of 10D multiple D$0$-branes~\eqref{eq:10D_SmD0=}. Notably, this action possesses all the properties expected for the description of mD$0$ system in flat target superspace. Specifically, it is doubly supersymmetric which means that it possesses both target space supersymmetry and worldline supersymmetry, the latter generalizing the $\kappa$-symmetry of the action of single D$0$-brane. This property ensures that the ground state of the dynamical system preserves half of the target space supersymmetry, and thus is a stable $1/2$ BPS state, the property expected for mD$0$-brane system. Moreover, as in the 3D counterpart, the generic candidate 10D mD$0$ action~\eqref{eq:10D_SmD0=} includes a positive definite function $\mathcal{M}(\mathcal{H}/\mu^6)$, and remarkably, the system remains doubly supersymmetric regardless of the choice of this $\mathcal{M}(\mathcal{H}/\mu^6)$. Curiously enough, the simplest candidate action proposed in~\cite{10D_mD0_Igor} can be recovered by choosing $\mathcal{M}(\mathcal{H}/\mu^6)= m$, with $m$ being the center of mass parameter which coincides with the mass parameter in a single D$0$-brane action use to describe the center of mass motion of mD$0$ system.

We have derived the complete set of equations of motion from this action for any $\mathcal{M}(\mathcal{H}/\mu^6)$, showing that the center of mass dynamics formally coincides with the  dynamics for single D$0$-brane, but with an effective mass 
\begin{equation*}
    \mathfrak{M} = m + \frac{2}{\mu^6} \frac{\mathcal{H}}{\mathcal{M}}
\end{equation*}
expressed in terms of $\mathcal{M}(\mathcal{H}/\mu^6)$. The gauge invariant relative motion equations are complicated, but we have shown that they imply $\text{d}\mathcal{H}=0$, which is the Noether identity manifesting the reparametrization symmetry (1d general coordinate invariance), ensuring $\mathfrak{M}$ is conserved, $\text{d}\mathfrak{M} = 0$.

Another interesting result obtained in this chapter~\ref{ch.10D_mD0} is that, in a suitable gauge (fixed on the fields from center of mass sector and on 1d SU$(N)$ gauge field), the relative motion equations simplify and reveal an interesting relation with the equations of $\mathcal{N}=16$ $\text{d}=1$ SU$(N)$ SYM theory. This relation, valid for any positive $\mathcal{M}(\mathcal{H}/\mu^6)$ is not a gauge equivalence; it links (equations of relative motion of the) models with different  $\mathcal{M}(\mathcal{H}/\mu^6)$, and thus different mass $\mathfrak{M}$, via Eq.~\eqref{Mass=}. 

Through this relation, we find that all supersymmetric bosonic solutions of the relative motion sector are in one-to-one correspondence with SUSY solutions of the SYM model. The associated BPS conditions reduce to a single equation $\mathcal{H}=0$, implying $ \mathfrak{M}=m$, which means that the effective mass $\mathfrak{M}$ of the configurations described by these supersymmetric solutions are given by center of mass parameter $m$. In other words, for any positive function  $\mathcal{M}(\mathcal{H}/\mu^6)$, the BPS spectrum of our mD$0$ model essentially coincides (in the relative motion sector) with the vacuum set of the maximally supersymmetric $\mathcal{N}=16$ $\text{d}=1$ SU$(N)$ SYM theory.

The above  correspondence also allows to study the non-supersymmetric solutions of the mD$0$ equations using known SYM solutions (see Appendix~\ref{sec:App_nonSUSY} for an example).

The origin of the unexpected multiplicity of massive $p=0$ supersymmetric objects founds in our study remains as an open question\footnote{Notice that the zoo of theses objects includes, besides our mD$0$ models with positive definite function $\mathcal{M}(\mathcal{H}/\mu^6)$, the elegant $0$-brane model of~\cite{Panda} with arbitrary function of relative motion variables $\bar{\mathcal{M}}= \bar{\mathcal{M}}(\mathbb{X}^i, {\boldsymbol{ \Psi}}_q )$.}. However, our study establishes a notable link between the gauge fixed equations of the mD$0$ model~\eqref{eq:10D_SmD0=} for any $\mathcal{M}(\mathcal{H}/\mu^6)$ and the maximally supersymmetric SYM equations, suggesting a possible reason behind this multiplicity.

In chapter~\ref{ch.11D_origin}, we revealed the $\text{D}=11$ origin of one representative of the family of candidates actions~\eqref{eq:10D_SmD0=} for the 10D mD$0$-brane system.  Specifically, we showed that the member of~\eqref{eq:10D_SmD0=} with the particular form of the positive definite function
\begin{equation*}
    \mathcal{M}(\mathcal{H}/\mu^6) = \dfrac{m}{2} + \sqrt{\dfrac{m^2}{4} + \dfrac{\mathcal{H}}{\mu^6}}
\end{equation*}
can be derived by the dimensional reduction of the 11D multiple M$0$-brane system.

Finally, in chapter~\ref{ch.quantization}, we have initiated the program of quantum description of the 10D mD$0$-brane system beginning by its quantization which should result in a superfield theory in superspace extended by bosonic and fermionic matrix coordinates. Since the quantization of the 10D system turned out to be quite complicated, we have begun by quantization of a toy model given by the simplest 3D counterpart of mD$0$ system, described by the action~\eqref{SmD0=3D} with the choice of positive definite function $\mathcal{M}(\mathcal{H}/\mu^6)=m$.

First we have developed the Hamiltonian formulation of the model. The calculation of canonical momenta for both the center of mass variables (including spinor moving frame) and the SU$(N)$ gauge field led to the identification of primary constraints. The SU$(N)$ gauge field, while carrying no physical degrees of freedom, serves as a Lagrange multiplier for the non-Abelian Gauss law, which appears as the only secondary constraint in this system. As expected from reparametrization invariance, the canonical Hamiltonian vanishes in the weak sense (i.e. when constraints are imposed). Then, we separated the constraints into first class (generating gauge symmetries) and second class ones (split on canonically conjugate pairs). However, the nontrivial Poisson brackets algebra of these first and second class constraints made powerful modern methods, like BRST quantization, difficult to apply. Surprisingly, even the Gupta-Bleuler approach failed for the bosonic sector, leaving the Dirac brackets method as the only way for quantizing this simplest 3D mD$0$ system.

In order to simplify the quantization, we moved to the so-called analytical coordinate basis
of the center of mass superspace, where the bosonic second class constraints are effectively resolved, allowing us to remove the ``unphysical'' coordinates from the phase space
without passing (explicitly) to the Dirac brackets. To deal with the fermionic second class constraints, we applied the Gupta-Bleuler (or more precisely, generalized conversion) method, reducing the system to a set of effective first class constraints that, upon quantization, can be imposed on the quantum state vector.

By imposing the quantum constraints on the state vector $\Xi$ in the (generalized) coordinate representation, we derived the field theory equations describing the simplest 3D counterpart of the quantum mD$0$-brane system. This theory is defined on the superspace extended by traceless $N\times N$ bosonic $\mathbb{Z} = (\bar{\mathbb{Z}})^\dagger$ and fermionic ${\boldsymbol{ \Psi}}$, $\bar{\boldsymbol{ \Psi}}$ matrices, together with the center of mass sector which can be identified with coordinates (coordinate functions) of Lorentz harmonic $\text{D}=3$ $\mathcal{N} = 2$ superspace. This superspace is parametrized by spinor moving frame variables and the usual coordinates $x^a,~\theta^{\alpha},~\bar{\theta}^{\dot{\alpha}}$, taken in the analytical basis (e.g., ${\rm x}^{(0)}= x^a \bar{w}\gamma_a w\, $, $\, \bar{\theta}{}^w=\bar{\theta}{}^\alpha w_\alpha $, etc.). Even for $N=1$, corresponding to the single D$0$-brane case, the resulting field theory has an unusual structure. We have discussed this case in detail, showing how the equations for the state vector superfield looks like in the central basis of Lorentz harmonic superspace, $\Xi=\Xi(x^a,\theta,\bar{\theta};\bar{w}, w)$, and how one can pass to the standard 3D field theory equations and solutions in spacetime. This central basis description is not available (not practical) for $N>1$.

For the general $N>1$ case, we have presented the formal solution of this system of
equations describing the dependence of the state vector on the fermionic center of mass coordinates. We have also investigated BPS-like solutions in the relative motion sector, where this sector decouples from the center of mass, describing configurations similar to the ground states of the BFSS matrix model at finite $N$, as studied in~\cite{Halpern, Graf, Frohlich, Hoppe_2, Hasler, Lin, Hoppe}. For $N=2$ case with SU$(2)$ gauge symmetry, we obtained field theory equations describing the ground state of the $\text{D}=3$ $\mathcal{N}=2$ SU$(2)$ SYM theory in Schrödinger-type representation\footnote{See~\cite{Nair} and refs. therein for studies of 3D YM model in Schrödinger representation.}. Following the approach of~\cite{Frohlich}, we also discussed the asymptotic behaviour of these solutions.

The results obtained in this thesis open the door to many directions for further research in the quest for a complete description of multiple D$p$-branes system. In particular, let us mention
\begin{itemize}
    \item{the construction of the Hamiltonian mechanics and quantization of the 10D mD$0$-brane system, starting from the single D$0$-brane case as well as a search for generalization of our mD$0$ action for the case of mD$p$-branes with $p \geq 1$, beginning from $p=1$.}
    \item{This (relatively) simplest case is intriguing in particular because it could help to identify which candidate in the 10D type IIA $0$-brane zoo truly describes the mD$0$ system, given that it should be related to mD$1$ by T-duality transformations. Such a study could also deepen our understanding of the nature and implications of T-duality.}
\end{itemize}
\newpage
In summary, the results and research directions outlined in this thesis aim to enhance the understanding of multiple D$p$-brane systems and help establish a foundation for uncovering deeper connections within the structure of String/M-theory.
 {\singlespacing\textcolor{cites}{\small{
\begin{changemargin}{6cm}{0cm}
\singlespacing\textcolor{cites}{ \small{
         \begin{flushright}The Road goes ever on and on down from the door\\ 
         where it began. Now far ahead the Road has gone,\\
         and I must follow, if I can, pursuing it with eager feet,\\
         until it joins some larger way where\\
         many paths and errands meet. And whither then?\\
         I cannot say.
         \end{flushright}
     \begin{flushright}     
          {\sffamily {\textit{The Lord of the Rings}}\\
         {by {J. R. R. Tolkien}. }}
     \end{flushright}}}
    \end{changemargin}}}
 }   

     \chapter*{Appendices}
\addcontentsline{toc}{chapter}{Appendices}
\thispagestyle{empty}
\vspace*{-1.5cm}

\begin{changemargin}{4cm}{0cm}
    \singlespacing\textcolor{cites}{ \small{
         \begin{flushright}But that is another story and shall be told another time.
         \end{flushright}
     \begin{flushright}     
         {\sffamily {\textit{The Neverending Story}}\\
         {by {Michael Ende}. }}
     \end{flushright}}}
    \end{changemargin}
    \vspace{12pt}

\appendix

    \chapter[Variation of the action of the 4D massless superparticle and its equations of
motion]{Variation of the action of the 4D massless superparticle and its equations of
motion}
\label{sec:eom4DN10}
\thispagestyle{empty}

In this appendix we obtain the equations of motion for the massless superparticle in $\text{D}=4$ ${\cal N}=1$ superspace by varying the action of its spinor moving frame formulation
\begin{equation*}
    S_0^{4\text{D}}= \int_{\mathcal{W}^1} \mathcal{L}_0^{4\text{D}} =  \int_{{\cal W}^1} \rho^{\#} \text{E}^{=} =  \int_{{\cal W}^1} \rho^\# \Pi^{\alpha\dot\beta}v_\alpha^-\bar{v}_{\dot{\beta}}^-\; ,
\end{equation*}
where $\Pi^{\alpha\dot\beta}$ is given in~\eqref{Pi=dtPi}.

The exterior derivative of the Lagrangian 1-form 
\begin{equation}
    \mathcal{L}_0^{4\text{D}} = \rho^\# \text{E}^= = \rho^\# \Pi^{\alpha\dot\beta}v_\alpha^-\bar{v}_{\dot{\beta}}^- ~
\end{equation}
reads
\be
{\rm d} \mathcal{L}_0^{4\text{D}} =- {\rm D} \rho^\#\wedge  \text{E}^= -4i\rho^\# \text{E}^-\wedge\bar{\text{E}}{}^-  + \rho^\#\text{E}^{+-}\wedge \Omega^{--} +\rho^\# \text{E}^{-+}\wedge \bar{\Omega}^{--} \; ,
\ee
where $\text{E}^=$, $\text{E}^{+-}$, $\text{E}^{-+}$, $\text{E}^-$ and $\bar{\text{E}}^-$ are defined in~\eqref{E--=}-\eqref{E-=} and $\Omega^{--}$, $\bar{\Omega}^{--}$ are Cartan forms~\eqref{eq:4D_Omega=}.
Using~\eqref{eq:delta},~\eqref{eq:iota} and \eqref{varL=D}, we find that
\begin{equation}
    \begin{array}{l}
         \begin{split}
             \delta \mathcal{L}_0^{4\text{D}} &= \iota_\delta {\rm D}\rho^\# \text{E}^= - {\rm D} \rho^\# \iota_\delta \text{E}^=   -4i\rho^\# \text{E}^-\iota_\delta\bar{\text{E}}{}^-  +4i\rho^\# \iota_\delta \text{E}^-\bar{\text{E}}{}^-  +\\
             &+ \rho^\# \text{E}^{+-} \iota_\delta\Omega^{--} - \rho^\#  \iota_\delta \text{E}^{+-} \Omega^{--} +\rho^\#  \text{E}^{-+} \iota_\delta\bar{\Omega}^{--} -\rho^\#  \iota_\delta \text{E}^{-+}\bar{\Omega}^{--}~,
         \end{split}
    \end{array}
\end{equation}
where we have used
\be\label{vPi}
{\rm d}\Pi^{\alpha\dot{\alpha}} = -4i {\rm d}\theta^\alpha \wedge d\bar{\theta}{}^{\dot\alpha} \qquad \Longrightarrow \qquad
\delta \Pi^{\alpha\dot{\alpha}} = +i \delta\theta^\alpha  {\rm d}\bar{\theta}{}^{\dot\alpha} -4i {\rm d}\theta^\alpha \delta \bar{\theta}{}^{\dot\alpha}+ {\rm d}(i_\delta \Pi^{\alpha\dot{\alpha}}) \; , \qquad \ee
as well as
\begin{equation}\label{vE++}
    \begin{array}{l}
         {\rm D}\text{E}^{\#}= -4i \text{E}^+\wedge\bar{\text{E}}{}^+  - \text{E}^{-+}\wedge \Omega^{++} - \text{E}^{+-}\wedge \bar{\Omega}^{++}  \qquad \Longrightarrow \\
          \qquad ~\Longrightarrow  \quad \delta \text{E}^{\#}= -4i \text{E}^+ \iota_\delta \bar{\text{E}}{}^++ 4i \iota_\delta \text{E}^+ \bar{\text{E}}{}^+ + \left(
\iota_\delta \text{E}^{-+} \Omega^{++}- \text{E}^{-+}\iota_\delta \Omega^{++}+ \rm c.c.
\right) \; ,
    \end{array}
\end{equation}
\begin{equation} \label{vE+}
    \begin{array}{l}
         ~~{\rm D}\text{E}^+= -\text{E}^-\wedge \Omega^{++} \qquad \Longrightarrow \qquad \delta \text{E}^+= \iota_\delta \text{E}^-\Omega^{++} -\text{E}^-\iota_\delta \Omega^{++}+ {\rm D}\iota_\delta \text{E}^+\; ,
    \end{array}
\end{equation}
\begin{equation} \label{vbE+}
    \begin{array}{l}
         ~~{\rm D}\bar{\text{E}}{}^+= -\bar{\text{E}}{}^-\wedge \bar{\Omega}{}^{++} \qquad \Longrightarrow \qquad \delta \bar{\text{E}}{}^+= \iota_\delta \bar{\text{E}}^-\bar{\Omega}{}^{++} -\bar{\text{E}}{}^-\iota_\delta \bar{\Omega}{}^{++}+ {\rm D}\iota_\delta \bar{\text{E}}{}^+\; ,
    \end{array}
\end{equation}
and
\begin{equation}\label{vE--}
    \begin{array}{l}
        {\rm D}\text{E}^{=}= -4i \text{E}^-\wedge\bar{\text{E}}{}^-  + \text{E}^{+-}\wedge \Omega^{--} + \text{E}^{-+}\wedge \bar{\Omega}^{--}  \qquad \Longrightarrow \\
          \qquad ~\Longrightarrow  \delta \text{E}^{=}= -4i \text{E}^- \iota_\delta \bar{\text{E}}{}^-+ 4i \iota_\delta \text{E}^- \bar{\text{E}}{}^- + \left(
-\iota_\delta \text{E}^{+-} \Omega^{--}+ \text{E}^{+-}\iota_\delta \Omega^{--}+ \rm c.c.
\right),
    \end{array}
\end{equation}
\begin{equation} \label{vE-}
    \begin{array}{l}
        ~~ {\rm D}\text{E}^-= \text{E}^+\wedge \Omega^{--} \qquad \Longrightarrow \qquad \delta \text{E}^-=- \iota_\delta \text{E}^+\Omega^{--} +\text{E}^+\iota_\delta \Omega^{--}+ {\rm D}\iota_\delta \text{E}^-\; ,
    \end{array}
\end{equation}
\begin{equation} \label{vbE-}
    \begin{array}{l}
     ~~~~~ {\rm D}\bar{\text{E}}{}^-= \bar{\text{E}}{}^+\wedge \bar{\Omega}{}^{--} \qquad \Longrightarrow \qquad \delta \bar{\text{E}}{}^-= -\iota_\delta \bar{\text{E}}^+\bar{\Omega}{}^{--}+\bar{\text{E}}{}^+\iota_\delta \bar{\Omega}{}^{--}+ {\rm D}\iota_\delta \bar{\text{E}}{}^-\; .
    \end{array}
\end{equation}
These expressions allow to easily find the following set of nontrivial equations of motion of the superparticle in the spinor moving frame formulation:
\bea\label{E--=0}
& \dfrac {\delta \mathcal{L}_0^{4\text{D}} }{ \iota_\delta {\rm D}\rho^\# } = \dfrac {\delta \mathcal{L}_0^{4\text{D}} }{ \delta \rho^\# } =0\qquad & \Longrightarrow \qquad  \text{E}^{=}=0\; , \qquad \\\nonumber
&\\
\label{E+-=0}
&  ~~~~\dfrac {\delta \mathcal{L}_0^{4\text{D}} }{ \iota_\delta \Omega^{--} }= v_\alpha^+ \dfrac {\delta \mathcal{L}_0^{4\text{D}} }{ \delta v_\alpha^{-}}=0\qquad & \Longrightarrow \qquad  \rho^\#  \text{E}^{+-}=0\; , \qquad \\ \nonumber
&\\ 
\label{E-+=0}
& ~~~~\dfrac {\delta \mathcal{L}_0^{4\text{D}} }{ \iota_\delta \bar{\Omega}{}^{--} }= \bar{v}_{\dot{\alpha}}^+ \dfrac {\delta \mathcal{L}_0^{4\text{D}} }{ \delta \bar{v}_{\dot{\alpha}}^{-}}=0\qquad & \Longrightarrow \qquad  \rho^\# \text{E}^{-+}=0\; , \qquad \\
\nonumber
&\\ 
 \label{Dr++=0}
& ~~~~~~~~~~~\dfrac {\delta \mathcal{L}_0^{4\text{D}} }{ \iota_\delta \text{E}^{=} }= \dfrac 1 2 u_{\mu}^\# \dfrac {\delta \mathcal{L}_0^{4\text{D}} }{ \delta x^{\mu}}=0\qquad & \Longrightarrow \qquad  {\rm D} \rho^\#  =0\; , \qquad \\
\nonumber
&\\ 
 \label{Om--=0}
& ~~~~~~~~~~~~\dfrac {\delta \mathcal{L}_0^{4\text{D}} }{ \iota_\delta \text{E}^{+-} }= \dfrac 1 2 u_{\mu}^{-+} \dfrac {\delta \mathcal{L}_0^{4\text{D}} }{ \delta x^{\mu}}=0\qquad & \Longrightarrow \qquad  \rho^\#  \Omega^{--}=0\; , \qquad  \\
\nonumber
&\\ \label{bOm--=0}
& ~~~~~~~~~~~~\dfrac {\delta \mathcal{L}_0^{4\text{D}} }{ \iota_\delta \text{E}^{-+} }= \dfrac 1 2 u_{\mu}^{+-} \dfrac {\delta \mathcal{L}_0^{4\text{D}} }{ \delta x^{\mu}}=0\qquad & \Longrightarrow \qquad  \rho^\#  \bar{\Omega}^{--}=0\; , \qquad 
\\\nonumber
&\\  \label{bE-=0}
& ~~~~~~~~~~~~~~\dfrac {\delta \mathcal{L}_0^{4\text{D}} }{ \iota_\delta \text{E}^{-} }= -v^{\alpha +}  \dfrac {\delta \mathcal{L}_0^{4\text{D}} }{ \delta \theta^{\alpha}}=0\qquad & \Longrightarrow \qquad  \rho^\#  \bar{\text{E}}^{-}=0\; , \qquad  \\
\nonumber
&\\ \label{E-=0}
& ~~~~~~~~~~~~~~\dfrac {\delta \mathcal{L}_0^{4\text{D}} }{ \iota_\delta\bar{\text{E}}^{-} }=-\bar{v}{}^{\dot{\alpha} +} \dfrac {\delta \mathcal{L}_0^{4\text{D}} }{ \delta \bar{\theta}{}^{\dot{\alpha}}}=0\qquad & \Longrightarrow \qquad  \rho^\#  \text{E}^{-}=0\; . \qquad
\eea
    \chapter{3D massless superparticle from dimensional reduction of 4D massless superparticle}
\label{sec:App_3Dderivatives}
\thispagestyle{empty}

In this appendix we discuss the derivation of the $\text{D}=3$ massless superparticle action from the action of $\text{D}=4$ massless superparticle in its spinor frame formulation. This serves as a useful and instructive warm up exercise, illustrating the general approach to dimensional reduction that have been applied to more complex systems in section~\ref{ch.3D_mD0}.

\section{Brink-Schwarz action}
To perform the dimensional reduction of the Brink-Schwarz action for massless superparticle, Eq.~\eqref{SBS=4D}, we can just isolate one of the momentum components (say, $p_2$, a choice is made here for convenience in later discussions), 
\be
p_\mu = (p_{\tilde{\mu}}, p_2)\; , \qquad {\tilde{\mu}}=0,1,3 \qquad \longleftrightarrow \qquad \mu= 0,1,2,3\; ,
\ee
and setting this component to zero in~\eqref{SBS=4D},
\be\label{p2=0}
p_2=0\; .
\ee
In such a way we obtain the $\text{D}=3$ massless superparticle action  \be\label{SBS=3D}
S_{\rm BS}^{3\rm D}= \int_{\mathcal{W}^1} \left( p_{\tilde{\mu}} \Pi^{\tilde{\mu}} +  \frac e 2  p_{\tilde{\mu}} p^{\tilde{\mu}}\right)\; .
\ee

\section{Spinor moving frame  formulation}
As in the spinor moving frame formulation the role of additional momentum variable are taken by bilinear of bosonic spinors, $v_\alpha^-$ and $\bar{v}_{\dot{\alpha}}^-$, it is natural to expect that similar dimensional reduction can be performed just by choosing a suitable ansatz for the $\text{D}=4$  harmonics in terms of $\text{D}=3$ ones. Actually the ansatz consists in imposing reality conditions
\be\label{v=v*=}
v_\alpha^-=\bar{v}_{\dot{\alpha}}^-={\rm v}_\alpha^- \; , \qquad v_\alpha^+=\bar{v}_{\dot{\alpha}}^+={\rm v}_\alpha^+\; ,
\ee
were we have introduced real 3D spinor moving frame variables ${\rm v}_\alpha^\pm$ which form the SL$(2,{\mathbb R})$ valued matrix, i.e. obey
\be\label{v-v+=1=3D*}
{\rm v}^{-\alpha}{\rm v}_\alpha^+=1 \; , \qquad ({\rm v}_\alpha^\pm)^*= {\rm v}_\alpha^\pm\; .
\ee
This choice is invariant under SL$(2,{\mathbb R})=\text{Spin}(1,2)$, subgroup of SL$(2,{\mathbb C})=\text{Spin}(1,3)$.

Using the split of the  $\text{D}=4$ relativistic Pauli matrices \eqref{Pauli=} as in
\eqref{s4d=s3d} and identifying the $\text{D}=3$ gamma matrices as in \eqref{g3=s4}, \eqref{tg3=ts4}, we find that the ansatz \eqref{v=v*=} and \eqref{VinSL}, \eqref{u--=v-sv-} implies
\be\label{u2--=0}
u_2^==0\; , \qquad u_2^\#=0\; , \qquad u_2^{-+} = - u_2^{+-}=(u_2^{+-})^*\; ,
\ee
while the real 3-vectors, obtained from $0,1,3$ components of the 4D moving frame vectors, are identified with the components of the 3D moving frame vectors
\be
 {\rm u}_{\tilde{\mu}}^= = {\rm v}^-\gamma_{\tilde{\mu}} {\rm v}^- =  {\rm v}^{-\alpha}\gamma_{\tilde{\mu}\, \alpha\beta} {\rm v}^{-\beta} \, , \qquad {\rm u}_{\tilde{\mu}}^\# =  {\rm v}^+\gamma_{\tilde{\mu}} {\rm v}^+ \, , \qquad
u_{\tilde{\mu}}^\perp =  {\rm v}^+\gamma_{\tilde{\mu}} {\rm v}^- \,   \qquad
\ee
as follows
\be
u_{\tilde{\mu}}^= = {\rm u}_{\tilde{\mu}}^=  \, , \qquad u_{\tilde{\mu}}^\# = {\rm u}_{\tilde{\mu}}^\#  \, , \qquad
u_{\tilde{\mu}}^{-+} = u_{\tilde{\mu}}^{+-}=u_{\tilde{\mu}}^\perp  \,  . \qquad
\ee

Now, let us turn to the 4D massless superparticle action \eqref{eq:S0D4==}. With the ansatz
\eqref{v=v*=}, which results in \eqref{u2--=0}, the component $\mu=2$ of the $\Pi^\mu$ 1-forms,
\be \label{Pi2==}
\Pi^2= \text{d}x^2-i {\rm d}\theta\sigma^2\bar{\theta}+i\theta\sigma^2 {\rm d}\bar{\theta}= {\rm d}x^2-\epsilon_{\alpha\beta} ( {\rm d}\theta^{\alpha} \bar{\theta}^{\beta} -\theta^{\alpha} {\rm d}\bar{\theta}^{\beta})\; ,
\ee
just disappears from the action. Then, the resulted expression can be identified as the 3D massless superparticle action in its spinor moving frame formulation
\be\label{S0D3=}
S_0^{3\rm D}= \int_{\mathcal{W}^1} \rho^{\#} {\rm v}_\alpha^-\bar{{\rm v}}_{\beta}^- \Pi^{\alpha\beta}=
\int_{\mathcal{W}^1} \rho^{\#} {\rm u}^=_{\tilde{\mu}}\Pi^{\tilde{\mu}}= \int_{\mathcal{W}^1} \rho^{\#} {\rm E}^{=}\; .
\ee
Here we have introduced 3D VA 1-forms
\be\label{VA=3d=}
\qquad
\Pi^{\tilde{\mu}}={\rm d}x^{\tilde{\mu}}-i {\rm d}\theta{\gamma}^{\tilde{\mu}}\bar{\theta} + i \theta{\gamma}^{\tilde{\mu}}{\rm d}\bar{\theta}\; ,
\ee
as well as one of 3D counterparts of the pull-backs of the 4D supervielbein forms \eqref{E--=}-\eqref{E+-=} adapted to the embedding,
\begin{equation}
    {\rm E}^{=}=  \Pi^{\tilde{\mu}}{\rm u}_{\tilde{\mu}}^= =  \frac 1 2  {\rm u}_{\alpha\beta}^= \Pi^{\alpha\beta} =
 \Pi^{\alpha\beta}{\rm v}_{\alpha}^-{\rm v}_{\beta}^-~,
\end{equation}
\begin{equation}
    {\rm E}^{\#}=   \Pi^{\tilde{\mu}}{\rm u}_{\tilde{\mu}}^\# =  \frac 1 2  {\rm u}_{\alpha\beta}^\# \Pi^{\alpha\beta} =
     \Pi^{\alpha\beta}{\rm v}_{\alpha}^+{\rm v}_{\beta}^+~,
\end{equation}
\begin{equation}
   ~~~~{\rm E}^{\perp}=  \Pi^{\tilde{\mu}}{\rm u}_{\tilde{\mu}}^\perp =  \frac 1 2  {\rm u}_{\alpha\beta}^\perp \Pi^{\alpha\beta} =
    \Pi^{\alpha \beta}{\rm v}_{(\alpha}^- {\rm v}_{\beta)}^+~. 
\end{equation}

\section[Some properties and applications of SL(2, \texorpdfstring{${\mathbb R}$}{R})/SO(2) Cartan forms]{Some properties and applications of SL(2,\boldmath\texorpdfstring{${\mathbb R}$}{R})/SO(2) Cartan forms}
Inverting the relations \eqref{u1du2} and \eqref{u0duI}with using the unity decomposition
\begin{equation}
    \delta_\mu^{~~\nu}= u_\mu^0u^{0\nu} - u_\mu^1u^{1\nu}  - u_\mu^2u^{2\nu} ~, 
\end{equation}
we find
\be
\left.\begin{matrix}{\rm d}{\rm u}^0_{\alpha\beta}= {\rm u}^1_{\alpha\beta}f^1 +{\rm u}^2_{\alpha\beta}f^2 \; ,  \cr
{\rm d}{\rm u}^1_{\alpha\beta}= {\rm u}^0_{\alpha\beta}f^1 +{\rm u}^2_{\alpha\beta}f^{qq}  , \cr  {\rm d}{\rm u}^2_{\alpha\beta}= {\rm u}^0_{\alpha\beta}f^2 -{\rm u}^1_{\alpha\beta}f^{qq} .  \cr
\end{matrix}\right\} \qquad  \Longleftrightarrow \qquad \begin{cases} \text{d}{\rm u}^0_{\alpha\beta}= {\rm u}^1_{\alpha\beta}f^1 +{\rm u}^2_{\alpha\beta}f^2 \; ,  \cr
{\rm D}{\rm u}^1_{\alpha\beta}={\rm d}{\rm u}^1_{\alpha\beta}-{\rm u}^2_{\alpha\beta}f^{qq}= {\rm u}^0_{\alpha\beta}f^1   , \cr  {\rm D}{\rm u}^2_{\alpha\beta}= {\rm d}{\rm u}^2_{\alpha\beta}+ {\rm u}^1_{\alpha\beta}f^{qq}= {\rm u}^0_{\alpha\beta}f^2 .  \cr
\end{cases}
\ee
and
\be
f^{11}=\frac 1 2 f^{qq}-\frac 1 2 f^{1}\; , \qquad f^{22}=\frac 1 2 f^{qq}+\frac 1 2 f^{1}\; , \qquad f^{12}=\frac 1 2 f^{2}\; .\qquad
\ee

Notice also that
\be
{\rm d}f^{qp}=-\epsilon_{q'p'}f^{qq'}\wedge f^{pp'}= f^{q}{}_{p'}\wedge f^{pp'}\; .
\ee
In particular this implies
\begin{equation}
    {\rm d}f^{qq}=-\epsilon_{q'p'}f^{qq'}\wedge f^{qp'}= 2(f^{11}-f^{22})\wedge f^{12}\; ,
\end{equation}
\begin{equation}
    {\rm d}(f^{11}-f^{22})=2f^{qq}\wedge f^{12}\; ,
\end{equation}
\begin{equation}
     {\rm d}f^{12}= f^{11}\wedge f^{22} = -\frac 1 2 f^{qq}\wedge (f^{11}-f^{22})\; .
\end{equation}
Equivalently, we can write this relations as
\be
{\rm D}f^1={\rm d}f^1 +f^{qq}\wedge f^2=0\; , \qquad {\rm D}f^2={\rm d}f^2 -f^{qq}\wedge f^1=0\; , \qquad {\rm d}f^{qq} = -f^{1}\wedge f^2\; . \qquad
\ee
In terms of covariant derivatives and SO$(2)$ Cartan forms the derivatives of 3D spinor harmonics are
\be\label{Dv1==}
{\rm D}{\rm v}_\alpha^1 ={\rm d}{\rm v}_\alpha^1+\frac 1 2 {\rm v}_\alpha^2f^{qq} = \frac 1 2 {\rm v}_\alpha^1f^{2} +
\frac 1 2 {\rm v}_\alpha^2f^{1}\; , \qquad {\rm D}{\rm v}_\alpha^2 ={\rm d}{\rm v}_\alpha^2-\frac 1 2 {\rm v}_\alpha^1f^{qq} = 
\frac 1 2 {\rm v}_\alpha^1f^{1} -\frac 1 2 {\rm v}_\alpha^2f^{2}  \; . \qquad
\ee

Using the above equations we can easily find
\be  {\rm D}{\cal E}^1 =\frac 1 2 {\cal E}^1\wedge f^{2}+ \frac 1 2 {\cal E}^2\wedge f^{1}\; , \qquad {\rm D}{\cal E}^2 =\frac 1 2 {\cal E}^1\wedge f^{1}- \frac 1 2 {\cal E}^2\wedge f^{2}\; , \qquad\ee
which imply
\begin{equation}
    \qquad {\rm D}({\cal E}^1-i{\cal E}^2)  =\frac 1 2 ({\cal E}^1+i{\cal E}^2)\wedge (f^{2}-if^{1})\; ,
\end{equation}
\begin{equation}
   ~~~~~~~~~{\rm D}(\bar{{\cal E}}^1+i\bar{{\cal E}}^2)  =\frac 1 2 (\bar{{\cal E}}^1-i\bar{{\cal E}}^2)\wedge (f^{2}+if^{1})\; .
\end{equation}
These equations and the Ricci identities for covariant derivatives of matrix fields~\eqref{DZ=3d},~\eqref{DPsi=3d}
\begin{equation}
   {\rm D}{\rm D}{\mathbb Z}= i f^1\wedge f^2 {\mathbb Z} + [{\mathbb F},{\mathbb Z}]\; , 
\end{equation}
\begin{equation}
  ~~~~  {\rm D}{\rm D}\bar{{\mathbb Z}}=- i f^1\wedge f^2\bar{{\mathbb Z}} + [{\mathbb F},\bar{{\mathbb Z}}]\; ,
\end{equation}
\begin{equation}
  ~~~~~~  {\rm D}{\rm D}{\Psi}=- \frac i 2 f^1\wedge f^2{\Psi} + [{\mathbb F},{\Psi}]\; ,
\end{equation}
\begin{equation}
   ~~ {\rm D}{\rm D}\bar{{\Psi}}=\frac i 2 f^1\wedge f^2\bar{{\Psi}} + [{\mathbb F},\bar{{\Psi}}]\; ,
\end{equation}
are useful to prove $\kappa$-symmetry of the 3D mD$0$ action. The same applies to
\be
{\rm d}{\rm E}^0 = {\rm E}^1\wedge f^1 +{\rm E}^2\wedge f^2 -2i ({\cal E}^1\wedge \bar{{\cal E}}^1+ {\cal E}^2\wedge \bar{{\cal E}}^2)\; ,
\ee
and
\be
{\rm d}\left({\rm d}\theta^\alpha \bar{\theta}_\alpha  -\theta^\alpha {\rm d}\bar{\theta}_\alpha \right)= -2 {\cal E}^1\wedge \bar{{\cal E}}^2 +2 {\cal E}^2\wedge \bar{{\cal E}}^1 \; . \qquad
\ee
The formal exterior derivative of the Lagrangian 1-form of this system reads
\begin{equation}
    \begin{array}{l}
         \begin{split}
             {\rm d}{\cal L}^{3\rm D}_{\rm D0}&= m {\rm d}\left({\rm E}^0 +{\rm d}\theta^\alpha \bar{\theta}_\alpha  -\theta^\alpha {\rm d}\bar{\theta}_\alpha \right) = \\
             &= -2im ({\cal E}^1 + i {\cal E}^2) \wedge (\bar{{\cal E}}^1-i\bar{{\cal E}}^2 )+ m{\rm E}^1\wedge f^1 +m{\rm E}^2\wedge f^2 = \\
             &= -2im ({\cal E}^2 - i {\cal E}^1) \wedge (\bar{{\cal E}}^2+i\bar{{\cal E}}^1 )+ m{\rm E}^1\wedge f^1 +m{\rm E}^2\wedge f^2 \; .
         \end{split}
    \end{array}
\end{equation}

    \chapter[Single D0-brane in spinor moving frame formulation and its \texorpdfstring{$\kappa$}{K}-symmetry]{Single D0-brane in spinor moving frame formulation and its \boldmath\texorpdfstring{$\kappa$}{K}-symmetry}
\label{sec:Appen:10D_kappa}
\thispagestyle{empty}

The action of the moving frame formulation of the 10D D$0$-brane  in flat type IIA superspace, which also appears as a part of the multiple D$0$-brane action \eqref{eq:10D_SmD0=} describing the center of mass dynamics of this system, reads \cite{superD0_Igor}
\begin{equation}\label{eq:L_D0}
S_{\text{D}0} = \int_{\mathcal{W}^1}{\cal L}_{\text{D}0}= m\int_{\mathcal{W}^1} \text{E}^0 -im \int_{\mathcal{W}^1} \left(\text{d} \theta^{1 \alpha}\theta_{\alpha}^{2} - \theta^{1 \alpha}\text{d}\theta_{\alpha}^{2} \right)~.
\end{equation}
Here $\text{d} =  \text{d}\tau\partial/\partial \tau =:\text{d}\tau \partial_\tau$, $\tau$ is proper time variable parametrizing the D$0$-brane  worldline
$\mathcal{W}^1$  defined as a line  in target $\text{D}=10$ type IIA superspace $\Sigma^{(10|32)}$ with $10$ bosonic and $16+16=32$ fermionic coordinates
\be\label{ZM} z^M=( x^\mu, \theta^{1 \alpha}, \theta_{\alpha}^2) \qquad  \ee
in terms of coordinate functions
\begin{equation}\label{ZMt}
     z^M(\tau)=( x^\mu(\tau), \theta^{1 \alpha}(\tau), \theta_{\alpha}^2 (\tau))\; ,
\end{equation}
\begin{equation}\label{cW1}
    \mathcal{W}^1 \in \Sigma^{(10|32)} : \qquad z^M=z^M(\tau)\; . 
\end{equation}
The constant $m$ entering both terms of \eqref{eq:L_D0} is the mass of D$0$-brane and $\text{E}^0$ is the contraction
\begin{equation}\label{E0=}
\text{E}^0 = \Pi^\mu u^0_\mu\qquad
\end{equation}
of the pull-back to the worldline of the 10D VA $1$-form
\begin{equation}
\Pi^\mu = \text{d}x^\mu -i\text{d}\theta^1\sigma^\mu \theta^1 -i\text{d}\theta^2\tilde{\sigma}^\mu \theta^2 \qquad
\end{equation}
with the vector field $u_\mu^0= u_\mu^0(\tau)$.
The pull-back of a differential form on target superspace is obtained by substituting the coordinate functions for coordinates; so that
Eq. \eqref{E0=} actually includes
 \begin{equation}
\Pi^\mu =\text{d}\tau \Pi_\tau^\mu = \text{d}x^\mu(\tau) -i\text{d}\theta^1(\tau)\sigma^\mu \theta^1(\tau) -i\text{d}\theta^2(\tau)\tilde{\sigma}^\mu \theta^2(\tau)\; .
\end{equation}
Notice that,  to simplify notation, below and along this thesis, we use the same symbols for the differential forms on the target superspace and their pull-backs to the worldline $\mathcal{W}^1$. The same applies to the superspace coordinates \eqref{ZM} and the coordinate functions \eqref{ZMt}. Particularly, in the second term of \eqref{eq:L_D0}  $\theta^{1 \alpha}$ and $\theta_{\alpha}^2$ denote  $\theta^{1 \alpha}(\tau)$ and $\theta_{\alpha}^2(\tau)$.

A very important property of the action  (\ref{eq:L_D0}) is that, besides manifest $\text{D}=10$ $\mathcal{N}=2$ spacetime supersymmetry, it is also invariant under the following local fermionic $\kappa$-symmetry transformations
\begin{equation}
\begin{array}{ccl}
~\delta_\kappa \theta^{1 \alpha}=   \kappa^q v_q^\alpha ~,~~~~~ \delta_\kappa \theta_{\alpha}^{2}= -  \kappa^q v_\alpha{}^q~,
\\ ~\delta_\kappa x^\mu =i\delta_\kappa\theta^1 \sigma^\mu \theta^1+i\delta_\kappa\theta^2 \tilde{\sigma}{}^\mu \theta^2~,\\
\qquad ~~~~~~~~~~\delta_\kappa v_\alpha{}^q=0 \qquad \Longrightarrow \qquad \delta_\kappa u^i_\mu = \delta_\kappa u^0_\mu = 0~~  ,
\end{array}
\label{eq:kappaD0=}
\end{equation}
where $\kappa^q =\kappa^q (\tau)$ with $q=1,\ldots,16$ are arbitrary  fermionic functions.

The Lagrangian 1-form  of the action \eqref{eq:L_D0} of single D$0$-brane in flat 10D type IIA superspace \cite{superD0_Igor}:
\begin{equation}
\mathcal{L}_{\text{D}0} = m\text{E}^0 - im (\text{d}\theta^1 \theta^2-\theta^1 \text{d}\theta^2)~.
\end{equation}

The formal exterior derivative  of  $\text{E}^0 = \Pi^\mu u^0_\mu$ in the first term of the Lagrangian form  is given by
\begin{equation}
\text{d}\text{E}^0 = \text{E}^i \wedge \Omega^i - i\left(\text{E}^{1q}\wedge \text{E}^{1q} +\text{E}^{2}_q \wedge \text{E}^{2}_q \right)~,
\label{eq_dE0}
\end{equation}
where
\begin{equation}\label{Ei=}
\text{E}^i = \Pi^\mu u^i_\mu\; , \qquad \text{E}^{1q}= \text{d}\theta^{1\alpha}\, v_\alpha{}^q\; ,   \qquad \text{E}_q^2= \text{d}\theta_\alpha^2 v_q{}^{\alpha}\; . \qquad
\end{equation}
To find~\eqref{eq_dE0} we have used
\begin{equation}
\text{d}\Pi^\mu = -i\text{d}\theta^1\sigma^\mu \wedge \text{d}\theta^1 -i\text{d}\theta^2\tilde{\sigma}^\mu \wedge \text{d}\theta^2~
\end{equation}
as well as Eqs. \eqref{u0s=vv} and  \eqref{Du0=}.

The derivative of the second, Wess-Zumino term of the D$0$-brane action is
\be
-2im \text{d}\theta^{1\alpha} \wedge \text{d}\theta_\alpha^2= -2im \text{E}^{1q}\wedge \text{E}_{q}^2\; ,
\label{eq:dthetas}
\ee
where we have used $\delta_\alpha^{~~\beta}= v_\alpha^q v_q^\beta$. Collecting~\eqref{eq_dE0} and~\eqref{eq:dthetas} we find that the formal exterior derivative of the Lagrangian form of single D$0$-brane can be written as
\begin{equation}
\text{d}{\cal L}_{\text{D}0}= m\text{E}^i\wedge \Omega^i -im (\text{E}^{1q}+\text{E}^2_{q})\wedge (\text{E}^{1q}+\text{E}^2_{q})~,
\end{equation}
where $\Omega^i$ is the covariant Cartan form defined in \eqref{Omi=}. Then, using the Lie derivative formula (\ref{varL=D}), we find
\bea\label{vcLD0=}
\delta \mathcal{L}_{\text{D}0} = m\left(\text{E}^i \iota_\delta \Omega^i - \iota_\delta \text{E}^i \Omega^i \right) - 2im\left(\text{E}^{1q} + \text{E}^2_q \right)\left(\iota_\delta \text{E}^{1q} + \iota_\delta \text{E}^2_q \right)~,
\eea
where $\iota_\delta \Omega^i$ defines essential variation of the spinor frame variable by
\begin{equation}
    \delta v_\alpha{}^q =\iota_\delta \text{D}v_\alpha{}^q
=  \frac 1 2\gamma^i_{qp} v_\alpha{}^p \iota_\delta \Omega^i~.
\end{equation}
This equation can be obtained from  the $\iota_\delta$ contraction of \eqref{Dv=vOm} by setting  $ \iota_\delta\Omega^{ij}=0$.

To conclude, let us note that the local fermionic $\kappa$-symmetry transformations \eqref{eq:kappaD0=} leaving invariant the D$0$-brane action  (\ref{eq:L_D0}) can be written in the following equivalent form
($\iota_\kappa {\rm d}:=\delta_\kappa$)
\begin{equation}
\begin{array}{c}
 \iota_\kappa \Pi^{\mu}=\delta_\kappa x^\mu -i\delta_\kappa\theta^1 \sigma^\mu \theta^1 - i\delta_\kappa\theta^2 \tilde{\sigma}{}^\mu \theta^2=0\qquad \Longrightarrow \qquad  \iota_\kappa \text{E}^0=0~, \qquad  \iota_\kappa \text{E}^i=0 ~,\\
\iota_\kappa\Omega^i=0~,
\qquad \iota_\kappa\Omega^{ij}=0~,\\
\iota_\kappa \text{E}^{1q}= - \iota_\kappa \text{E}^{2}_{q}= \kappa^q \qquad \Longrightarrow \qquad \iota_\kappa (\text{E}^{1q}+ \text{E}^{2}_{q})=0 \; .
\end{array}
\label{eq:kappaD0}
\end{equation}
Indeed, it is not difficult to check that substituting the above defined $\iota_\kappa$ for $\iota_\delta$ in \eqref{vcLD0=}, we find $\delta_\kappa {\cal L}_{\text{D}0}=0$.

    \chapter{Multiple D0-brane action and its worldline supersymmetry}
\label{sec:Append:10D_mD0}
\thispagestyle{empty}

In this appendix we present some details of the proof of the worldline supersymmetry of the candidate mD$0$ action \eqref{eq:10D_SmD0=}.

\section{Formal exterior derivative of the  Lagrangian form of the mD0 action}
The first stage is to calculate the formal exterior derivative of the Lagrangian form of the action \eqref{eq:10D_SmD0=},
\begin{equation}\label{LmD0==}
\begin{array}{c}
\begin{split}
\mathcal{L}_{\text{mD}0} &= m\text{E}^0 - im (\text{d}\theta^1 \theta^2-\theta^1 \text{d}\theta^2)~+\\
&+ \frac 1 {\mu^6}  \left[  \text{tr}\left({\mathbb P}^i \text{D} {\mathbb X}^i + 4i  {\boldsymbol{ \Psi}}_q \text{D}
 {\boldsymbol{ \Psi}}_q  \right) +  \frac 2 {\cal M} \text{E}^{0} {\cal H}-   \frac {\text{d} {\cal M}}{ {\cal M} } \text{tr} ({\mathbb P}^i{\mathbb X}^i)~+\right.\\
& \left. +\frac 1 {\sqrt{2{\cal M}}}(\text{E}{}^{q1}-\text{E}{}^{2}_{q}) \text{tr} \left(-4i (\gamma^i  {\boldsymbol{ \Psi}})_q  {\mathbb P}^i + {1\over 2}
(\gamma^{ij}  {\boldsymbol{ \Psi}})_q  [{\mathbb X}^i, {\mathbb X}^j]  \right)\right]~.
\end{split}
\end{array}
\end{equation}
Let us recall that in it ${\cal H}$ is given in Eq. \eqref{HmM0==}, the covariant derivatives D of the bosonic and fermionic Hermitian traceless $N \times N$ matrix fields are defined in \eqref{DXi=_derivatives} and \eqref{DPsi=_derivatives}
 with the use of $1$d gauge field 1-form $\mathbb{A}= \text{d}\tau \mathbb{A}_\tau $ and Cartan forms (\ref{Omi=}).
 
 To compute the exterior derivative of \eqref{LmD0==}, we have to use the Ricci identities
\bea
\text{DD}{\mathbb X}^i= \Omega^i\wedge \Omega^j \, {\mathbb X}^j + [{\mathbb F}, {\mathbb X}^i]~, \qquad
\text{DD} {\boldsymbol{ \Psi}}_q = \frac 1 4 \, \Omega^i\wedge \Omega^j\, (\gamma^{ij} {\boldsymbol{ \Psi}})_q  + [{\mathbb F},  {\boldsymbol{ \Psi}}_q]~, \label{Ricci}
\eea
where $\mathbb{F} = \text{d}\mathbb{A} - \mathbb{A} \wedge \mathbb{A}$ is the formal 2-form field strength of the 1d gauge field $\mathbb{A}$. Eqs. \eqref{Ricci} are obtained using the Maurer-Cartan equations \eqref{MC=Eq}.

After some algebra, the exterior derivative of the multiple D$0$-branes Lagrangian form \eqref{LmD0==} can be found to be
\begin{equation}\label{dLmD0==}
\begin{array}{c}
\begin{split}
\mu^6\text{d}{\cal L}_{\text{mD}0}&= \mu^6 m \text{E}^i\wedge \Omega^i -i\mu^6 m (\text{E}^{1q}+\text{E}^2_{q})\wedge (\text{E}^{1q}+\text{E}^2_{q})
+ \Omega^i\wedge \Omega^j \, {\rm tr} ({\mathbb P}^i {\mathbb X}^j+ i\boldsymbol{\Psi}\gamma^{ij}\boldsymbol{\Psi})~+\\
&+{\rm tr} \left[{\mathbb F} \left(\left[{\mathbb X}^i,{\mathbb P}^i\right]- 4i \{\boldsymbol{\Psi}_q,\boldsymbol{\Psi}_q\}\right)\right]  -  {\rm tr} (\text{D}{\mathbb P}^i \wedge \text{D}{\mathbb X}^i) -4i {\rm tr} (\text{D} {\boldsymbol{ \Psi}}_q \wedge \text{D} {\boldsymbol{ \Psi}}_q)~+\\
&+ \frac 2 {{\cal M}} \text{E}^i\wedge \Omega^i {\cal H} -i \frac 2 {{\cal M}}  ( \text{E}^{1q}\wedge \text{E}^{1q}+ \text{E}^2_{q}\wedge \text{E}^2_{q}) {\cal H}  -  \frac 1 {2\sqrt{2{\cal M}}}  (\text{E}^{1q}+\text{E}_q^2) \gamma^i_{qp}i\nu_p \wedge \Omega^i~+\\
& + \frac 2 {{\cal M}} \left(1- \frac 1 {\mu^6}  \frac {{\cal M}^\prime}{{\cal M}}{\cal H} \right) \text{E}^0\wedge \text{d}{\cal H}  + \frac 1 {\sqrt{2{\cal M}}} (\text{E}^{1q}-\text{E}_q^2)\wedge i\text{D}\nu_q + \frac 1 {\mu^6}  \frac {{\cal M}^\prime}{{\cal M}}  \text{d}{\cal K} \wedge \text{d}{\cal H}~+\\
&+ \frac 1 {\mu^6} \, \frac 1 {2\sqrt{2{\cal M}}} \frac {{\cal M}^\prime}{{\cal M}} (\text{E}^{1q}-\text{E}_q^2) i\nu_q \wedge \text{d}{\cal H}~,
\end{split}
\end{array}
\end{equation}
where ${\cal K}:= {\rm tr} ({\mathbb X}^i{\mathbb P}^i)$, $\nu_q$ and ${\cal H}$ are defined in \eqref{inu=} and \eqref{HmM0==}, respectively. The derivatives of these ``blocks'', which also enter \eqref{dLmD0==},  read
\begin{equation}
    \text{d}{\cal K}= {\rm tr} (\text{D}{\mathbb X}^i\,{\mathbb P}^i+{\mathbb X}^i\text{D}{\mathbb P}^i)\; ,
\end{equation}
\begin{equation}
     i\text{D}\nu_q = {\rm tr} \left(-4i (\gamma^i \boldsymbol{\Psi})_q  \text{D}{\mathbb P}^i -4i (\gamma^i \text{D}\boldsymbol{\Psi})_q  {\mathbb P}^i -
     \text{D}{\mathbb X}^i[(\gamma^{ij} \boldsymbol{\Psi})_q  , {\mathbb X}^j] + {1\over 2}
     (\gamma^{ij} \text{D}\boldsymbol{\Psi})_q  [{\mathbb X}^i, {\mathbb X}^j]  \right)~,
\end{equation}
\begin{equation}\label{dHmM0==}
~~~~~~   \text{d}{\cal H} = \text{tr}\left( {\mathbb P}^i \text{D} {\mathbb P}^i+ \frac 1 {16} \text{D} {\mathbb X}^i [[ {\mathbb X}^i, {\mathbb X}^j], {\mathbb X}^j] - \text{D} {\mathbb X}^i \gamma^i_{pq} \{ {\boldsymbol{ \Psi}}_p,  {\boldsymbol{ \Psi}}_q\}    - 2\, \text{D} {\boldsymbol{ \Psi}}_q  [(\gamma^i  {\boldsymbol{ \Psi}})_q , {\mathbb X}^i]\, \right)~.
\end{equation}

\section[Worldline supersymmetry (\texorpdfstring{$\kappa$}{K}-symmetry) transformations of the center of mass variables]{Worldline supersymmetry (\boldmath\texorpdfstring{$\kappa$}{K}-symmetry) transformations of the center of mass variables}
The previous experience with lower dimensional counterparts of the mD$0$ system (see chapter~\ref{ch.3D_mD0}) suggested to assume that the worldline supersymmetry acts on the center of mass variables of the mD$0$ system (i.e. on the superspace coordinate functions and spinor frame variables)
as the $\kappa$-symmetry of the single D$0$-brane action (see appendix~\ref{sec:Appen:10D_kappa}) acts on their single brane counterparts. Namely, we set
\footnote{The re-scaling of fermionic function ${\kappa^q}\mapsto \dfrac {\kappa^q} {\sqrt{2}}$ is performed to simplify the worldline supersymmetry transformation rules of the matrix fields.}
\begin{equation}\label{ikPi=0}
\begin{array}{ccl}
 \qquad \qquad  \qquad ~~~~~ \iota_\kappa \Pi^{\mu}=0 & \Longrightarrow & \qquad \iota_\kappa \text{E}^0=0\; , \qquad  \iota_\kappa \text{E}^i=0\; , \qquad \\
\iota_\kappa\Omega^i=0~, \qquad \iota_\kappa\Omega^{ij}=0& \Longrightarrow & ~~~~~~~~\delta_\kappa u^0_\mu = 0\; , \qquad \delta_\kappa u^i_\mu = 0 \; , \qquad \delta_\kappa v_\alpha{}^q=0\; ,
\end{array}
\end{equation}
and
\begin{equation}\label{ikEq1=}
\iota_\kappa \text{E}^{1q}= - \iota_\kappa \text{E}^{2}_{q}= \frac {\kappa^q} {\sqrt{2}}\qquad \Longrightarrow \qquad \delta_\kappa \theta^{1 \alpha}=  \frac {\kappa^q} {\sqrt{2}}v_q^\alpha ~, \qquad \delta_\kappa \theta_{\alpha}^{2}= - \dfrac {\kappa^q} {\sqrt{2}}v_\alpha{}^q~.
\end{equation}
These expressions are equivalent to \eqref{kappa=}, but they are more convenient to use in our method of calculation of the variation of Lagrangian form.

Then, using the Lie derivative formula with \eqref{ikPi=0}, \eqref{ikEq1=} and furthermore identifying in it
\begin{equation}
\iota_\kappa \text{D}=\delta_\kappa~, \qquad \iota_\kappa {\mathbb F}=\delta_\kappa {\mathbb A}~, \qquad \iota_\kappa {\mathbb A}=0~,
\end{equation}
we find that, modulo total derivative, the variation $\delta_\kappa$ of the Lagrangian form ${\cal L}_{\text{mD}0}$
reduces to
\begin{equation}
\begin{array}{c}
\begin{split}
\mu^6\delta_\kappa {\cal L}_{\text{mD}0}&=  \text{tr}\left[\delta_\kappa \mathbb{A} \left(\left[{\mathbb X}^i,{\mathbb P}^i\right]- 4i \{\boldsymbol{\Psi}_q,\boldsymbol{\Psi}_q\}\right)\right] + {\rm tr} \left(\delta_\kappa{\mathbb P}^i \text{D}{\mathbb X}^i - \text{D}{\mathbb P}^i \delta_\kappa {\mathbb X}^i-8i  \text{D} {\boldsymbol{ \Psi}}_q \delta_\kappa  {\boldsymbol{ \Psi}}_q\right) ~-\\
&- i \frac {2\sqrt{2}} {{\cal M}}  ( \text{E}^{1q}- \text{E}^2_{q})\kappa^{q}{\cal H} + \frac 2 {{\cal M}} \left(1- \frac {{\cal H}} {\mu^6} \, \frac {{\cal M}^\prime}{{\cal M}}\right) \text{E}^0\delta_\kappa {\cal H}+  \frac 1 {\mu^6} \,  \frac {{\cal M}^\prime}{{\cal M}}  \text{d}{\cal K} \delta_\kappa {\cal H}~+\\
&+ \frac 1 {\mu^6} \, \frac 1 {2\sqrt{2{\cal M}}} \frac {{\cal M}^\prime}{{\cal M}} \,  (\text{E}^{1q}-\text{E}_q^2) i\nu_q \delta_\kappa {\cal H} - \frac 1 {\mu^6} \,  \frac {{\cal M}^\prime}{{\cal M}} \,  \delta_\kappa{\cal K}\,  \text{d}{\cal H}~-\\
&-\frac 1 {\mu^6} \, \frac 1 {2\sqrt{{\cal M}}} \frac {{\cal M}^\prime}{{\cal M}} \, \kappa^q i\nu_q\,  \text{d}{\cal H} - \frac 1 {\sqrt{{\cal M}}} \kappa^q i\text{D}\nu_q
 + \frac 1 {\sqrt{2{\cal M}}} (E^{1q}-E_q^2)\, i\delta_\kappa\nu_q~.
\end{split}
\end{array}
\label{susy=cLmD0}
\end{equation}
The worldline supersymmetry transformation rules of the matrix fields can be found by requiring this variation to vanish.
As this calculation is a bit subtle, we present below some details.

\subsection{Worldline supersymmetry transformations of the matrix matter fields }

To find the  supersymmetry transformation leaving invariant $S_{\text{mD}0}= \int_{\mathcal{W}^1} {\cal L}_{\text{mD}0}$, i.e. obeying $\delta_\kappa {\cal L}_{\text{mD}0}=0$ (modulo total derivative),   we have to set equal to zero the coefficients for all the independent 1-forms in \eqref{susy=cLmD0}. Requiring to vanish the terms proportional to D${\mathbb P}^i$, D${\mathbb X}^i$ and  D$ {\boldsymbol{ \Psi}}_q$, we find the  set of {\it equations} for the worldline supersymmetry transformations of the matrix ``matter'' fields of the form of relations \eqref{susy=X}, \eqref{susy=P} and \eqref{susy=Psi}. Let us stress that, at this stage, these are equations because their r.h.s.-s contain $\Delta_\kappa {\cal K}$ from and $\delta_\kappa {\cal H}$ which, in their turn, are expressed in terms of $\delta_\kappa {\mathbb X}^i$, $\delta_\kappa {\mathbb P}^i$ and $\delta_\kappa  {\boldsymbol{ \Psi}}_q$.

To solve these equations it is convenient to compute formally the variations of composite quantities
$\delta_\kappa {\cal H}$ and $\Delta_\kappa {\cal K}$ with \eqref{susy=X}-\eqref{susy=Psi}.
On this way we find the following equations
\begin{equation}
    \Delta_\kappa {\cal K} = \frac 1{ 2\sqrt{{\cal M}}}{\rm tr} \left(4i (\kappa\gamma^i \boldsymbol{\Psi}) {\mathbb P}^i + {5\over 2}(\kappa\gamma^{ij}\boldsymbol{\Psi}) [{\mathbb X}^i, {\mathbb X}^j]  \right) - \frac 1 {\mu^6} \,  \frac {{\cal M}^\prime}{{\cal M}} \,\Delta_\kappa {\cal K}\, {\frak H}~,
\end{equation}
\begin{equation} \label{kappaH==}
    \delta_\kappa {\cal H} =  \frac 1{ 2\sqrt{{\cal M}}}{\rm tr} \left(\kappa^q {\boldsymbol{ \Psi}}_q\left( [{\mathbb X}^i, {\mathbb P}^i] -4i\{ {\boldsymbol{ \Psi}}_p,  {\boldsymbol{ \Psi}}_p\}\right) \right) - \frac 1 {\mu^6} \,  \frac {{\cal M}^\prime}{{\cal M}} \,\delta_\kappa {\cal H}\, {\frak H}~,
\end{equation}
where  $\frak{H}$ is given in Eq. \eqref{frakH=} \footnote{To obtain \eqref{kappaH==} one has to use the identity
$\gamma^i_{s(r}\gamma^i_{pq)}\equiv \delta_{s(r}\delta_{pq)}$ and also notice that
 $ {\rm tr} (  {\boldsymbol{ \Psi}}_r\{  {\boldsymbol{ \Psi}}_p,   {\boldsymbol{ \Psi}}_q\})= {\rm tr} (  {\boldsymbol{ \Psi}}_{(r}\{  {\boldsymbol{ \Psi}}_p,   {\boldsymbol{ \Psi}}_{q)}\})$ is completely symmetric with respect to $(rpq)$ indices while   ${\rm tr} ( [(\gamma^i  {\boldsymbol{ \Psi}})_q, {\mathbb X}^i]\,[(\gamma^j  {\boldsymbol{ \Psi}})_q, {\mathbb X}^j])=0$ vanishes.
}. These equations  are solved by \eqref{kappaK=} and \eqref{kappaH=}.

Thus, worldline supersymmetry transformations of the matrix matter fields are given by \eqref{susy=X}-\eqref{susy=Psi} with
\eqref{kappaK=} and \eqref{kappaH=}.

\subsection{Worldline supersymmetry transformations of the worldvolume gauge field}
Taking into account the above results for supersymmetry transformations of the matrix matter fields, we find that the remaining variation of the Lagrangian form \eqref{susy=cLmD0} can be written as
\begin{equation}
\begin{array}{c}
\begin{split}
\mu^6\delta_\kappa {\cal L}_{\text{mD}0} &= {\rm tr} \left[\delta_\kappa {\mathbb A}\left([{\mathbb X}^i,{\mathbb P}^i]- 4i \{ \boldsymbol{\Psi}_q,\boldsymbol{\Psi}_q \}\right) \right] + \frac 2 {{\cal M}} \left(1- \frac {{\cal H}} {\mu^6} \, \frac {{\cal M}^\prime}{{\cal M}}\right) \text{E}^0\delta_\kappa {\cal H}~+\\
&+ \frac 1 {\sqrt{2{\cal M}}} (\text{E}^{1q}-\text{E}_q^2)\, \left(  i\delta_\kappa\nu_q -\frac { 4i} {\sqrt{{\cal M}}}\kappa^{q}\, {\cal H}    + \frac 1 {\mu^6} \, \frac 1 {2} \frac {{\cal M}^\prime}{{\cal M}} \, i \nu_q \delta_\kappa {\cal H} \right)~.
\end{split}
\end{array}
\label{susy=cLmD0-1}
\end{equation}
To proceed further, we calculate $ i\delta_\kappa\nu_q $ which reads
\begin{equation}
    \begin{array}{l}
         \begin{split}
              i\delta_\kappa\nu_q &= -  \frac {1} {\sqrt{{\cal M}}}\, (\kappa\gamma^i)_q \, {\rm tr} \left[{\mathbb X}^i \, ([{\mathbb X}^j,{\mathbb P}^j]- 4i \{ {\boldsymbol{ \Psi}}_r, {\boldsymbol{ \Psi}}_r\} ) \right]+ \frac { 4i} {\sqrt{{\cal M}}}\kappa^{q}\, {\cal H} +\\
              &+ \frac 1 {\mu^6} \, \frac {{\cal M}^\prime}{{\cal M}} \,   {\rm tr}\left(4i (\gamma^i  {\boldsymbol{ \Psi}})_q  {\mathbb P}^i +
              (\gamma^{ij} {\Psi})_q  [{\mathbb X}^i, {\mathbb X}^j]  \right)\, \delta_\kappa {\cal H} -\\
              &- \frac 1 {\mu^6} \, \frac {{\cal M}^\prime}{{\cal M}} \, \, \Delta_\kappa {\cal K}\, {\rm tr} \left[ {\boldsymbol{ \Psi}}_q ([{\mathbb X}^i,{\mathbb P}^i]- 4i \{ {\boldsymbol{ \Psi}}_q, {\boldsymbol{ \Psi}}_q \})\right]
         \end{split}
    \end{array}
\end{equation}
and substitute it to \eqref{susy=cLmD0-1} thus arriving at
\begin{equation}\label{susy=cLmD0-2}
    \begin{array}{l}
         \begin{split}
             &\mu^6\delta_\kappa {\cal L}_{\text{mD}0} =  {\rm tr} \left\lbrace([{\mathbb X}^i,{\mathbb P}^i]- 4i \{ {\boldsymbol{ \Psi}}_r, {\boldsymbol{ \Psi}}_r\} ) \, \left[\delta_\kappa {\mathbb A}   + \dfrac 2 {{\cal M}\sqrt{{\cal M}}}\, \text{E}^0\,  (\kappa^q {\boldsymbol{ \Psi}}_q) \dfrac {\left(1- \frac 1 {\mu^6} \, \frac {{\cal M}^\prime}{{\cal M}}{\cal H}\right)}{\left(1+ \frac 1 {\mu^6} \, \frac {{\cal M}^\prime}{{\cal M}}\, {\frak H}\right)}  \right. \right. \\
             & -\frac 1 {\sqrt{2}{\cal M}} \, (\text{E}^{1q}-\text{E}_q^2)(\gamma^i\kappa)_q \,  {\mathbb X}^i~+ \\
             &\left.\left.+  (\text{E}^{1q}-\text{E}_q^2)\,  \frac 1 {\mu^6} \, \frac {{\cal M}^\prime}{2{\cal M}\sqrt{2{\cal M}}} \, \left(  -2 \Delta_\kappa {\cal K}\,  {\boldsymbol{ \Psi}}_q  + \frac  {\kappa^p {\boldsymbol{ \Psi}}_p} {\sqrt{{\cal M}}}\; \frac {  {\rm tr}\left(4i (\gamma^i  {\boldsymbol{ \Psi}})_q  {\mathbb P}^i +\frac 5 2
             (\gamma^{ij} \boldsymbol{\Psi})_q  [{\mathbb X}^i, {\mathbb X}^j]  \right)}{\left(1+ \frac 1 {\mu^6} \, \frac {{\cal M}^\prime}{{\cal M}}\, {\frak H}\right)}\, \right)\right]
 \right\rbrace~.
         \end{split}
    \end{array}
\end{equation}
The above expression vanishes if the SU$(N)$ gauge field transforms under worldline supersymmetry as

\begin{equation}
    \begin{array}{l}
         \begin{split}
             \delta_\kappa {\mathbb A} &=  - \frac 2 {{\cal M}\sqrt{{\cal M}}}\, \text{E}^0\,  (\kappa^q {\boldsymbol{ \Psi}}_q) \frac {\left(1- \frac 1 {\mu^6} \, \frac {{\cal M}^\prime}{{\cal M}}{\cal H}\right)}{\left(1+ \frac 1 {\mu^6} \, \frac {{\cal M}^\prime}{{\cal M}}\, {\frak H}\right)}  +\frac 1 {\sqrt{2}{\cal M}} \, (\text{E}^{1q}-E_q^2)(\gamma^i\kappa)_q \,{\mathbb X}^i~+\\
             &~\\
             &+(\text{E}^{1q}-\text{E}_q^2)\,  \frac 1 {\mu^6} \, \frac {{\cal M}^\prime}{2{\cal M}\sqrt{2{\cal M}}} \, \left(  2 \Delta_\kappa {\cal K}\,  {\boldsymbol{ \Psi}}_q  - \frac  {\kappa^p {\boldsymbol{ \Psi}}_p} {\sqrt{{\cal M}}}\; \frac {{\rm tr}\left(4i (\gamma^i \boldsymbol{\Psi})_q  {\mathbb P}^i +\frac 5 2
             (\gamma^{ij} \boldsymbol{\Psi})_q  [{\mathbb X}^i, {\mathbb X}^j]  \right) }{\left(1+ \frac 1 {\mu^6} \, \frac {{\cal M}^\prime}{{\cal M}}\, {\frak H}\right)} \right)~.
         \end{split}
    \end{array}
\end{equation}
Finally, substituting the expression \eqref{kappaK=} in the above result, we obtain Eq.~\eqref{susy=A} from the main text.

    \chapter{A non-supersymmetric solution of mD0 equations}
\label{sec:App_nonSUSY}
\thispagestyle{empty}
As an example of using the correspondence between the relative motion equations of the mD$0$ system and 1d SYM to find non-supersymmetric solutions, let us consider the SYM solution employed in \cite{Brahma} to explore cosmological scenarios within the BFSS matrix model framework \cite{BFSS}. This approach is based on the ansatz 
\be \label{Xi=aY}
{\mathbb X}^i (\tau) = a(\tau)  {\mathbb Y}^i \,
\ee
where ${\mathbb Y}^i$ are nine constant traceless $N\times N$ matrices satisfying
\be
[[{\mathbb Y}^i, {\mathbb Y}^j],{\mathbb Y}^j] =16\lambda {\mathbb Y}^i
\ee
for some constant $\lambda$. Our goal is to find a solution of the mD$0$ equations of motion with this ansatz. From Eq. \eqref{DXi==gf}, and fixing the gauge $\dot{x}^{{\bf 0}}=1$, we deduce
\be\label{Pi=bY}
 {\mathbb P}^i (\tau) = b(\tau)  {\mathbb Y}^i
\ee
where the function $b(\tau)$ should be determined from solving Eq.~\eqref{DXi==gf} with \eqref{Xi=aY} and \eqref{Pi=bY}.
Then, with the above ansatz
\be\label{H=anstaz}
{\cal H}=c\left(b^2+\frac \lambda {2} a^4\right) \; , \qquad {\frak H}=2{\cal H} +\frac 3 {32}{\rm tr} ([{\mathbb X}^i , {\mathbb X}^i ]^2 )=2 {\cal H} - 3c\lambda a^4 = 2c\left(b^2-\lambda  a^4\right) \; , \qquad
\ee
with 
\be
c=\frac 1 2 {\rm tr}({\mathbb Y}^i{\mathbb Y}^i)\; ,
\ee
so that the straightforward approach to  \eqref{DXi==gf} does not look too promising. However, at this stage we can use the equation \eqref{dH=0} stating that on the mass shell
${\cal H}$ is constant and conclude from \eqref{H=anstaz} that
\be\label{b2=}
 b^2= \frac {{\cal H}} c- \frac \lambda {2} a^4 \; , \qquad {\frak H}=2{\cal H} -\frac {3\lambda} {2} a^4 \; . \qquad
\ee 
Following the method of \cite{Brahma}, we consider a field-dependent redefinition of the time variable (similar to Eq.~\eqref{dt=})
\be\label{dt=2}
\text{d}t = \text{d}\tau  \,  \dfrac{2}{\mathcal{M}}\dfrac {\left(1 - \dfrac{1}{\mu^6}\dfrac{\mathcal{M'}}{\mathcal{M}}\mathcal{H} \right)} { \left[1 + \dfrac{1}{\mu^6} \dfrac{\mathcal{M'}}{\mathcal{M}}\left(2{\cal H} -\dfrac {3\lambda} {2} a^4(\tau)\right)  \right]} \qquad
\ee
and obtain in such a way the equations
\be
\frac{\text{d}}{\text{d}t}a=- b\; , \qquad \frac{\text{d}^2 a}{\text{d}t^2}=-\frac \lambda  a^3\; . \qquad
\ee
Following \cite{Brahma}, we can now multiply the second equation by $\frac{\text{d}a}{\text{d}t}$ and integrate it over $\text{d}t$ thus arriving at
\be
\frac 1 2 \left(\frac{\text{d}a}{\text{d}t}\right)^2= - \frac \lambda 4 a^4 +k
\ee
with some integration constant $k$. Introducing a further time redefinition,
\be\label{dtt}
\text{d}\tilde{t}= \text{d}t a(t)
\ee
we obtain a Friedmann-like equation
\be
\frac 1 {a^2} \left(\frac{\text{d}a}{\text{d} \tilde{t}}\right)^2= - \frac \lambda 2 +\frac {2k} {a^4}\; .
\ee
This is the basis for the cosmological interpretation of the matrix model proposed in \cite{Brahma}.

Although this final time redefinition \eqref{dtt} is is structurally similar to our earlier \eqref{dt=} in our model with non-constant $\mathcal{M}$, the complete equation for the change of time variable is much more complicated
\be\label{dtt=}
\text{d}\tilde{t}=  \text{d}\tau  \,  \cfrac{2a(\tau) }{\mathcal{M}}\dfrac {\left(1 - \cfrac{1}{\mu^6}\cfrac{\mathcal{M'}}{\mathcal{M}}\mathcal{H} \right)} { \left[1 + \dfrac{1}{\mu^6} \dfrac{\mathcal{M'}}{\mathcal{M}}\left(2{\cal H} -\dfrac {3\lambda} {2} a^4(\tau)\right)  \right]}\;.   \qquad
\ee
 As a result, to find the dynamics in terms of proper time $\tilde{t}$ becomes a more challenging task.
    \chapter{Convenient representations for 11D gamma matrices}
\label{sec:App_Gamma}
\thispagestyle{empty}
In this appendix, we summarize two useful representations of the 11-dimensional gamma matrices $\Gamma^{\underline{\mu}}$ and the charge conjugation matrix $C_{\underline{\alpha}\underline{\beta}}$ relevant for spinor moving frame (Lorentz harmonics) formalism for M$0$-brane and for reduction of its action to 10D which produces D$0$-brane action.

\section{SO(1,9) covariant representation}
The first representation is manifestly covariant under the 10D Lorentz group $\text{SO}(1,9) \subset \text{SO}(1,10)$. In it
\be\label{G11=s10}
~~~~~~~~\Gamma^\mu_{\underline{\alpha}\underline{\beta}} = \left(\begin{matrix} \sigma^{\mu}_{\alpha\beta} & 0\cr
 0 & \tilde{\sigma}{}^{\mu\alpha\beta}
\end{matrix}\right)\; , \qquad ~~~~~~\Gamma^*_{\underline{\alpha}\underline{\beta}} = \left(\begin{matrix} 0 & - \delta_{\alpha}{}^{\beta} \cr
- \delta_{\beta}{}^{\alpha}  & 0
\end{matrix}\right)\; , \qquad
\ee
\be\label{tG11=s10}
\tilde{\Gamma}^{\mu \underline{\alpha}\underline{\beta}} = \left(\begin{matrix}  \tilde{\sigma}{}^{\mu\alpha\beta}& 0\cr
 0 & \sigma^{\mu}_{\alpha\beta}
\end{matrix}\right)\; , \qquad \tilde{\Gamma}^{*\underline{\alpha}\underline{\beta}} = \left(\begin{matrix} 0 & \delta_{\beta}{}^{\alpha} \cr
\delta_{\alpha}{}^{\beta} & 0
\end{matrix}\right)\; , \qquad
\ee
where $\mu = 0, \ldots, 9$ is a 10D vector index, $\underline{\alpha}, \underline{\beta} = 1, \ldots, 32$ and $\alpha, \beta = 1, \ldots, 16$ label 11D Majorana and 10D Majorana-Weyl spinor indices, respectively. 

In this representation, 11D charge conjugation matrix takes the form
\be\label{C11=10}
C_{\underline{\alpha}\underline{\beta}} = i\,\left(\begin{matrix} 0 &\delta_{\alpha}{}^{\beta} \cr
 -\delta_{\beta}{}^{\alpha}  & 0
\end{matrix}\right)\; , \qquad C^{\underline{\alpha}\underline{\beta}} = i\,\left(\begin{matrix} 0 & \delta_{\beta}{}^{\alpha} \cr
-\delta_{\alpha}{}^{\beta} & 0
\end{matrix}\right)\; , \qquad
\ee
and it is easy to check that
\be
\tilde{\Gamma}^{\underline{\mu}}=C {\Gamma}^{\underline{\mu}}C\; .
\ee

\section[SO(1,1)\texorpdfstring{$\otimes$}{x}SO(9) covariant representation]{SO(1,1)\boldmath\texorpdfstring{$\otimes$}{x}SO(9) covariant representation}
The second representation is adapted to the light-cone decomposition of 11D spacetime and is covariant under the subgroup $\text{SO}(1,1) \otimes \text{SO}(9)$. It is particularly useful in spinor moving frame formulation of massless 11D superparticle:
\begin{eqnarray}
 \label{11DGC=} & (\Gamma ^{\#})_{\underline{\alpha}\underline{\beta}}
 = \left( \begin{matrix}2\delta_{pq} & 0
 \cr 0 & 0 \end{matrix}  \right) =  \tilde{\Gamma}{}^{=\underline{\alpha}\underline{\beta}} \;  , \qquad  (\Gamma ^{=})_{\underline{\alpha}\underline{\beta}}
 = \left( \begin{matrix} 0 & 0
 \cr 0 & 2\delta_{pq}  \end{matrix}  \right) =  \tilde{\Gamma}{}^{\# \underline{\alpha}\underline{\beta}}  \;  , \qquad \\
  & (\Gamma ^{i})_{\underline{\alpha}\underline{\beta}}
 = \left( \begin{matrix} 0 & \gamma^{i}_{pq}
 \cr \gamma^{i}_{pq} & 0 \end{matrix}  \right)=-  \tilde{\Gamma}{}^{i \underline{\alpha}\underline{\beta}} \;  , 
\end{eqnarray}
where $i = 1, \dots, 9$ are spatial SO(9) vector indices and $p,q = 1,\dots,16$ are SO(9) spinor indices. The charge conjugation matrix in this representation is
\begin{eqnarray}
\label{11DC=} & C_{\underline{\alpha}\underline{\beta}}  =- C_{\underline{\beta}\underline{\alpha}}
= \left( \begin{matrix} 0 & i\delta_{pq}
 \cr -i\delta_{pq} & 0
\cr\end{matrix} \right)=  (C^{-1}){}^{\underline{\alpha}\underline{\beta}}=:  C^{\underline{\alpha}\underline{\beta}}\;  . \qquad
\end{eqnarray}

    \chapter{Useful Poisson bracket relations for matrix fields, constraints and currents of 3D mD0}
\label{App=PB}
\thispagestyle{empty}
In this appendix we collect the Poisson bracket relations for matrix variables, SU$(N)$ constraints and currents of 3D mD$0$ system.

Let us begin by presenting the Poisson brackets of the Gauss law constraint
$\mathbb{G}$  \eqref{bbG:=}
with the matrix variables. It is convenient  to encode these in the brackets of the trace of the product of $\mathbb{G}$ with some ``reference'' traceless matrix $\mathbb{Y}$, i.e. of $\text{tr}(\mathbb{Y}\mathbb{G}) = \mathbb{Y}_i^j \mathbb{G}^i_j$. In such a way we arrive at simple and transparent expressions
\begin{equation}
\begin{array}{ccc}
[\text{tr}(\mathbb{Y}\mathbb{G}), \mathbb{Z}]_{\text{PB}} = [\mathbb{Z}, \mathbb{Y}]~,&~&[\text{tr}(\mathbb{Y}\mathbb{G}), \bar{\mathbb{Z}}]_{\text{PB}} = [\bar{\mathbb{Z}}, \mathbb{Y}]~,
\end{array}
\end{equation}
\begin{equation}
\begin{array}{ccc}
[\text{tr}(\mathbb{Y}\mathbb{G}), \mathbb{P}]_{\text{PB}} = [\mathbb{P}, \mathbb{Y}]~,&~&[\text{tr}(\mathbb{Y}\mathbb{G}), \bar{\mathbb{P}}]_{\text{PB}} = [\bar{\mathbb{P}}, \mathbb{Y}]~,
\end{array}
\end{equation}
\begin{equation}
\begin{array}{ccc}
[\text{tr}(\mathbb{Y}\mathbb{G}), {\boldsymbol{ \Psi}}]_{\text{PB}} = [{\boldsymbol{ \Psi}}, \mathbb{Y}]~,&~&[\text{tr}(\mathbb{Y}\mathbb{G}), \bar{{\boldsymbol{ \Psi}}}]_{\text{PB}} = [\bar{{\boldsymbol{ \Psi}}}, \mathbb{Y}]
\end{array}
\end{equation}
which clearly represent the infinitesimal SU$(N)$ transformations.

These relations allow, in particular, to obtain the $\mathfrak{su}(N)$ algebra \eqref{GG=G} generated by the Gauss constraints
\begin{equation}
[\text{tr}(\mathbb{Y}\mathbb{G}), \text{tr}(\mathbb{Y}^\prime \mathbb{G})]_{\text{PB}} = \text{tr}\left([\mathbb{Y}, \mathbb{Y}^\prime]\mathbb{G}\right)~.
\end{equation}
It is also easy to check that the Poisson brackets of the Gauss law with different currents that appear in our constraints,
\begin{equation}
\begin{array}{ccccc}
{\nu} := \text{tr}({\boldsymbol{ \Psi}} \mathbb{P} + \bar{{\boldsymbol{ \Psi}}}[\mathbb{Z}, \bar{\mathbb{Z}}])~,&~&{\bar{\nu}} := \text{tr}( \bar{{\boldsymbol{ \Psi}}} \bar{\mathbb{P}} + {\boldsymbol{ \Psi}}[\mathbb{Z}, \bar{\mathbb{Z}}])~,&~& \mathcal{B}:= \text{tr}\left(\bar{\mathbb{P}}\mathbb{Z} - \mathbb{P}\bar{\mathbb{Z}} + \dfrac i {8}{{\boldsymbol{ \Psi}}} \bar{{\boldsymbol{ \Psi}}} \right),
\label{eq:currentsMatrix}
\end{array}
\end{equation}
and
\begin{equation}\label{cH==B}
\begin{array}{ccc}
\mathcal{H} =    {\rm tr}\left( {\mathbb P} \bar{\mathbb P} +  [{\mathbb Z},  \bar{\mathbb Z}]^2 -
\dfrac{i}{2} {\mathbb Z}{ {\boldsymbol{ \Psi}}}{ {\boldsymbol{ \Psi}}} + \dfrac{i}{2} \bar{\mathbb Z} \bar{{\boldsymbol{ \Psi}}}  \bar{{\boldsymbol{ \Psi}}} \right)~,
\end{array}
\end{equation}
vanish,
\begin{eqnarray}
&& [\text{tr}(\mathbb{Y}\mathbb{G}), \nu]_{\text{PB}} = 0~, \qquad [\text{tr}(\mathbb{Y}\mathbb{G}), \bar{\nu}]_{\text{PB}} = 0~,\qquad [\text{tr}(\mathbb{Y}\mathbb{G}), \mathcal{B}]_{\text{PB}} = 0~, \qquad  [\text{tr}(\mathbb{Y}\mathbb{G}), \mathcal{H}]_{\text{PB}} = 0~.\qquad
\end{eqnarray}
This is just the reflection of SU$(N)$ invariance of these objects.

The Poisson brackets of the fermionic and bosonic currents \eqref{eq:currentsMatrix} among themselves and with the ``relative motion Hamiltonian'' \eqref{cH==B} are
\begin{equation}
\begin{array}{ccccc}
\{\nu, \nu \}_{\text{PB}} =  -8i(\mu^{6})^2 {\rm tr} ({\mathbb Z}{\mathbb G})~,&~& \{\bar{\nu}, \bar{\nu} \}_{\text{PB}} =8i(\mu^{6})^2 {\rm tr} (\bar{{\mathbb Z}}{\mathbb G})~,&~& \{\nu, \bar{\nu} \}_{\text{PB}} =-4i \mu^6 \mathcal{H}~, \\ \\ {}
[\nu, \mathcal{B}]_{\text{PB}}= \dfrac 1 2 \mu^6 \nu,&~&[\bar{\nu}, \mathcal{B}]_{\text{PB}}=  -\dfrac 1 2 \mu^6 \bar{\nu}~,&~&
\end{array}
\end{equation}
and
\begin{equation}
\begin{array}{ccccc}
[\nu, \mathcal{H}]_{\text{PB}} = (\mu^{6})^2 {\rm tr}(\bar{ {\boldsymbol{ \Psi}}}{\mathbb G})~,&~&[\bar{\nu}, \mathcal{H}]_{\text{PB}} = -(\mu^{6})^2 {\rm tr}( {\boldsymbol{ \Psi}}{\mathbb G})~,&~&[\mathcal{B}, \mathcal{H}]_{\text{PB}} = 0~. \\ {}
\end{array}
\end{equation}
In terms of renormalized currents and matrix fields \eqref{tcH:=}, \eqref{tB=} and \eqref{tG=G/m}, the above relations simplify to
\begin{equation}
\begin{array}{ccccc}
\{\tilde{\nu}, \tilde{\nu} \}_{\text{PB}} =  -{2i}{\rm tr} ({\mathbb Z}\tilde{{\mathbb G}})~,&~& \{\bar{\tilde{\nu}}, \bar{\tilde{\nu}} \}_{\text{PB}} =2i{\rm tr} (\bar{{\mathbb Z}}\tilde{{\mathbb G}})~,&~& \{\tilde{\nu}, \bar{\tilde{\nu}} \}_{\text{PB}} =-i \tilde{\mathcal{H}}~, \\ \\ {}
[\tilde{\nu}, \tilde{\mathcal{B}}]_{\text{PB}}= \dfrac 1 2\tilde{ \nu},&~&[\bar{\tilde{\nu}}, \tilde{\mathcal{B}}]_{\text{PB}}=  -\dfrac 1 2 \bar{\tilde{\nu}}~,&~&[\tilde{\mathcal{B}}, \tilde{\mathcal{B}}]_{\text{PB}} \equiv 0~,
\\ \\
{}[\tilde{\nu}, \tilde{\mathcal{H}}]_{\text{PB}} = \dfrac 1 2  {\rm tr}(\tilde{\bar{ {\boldsymbol{ \Psi}}}}\tilde{{\mathbb G}})~,&~&[\bar{\tilde{\nu}}, \tilde{\mathcal{H}}]_{\text{PB}} = -\dfrac 1 2 {\rm tr}(\tilde{{ {\boldsymbol{ \Psi}}}}\tilde{{\mathbb G}})~,&~&[\tilde{\mathcal{B}}, \tilde{\mathcal{H}}]_{\text{PB}} = 0~. \\ {}
\end{array}
\end{equation}
The above relations were obtained by using the brackets of currents with matrix fields which are
\begin{equation}
\begin{array}{lllll}
[\nu, \mathbb{Z}]_{\text{PB}}=0~,&~&[\nu, \mathbb{P}]_{\text{PB}}=0~,&~&\{ \nu, {\boldsymbol{ \Psi}} \}_{\text{PB}}= -{4i \mu^6 [\mathbb{Z}, \bar{\mathbb{Z}}]}~,
\\ \\ {}
[\nu, \bar{\mathbb{Z}}]_{\text{PB}}=- {\mu^6 {\boldsymbol{ \Psi}}}~,&~&[\nu, \bar{\mathbb{P}}]_{\text{PB}}=0~,&~& \{\nu, \bar{{\boldsymbol{ \Psi}}} \}_{\text{PB}}=-4i \mu^6 \mathbb{P},
\\ \\ {}
[\bar{\nu}, \mathbb{Z}]_{\text{PB}}=-{\mu^6 \bar{{\boldsymbol{ \Psi}}}}~,&~&[\bar{\nu}, \mathbb{P}]_{\text{PB}}=0~,&~&\{\bar{\nu}, {\boldsymbol{ \Psi}} \}_{\text{PB}}=-4i \mu^6 \bar{\mathbb{P}} ~,
\\ \\ {}
[\bar{\nu}, \bar{\mathbb{Z}}]_{\text{PB}}=0~,&~&[\bar{\nu}, \bar{\mathbb{P}}]_{\text{PB}}=0~,&~&\{\bar{\nu}, \bar{ {\boldsymbol{ \Psi}}}\}_{\text{PB}}= {-4i \mu^6 [\mathbb{Z}, \bar{\mathbb{Z}}]}~,
\\ \\ {}
[\mathcal{B}, \mathbb{Z}]_{\text{PB}}=-\mu^6 \mathbb{Z}~,&~&[\mathcal{B}, \mathbb{P}]_{\text{PB}}=-\mu^6 \mathbb{P}~,&~&[ \mathcal{B},  {\boldsymbol{ \Psi}}]_{\text{PB}}=\;  \dfrac 1 2 \mu^6 {\boldsymbol{ \Psi}} ~,
\\ \\ {}
[\mathcal{B}, \bar{\mathbb{Z}}]_{\text{PB}}=\mu^6 \bar{\mathbb{Z}}~,&~&[\mathcal{B}, \bar{\mathbb{P}}]_{\text{PB}}=\mu^6 \bar{\mathbb{P}}~,&~&[ \mathcal{B}, \bar{{\boldsymbol{ \Psi}}} ]_{\text{PB}}= -\dfrac 1 2 \mu^6 \bar{{\boldsymbol{ \Psi}}} ~.
\end{array}
\end{equation}
    \chapter{Problems with Gupta-Bleuler quantization scheme for massive particle in moving frame formulation}
\label{GB-failed}
\thispagestyle{empty}

Let us consider the system of first class constraints \eqref{Phi0=0}, \eqref{U0=0} and second class constraints  \eqref{eqs:frak_d}, \eqref{eqs:Phi} and set to zero all fermionic and matrix fields. Under these conditions, the system of constraints describes, within in the frame of Hamiltonian approach, just a massive bosonic particle in its spinor moving frame formulation. 

In this appendix we show that, somewhat unexpectedly, the quantization of such a simple system by the Gupta-Bleuler method in its canonical form fails\footnote{Of course, the quantization of the same system in the frame of standard formulation, using the bosonic limit of the action \eqref{SD0=A+L}, can be easily performed and gives an expected result: the theory of free scalar field obeying the Klein-Gordon equation. However, our interest lies in quantization of this simple system in its moving frame formulation, since this arises as a pure bosonic limit of our 3D mD$0$ system when we set to zero all the matrix fields.}.

Imposing the quantum versions of the first class constraints \eqref{Phi0=0}, \eqref{U0=0} and two of second class constraints \eqref{eqs:frak_d}, \eqref{eqs:Phi} which commute and are not related by Hermitian  conjugation among themselves, and choosing a coordinate representation for the state vector, $\Xi_0=\Xi_0(x^a,\bar{w},w)$, we arrive at the following system of differential equations
\bea
\label{u0dxXi0=}
(iu^{(0)a}\partial_{a}+m)\Xi_0 =0  \; , \qquad  \\ 
\label{D0Xi0=}
\left({\mathbb D}^{(0)}  -q\right) \Xi_0=0 \; , \qquad
\\ 
\label{barbbDXi0=}
\bar{{\mathbb D}}  \Xi_0=0 \; , \qquad
\\  \label{budxXi0=0}
\bar{u}{}^{a}\partial_{a}\Xi_0 =0  \; . \qquad  
\eea
The $x^a$-dependence of the state vector is fixed by Eqs.~\eqref{u0dxXi0=} and \eqref{budxXi0=0} to be
\be\label{Xi0=exp.chi}
\Xi_0 = e^{imx^a{u}_a^{(0)} } \, \chi_0 ( x^a\bar{u}_a, \bar{w}_\alpha,  w_\alpha) \; . \qquad
\ee
Substituting this into \eqref{barbbDXi0=}, we find that $\chi_0 (x^a \bar{u}_a, \bar{w}_\alpha, w_\alpha)$ satisfies
\be
\bar{{\mathbb D}} \chi_0 ( x^a\bar{u}_a, \bar{w}_\alpha,  w_\alpha):=
\bar{w}_\alpha \frac {\partial} {\partial {w}_\alpha}
\chi_0 ( x^a\bar{u}_a, \bar{w}_\alpha,  w_\alpha) \; =- imx^a\bar{u}_a\, \chi_0 ( x^a\bar{u}_a, \bar{w}_\alpha,  w_\alpha) \; .
\ee
The second form of the l.h.s. of this equation makes manifest that its nontrivial solution, if existed, would be $\chi_0 = e^{-imx^a{u}_a^{(0)}}\tilde{\chi}_0 ( x^a\bar{u}_a, \bar{w}_\alpha, w_\alpha)$, which contradicts the fact that, in \eqref{Xi0=exp.chi}, $\chi_0$ is independent of $x^a u_a^{(0)}$.

This example (actually the only presently known to the authors) indicates that the selfconsistency of the Gupta-Bleuler quantization method is not guaranteed and thus its should be used with precaution.

Curiously, if we modify the above procedure by imposing, besides \eqref{u0dxXi0=} and \eqref{D0Xi0=}, the complex (Hermitian) conjugate but not canonically conjugate pair of second class constraints \eqref{eqs:frak_d}, which commute among themselves, we obtain the system
\bea
\label{u0dxXi0==}
(iu^{(0)a}\partial_{a}+m)\Xi_0 =0  \; , \qquad  \\ 
\label{D0Xi0==}
\left({\mathbb D}^{(0)}  -q\right) \Xi_0=0 \; , \qquad
\\ 
\label{udxXi0==0}
u^{a}\partial_{a}\Xi_0 =0  \; , \qquad
\\ \label{budxXi0==0}
\bar{u}{}^{a}\partial_{a}\Xi_0 =0  \qquad \;.
\eea
This system does have a nontrivial solution which is equivalent to the solution of the massive Klein-Gordon equation.

While this modification falls outside the standard Gupta-Bleuler prescription, as we show in the main text and also in Appendix~\ref{App=D0}, its result is equivalent to the quantization with the method of Dirac brackets, which is further simplified and reduced to the explicit resolution of the constraints in the analytical basis. Such Dirac brackets/analytical basis quantization gives a consistent field theory with a nontrivial solution which, in the simplest case considered here, reproduces the result of quantization in the standard formulation of the massive particle.
    \chapter{Hamiltonian formalism of single 3D D0-brane in the so-called analytical basis}
\label{App=D0}
\thispagestyle{empty}

In this appendix we provide the details of the Hamiltonian formalism for a single D$0$-brane in the analytical basis.

The Lagrangian of this 3D counterpart of the D$0$-brane in the spinor moving frame formulation ({\it cf.} \eqref{SD0=A+L}) reads
\begin{equation}\label{LD0=}
 \mathcal{L}_{\text{D}0} =\text{d} \tau \mathcal{L}_{\tau \text{D}0} = \, m  {\rm E}^{0} + m (\text{d}\theta^\alpha \bar{\theta}_\alpha - \text{d}\bar{\theta}^\alpha \theta_\alpha)
\end{equation}
where ${\rm E}^{(0)}$ is defined in \eqref{E0=} with  \eqref{Ea=}. The dynamical variables in this Lagrangian 1-form  are coordinate functions that parametrize the embedding of the worldline in flat superspace
\begin{equation}
\begin{array}{ccccccc}
\mathcal{W}^1  \subset \Sigma^{\left(3\left. \right| 4 \right)}: &~& x^a = x^a (\tau), &~&  \theta^\alpha= \theta^\alpha(\tau),&~& \bar{\theta}^\alpha =  \bar{\theta}^\alpha(\tau)
\end{array}~
\end{equation}
as well as the spinor frame variables $w_\alpha(\tau)$ and its complex conjugate $\bar{w}_\alpha(\tau)$, satisfying \eqref{bww=i}. In the terminology of \cite{Igor_Russian}, originating in the seminal works on Harmonic superspaces \cite{Galperin2, Galperin3, harmonic, lightcone}, these spinor frame variables can be referred to as Lorentz harmonics, reflecting the fact that they parametrize the double cover of the 3D Lorentz group SL$(2,{\mathbb R})$. Upon imposing invariance under U$(1)$ gauge transformations acting on $w$ and $\bar{w}$, they serve as homogeneous coordinates on the coset  SL$(2,{\mathbb R})/\text{U}(1)$. Furthermore, the configuration space of this dynamical system (the target space of the spinor moving frame formulation of the 3D D$0$-brane) can be called, following the same terminology, {\it Lorentz harmonic superspace} which we denoted $\Sigma^{(3+3|4)}$. This is the superspace with coordinates
\be
\Sigma^{(3+3|4)} = \{ (x^a,\theta^\alpha , \bar{\theta}^\alpha , w_\alpha , \bar{w}_\alpha )\} =: \{ (Z^M , w_\alpha , \bar{w}_\alpha )\} \; , \qquad \bar{w}^\alpha{w}_\alpha =i\;.
\ee
In the spinor moving frame approach we are considering D$0$-brane as a particle moving in this extended and enlarged $\text{D}=3$ superspace. Here ``extended'' refers to ${\cal N}=2$ supersymmetry and ``enlarged'' reflects the presence of additional bosonic spinor (spinor moving frame or Lorentz harmonic) coordinates.

This perspective is convenient, in particular, because it suggests the possibility to choose an alternative coordinate basis of the Lorentz harmonic superspace, \textit{the analytical coordinate basis} which is given by
\be
\Sigma^{(3+3|4)} = \{ ({\rm x}{}^{(0)},{\rm x}_A,\bar{{\rm x}}_A\; ,  \theta^w , {\theta}^{\bar{w}} , \bar{\theta}^{{w}} ,\bar{\theta}^{\bar{w}}  ; w_\alpha , \bar{w}_\alpha )\} =:\{ ({\rm Z}{}^{(M)}_{An}, \; w_\alpha , \bar{w}_\alpha )\} \; , \qquad \bar{w}^\alpha{w}_\alpha =i\;
\ee
where (see~\cite{Igor_Russian, harmonic, lightcone})
\begin{equation}\label{xA=D0}
\begin{array}{c}
     {\rm x}^{(0)}:=x^au_a^{(0)}\; ,  \\
      \qquad  \fbox{${\rm x}_A := {\rm x}-2i \theta^w \bar{\theta}^{{w}}$}=x^au_a-2i \theta^w \bar{\theta}^{{w}}\; , \qquad \fbox{$\bar{{\rm x}}_A:= \bar{{\rm x}} +2i{\theta}^{\bar{w}} \bar{\theta}^{\bar{w}}$}=x^a \bar{u}_a+2i {\theta}^{\bar{w}} \bar{\theta}^{\bar{w}}\; ,~~~~~~~~~~~~~
\end{array}
\end{equation}
\begin{equation}\label{thw=D0}
    \theta^w := \theta^\alpha w_\alpha~, \qquad \theta^{\bar{w}}:= \theta^\alpha \bar{w}_\alpha~,
\end{equation}
\begin{equation}\label{bthw=D0}
    \bar{\theta}^w := \bar{\theta}^\alpha w_\alpha~, \qquad  \bar{\theta}^{\bar{w}}:= \bar{\theta}^\alpha \bar{w}_\alpha~. \qquad
\end{equation}
The definition of ${\rm x}_A=(\bar{{\rm x}}_A)^*$ in~\eqref{xA=D0} is designed in such a way that, in the analytical basis, the Lagrangian form \eqref{LD0=} simplifies to
\begin{equation}
    \begin{array}{l}
         \begin{split}
              \mathcal{L}_{\text{D}0} &= m \text{d}{\rm x}^{(0)} - im f\; \bar{{\rm x}}_A  + im \bar{f}\; {\rm x}_A  -4ma\theta^{\bar{w}} \bar{\theta}^w - 2im (\text{d}\theta^{\bar{w}} \bar{\theta}^{w} + \text{d}\bar{\theta}^{w} \theta^{\bar{w}})=\\
              &=  m \text{d}{\rm x}^{(0)} - 2im (\text{d}\theta^{\bar{w}}-ia\theta^{\bar{w}})\, \bar{\theta}^{w}  - 2im (\text{d}\bar{\theta}^{w}+ia\bar{\theta}^{w} )\, \theta^{\bar{w}}  - im f\; \bar{{\rm x}}_A  + im \bar{f}\; {\rm x}_A \; .
         \end{split}
    \end{array}
\end{equation}
In the main text we also needed explicit expressions for the kinetic and WZ terms in the single 3D D$0$ action,
\begin{equation}\label{E0=an}
    \begin{array}{l}
         \begin{split}
             {\rm E}^{0} &= \text{d}{\rm x}^{(0)} - i(\text{d}\theta^{{w}}+ia\theta^{{w}})\, \bar{\theta}^{\bar{w}}- i(\text{d}\theta^{\bar{w}}-i (\text{d}\bar{\theta}^{w}+ia\bar{\theta}^{w}) \theta^{\bar{w}} -ia\theta^{\bar{w}})\, \bar{\theta}^{w} -\\
             &-i (\text{d}\bar{\theta}^{\bar{w}}-ia\bar{\theta}^{\bar{w}}) \theta^{w}+i f\; \bar{{\rm x}} -i \bar{f}\; {\rm x}  \; ,
         \end{split}
    \end{array}
\end{equation}
\begin{equation}\label{WZ=an}
    \begin{array}{l}
         \begin{split}
             \text{d}\theta^\alpha \bar{\theta}_\alpha -\text{d}\bar{\theta}^\alpha {\theta}_\alpha   &= 2f\;\theta^{\bar{w}} \bar{\theta}^{{\bar{w}}}  + 2 \bar{f}\; \theta^{{w}} \bar{\theta}^{{w}} + i(\text{d}\theta^{{w}}+ia\theta^{{w}})\, \bar{\theta}^{\bar{w}}- i(\text{d}\theta^{\bar{w}}-ia\theta^{\bar{w}})\, \bar{\theta}^{w} - \\
             &  -i (\text{d}\bar{\theta}^{w}+ia\bar{\theta}^{w}) \theta^{\bar{w}} +i (\text{d}\bar{\theta}^{\bar{w}}-ia\bar{\theta}^{\bar{w}}) \theta^{w} \; .
         \end{split}
    \end{array}
\end{equation}
The canonical Hamiltonian $H_0$, expressed in terms of the analytical basis coordinates, is defined by
\begin{equation}
\text{d}\tau H_0 = \text{d}\text{x}^{(0)}p^{(0)} + \text{d}{\rm x}_A \bar{p} + \text{d}\bar{\text{x}}_A p + \text{d}\theta^w \Pi^\theta_w + \text{d}\theta^{\bar{w}} \Pi^\theta_{\bar{w}} + \text{d} \bar{\theta}^w \bar{\Pi}^{\bar{\theta}}_w + \text{d} \bar{\theta}^{\bar{w}} \bar{\Pi}^{\bar{\theta}}_{\bar{w}} + ia \tilde{\mathfrak{d}}^{(0)} + i f \bar{\tilde{\mathfrak{d}}} - i \bar{f} \tilde{\mathfrak{d}} - \mathcal{L}_{\text{D}0}~.
\end{equation}
The momenta conjugate to the coordinate functions, i.e. having the nonvanishing Poisson brackets
\begin{equation}
\begin{array}{ccccc}
\left[p^{(0)}, \text{x}^{(0)} \right]_{\text{PB}} = -1~,&~&\left[p, \bar{{\rm x}}_A \right]_{\text{PB}} = -1~,&~&\left[\bar{p}, {\rm x}_A \right]_{\text{PB}} = -1~,
\end{array}
\end{equation}
are defined by \eqref{up=-2p},
\begin{equation}\label{p-pAnA}
\begin{array}{ccccc}
p^{(0)}:= u^{a(0)}p_a~,&~&p := -\dfrac{1}{2}u^a p_a~,&~&\bar{p} := -\dfrac{1}{2}\bar{u}^a p_a~.
\end{array}
\end{equation}
Similarly, the momenta conjugate to the fermionic coordinate functions satisfy
\begin{equation}
\begin{array}{ccccccc}
{}\{\Pi^\theta_w , \theta^w\}_{\text{PB}} = -1~,&~& \{\Pi^\theta_{\bar{w}} , \theta^{\bar{w}}\}_{\text{PB}} = -1~, &~& \{\bar{\Pi}^{\bar{\theta}}_w , \bar{\theta}^w\}_{\text{PB}} = -1~, &~& \{\bar{\Pi}^{\bar{\theta}}_{\bar{w}} , \bar{\theta}^{\bar{w}}\}_{\text{PB}} = -1~.
\end{array}
\end{equation}
and are related to the ones of the central basis by \eqref{Pi-PiAn}
\begin{equation}\label{Pi-PiAnA}
\begin{array}{ccc}
\Pi^\theta_{w}:= -i\bar{w}^\alpha \Pi_\alpha+2i \bar{\theta}^w\bar{p}~,&~& \Pi^\theta_{\bar{w}}:= iw^\alpha \Pi_\alpha - 2i \bar{\theta}^{\bar{w}}p~,\\
~\\
\bar{\Pi}^{\bar{\theta}}_{w}:= -i\bar{w}^\alpha \bar{\Pi}_\alpha -2i \theta^w \bar{p} ~,&~& \bar{\Pi}^{\bar{\theta}}_{\bar{w}}:= iw^\alpha \bar{\Pi}_\alpha+ 2i \theta^{\bar{w}}p~.
\end{array}
\end{equation}
The covariant momenta in the analytical basis are related to those in the central basis as
\begin{equation}
\begin{array}{lll}
&~& \mathfrak{d}  := \tilde{\mathfrak{d}} + (\text{x}_A+2i \theta^{w}  \bar{\theta}^{\bar{w}} ) p^{(0)} + 2(\text{x}^{(0)}+i \theta^{w}  \bar{\theta}^{\bar{w}}+i \theta^{\bar{w}}  \bar{\theta}^{{w}}) p + \theta^{w} \Pi_{\bar{w}}^\theta +  \bar{\theta}^{w} \bar{\Pi}_{\bar{w}}^{\bar{\theta}}~,\\
~\\ &~& \bar{\mathfrak{d}} := \bar{\tilde{\mathfrak{d}}} +(\bar{\text{x}}_A -2i{\theta}^{\bar{w}}\bar{\theta}^{\bar{w}}) p^{(0)} + 2(\text{x}^{(0)}-i{\theta}^{{w}}\bar{\theta}^{\bar{w}}-i{\theta}^{\bar{w}}\bar{\theta}^{{w}})\bar{p} + \theta^{\bar{w}} \Pi_{w}^\theta +  \bar{\theta}^{\bar{w}} \bar{\Pi}_{w}^{\bar{\theta}}~,\\
~\\
&~&{\mathfrak{d}}^{(0)} := \tilde{\mathfrak{d}}^{(0)}- 2\text{x}_A \bar{p} + 2\bar{\text{x}}_Ap+ \theta^{\bar{w}} \Pi_{\bar{w}}^\theta + \bar{\theta}^{\bar{w}} \bar{\Pi}_{\bar{w}}^{\bar{\theta}}- \theta^{w} \Pi^{\theta}_w - \bar{\theta}^w \bar{\Pi}^{\bar{\theta}}_w~.
\end{array}
\end{equation}
This implies that
\begin{equation}
\begin{array}{lll}
{}[\tilde{\mathfrak{d}}^{(0)} , {\rm Z}_{An}^{(M)}]_{\text{PB}} =0 \; , \qquad {}[\tilde{\mathfrak{d}}, {\rm Z}_{An}^{(M)}]_{\text{PB}} =0 \; , \qquad  {}[\bar{\tilde{\mathfrak{d}}}, {\rm Z}_{An}^{(M)}]_{\text{PB}} =0 \; , \qquad
\end{array}
\end{equation}
while
\begin{equation}
\begin{array}{lll}
{}[{\mathfrak{d}}^{(0)} , Z^{M}]_{\text{PB}} =0 \; , \qquad {}[{\mathfrak{d}},  Z^{M}]_{\text{PB}} =0 \; , \qquad  {}[\bar{{\mathfrak{d}}}, Z^{M}]_{\text{PB}} =0 \; . \qquad
\end{array}
\end{equation}
The calculation of all the canonical and covariant momenta results in the (primary) constraints
\begin{eqnarray}\label{Phi=An}
\Phi^{(0)}:= p^{(0)} - m \approx 0\; ,\qquad & \qquad    \bar{\Phi}:= \bar{p}  \approx 0\; ,\qquad & \Phi:= p  \approx 0\; ,\qquad \\ \nonumber  \\ \label{frakd=An}
\tilde{\mathfrak{d}}^{(0)} -4im \theta^{\bar{w}} \bar{\theta}^w   \approx 0~, \qquad   &  \qquad
\tilde{\mathfrak{d}}+ m {\rm x}_A  \approx 0~,   \qquad
 & \bar{\tilde{\mathfrak{d}}}+ m \bar{{\rm x}}_A  \approx 0~,
 \\ \nonumber
 \\ \label{dw=An} d_w:= \Pi^\theta_w \approx 0~, \qquad & \qquad d_{\bar{w}}:= \Pi^\theta_{\bar{w}}+2i m \bar{\theta}^w \approx 0~, \qquad & \\ \nonumber
 \\ \label{bdw=An} \bar{d}_{\bar{w}} := \bar{\Pi}^{\bar{\theta}}_{\bar{w}} \approx 0~,  \qquad
  & \qquad \bar{d}_w := \bar{\Pi}^{\bar{\theta}}_w + 2im\theta^{\bar{w}} \approx 0~. \qquad  &  \\ \nonumber
\end{eqnarray}
We have written these in three columns in such a way that the rows correspond to different sectors of dynamical  variables. From  the second and third columns one can read the pairs of conjugate second class constraints (which, in the case of bosonic constraints, are explicitly solved), while the first column contains a prototypes of the first class constraints.

The canonical Hamiltonian vanishes on the surface of primary constraints, this is
\be
H_0\approx 0\; .
\ee
The true first class constraints,
\be\label{1st=An}
 \Phi^{(0)}\approx 0, \qquad d_w,  \approx 0, \qquad \bar{d}_{\bar{w}} \approx 0, \qquad  \text{and}\qquad \tilde{\tilde{U}}{}^{(0)} \approx 0, \qquad
\ee
are given by the first equations in \eqref{Phi=An}, \eqref{dw=An}, \eqref{bdw=An} and by the sum of the first equation in \eqref{frakd=An} with certain linear combination of the second class constraints,
\bea\label{U0=An}
\tilde{\tilde{U}}{}^{(0)} = \tilde{\mathfrak{d}}^{(0)} -4im \theta^{\bar{w}} \bar{\theta}^w  -2\text{x}_A\bar{p}+2\bar{\text{x}}_Ap-\bar{\theta}^{w}\bar{d}_w+ {\theta}^{\bar{w}}d_{\bar{w}} =\qquad  \nonumber \\ \nonumber \\ = \tilde{\mathfrak{d}}^{(0)}   -2\text{x}_A\bar{p}+2\bar{\text{x}}_Ap-\bar{\theta}^{w}\bar{\Pi}^{\bar{\theta}}_w+ {\theta}^{\bar{w}}{\Pi}^{{\theta}}_{\bar{w}} \approx 0 .  \;
\eea
Also the set of second class constraints, in order to put their algebra in the canonical form, should be redefined a bit and written as
\be\label{2nd=class}
{\text{2nd class}}: \qquad \qquad \begin{cases} \bar{\Phi}=\bar{p}\approx 0 \cr \tilde{\tilde{\mathfrak{d}}}\approx 0\end{cases}\; , \qquad \begin{cases} {\Phi}={p}\approx 0 \cr \bar{\tilde{\tilde{\mathfrak{d}}}} \approx 0\end{cases}\; ,
\qquad \begin{cases} d_{\bar{w}}\approx 0 \cr \bar{d}_w\approx 0\end{cases}\; , \qquad  \ee
 where
\be\label{ttfrakd=}
\tilde{\tilde{\mathfrak{d}}}:= \tilde{\mathfrak{d}}+ m {\rm x}_A+2i {\theta}^{\bar{w}} \bar{\theta}^{w}p\; , \qquad \bar{\tilde{\tilde{\mathfrak{d}}}}:= \bar{\tilde{\mathfrak{d}}}+ m \bar{{\rm x}}_A -2i {\theta}^{\bar{w}} \bar{\theta}^{w}\bar{p} ~.   \qquad
\ee
The algebra of the first class constraints \eqref{1st=An} is Abelian (this is to say all the Poisson brackets of the first class constraints vanish). The only nonvanishing brackets between first class constraints and second class constraints appear when the first class constraint is given by the U$(1)$ generator \eqref{U0=An},
\be
{}[ \tilde{\tilde{U}}{}^{(0)} \, , \, \left(\begin{matrix}\bar{p}\cr \tilde{\tilde{\mathfrak{d}}}\cr  {p} \cr \bar{\tilde{\tilde{\mathfrak{d}}}}
\cr  d_{\bar{w}} \cr \bar{d}_w \end{matrix}\right)]_{{\text{PB}}} =  \, \left(\begin{matrix}-2\bar{p}\cr 2\tilde{\tilde{\mathfrak{d}}}\cr  2{p} \cr -2\bar{\tilde{\tilde{\mathfrak{d}}}}
\cr d_{\bar{w}} \cr -\bar{d}_w \end{matrix}\right)\, .
\ee
The coefficients in the r.h.s. represent the charges of second class constraints with respect to U$(1)$ gauge symmetry.

Finally, the nonvanishing brackets of the second class constraints are given by
\be
 {}[\bar{p}\, , \,  \tilde{\tilde{\mathfrak{d}}} ]_{{\text{PB}}}= m\; , \qquad  [ {p} \, , \,  \bar{\tilde{\tilde{\mathfrak{d}}}}
]_{{\text{PB}}}=m \; , \qquad \{ d_{\bar{w}},  \bar{d}_w \}_{{\text{PB}}}=4im\; ,
\ee
reflecting their second class nature, and the weakly vanishing brackets
\bea
 {}[ \tilde{\tilde{\mathfrak{d}}}\, , \,  \bar{\tilde{\tilde{\mathfrak{d}}}} ]_{{\text{PB}}}= \tilde{\tilde{U}}{}^{(0)} +2\text{x}_A\bar{p}-2\bar{\text{x}}_Ap+\bar{\theta}^{w}\bar{d}_w- {\theta}^{\bar{w}}d_{\bar{w}}\;\approx 0 ,  \qquad \\ \nonumber \\
  {}[ \tilde{\tilde{\mathfrak{d}}}\, , \, d_{\bar{w}} ]_{{\text{PB}}}= 2i\bar{\theta}^{w}p  \approx 0 \; , \qquad {}[ \tilde{\tilde{\mathfrak{d}}}\, , \, d_{\bar{w}} ]_{{\text{PB}}}= -2i {\theta}^{\bar{w}}{p}  \approx 0  \; , \qquad
   \\ \nonumber \\
  {}[ \bar{\tilde{\tilde{\mathfrak{d}}}}\, , \, d_{\bar{w}} ]_{{\text{PB}}}= -2i\bar{\theta}^{w}\bar{p}  \approx 0 \; ,  \qquad  {}[ \bar{\tilde{\tilde{\mathfrak{d}}}}\, , \, d_{\bar{w}} ]_{{\text{PB}}}= 2i {\theta}^{\bar{w}}\bar{p}  \approx 0 \; . \qquad
\eea
In summary, the algebra of the constraints is presented in Table \ref{table:singleD0}
\begin{table}[h!]
\resizebox{\textwidth}{!}{
\begin{tabular}{c||cccc||cccccc}
 $[...,... \}_{\text{PB}}$
& $\Phi^{(0)}$&  $\tilde{\tilde{U}}{}^{(0)}$ & $d_w$ & $\bar{d}_{\bar{w}}$ &
$\bar{p}$ & $\tilde{\tilde{\mathfrak{d}}}$ & ${p}$ & $\tilde{\bar{\tilde{\mathfrak{d}}}}$ & $d_{\bar{w}}$ & $ \bar{d}_w$ \\
 \hline \hline
$\Phi^{(0)}$& 0 & 0 & 0 & 0 & 0 & 0 & 0 & 0 & 0 & 0 \\
 $\tilde{\tilde{U}}{}^{(0)}$ & 0 & 0 & 0 & 0 & $-2\bar{p}$ & $2\tilde{\tilde{\mathfrak{d}}}$ & $2{p}$ & $-2\bar{\tilde{\tilde{\mathfrak{d}}}}$ & $d_{\bar{w}}$ & $- \bar{d}_w$\\
 $d_w$ &  0 & 0 & 0 & 0 & 0 & 0 & 0 & 0 & 0 & 0 \\ $\bar{d}_{\bar{w}}$ &  0 & 0 & 0 & 0 & 0 & 0 & 0 & 0 & 0 & 0 \\
 \hline \hline
$\bar{p}$ &  0 & 2$\bar{p}$ & 0 & 0 & 0 & $m$ & 0 & 0 & 0 & 0 \\
$\tilde{\tilde{\mathfrak{d}}}$  &  0 & $-2\tilde{\tilde{\mathfrak{d}}}$ & 0 & 0 & $-m$  & 0 & 0 & \fbox{$\begin{matrix}\tilde{\tilde{U}}{}^{(0)} +\bar{\theta}^{w}\bar{d}_w - {\theta}^{\bar{w}}d_{\bar{w}}+ \cr+2 \text{x}_A{p} -2\bar{\text{x}}_Ap\end{matrix}$} & $2i\bar{\theta}{}^w\bar{p}$ & $-2i{\theta}{}^{\bar{w}}{p} $
\\ ${p}$ &  0 & -2$p$ & 0 & 0 & 0 & 0 & 0 & m & 0 & 0  \\ $\bar{\tilde{\tilde{\mathfrak{d}}}}$ &  0 & $2\bar{\tilde{\tilde{\mathfrak{d}}}}$ &0 & 0 & 0 &  \fbox{$\begin{matrix}-\tilde{\tilde{U}}{}^{(0)} -\bar{\theta}^{w}\bar{d}_w +{\theta}^{\bar{w}}d_{\bar{w}}- \cr -2\text{x}_A\bar{p} +2\bar{\text{x}}_Ap\end{matrix}$} & -m & 0 & $-2i\bar{\theta}{}^w\bar{p}$ & $2i{\theta}{}^{\bar{w}}\bar{p}$ \\ $d_{\bar{w}}$ &  0 & -$d_{\bar{w}}$ & 0 & 0 & 0 & 0 & 0 & 0 & 0 & $-4im$\\ $ \bar{d}_w$ &  0 &  $ \bar{d}_w$& 0 & 0 & 0 & 0 & 0 & 0 & $-4im$ & 0\\
 \hline \hline
\end{tabular}}
\caption{Closed algebra of the first and second class constraints of single D$0$-brane system.}
\label{table:singleD0}
\end{table}

The bosonic second class constraints in \eqref{2nd=class} are now explicitly solved with respect to ${\rm x}_A$, $\bar{{\rm x}}_A$ and their conjugate momenta. Thus a consistent way is to use these to reduce the phase space of our dynamical system before quantization. An equivalent approach is provided by the (classical counterpart of) generalized/deformed Gupta-Bleuler procedure which implies that we impose only half of the second- lass constraints on the quantum system $p\approx 0$ and $\bar{p}\approx 0$. At the classical level, this implies omitting the conjugate constraints $\bar{\tilde{\tilde{\mathfrak{d}}}}\approx 0$ and $\tilde{\tilde{{\mathfrak{d}}}}\approx 0$ from consideration, thus converting their conjugate constraints $p\approx 0$ and $\bar{p}\approx 0$ into first class constraints\footnote{Notice that this is a deformed version of Gupta-Bleuler quantization. As we have already written in the main text and shown in appendix \ref{GB-failed}, the canonical Gupta-Bleuler approach to the bosonic second class constraints of our system fails to produce the correct result equivalent to the Dirac bracket quantization.}.

We prefer this latter treatment of the bosonic second class constraints because it aligns with the (canonical) Gupta-Bleuler treatment of the fermionic second class constraints in \eqref{2nd=class}. These are complex conjugate pairs and, after quantization, we will impose on the state vector the quantum counterpart of only one of two fermionic second class constraints, $\bar{d}_w\approx 0$. This is equivalent to omitting, at the classical level, the conjugate $d_{\bar{w}}\approx 0$ constraint, which converts $\bar{d}_w\approx 0$ into a first class constraint. The resulting algebra of the effective first class constraints appearing as a result of the implementation of the above classical counterpart of the generalized/deformed Gupta-Bleuler procedure is presented in Table \ref{table:singleD0eff}.
\begin{table}[h!]
\hspace{11.0em}
\begin{tabular}{c||ccccccc}
 $[...,... \}_{\text{PB}}$
& $\Phi^{(0)}$&  $\tilde{\tilde{U}}{}^{(0)}$ & $d_w$ & $\bar{d}_{\bar{w}}$ &
$\bar{p}$ &  ${p}$ &  $ \bar{d}_w$ \\
 \hline \hline
$\Phi^{(0)}$& 0 & 0 & 0 & 0 &  0 & 0 & 0  \\
 $\tilde{\tilde{U}}{}^{(0)}$ & 0 & 0 & 0 & 0 & $-2\bar{p}$ & $2{p}$ &  $- \bar{d}_w$\\
 $d_w$ &  0 & 0 & 0 & 0 & 0 & 0 & 0 \\ $\bar{d}_{\bar{w}}$ &  0 & 0 & 0 & 0 & 0 & 0 & 0 \\
 \hline \hline
$\bar{p}$ &  0 & 2$\bar{p}$ & 0 & 0 & 0 & 0 & 0  \\
 ${p}$ &  0 &  $-2p$ & 0 & 0 & 0 & 0 & 0 \\  $ \bar{d}_w$ &  0 &  $ \bar{d}_w$& 0 & 0 & 0 & 0 & 0 \\
 \hline \hline
\end{tabular}
\caption{Closed algebra of the effective first  class constraints of single D$0$-brane system.}
\label{table:singleD0eff}
\end{table}
To quantize our dynamical system in the supercoordinate representation we represent our effective first class constraints by differential operators,
\begin{equation}\label{hPhi0=}
    \hat{\Phi}^{(0)}= -i\partial_{\text{x}^{(0)}}-m \; , 
\end{equation}
\begin{equation} \label{hPhi=}
    \hat{\bar{\Phi}}= -i\partial_{\text{x}_A}\; , \qquad  \hat{{\Phi}}= -i\bar{\partial}_{\bar{\text{x}}_A}\;,
\end{equation}
\begin{equation}\label{hU0=}
    \begin{array}{l}
         \begin{split}
             \hat{\tilde{\tilde{U}}}{}^{(0)} &=-i\left({\mathbb D}^{(0)}+4m\theta^{\bar{w}} \bar{\theta}^w  -2\text{x}_A\partial_{\text{x}_A}+2\bar{\text{x}}_A \partial_{\bar{\text{x}}_A}-\bar{\theta}^{w} \bar{d}_w+ {\theta}^{\bar{w}}d_{\bar{w}} -q\right) =\\
             &= -i\left({\mathbb D}^{(0)}-2\text{x}_A\partial_{\text{x}_A}+2\bar{\text{x}}_A \partial_{\bar{\text{x}}_A}-\bar{\theta}^{w}\partial_{\bar{\theta}^{{w}}}+ {\theta}^{\bar{w}}\partial_{{\theta}^{\bar{w}}}-q \right) \; ,
         \end{split}
    \end{array}
\end{equation}
\begin{equation}\label{hdw=}
    \hat{d}_w=-i {\partial}_{{\theta}{}^{w}}  \; ,
\end{equation}
\begin{equation}\label{hbdbw=}
    \hat{\bar{d}}_{\bar{w}}=-i \bar{\partial}_{\bar{\theta}{}^{\bar{w}}} \; , 
\end{equation}
\begin{equation}\label{hbdw=}
\hat{\bar{d}}_w=-i\left(\bar{\partial}_{\bar{\theta}{}^{w}}  -2m{\theta}^{\bar{w}}\right)
\end{equation}
and impose them on the state vector. In \eqref{hPhi0=}-\eqref{hbdw=} the derivatives and covariant derivatives are defined as follows
\bea
\partial_{\text{x}^{(0)}} =\frac {\partial} {\partial \text{x}^{(0)}} \; , \qquad \partial_{{\text{x}_A}} =\frac {\partial} {\partial {\text{x}_A}} \; , \qquad \partial_{\bar{\text{x}}_A} =\frac {\partial} {\partial \bar{\text{x}}_A} \; , \qquad \\
\nonumber \\
{\mathbb D}^{(0)}= \bar{w}_\alpha \frac {\partial} {\partial \bar{w}_\alpha} - {w}_\alpha \frac {\partial} {\partial {w}_\alpha}  \; , \qquad
{\mathbb D} = w_\alpha \frac {\partial} {\partial \bar{w}_\alpha}  \; , \qquad  \bar{{\mathbb D}}= \bar{w}_\alpha\frac {\partial} {\partial {w}_\alpha}  \; . 
\eea
Notice the appearance of the ordering constant in the homogeneous differential operator \eqref{hU0=}.

To understand the structure of the fermionic constraint \eqref{hbdw=} it is useful to keep in mind
the relation between derivatives and covariant derivatives in the central and analytical basis. This relationship can be derived from the identity
\begin{equation}
    \begin{array}{l}
         \begin{split}
             \text{d}&= \text{d}x^a\partial_a + \text{d}\theta^\alpha \partial_\alpha + \text{d}\bar{\theta}{}^\alpha \bar{\partial}_\alpha + ia {\mathbb{D}}^{(0)} + i f \bar{{\mathbb{D}}} - i \bar{f} {\mathbb{D}}= \\
             & =\Pi^a\partial_a + \text{d}\theta^\alpha \text{D}_\alpha + \text{d}\bar{\theta}{}^\alpha \bar{\text{D}}_\alpha + ia {\mathbb{D}}^{(0)} + i f \bar{{\mathbb{D}}} - i \bar{f} {\mathbb{D}}~= \\
             &= \text{d}{\rm x}^{(0)} \partial_{{\rm x}^0 } + \text{d}{\rm x}_A \partial_{{\rm x}_A} + \text{d}\bar{{\rm x}}_A \partial_{\bar{{\rm x}}_A } + \text{d}\theta^w \partial_{\theta^w } +\text{d}\theta^{\bar{w}} \partial_{\theta^{\bar{w}} } +\\
             &  \qquad \qquad \qquad \qquad \qquad \quad ~~~~~~~~~~~~~~~~~~~~ + \text{d}\bar{\theta}{}^w \partial_{\bar{\theta}{}^w} + \text{d}\bar{\theta}{}^{\bar{w}} \partial_{\bar{\theta}{}^{\bar{w}}} + ia \tilde{\mathbb{D}}^{(0)} + i f \bar{\tilde{\mathbb{D}}} - i \bar{f} \tilde{\mathbb{D}}\, .
         \end{split}
    \end{array}
\end{equation}
This identity implies, in particular,
\bea
\partial_a= u_a^{(0)} \partial_{{\rm x}^{(0)} } + u_a \partial_{{\rm x}_A} + \bar{u}_a \partial_{\bar{{\rm x}}_A}\; , \qquad  \\ 
\text{D}_\alpha = \partial_\alpha + i(\gamma^a\bar{\theta})_\alpha =
w_\alpha \left(\partial_{\theta^w } +i\bar{\theta}{}^{\bar{w}}  \partial_{{\rm x}^{(0)} }  \right) + \bar{w}_\alpha \left(\partial_{\theta^{\bar{w}} } +4i\bar{\theta}{}^{\bar{w}} \partial_{\bar{\rm x}_A} +i\bar{\theta}{}^{{w}}\partial_{{\rm x}^{(0)} }  \right)\; ,  \\  \bar{\text{D}}_\alpha = \bar{\partial}_\alpha + i(\gamma^a{\theta})_\alpha =
w_\alpha \left(\bar{\partial}_{\bar{\theta}{}^w} +4i{\theta}{}^{w} \partial_{{\rm x}_A} +i{\theta}{}^{\bar{w}}\partial_{{\rm x}^{(0)} } \right) +  \bar{w}_\alpha \left(\bar{\partial}_{\bar{\theta}^{\bar{w}}} +i{\theta}{}^{w}  \partial_{{\rm x}^{(0)} }  \right)\; .
\eea
The last equation can be used to write \eqref{hbdw=} in the form
\be\label{hbdw==}
\hat{\bar{d}}_w=-\bar{w}\bar{\text{D}}+im{\theta}^{\bar{w}}-4i\theta^w\hat{\Phi}-i\theta^{\bar{w}}\hat{\Phi}{}^{(0)}\; .
\ee

    \chapter[Born-Oppenheimer-like approach to  asymptotic form of a solution for \boldmath\texorpdfstring{$N$}{N}=2 3D mD0 system]{Born-Oppenheimer-like approach to asymptotic form of a solution for \boldmath\texorpdfstring{$N$}{N}=2 3D mD0 system}
\label{AppBornOppenhemer}
\thispagestyle{empty}

Following \cite{Frohlich}, we can search for an asymptotic form of the solution of Eqs.  \eqref{Psi0}-\eqref{Psi2=} using the so-called \textit{tabular} or \textit{endpoint coordinates} (see \cite{Frohlich} for references)
\be\label{Z=X+z}
Z^I= X^I e^{i\beta} + |X|^{-1/2} z^I \; , \qquad \bar{Z}{}^I= X^I e^{-i\beta} + |X|^{-1/2} \bar{z}{}^I \; , \qquad
\ee
 in the neighbourhood of the classical vacuum configuration
 \begin{equation}
 Z^I= X^I e^{i\beta}~, \qquad\bar{Z}{}^I= X^I e^{-i\beta}~,
 \end{equation}
 which represents a general solution to the condition for vanishing YM potential which reduces to $\epsilon_{IJK}  Z^J\bar{Z}{}^K=0$. In~\eqref{Z=X+z} $X^I$ is real, $X^I=(X^I)^*$, with $|X|= \sqrt{X^IX^I}$, and the complex variables $z^I=(\bar{z}{}^I)^*$ are orthogonal to $X^I$. Additionally, they satisfy two constraints, which we choose as $\Re{\rm e} (z^I e^{-i\beta})=0$,
\be\label{XIzI=0}
X^I z^I=0 = X^I\bar{z}{}^I \; \; , \qquad z^I e^{-i\beta}+ \bar{z}^I e^{i\beta}=0\; .
\ee
The asymptotic regime corresponds to large $|X|$, formally
\be\label{r-infty}
r=|X| =\sqrt{X^IX^I} \mapsto \infty \; ,
\ee
with $z^I\bar{z}{}^I$ kept to be finite. In our case, it is convenient to solve \eqref{XIzI=0} by setting $z^I= iy^I  e^{i\beta}$ in terms of a real vector $y^I$, yielding $\bar{z}^I= -iy^I  e^{-i\beta}$. Thus, the above coordinate system is finally described by
\bea\label{Z=X+y}
Z^I= (rn^I  + ir^{-1/2} y^I)e^{i\beta} \; , \qquad \bar{Z}{}^I=( rn^I-i r^{-1/2} y{}^I)e^{-i\beta} \; , \qquad \\  \label{r=|X|}
r:= |X| =\sqrt{X^IX^I}\; , \qquad n^I = X^I/r=X^I/|X|\quad \Rightarrow \quad n^In^I=1 \; , \qquad  \\  \label{yn=0} y^I = (y^I)^* \; , \qquad n^I y^I=0\; . \qquad
\eea
To proceed, it is convenient to introduce complex null-vectors that complete the unit vector $n^I$ till a nondegenerate SO$(3)$ frame, i.e. vectors $U^I$ and $\bar{U}^I = (U^I)^*$ which obeys
\be
n^IU^I=0\; , \qquad n^I\bar{U}^I=0\; , \qquad U^IU^I=0\; , \qquad\bar{U}^I\bar{U}^I=0\; , \qquad U^I\bar{U}^I=1\; .
\ee
Then,
\be\label{I=nn+}
\delta^{IJ}= n^In^J+U^I\bar{U}^J + \bar{U}^IU^J\; \qquad
\ee
and the constraint \eqref{yn=0} can be solved by
\be\label{yI=}
y^I= y_U\bar{U}^I+y_{\bar{U}} U^I\; , \qquad {\text{so that}}\qquad y^Iy^I=|y_U|^2 \; .
\ee
We fix the orientation of the SO$(3)$ frame by setting $\epsilon_{IJK}\bar{U}^IU^Jn^K=i$, which implies
\be\label{eUn=iU}
\epsilon_{IJK}\bar{U}^Jn^K=-i\bar{U}^I\; , \qquad  \epsilon_{IJK}{U}^Jn^K=i{U}^I\; . \qquad
\ee
Using these relations, we find
\be\label{eZbZ=}
\epsilon_{IJK}Z^J\bar{Z}^K= 2\sqrt{r} (y_U\bar{U}{}^I -y_{\bar{U}}U^I )=2i\sqrt{r} \epsilon_{IJK}y^Jn^K \; .
\ee
The derivatives of the constrained  vectors can be expressed in terms of three Cartan forms, defines as
\be\label{Om1=}
\Omega_1=U^I\text{d}n^I\, , \qquad \bar{\Omega}_1 = \bar{U}^I\text{d}n^I \qquad \text{ and} \qquad \Omega_1^{(0)}= \bar{U}^I\text{d}{U}^I\;. 
\ee
These allow us to write
\be\label{dnI=}
\text{d}n^I= U^I \bar{\Omega}_1 + \bar{U}^I\Omega_1 \; ,  \qquad  \text{d}{U}^I= {U}^I\Omega_1^{(0)} - n^I\Omega_1  \; ,  \qquad  \text{d}\bar{U}^I = -\bar{U}^I\Omega_1^{(0)} -n^I  \bar{\Omega}_1\; .  \qquad
\ee
Using these, the derivative of the coordinates $y^I$ can be expressed as
\be\label{dyI=}
\text{d}y^I=  (\text{d}y_U-y_U\Omega_1^{(0)}) \bar{U}^I+(\text{d}y_{\bar{U}}+y_{\bar{U}}\Omega_1^{(0)}) U^I-n^I (y_U \bar{\Omega}_1 + y_{\bar{U}}\Omega_1)\; .
\ee
Now, using equations \eqref{Z=X+y}, \eqref{yI=}, \eqref{dnI=}, \eqref{dyI=} we can compute d$Z^I$ and d$\bar{Z}{}^I$,  and subsequently decompose the differential operator
\be
\text{d}=\text{d}Z^I\partial_I +\text{d}\bar{Z}{}^I\bar{\partial}_I =\text{d}r \partial_r +\text{d}\beta \partial_{(\beta)} + \text{d}y^I\partial^y_I +\text{d}n^I\partial_{n^I }
\ee
on the basis of six independent 1-forms: $\text{d}r$, $\text{d}\beta$, $\text{d}y_U$, $\text{d}y_{\bar{U}}$, $\Omega_1$, $\bar{\Omega}_1$ \footnote{$\Omega^{(0)}$ also appears in intermediate expressions; however, its coefficient does not produce independent equations, reflecting the underlying U$(1)$ gauge symmetry of the construction.}. In this way, we obtain the following expressions for derivatives in the tabular (endpoint) coordinates
\bea\label{d-dr=}
\partial_r= n^I (e^{i\beta}\partial_I + e^{-i\beta}\bar{\partial}_I ) - \frac i 2\,\frac 1 { r^{3/2}}\, (y_U \bar{U}^I+y_{\bar{U}} U^I)\, (e^{i\beta}\partial_I - e^{-i\beta}\bar{\partial}_I )\; , \qquad
\\ \nonumber \\ \label{d-dbeta=} \partial_{(\beta)}=ir n^I (e^{i\beta}\partial_I - e^{-i\beta}\bar{\partial}_I ) - \frac 1 { r^{1/2}}\, (y_U \bar{U}^I+y_{\bar{U}} U^I)\, (e^{i\beta}\partial_I + e^{-i\beta}\bar{\partial}_I )\; , \qquad
\\ \nonumber \\  \label{bUd-dy=}
 \bar{U}^I\left(\partial^y_{I }-\frac i { r^{1/2}}(e^{i\beta}\partial_I - e^{-i\beta}\bar{\partial}_I )\right) =0\, , \qquad
\\ \nonumber \\  \label{Ud-dy=}  {U}^I\left(\partial^y_{I }-\frac i { r^{1/2}}(e^{i\beta}\partial_I - e^{-i\beta}\bar{\partial}_I )\right) =0\, , \qquad
\\ \nonumber \\ \label{bUd-dn=}
 \bar{U}^I\left(\partial_{n^I }-r(e^{i\beta}\partial_I + e^{-i\beta}\bar{\partial}_I )\right)
- y_{\bar{U}}n^I\left(\partial^y_{I } -\frac i { r^{1/2}}(e^{i\beta}\partial_I - e^{-i\beta}\bar{\partial}_I )\right) =0\, , \qquad
\\ \nonumber \\  \label{Ud-dn=}{U}^I\left(\partial_{n^I }-r(e^{i\beta}\partial_I + e^{-i\beta}\bar{\partial}_I )\right)
- y_{U}n^I\left(\partial^y_{I } -\frac i { r^{1/2}}(e^{i\beta}\partial_I - e^{-i\beta}\bar{\partial}_I )\right) =0\, . \qquad
\eea
These relations are exact. However, to solve them for ${\partial}_I $ and $\bar{\partial}_I $, we need to use the condition $r\mapsto  \infty$ of asymptotic regime. To this end let us first rewrite  Eqs.~\eqref{d-dr=}-\eqref{Ud-dn=} in the form convenient for perturbative solution of the system for
\be \label{d+-I=}
\partial^\pm_I= e^{i\beta}\partial_I \pm e^{-i\beta}\bar{\partial}_I \; . \qquad 
\ee
Firstly, from \eqref{d-dbeta=}, \eqref{bUd-dy=} and \eqref{Ud-dy=}, we find
\bea
\label{nId-I=}    n^I \partial^-_I=-\frac i r  \partial_{(\beta)}- \frac i {r\sqrt{r}}  (y_U \bar{U}^I+y_{\bar{U}} U^I)\, \partial^+_I\; , \qquad
\\ \nonumber \\ \label{Ud-I=}\bar{U}^I\partial^-_I = -i\sqrt{r} \bar{U}^I \partial^y_{I }\; , \qquad {U}^I\partial^-_I = -i\sqrt{r} {U}^I \partial^y_{I }\; . \qquad
\eea
Using \eqref{I=nn+}, these can be unified in one equation
\be
\partial^-_I=-i\sqrt{r} (\delta^{IJ} - n^In^J)\partial^y_{J}
-\frac i r  n^I \partial_{(\beta)}- \frac i {r\sqrt{r}} n^I (y_U \bar{U}^J+y_{\bar{U}} U^J)\, \partial^+_J\; . \qquad
\ee
Now, using these expressions, we can solve formally Eqs.~\eqref{d-dr=}, \eqref{bUd-dn=} and  \eqref{Ud-dn=} with respect to projections of $\partial^+_I$:
\bea \label{nId+I=} n^I\partial^+_I=
\partial_r+ \frac 1 {2r}\, (y_U \bar{U}^J+y_{\bar{U}} U^J)\, \partial^y_{J} \; , \qquad
\\ \nonumber \\ \label{bUId+I=} \bar{U}^I \partial^+_I=\frac 1 {r}\,  \bar{U}^I\partial_{n^I }- \frac 1 {r}\,  y_{\bar{U}}  n^I\partial^y_{I } + \frac 1 {r^2\sqrt{r}}  y_{\bar{U}} \, \partial_{(\beta)} + \frac 1 {r^3}  y_{\bar{U}} \, (y_U \bar{U}^J+y_{\bar{U}} U^J)\, \partial^+_{J} \, , \qquad
\\ \nonumber \\ \label{UId+I=} {U}^I \partial^+_I=\frac 1 {r}\,  {U}^I\partial_{n^I }- \frac 1 {r}\,  y_{U}  n^I\partial^y_{I } + + \frac 1 {r^2\sqrt{r}}  y_{{U}} \, \partial_{(\beta)}  + \frac 1 {r^3} y_{{U}}  \, (y_U \bar{U}^J+y_{\bar{U}} U^J)\, \partial^+_{J} \, .\qquad
\eea
Note that, according to \eqref{bUId+I=} and \eqref{UId+I=}, $(y_U \bar{U}^J+y_{\bar{U}} U^J)\, \partial^+_J \propto \frac 1 r $ and the derivative $\partial_r$ only appears  in the first term of \eqref{nId+I=}. Using these observations and repeatedly applying \eqref{I=nn+}, we obtain the following approximate solution
\bea \label{d+I=} \partial^+_I= \frac 1 r \left[ n^I \left(r\, \partial_r  + \frac 12 y^J \partial^y_{J} \right) +(\delta^{IJ}- n^I n^J) \, \partial_{n^J} + y^I n^J\partial^y_{J}
\right] +{\cal O}(r^{-5/2})  \; , \qquad
\\ \nonumber
\\ \label{d-I=} \partial^-_I=-i\sqrt{r} (\delta^{IJ} - n^In^J)\partial^y_{J}
-\frac i r  n^I \partial_{(\beta)}+{\cal O}(r^{-5/2}) \;  \qquad
\eea
in which the ${\cal O}(r^{-5/2})$ terms do not involve $\partial_r$ derivative.

Finally, from \eqref{d+-I=} we easily find
\begin{equation}\label{d-dZI=}
    \begin{array}{l}
         \begin{split}
             \partial_I&= \frac 1 2 e^{-i\beta} \Bigg[ -i\sqrt{r} (\delta^{IJ} - n^In^J)\partial^y_{J} +\frac 1 r \left( n^I \left(r\, \partial_r  -i \partial_{(\beta)}  + \frac 12  y^J\partial^y_{J} \right) + \right.\\
             &  + (\delta^{IJ}- n^I n^J) \, \partial_{n^J} + y^I n^J\partial^y_{J}
             \biggr) \Bigg] +{\cal O}(r^{-5/2})  \; ,
         \end{split} 
    \end{array}
\end{equation}
\begin{equation}\label{d-dbZI=}
    \begin{array}{l}
         \begin{split}
             \bar{\partial}_I &=  \frac 1 2 e^{i\beta} \Bigg[ +i\sqrt{r} (\delta^{IJ} - n^In^J)\partial^y_{J} +\frac 1 r \left( n^I \left(r\, \partial_r   +i \partial_{(\beta)} + \frac 12  y^J \partial^y_{J} \right) + \right.\\
             &  + (\delta^{IJ}- n^I n^J) \, \partial_{n^J} + y^I n^J\partial^y_{J}
            \biggr) \Bigg] +{\cal O}(r^{-5/2})   \; .
         \end{split} 
    \end{array}
\end{equation}
Notice that in the asymptotic region \eqref{r-infty}, $r\mapsto \infty$,  the leading order in the above decomposition of the derivatives is $\propto \sqrt{r}$,
\bea \label{d-dZI=r1-2} \partial_I= - \frac i 2 e^{-i\beta} \sqrt{r} (\delta^{IJ} - n^In^J)\partial^y_{J} +{\cal O}(r^{-1})  \; , \qquad
\\ \nonumber
\\ \label{d-dbZI=r1-2} \bar{\partial}_I=  \frac i 2 e^{i\beta}\sqrt{r} (\delta^{IJ} - n^In^J)\partial^y_{J} +{\cal O}(r^{-1})  \; , \qquad
\\ \nonumber
\eea
so that
\bea \label{nId-dZI=0} n^I \partial_I= 0+{\cal O}(r^{-1})  \; , \qquad
\\ 
\label{nId-dbZI=r1-2} n^I\bar{\partial}_I= 0 +{\cal O}(r^{-1})  \; . \qquad
\eea
According to \eqref{eZbZ=}, the factor $\sqrt{r}$ also appears in all nonvanishing r.h.s.'s of Eqs~\eqref{Psi0}-\eqref{fIJ=dIfJ} and \eqref{Psi0=}-\eqref{Psi2=}, which can be written in the form
\bea\label{Psi0=1}
&& 2i \sqrt{r}\epsilon_{IJK} y^Jn^K {\mathfrak{f}}_I^{(q-1)}=0\qquad \Longleftrightarrow \qquad y_U\bar{U}^I{\mathfrak{f}}_I^{(q-1)}=y_{\bar{U}}U^I{\mathfrak{f}}_I^{(q-1)}\; , \qquad
\\ \nonumber \\ \label{Psi1=1}
&& \bar{\partial}_{I} {\mathfrak{f}}^{(q)}=8i\sqrt{2}\sqrt{r}
\epsilon_{JKL} y^Jn^K  \bar{\partial}_{[L} {\mathfrak{f}}_{I]}{}^{(q)} \; , \qquad
\\ \nonumber \\ \label{Psi2=1}
&&  \bar{\partial}_{[I} {\mathfrak{f}}_{J]}{}^{(q-1)}= 8i\sqrt{2}\sqrt{r}
y^{[I}n^{J]}  {\mathfrak{f}}^{(q-3)} \; \qquad
\eea
and
\bea \label{Psi0==1}
&& \partial_{I}  {\mathfrak{f}}^{(q-1)}_{I} =0 \; , \qquad
\\ \nonumber \\ \label{Psi1==1}
&& \partial_{J}  \bar{\partial}_{[I} {\mathfrak{f}}^{(q)}_{J]}=- \frac i {\sqrt{2} \mu^{12}}\, \sqrt{r}\, \epsilon_{IJK}y^J n^{K}f^{(q)}\; ,
\\
 \label{Psi2=11}
 &&  \partial_{I}  {\mathfrak{f}}^{(q-3)} = - \frac  {i\sqrt{2}}{\mu^{12}} \, \sqrt{r}\, y^{[I}n{}^{J]} {\mathfrak{f}}^{(q-1)}_{J}\; .  \qquad
\eea
The algebraic equation \eqref{Psi0=1} can be easily solved by
\be\label{ffIq-1=}
{\mathfrak{f}}^{(q-1)}_{I}= n^I {\mathfrak{f}}^{(q-1)}_{(n)}+ y^I {\mathfrak{f}}^{(q-1)}_{(y)}\; . \qquad
\ee
We can now decompose the ``wavefunctions'' into inverse powers of $r\sqrt{r}$ and try to solve the equations order by order. To be consistent, we set
\be
 {\mathfrak{f}}^{(q\ldots)}_{\ldots}= r^{-k} \left( {\mathfrak{f}}^{(q\ldots)}_{\ldots[0]} + \frac 1 {r\sqrt{r}}   {\mathfrak{f}}^{(q\ldots)}_{\ldots[1]} +  \frac 1 {r^3} {\mathfrak{f}}^{(q\ldots)}_{\ldots[2]}+\ldots\right)\; ,
\ee
with some $k$ to be determined, for all but the leading components of the state vector superfield. For the leading component, we instead assume
\be
 {\mathfrak{f}}^{(q)}= \sqrt{r}{\mathfrak{f}}^{(q)}_{[0]} + \frac 1 {r}   {\mathfrak{f}}^{(q)}_{[1]} +  \frac 1 {r^2} {\mathfrak{f}}^{(q)}_{[2]}+\ldots\; 
\ee
Here we focus on the zeroth order, where (following the terms $\propto \sqrt{r}$) we obtain the equations
\bea\label{Psi1=1-0}
&& (\delta^{IJ} - n^In^J)\partial^y_{J} {\mathfrak{f}}^{(q)}_{[0]}=8i\sqrt{2}
\epsilon_{JKL} y^Jn^K(\delta^{[L|P} - n^{[L}n^P)\partial^y_{P} {\mathfrak{f}}_{I][0]}{}^{(q)} \; , \qquad
\\ \nonumber \\ \label{Psi2==1}
&&   (\delta^{[I|K} - n^{[I|}n^K)\partial^y_{K} {\mathfrak{f}}_{J][0]}{}^{(q-1)}=16\sqrt{2}e^{-i\beta}
y^{[I}n^{J]}  {\mathfrak{f}}^{(q-3)}_{[0]} \; ,  \qquad
\\ \nonumber \\  \label{Psi0==1-0}
&&   (\delta^{IJ} - n^{I}n^J)\partial^y_{J}  {\mathfrak{f}}^{(q-1)}_{I[0]} =0 \; , \qquad
\\ \nonumber \\ \label{Psi1==1-0}
&& (\delta^{JK} - n^{J}n^K)\partial^y_{K} \, (\delta^{[I|L} - n^{[I|}n^L)\partial^y_{L}    {\mathfrak{f}}^{(q)}_{|J][0]}=-\frac {2\sqrt{2}\, i }  {(\mu^{6})^2}\,  \,\epsilon_{IJK}y^J n^{K}f^{(q)}_{[0]}\; ,
\\
 \label{Psi2==1-0}
 &&  (\delta^{IJ} - n^{I}n^J)\partial^y_{J} {\mathfrak{f}}^{(q-3)}_{[0]} =  \frac  {2\sqrt{2}}{(\mu^{6})^2} \, e^{i\beta}  \, y^{[I}n{}^{J]} {\mathfrak{f}}^{(q-1)}_{J[0]}\; .  \qquad
\eea
 Observe that this system of equations involves only derivatives with respect to the variables $y^I$. With this in mind, one immediately deduces that \eqref{Psi1=1-0} implies
\be\label{Psi1=1-0n}
\epsilon_{JKL} y^Jn^K(\delta^{LP} - n^{L}n^P)\partial^y_{P} (n^I {\mathfrak{f}}_{I[0]}{}^{(q)})=0 \; , \qquad
\ee
which indicates that $n^I {\mathfrak{f}}_{I[0]}{}^{(q)}$ depends on $y^J$ only through its square, $|y|^2=y^Jy^J$.

Eq.~\eqref{Psi0==1-0}, together with leading order term of \eqref{ffIq-1=},
\be\label{ffIq-1=1}
{\mathfrak{f}}^{(q-1)}_{I[0]}= n^I {\mathfrak{f}}^{(q-1)}_{(n)[0]}+ y^I {\mathfrak{f}}^{(q-1)}_{(y)[0]}
\ee
does not impose any condition of ${\mathfrak{f}}^{(q-1)}_{(n)[0]}$, but requires ${\mathfrak{f}}^{(q-1)}_{(y)[0]}$ to satisfy $(y^I\partial^y_{I} +2){\mathfrak{f}}^{(q-1)}_{(y)[0]}=0$. Nevertheless, as it will become clear shortly, we have to choose the trivial solution of this equation. Indeed, substituting \eqref{ffIq-1=1} into  \eqref{Psi2==1-0}, we find
\bea
 \label{Psi2==1-01}
 &&  (\delta^{IJ} - n^{I}n^J)\partial^y_{J} {\mathfrak{f}}^{(q-3)}_{[0]} =  \frac  {\sqrt{2}}{(\mu^{6})^2} \, e^{i\beta}  \, \left(y^{I} {\mathfrak{f}}^{(q-1)}_{(n)[0]}-n{}^{I} (y^{J}y^{J}){\mathfrak{f}}^{(q-1)}_{y[0]}\right)\; .  \qquad
\eea
Since both l.h.s. and the first term on the r.h.s. of this equation vanish when contracted with $n^I$, the last term must vanish independently,
\be
{\mathfrak{f}}^{(q-1)}_{y[0]}=0\qquad \Longrightarrow \qquad {\mathfrak{f}}^{(q-1)}_{I[0]}= n^I {\mathfrak{f}}^{(q-1)}_{(n)[0]}\; .
\ee
Then, Eq.~\eqref{Psi2==1-01} reduces to
\bea
 \label{Psi2==1-02}
 &&  (\delta^{IJ} - n^{I}n^J)\partial^y_{J} {\mathfrak{f}}^{(q-3)}_{[0]} =  \frac  {\sqrt{2}}{(\mu^{6})^2} \, e^{i\beta}  \,y^{I} {\mathfrak{f}}^{(q-1)}_{(n)[0]}\; .  \qquad
\eea
This equation implies that both $ {\mathfrak{f}}^{(q-3)}_{[0]}$ and $ {\mathfrak{f}}^{(q-1)}_{(n)[0]}$ depend on $y^I$ vector only through its length $|y|=\sqrt{y^Iy^I}$. Therefore,
\be\label{dy2fq-3=} \partial_{|y|^2} {\mathfrak{f}}^{(q-3)}_{[0]}= \frac  {1}{(\mu^{6})^2\sqrt{2}} \, e^{i\beta}  \,{\mathfrak{f}}^{(q-1)}_{(n)[0]}\; . \qquad 
\ee
On the other hand, with the above conclusion on dependence on $y^I$ vector only through its length $|y|=\sqrt{y^Iy^I}$, Eq.~\eqref{Psi2=1} reduces to
\begin{equation}
    \partial_{|y|^2}{\mathfrak{f}}^{(q-1)}_{(n)[0]}= 8\sqrt{2} \, e^{-i\beta}  \, {\mathfrak{f}}^{(q-3)}_{[0]}~.
\end{equation}
Using this and \eqref{dy2fq-3=}, we find a simple equation
\be\label{dy22fq-1=} \frac {\partial}{\partial |y|^2} \frac {\partial}{\partial |y|^2} {\mathfrak{f}}^{(q-1)}_{(n)[0]}= \frac  {8}{(\mu^{6})^2}   \,{\mathfrak{f}}^{(q-1)}_{(n)[0]}\;  \qquad \ee
which is solved by
\be\label{fq-10=exp}
 {\mathfrak{f}}^{(q-1)}_{(n)[0]}= {\mathfrak{h}}^{(q-1)}(r,\beta, \vec{n})\; {\rm exp}\left(-2\sqrt{2}\frac{|y|^2}{\mu^{6}} \right)\;
\ee
with some function ${\mathfrak{h}}^{(q-1)}(r,\beta, \vec{n})$ depending on the remaining coordinates.

Similarly one can study the case of other components of the state vector superfield, whose statistics (bosonic if the superfield is bosonic) are opposite to the components discussed above (fermionic if the superfield is bosonic), which obey Eqs. \eqref{Psi1=1-0} and  \eqref{Psi1==1-0}. Since the study of this case is somewhat more involved than previous one, we simplify the analysis by choosing a special frame in which $n^I=(0,0,1)=\delta^I_3$, and consequently $y^I= (y^1,y^2, 0)= \delta^I_r y^r$.

In this SO$(3)$ frame, the solution to \eqref{Psi1=1-0n} can be written in the form ${\mathfrak{f}}_{3[0]}{}^{(q)}(y^J)=  {\mathfrak{f}}_{3[0]}{}^{(q)}(|y|^2)$, and the remaining components of  \eqref{Psi1=1-0} and \eqref{Psi1==1-0} simplify to
\bea\label{Psi1=1-00}
&& \partial^y_{r} {\mathfrak{f}}^{(q)}_{[0]}=8i\sqrt{2}
\epsilon_{st} y^t\partial^y_{[s} {\mathfrak{f}}_{r][0]}{}^{(q)} \; , \qquad
\\ \nonumber \\ \label{Psi1==1-00}
&& \partial^y_{s} \, \partial^y_{[r}    {\mathfrak{f}}^{(q)}_{s][0]}=-\frac {i2\sqrt{2} }  {(\mu^{6})^2}\,  \,\epsilon_{rs}y^s{\mathfrak{f}}^{(q)}_{[0]}\; \; .  \qquad
\eea
Furthermore, since $r,s,t=1,2$, the antisymmetric derivative can be expressed as $\partial^y_{[s} {\mathfrak{f}}_{r][0]}{}^{(q)}=\epsilon_{sr}\tilde{{\mathfrak{f}}}^{(q)}_{[0]}$, where
$\tilde{{\mathfrak{f}}}^{(q)}_{[0]}= \frac 1 2 \epsilon_{sr} \partial^y_{[s} {\mathfrak{f}}_{r][0]}{}^{(q)}$ and the above equations reduce to
\be\label{Psi1=1-00=}
 \partial^y_{r} {\mathfrak{f}}^{(q)}_{[0]}=8i\sqrt{2}y^r \tilde{{\mathfrak{f}}}^{(q)}_{[0]}\; , \qquad
 \partial^y_{s} \, \tilde{{\mathfrak{f}}}^{(q)}_{[0]}=-\frac{2\sqrt{2} i}{(\mu^{6})^2}\, y^s  \,{\mathfrak{f}}^{(q)}_{[0]}\; .  \qquad
\ee
It follows immediately that both ${\mathfrak{f}}^{(q)}_{[0]}$ and $\tilde{{\mathfrak{f}}}^{(q)}_{[0]}$ depend on $y^I$ only through its square $|y|^2=y^Iy^I$, and that they obey the second order linear differential equation in $|y|^2$,
\be
\partial_{|y|^2}\partial_{|y|^2}{\mathfrak{f}}^{(q)}_{[0]}= \frac 8 {(\mu^{6})^2}{\mathfrak{f}}^{(q)}_{[0]}\;,
\ee
which is solved by
\be\label{fq0=exp}
{\mathfrak{f}}^{(q)}_{[0]}= {\mathfrak{h}}^{(q)}_{[0]}(r,\beta,\vec{n}) \exp \left(  -\frac {2\sqrt{2}} {\mu^{6}}\, |y|^2\right)\; ,
\ee
where ${\mathfrak{h}}^{(q)}_{[0]}(r,\beta,\vec{n})$ is a function of the remaining variables.

We have chosen the negative sign in the exponent in both  \eqref{fq-10=exp} and \eqref{fq0=exp} to ensure that the wavefunction is convergent in $y^I$. 

We will not consider higher orders of the Born-Oppenheimer-like approximation, since this goes beyond the scope of the present thesis; we refer the reader to \cite{Frohlich} for a detailed treatment of such corrections.  It is worth noting that in that work, besides the case of Matrix model obtained by dimensional reduction of $\text{D}=3$ SYM model to $\text{d}=1$, Matrix models from reductions of higher dimensional SYM were also studied, with particular attention drawn to the exceptional properties of $\text{D}=10$ case. This will appear in a setup similar to \eqref{nuXi=0}-\eqref{cHXi=0} from the field theory of our 10D mD$0$ model which we aim to quantize in future work.

\renewcommand\thechapter{\Alph{chapter}}

    \renewcommand{\bibname}{Bibliography}
\clearpage
\thispagestyle{empty}

\renewcommand{\bibpreamble}{\begin{changemargin}{8cm}{0cm}
\vspace*{-0.5cm}
    \singlespacing\textcolor{cites}{ \small{
\hspace*{-3pt}
~~~~~~~~~~We are more than the parts that form us. 
    \begin{flushright}
         {\sffamily {\textit{The Kingkiller Chronicle, Day One:\\ The Name of the Wind}}\\{by {Patrick Rothfuss}.}}
     \end{flushright}}}
    \end{changemargin}
    \vspace{12pt}
}


 \backmatter
 
     \chapter*{}
     \thispagestyle{empty}
     \includepdf{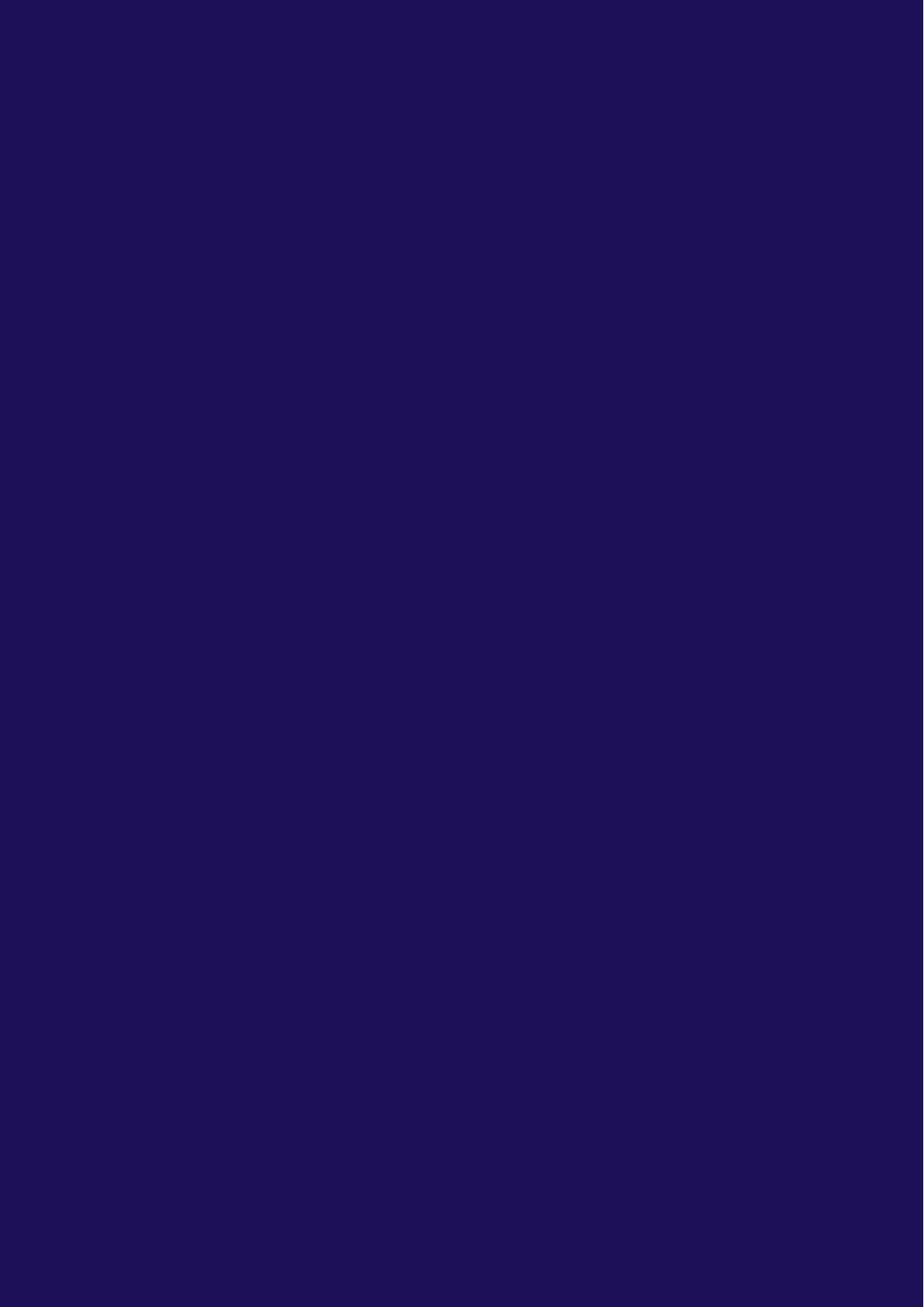} 

\end{document}